\documentclass[11pt,towside,a4paper]{report}
\usepackage{latexsym}
\usepackage{amsmath}
\usepackage{fancyhdr} 
\usepackage{amssymb}
\usepackage{dcolumn}
\usepackage{bm}
\usepackage{color}
\usepackage{url}
\usepackage{amsmath,amssymb,graphicx}
\usepackage{amsmath}
\usepackage{amssymb}
\usepackage{mathrsfs}
\usepackage{accents}
\usepackage{epsfig}
\usepackage[lofdepth,lotdepth,caption=false]{subfig}

\usepackage{setspace} 
\usepackage{fancyhdr} %
\usepackage{mathrsfs}
\usepackage{slashed}
\usepackage{braket}
\newcommand{\nc}{\newcommand}
\nc{\ba}{\begin{eqnarray}} \nc{\ea}{\end{eqnarray}}
\newcommand\be{\begin{equation}}
	\newcommand\ee{\end{equation}}
\nc{\x}{{\bf{x}}}
\nc{\e}{{\bf{e}}}
\nc{\bfk}{{\mathbf{k}}}
\nc{\bfq}{{\mathbf{q}}}
\nc{\bfp}{{\mathbf{p}}}
\nc{\bk}{{\mathbf{k}}}
\nc{\bq}{{\mathbf{q}}}
\nc{\bp}{{\mathbf{p}}}
\nc{\bn}{{\mathbf{n}}}
\nc{\calR}{{\cal R}}
\nc{\calP}{{\cal P}}

\nc{\bfx}{{\bf{x}}}

\nc{\eH}{{\epsilon_H}}

\nc{\im}{{ \mathrm{Im} } }

\nc{\kk}{{\bf{k}}}
\nc{\pp}{{\bf{p}}}
\nc{\cM}{{\cal{M}}}

\begin{document}
\marginparsep 0pt
\textwidth 15.0 truecm


\setlength{\baselineskip}{0.780cm}
\begin{center}
	\includegraphics{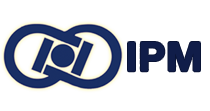}
\end{center}

\pagestyle{empty}
\begin{center}
\Large{ \bf Institute for Research in Fundamental Sciences} \\
	(School of Physics)
\end{center}
\vspace{6mm}

\vspace{12mm}

\begin{center}
	\Large{ Ph.D Thesis}
\end{center}
\vspace{12mm}

\begin{center}
	\Large{
		\bf Anisotropic Inflation and Cosmological Observations}
\end{center}

\vspace{21mm}
\begin{center}
\Large{ \bf By: Razieh Emami Meibody}\\
\end{center}

\vspace{24mm}
\begin{center}
Supervisor\\
{\bf Hassan Firouzjahi}
\end{center}

\vspace{12mm}
\begin{center}
21 July 2015
\end{center}









\newpage

\noindent
\vspace{5 cm}
\begin{center}
\Huge {\textbf{\textit{To my parents}}}
\end{center}

\newpage
\setcounter{page}{1}
\pagestyle{myheadings}
\markright{}

\noindent
{\Huge {\bf Abstract}}
\vspace{1.2cm}

\noindent
Recent observations opened up a new window on the inflationary model building. As it was firstly reported by the WMAP data, there may be some indications of statistical anisotropy on the CMB map, although the statistical significance of these findings are under debate. Motivated by these observations, people begun considering new inflationary models which may lead to statistical anisotropy. The simplest possible way to construct anisotropic inflation is to introduce vector fields. During the course of this thesis, we study models of anisotropic inflation and their observational implications such as power spectrum, bispectrum etc. \\ 
Firstly we build a new model, which contain the gauge field which breaks the conformal invariance while preserving the gauge invariance. We show that in these kind of models, there is an attractor phase in the evolution of the system when the back-reaction of the gauge field becomes important in the evolution of the inflaton field. We then study the cosmological perturbation theory in these kind of models. More specifically, we calculate the anisotropic corrections due to the presence of the vector field. \\
We then generalize the separate universe formalisms to our anisotropic set up and use this formalism in some specific examples of anisotropic inflation. \\
Finally, we connect the primordial anisotropies to the specific examples and to CMB observations. We calculate the TT, TE, TB, EB and BB correlation in the model of charged scalar field model and look for the unique signatures that the anisotropic inflation can have on the CMB map. Any future detection of these statistical anisotropies would rule out the isotropic FRW models.   

\vspace{7mm}

\newpage
\begin{center}
{\large {\bf ACKNOWLEDGMENTS}}
\end{center}

\vspace{5mm}
These days that I am finalizing my PhD thesis have made me to look back and review the whole of my very joyous life as a graduate student.  As I am coming into my own way as a physicist, I see that my PhD period was incredibly important to give me a real running start. It was indeed the best scientific years for me to push myself forward and move into the next stage of my career and I thus have a lot of people to thank for.  \\
First and foremost, I am extremely grateful to my PhD supervisor, Hassan Firouzjahi, for truly guiding me on my path from being interested in physics to be a scientist. At first, he provided a very active and peaceful scientific environment for me to understand how to tackle the problems in science and how to pick interesting topics and very good collaborators to work
with. Indeed I learned tremendous amounts from working with him. Next, he opened up the window of scientific world to me through exposing me abroad. In the time that nobody knew me, he was my best support to introduce and expose me to several wonderful scientific groups throughout the world from Europe, USA and recently Asia for extending my expertises through working with theses international groups.
He knew how important it was for me to be familiar with different topics, beyond my PhD thesis, and he thus gave me considerable freedom in working on whatever I found interesting and whoever I was eager to work with. These helped me to build up new international collaborations with expert cosmologists around the world. 
Still, even in the time that I was very far away from him, he was always present when I dealt with new difficult conditions and I have always benefited from his supports. 
Finally, I learned from him how to behave nicely in the scientific collaborations. He was indeed my shedding light during the whole of my PhD career and I thus really appreciate all of his considerations. \\
Next, it is my great pleasure to express my very deep gratitude to Marc Kamionkowski, a true leader of theoretical cosmology, and astronomy all around the world, for kindly accepting to be my advisor during my long term visit to Johns Hopkins University through the Oxford-JHU Balzan fellowship and for making my time there the best visit that I could ever image. He kindly opened up to me a deep and completely new window of research which was very far from what I have done before. Especially I learned from him how to do the professional numerical analysis which are incredibly important for the young researchers in any fields particularly in cosmology and astronomy. He kindly introduced me to his scientific group and provided me a very nice scientific environment working with several of senior postdocs, J. Chluba,  E. Dimastrogiovanni, D. Grin, A. Raccanelli and Josef Pradler for several amazing collaborations which some of them are in progress which I thank all of them as well as other postdocs, E. Kovetz, Y.A. Haimoud for the many insightful discussions. Marc was indeed my shedding light during my visit and I have always benefited from the fruitful discussions with him about new topics to work on. I am also grateful to all of the faculty members of JHU, particularly Daniel H.Reich, the head of the Department of Physics and Astronomy, and all of my nice friends there for their kindness and supports during my presence there which made me to never feel that I am abroad!
It is my pleasure to thank Oxford University especially Joseph Silk for the financial support of my visit. \\
I am also really grateful to Xingang Chen and Yi Wang for many insightful discussions and several collaborations during the course of this thesis and beyond that. \\
I am also thankful to Paolo Creminelli for being my host during my long term visit to ICTP during which we built up a new collaboration on new topic as a result of interesting discussions. 
Thanks also to his group members, postdocs and all of the faculty members of ICTP especially Fernando Quevedo, director of ICTP, for their kindness during my visit there. \\
In addition, I am also thankful to R. Asgari the former head of school of Physics for all of his efforts to make IPM an amazing place to work as well as all of his kindness  during the whole of my PhD and supporting my visits to ICTP and JHU.  I am also really thankful to M. Alishahiha as the IPM deputy research who facilitated my visits to ICTP and JHU. I am also deeply grateful to S. Sheikh Jabbari (the new head of the school of physics) for all the nice things that I have learned during the whole of my study in IPM from discussing with him. \\
Thanks also to all of my professors during the whole of my PhD study including Y. Farzan, A. Naji, H. Seyed-Allaei, S. Sheikh Jabbari for kindly teaching us different courses and skills during the first years of my PhD which were incredibly important to prepare myself for doing research. I am also really grateful to the whole of the honest staffs in IPM, J. Aliabadi, M. Babanzadeh, S. Jam, H. Zarei Khatuni, A. Mousavi and especially N. Pileroudi for kindly helping us to have a wonderful environment for study and by taking care of every administrative processes. 
Thanks so much to all of my friends and colleagues at IPM for making the whole of my PhD to be the most enjoyable time. I learned a lot from each of them. 
It is also a pleasure to thank
A.A. Abolhasani, M. Akhshik, N.Bartolo, S. Baghram, D. Baumann, L. Dai,  E. Komatsu, A. Malek Nejad,  S. Matarrese, S.M.S. Movahed, M. Noorbala, M. Peloso, L. Senatore, S. Sheikh Jabbari, J. T. Firouzjaee and M. Zarei for the very insightful discussions and collaborations. \\
Last but not least, I want to express my deep gratitude to my family, especially my parents for honestly furnishing me any possible opportunities to discover and pursue higher studies that was my passion. Indeed without their continuous supports and encouragements, it was hardly possible for me to make any success in any steps of my education. I thus extremely appreciate their patience and supports during every moments of my scientific career and my life. They are indeed the meaning of my life!

\newpage
{\large {\tableofcontents}}

\listoffigures
\chapter{Introduction}

Cosmology deals with understanding the physical contents and the evolution of the universe as a whole. Modern cosmology is born in 1917 with the invention of general relativity by Albert Einstein. Although at first the universe was supposed to be static, this picture has changed by the discovery of the expansion of the universe by Edwin Hubble in 1929. Subsequent developments in both theoretical and observational cosmology led to a very consistent theoretical model, the so called standard model of cosmology. This is a very rich framework for describing the evolution of the universe and its history. Moreover, this model has a predictive power which transformed cosmology from a largely speculative science into a predictive subject which can be tested observationally. There have been several advanced observations to test this model and to find the parameters of this model precisely. \\
Although a very good picture of universe is achieved by this model, there are some shortcomings in explaining the observed universe,
unless we impose some fine tunings in the initial conditions of the universe, such as the horizon problem and the flatness problems. The horizon problem refers to the fact that the observed cosmic microwave background radiation seems to be to a very good approximation isotropic over the sky. However, back in time, these regions were outside the causal contact by the time of recombination. Indeed, there were about $10^{4}$ casually disconnected regions at the recombination surface and it is thus really surprising how these regions have a very uniform CMB temperature today. One expects that the modern cosmology would be able to explain this large scale uniformity without resorting to the initial fine-tuning. Next is the flatness problem in the following way. The current observations confirm that our universe can be approximated with a nearly flat universe. However, this flat phase is not an attractor phase which means that in the past it must have been much more flat than it is now. So we need a fine-tuning to end up into our current observed universe. \\
Both of these puzzles can be solved if we consider a new phase in the evolution of the universe, called as the cosmic inflation which was first proposed by Alan Guth in 1981. Cosmological inflation, an early period of accelerated expansion, allows our universe to arise from generic initial conditions. Beside solving the above puzzles, inflation provides a natural explanation about the origin of the structures in the observed universe. According to inflation, the quantum fluctuations during this phase are stretched and become classical fluctuations and thus seed the large scale inhomogeneities. The suitable framework for investigating these inhomogeneities is called the cosmological perturbation theory. \\ 
Due to the importance of the inflation in addressing the big-bang puzzles as well as the seeds for the structure formations, there was a lot of efforts in the community trying to build different inflationary models and figure out which one is consistent with the observations such as WMAP and more recently with the Planck data. Currently there are a lot of interesting inflationary models in the market in which most of them are based on scalar fields mainly due to their simplicity. However, recently there have been considerable interests on primordial anisotropies in models of anisotropic inflation. Observationally, there may be some indications of the statistical anisotropy on CMB, although the statistical significance of these findings are not significant. On the theoretical sides, there have been many attempts to construct models of inflation with vector fields or gauge fields which can create sizable amount of anisotropy on curvature perturbations.  \\
In this thesis, we mainly focus on the theoretical and observational signatures of anisotropic inflation. We present a variety of different anisotropic inflationary models. We then consider the specific predictions of these models. As we will see, it is required to extend some of the existing formalisms in the literature to models of anisotropic inflation. We finally connect the primordial anisotropies to what is actually observed in the CMB.\\
The rest of the thesis is organized as follows. In chapter 2, we review the standard model of cosmology. We present the main equations which govern the evolution of the universe. We then present the big-bang puzzles and introduce inflation as a solution of these problems. In chapter 3, we study the linear perturbation theory. We then try to build up some physical observables which can be checked observationally . In chapter 4, we review the models of anisotropic inflation. We consider different models of inflation and show that there is an attractor solutions where the anisotropies produced during inflation becomes comparable to the slow-roll parameters. In chapter 5, we consider the curvature perturbation in the anisotropic model with a complex scalar field charged under a $U(1)$ gauge field in Bianchi 1 universe. We identify the dominant contribution into the statistical anisotropy. We obtain an upper bound on gauge coupling in order to satisfy the observational constraints on curvature perturbations anisotropies. In chapter 6, we present a consistent $\delta N$ formalism for calculating the curvature perturbations in anisotropic backgrounds. We then employ our formalism to calculate the power spectrum, bispectrum and the trispectrum in models of anisotropic inflation. In chapter 7, we study the primordial anisotropies generated in the model of gauged hybrid inflation in which the complex waterfall field is charged under a $U(1)$ gauge field. In this model primordial anisotropies are generated either actively during inflation or from inhomogeneities modulating the surface of end of inflation. We show that the primordial anisotropies generated at the end of inflation does not depend on the number of e-folds and thus do not produce IR divergences. Interestingly, the gauge field fluctuations induce a red tilted power spectrum so the averaged power spectrum from the gauge field can change the total power from blue to red. In chapter 8, we study the TT, TB, EB and BB correlations associated with the B-mode polarization of the CMB map. We show that the asymmetry in the tensor power spectrum is a very sensitive probe of the gauge coupling. In addition, the TT correlation receives an anisotropic contribution from the tensor sector which naturally decays after $l> 100$. In chapter 9, we examine the impact of different anisotropic relics on inflation. There are two different types of background relics, one coming from the matter while the other is due the metric. Although the angular dependence of the statistical anisotropies are the same in both cases, the scale dependence are observationally distinctive. \\
Finally we conclude and leave numerous technicalities into the appendices.

\newpage


\chapter{Standard Model of Cosmology} 

\label{Chapter1} 


\vspace{0.5cm}
\hrule \vspace{0.3cm}

\begin{quote}
\textbf{Abstract:}	The purpose of the following chapter is to review big-bang cosmology. We start by presenting the kinematic and the dynamical equations which govern the evolution of the universe. We then briefly discuss the thermal history of the universe. After that, we show that conventional big-bang theory leads to the initial condition problems in the sense that our current universe can only arise from a very specific initial conditions. We then introduce the inflationary phase as a solution for the above problems. We leave the detailed analysis of inflation for the next chapter.
\end{quote}

\vspace{0.1cm}  \hrule
\vspace{0.5cm}
\section{The Homogeneous and Isotropic Universe}
\textbf{\textit{Overview:}}
The goal of the cosmology is to describe the structure and the evolution of the universe on the largest scales. Modern cosmology is based on the assumption that our universe to a very good approximation is homogeneous and isotropic on sufficiently large scales, where by the large scale we mean scales larger than $100$ Mpc. This assumption is somehow the generalization of the Copernican Principle which states that not only our place inside the solar system is not a very special place, but also the position of the Milky Way in the universe should not be statistically distinguishable from other galaxies. Moreover, no direction should be distinguishable. 

However, the small deviations from the homogeneity and isotropy in the CMB are extremely important. The reason is that they generate the seeds which later on via gravitational instability have led to the formation of the large scale structure in the universe. 

Furthermore, we assume that the above mentioned deviations from the homogeneity and isotropy have been created from tiny quantum fluctuations which have been amplified during a period of inflationary expansion of the universe. \\
The aim of this chapter is to review of the standard model of cosmology. A more extensive review can be found in \cite{Mukhanov, Weinberg:2008zzc, Baumann:2009ds, Gorbunov:2011zzc, Gorbunov:2011zz, Ruth Durrer}.
\subsection{FRW Spacetime}
As we have already discussed, our universe is to a good approximation homogeneous and isotropic on very large scales. This leads to the Friedmann-Robertson-Walker (FRW) metric, with the following form, 
\begin{equation}
	\label{equ:FRW}
	d s^2 = - d t^2 + a^2(t)  \left(\frac{d r^2}{1-k r^2} + r^2 (d \theta^2 + \sin^2 \theta d \phi^2) \right).
\end{equation}
Here, $a(t)$ is the scale factor and characterizes the relative size of the space-like hyper-surface at different times. In addition, the curvature constant parameter, i.e. $k$, is positive for the closed universes, $0$ for the flat and negative for the open universe. In the above metric, $r, \theta$ and $\phi$ are the comoving coordinates. They are fixed in the absence of the peculiar motion. The corresponding physical distance is obtained by multiplying $r$ with the scale factor, 
$R = a(t)r$. $R$ is time dependent even for the objects with vanishing peculiar velocities. \\
From the FRW metric, it is transparent that the evolution of the universe at large scales is given by a single function, $a(t)$ where the evolution of the scale factor can be read from the Einstein field equations. A very important quantity which characterizes the FRW spacetime is the expansion rate, $H$. This sets the fundamental scale of the FRW universe and is given by, 
\begin{equation} 
	\label{hubble}
	H \equiv \frac{\dot{a}}{a}
\end{equation}
In addition, the age of the universe to reasonable approximation is scaled as $t_0 \simeq \frac{1}{H_0}$ in which the subscript 0 indicates the values of the corresponding quantities at the current time. 
\subsection{Kinematic of the Universe}
Having presented the metric for the homogeneous spacetime, we can now study the kinematics of the universe. 
\subsubsection{Conformal Time and Null Geodesics}
\begin{figure}[htbp]
		\centering
	\includegraphics[width=.4\textwidth]{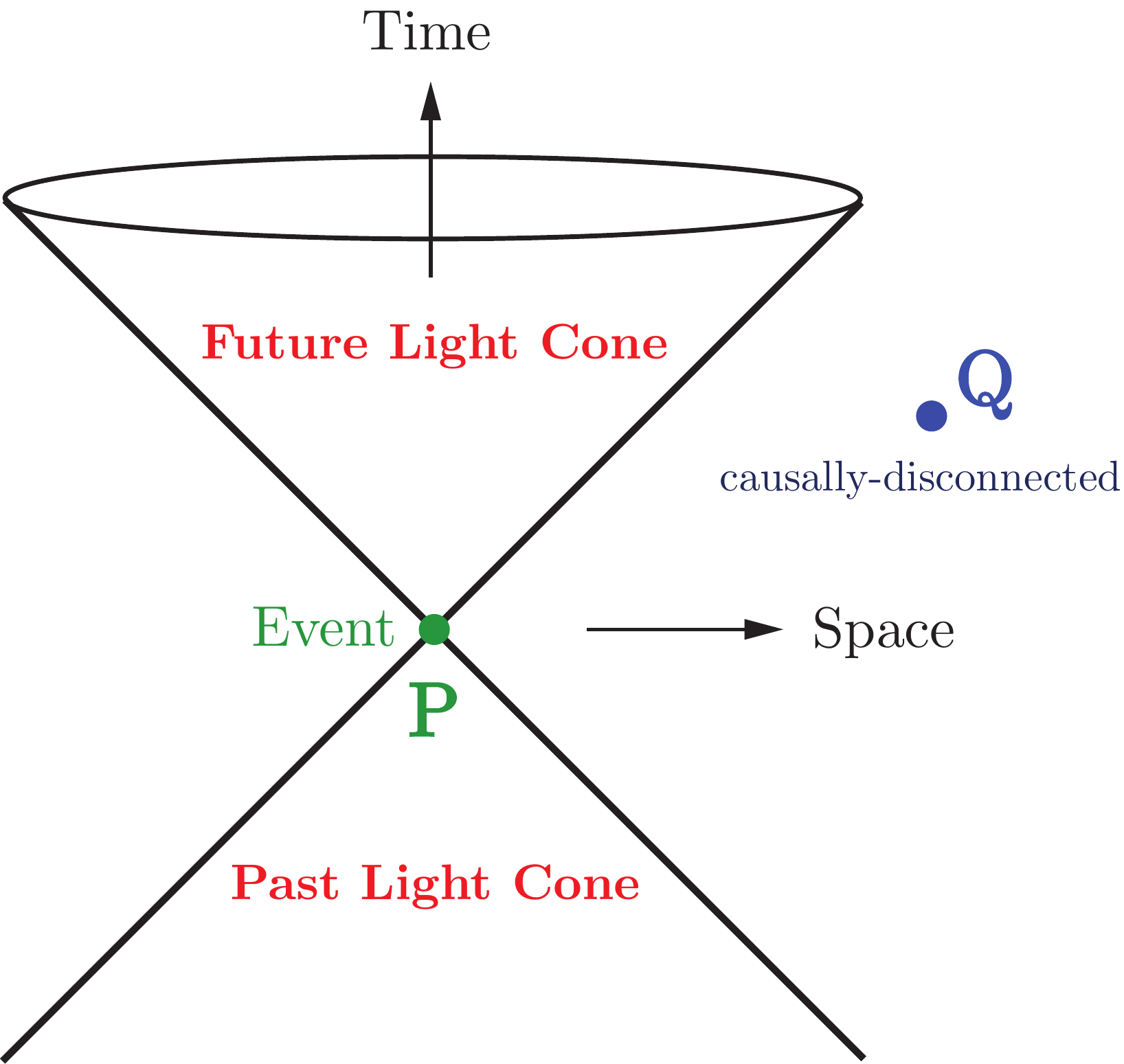}
	\caption{Photons travel along the {\it null geodesics}, $d s^2 =0$.
	 On the other hand, massive particles travel along the {\it timelike geodescis}, $d s^2 > 0$. Causally disconnected regions of spacetime are separated by {\it spacelike} intervals, $d s^2 < 0$. A set of the all null geodesics which are passing through a given point (or {\it event}) in spacetime is called the {\it light cone}. The interior of light cone, consisting of all null and timelike geodesics, would define the region of spacetime causally related to that event.}
	\label{fig:lightcone}
\end{figure}

The causal structure of the universe is determined by the propagation of light in the FRW spacetime (\ref{equ:FRW}).  (Massless) photons follow the null geodesics, $d s^2 = 0$. Their trajectories are studied most easily if we define conformal time,
\begin{equation} 
	\label{tau}
	\tau = \int \frac{d t}{a(t)} ,
\end{equation}
for which the FRW metric becomes,
\begin{equation} 
	d s^2 =a(\tau)^2 \left[- d \tau^2 +  \left(d \chi^2 + \Phi_k(\chi^2) (d \theta^2 + \sin^2 \theta d \phi^2) \right) \right].
\end{equation}
Where $\Phi_k$ is given by,
\begin{equation} 
r^2 = \Phi_k(\chi^2) \equiv  \left\{
\begin{array}{c} \sinh^2 \chi \\ \chi^2 \\ \sin^2 \chi \\
\end{array} \right. \quad \begin{array}{l} k=-1 \\ k=0 \\ k=+1 \end{array}\, .
\end{equation}
In an isotropic universe, due to the presence of the spherical symmetry, it is convenient to consider the radial propagation of light which is determined by the two dimensional line element, 
\begin{equation} 
	\label{radial line element}
	d s^2 = a(\tau)^2 \left[ - d \tau^2 + d \chi^2\right]. 
\end{equation}
As we see, in this way, the metric is conformal to static Minkowski background with the conformal factor $a(\tau)$. Then the radial null geodesics of the light in FRW background in terms of the conformal time would be given by,
\begin{equation} 
	\label{null geodesics}
	\chi(\tau) = \pm \tau\, +\, {\rm const.}\, ,
\end{equation}
They corresponds to straight lines at angles $\pm 45^{o}$ in the $\tau-\chi$ plane, see Fig. \ref{fig:lightcone}.
\subsubsection{Particle Horizon}
The maximum comoving distance that light can propagate between an initial time $t_{\rm i}$ and a later time $t$ is
\begin{equation} 
\label{particleH}
	\chi_p(\tau) = \tau - \tau_{\rm i} = \int_{t_{\rm i}}^t \frac{d t}{a(t)}\, .
\end{equation}
This is called the (comoving) particle horizon. The initial time $t_{\rm i}$ is often chosen to be the `origin of the universe', i.e. $t_{\rm i} \equiv 0$, defined by the initial singularity, i.e. $a(t_{\rm i} \equiv 0) \equiv 0$.\footnote{Whether $t_{\rm i} = 0$ also corresponds to $\tau_{\rm i} = 0$ depends on the evolution of the scale factor $a(t)$; {\it e.g.}~for inflation $t_{\rm i}=0$ will {\it not} be $\tau_{\rm i}=0$.}
The physical size of the particle horizon is 
\begin{equation} 
	d_p(t) = a(t) \chi_p\, .
\end{equation}
The particle horizon is very importance in order to understand the ``causal structure" of the universe and it will be fundamental for our discussion of inflation.
As we will see later, the conventional Big Bang model begins at some finite time in the past and at any time the particle horizon is finite, limiting the distance over which spacetime region could have been in the causal contact.
This feature is at the heart of the `Big Bang puzzles'.
\subsection{Dynamics of Cosmological Expansion}
\label{sec:EinsteinEqns}
The Standard model of Cosmology is based on the Einstein theory of general relativity. So in order to determine the evolution of the scalar factor, we use the Einstein equations,
\begin{eqnarray}
	\label{Einstein}
	G_{\mu \nu} \equiv R_{\mu \nu} - \frac{1}{2} g_{\mu\nu} R
	= 8\pi G\, T_{\mu \nu}  .
\end{eqnarray}
Where $R_{\mu \nu}$ is the Ricci tensor given in terms of Christofel symbol via, 
\begin{equation} 
R_{\mu \nu} = \partial_{\lambda} \Gamma^{\lambda}_{\mu \nu} - \partial_{\mu} \Gamma^{\lambda}_{\nu \lambda} + \Gamma^{\lambda}_{\mu \nu} \Gamma^{\sigma}_{\lambda \sigma} - 
\Gamma^{\lambda}_{\mu \sigma} \Gamma^{\sigma}_{\nu \lambda} ,
\end{equation}
where $R$ is the Ricci scalar which is given by, $R = R_{\mu \nu} g^{\mu \nu}$. \\
It is convenient to define the reduced Planck mass as,  
\begin{equation} 
M_{Pl}^2 \equiv \frac{1}{8\pi G}
\end{equation}
In addition, the energy momentum tensor, i.e. $T_{\mu \nu}$, specify the matter content of the universe. At the cosmological scales, $T_{\mu \nu}$ can be described as a `perfect fluid' with the energy density $\rho(t)$, pressure $p(t)$ as well as the velocity $u^{\mu}(t) \equiv \frac{dx^{\mu}}{ds}$, 
\begin{equation} 
\label{perfect fluid}
T_{\mu \nu} = \left( \rho + p \right) u_{\mu} u_{\nu} - p g_{\mu \nu}
\end{equation}
Moving to comoving frame, we have $u^{\mu} = (1, 0, 0, 0)$. \\
Plugging back Eq. (\ref{perfect fluid}) into Eq. (\ref{Einstein}), we obtain the dynamical equations governing the evolution of the scale factor, 
\begin{eqnarray}
\label{Friedmann}
\left(\frac{\dot{a}}{a}\right)^2 = \frac{1}{3M_{Pl}^2} \rho - \frac{k}{a^2} ~~~~~~,~~~~~~ \frac{\ddot{a}}{a} = -\frac{1}{6 M_{Pl}^2} \left(\rho + 3 p \right)
\end{eqnarray}
where as we mentioned before, $k$ is the curvature constant. However, since observations give us a very small value for contribution of curvature, corresponding to a nearly flat universe, it is convenient to set $k$ equal to zero. 

To close the system of equations determining the dynamics of the universe, we would also need to specify the `equation of state' for the matter, 
\begin{equation} 
\label{equation of state}
 p = p(\rho)
\end{equation}
We should notice that Eq. (\ref{equation of state}) is not a consequence of the Einstein gravity and as we will see in the following, it depends on the matter content of the universe.
In addition to the above equations, by using the covariant conservation of the energy-momentum tensor, we can get another equation. Although this equation does not contain any further information beyond the above equations, it represents the energy conservation in more transparent way. We call this equation the `continuity equation' and is given by,
\begin{equation} 
\label{continuity}
\dot{\rho} + 3 H \left(\rho + p \right) = 0
\end{equation}
From this equation, we can infer the evolution of $\rho(t)$ as,
\begin{equation} 
\label{rho dynamic}
\rho \propto a^{-3(1 + w)} ~~~, ~~~ p \equiv w \rho
\end{equation} 
Having presented the dynamical equations for the universe, we now determine the evolution of the scale factor and the energy density in specific some examples.

\textbf{Relativistic particles: } \\
In this case, the equation of state is given by,
\begin{equation} 
w = \frac{1}{3}
\end{equation}
Plugging this back inside the above equations, we obtain
\begin{equation} 
\rho(t) \propto a^{-4} ~~~~~,~~~~~ a(t) \propto t^{1/2} 
\end{equation} 
\textbf{Non-Relativistic particles: } \\
In this case, the equation of state is given by,
\begin{equation} 
w = 0
\end{equation}
So by plugging back this equation, we obtain,
\begin{equation} 
\rho(t) \propto a^{-3} ~~~~~,~~~~~ a(t) \propto t^{2/3} 
\end{equation} 
\textbf{Vacuum:} \\
In this case, the energy momentum tensor is given by $T_{\mu \nu} = \rho_{vac} \eta_{\mu \nu}$. This means that the equation of state is given by,
\begin{equation} 
w = -1
\end{equation}
Plugging this back inside the above equations, we obtain
\begin{equation} 
\rho(t) = const = \rho_{vac}  ~~~~~,~~~~~ a(t) \propto  exp{(H_{ds}t)}
\end{equation} 
Where $H_{ds}$ is given by,
\begin{equation} 
H_{ds} = \sqrt{\frac{\rho_{vac}}{3 M_{Pl}^2}}
\end{equation} 
We should notice that the vacuum, or the cosmological constant, is a very specific example of a wide class of theories called Dark Energy in which $-1 \leqslant w \leqslant -1/3$. 

So far we considered the contributions of different type of matter in the energy density separately. It is straightforward to generalize the above picture and consider the case where more than one matter species (e.g. photons, baryons, neutrinos, dark matter, dark energy, curvature, etc) contributes in the energy density and the pressure of the universe. So we have,
\begin{equation} 
\label{energy-pressure}  
 \rho \equiv \sum_{i} \rho_i ~~~~~~~~~~,~~~~~~~~~~ p\equiv \sum_{i} p_i 
\end{equation} 
For the next references, we may now define the critical energy density, defined to be the current total energy density of the universe,
\begin{equation} 
\label{critical}
\rho_{crit} \equiv \frac{3H_0^2}{8\pi G} = 3M_{Pl}^2 H_0^2
\end{equation} 
Here and in the following, the subscript `0' means the value of a quantity at the present time, $t_0$. 

For each matter pieces, we can define the present ratio of its energy density relative to the critical energy density which defined above,
\begin{equation} 
\label{rational energy}
\Omega_{i} \equiv \frac{\rho_{i0}}{\rho_{crit}}
\end{equation} 
So we can rewrite the energy density as,
\begin{equation} 
\label{total rho}
\rho = \rho_{crit} \left(\Omega_{\Lambda} + \Omega_{M} \left(\frac{a}{a_0}\right)^3 + \Omega_{R} \left(\frac{a}{a_0}\right)^4 + \Omega_{k} \left(\frac{a}{a_0}\right)^2 \right)
\end{equation} 
where $\Omega_{\Lambda}$, $\Omega_{M}$, $\Omega_{R}$ and $\Omega_{k}$ denotes the relative energy density for the Dark Energy (DE), matter component, i.e. baryons plus the cold dark matter, the radiation and the curvature of the universe, respectively. 

\begin{figure}[t!]
	\centering
	\includegraphics[width=0.7\textwidth]{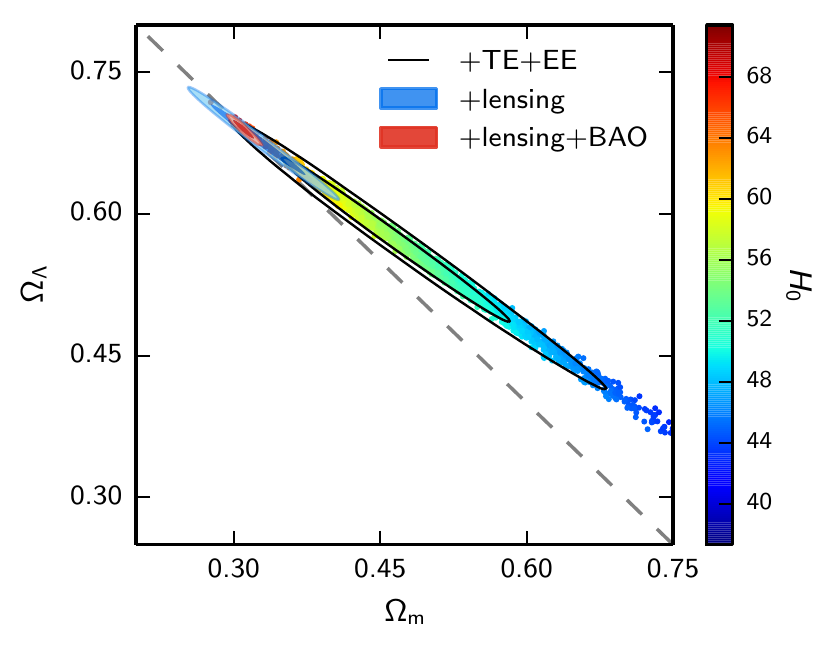}
	\label{fig:omegam-l}
	\caption{ constraints on the value of $\Omega_m$ and $\Omega_{\Lambda}$ using the Planck data, \cite{Planck:2015xua}.}
\end{figure}

By using the observations of the cosmic microwave background and the large-scale structure, we can find the above parameters. Fig. \ref{fig:omegam-l} presents the constraints on the value of $\Omega_m$ and $\Omega_{\Lambda}$ using the Planck data, \cite{Planck:2015xua}. 

\section{Thermal History of the Universe}
The goal of this section is to give a short summary about the thermal history of the universe and then focusing on the most relevant eras of the cosmic expansion for this thesis, i.e. the recombination, last scattering surface as well as the Inflationary universe. \\
Before getting started with the thermal history of the universe, it is worth to mention that there are two main principles characterizing the physics in an expanding universe:
\begin{itemize}
\item Interactions between particles freeze out as soon as their interaction rate drops below the expansion rate of the universe.\\ 	
\item It is possible that the broken  symmetries in the laws of physics are restored at very high energies.	
\end{itemize}
Having presented the main principles in the expanding universe, let us now focus on the thermal history of the universe. We have listed it in table \ref{tab:timeline} and in what follows, first of all, we give a qualitative summary of these eras in the evolution of the universe and then we will focus on some of the main eras in the thermal history of the universe. 

While there are many different and potentially interesting eras in the evolution of the universe, from the observational point of view, all can be divided into two main categories, namely 
`Upper Collider'
and `Below Collider'
physics. Where by the collider physics, we mean the energies of order 1TeV, corresponding to $t \simeq 10^{-10}$ seconds after the Big Bang. 

As it is well known, the fundamental laws of the high energy physics are well determined up to energies of order 1TeV which are accessible by the current particle accelerators like Large Hadron Collider. 
Up to this energy, the physics can be very well described and tested by the Standard Model of the particle physics, with the general relativity and fluid physics. And, as we see briefly in the following, we are confident about the physical processes which happen in this era. \\
However, it is not the case for the upper collider physics, $t < 10^{-10}$ seconds. In this case, the energy of the universe exceeds 1TeV. Therefore we can not have a direct test of the physical laws in this era. \\
Having introduced two main classes in the cosmic evolution, let us now consider them in some more details, \\
{\color{blue}{\textbf{ $\bigstar$ Upper Collider Physics:}}}\\ As we briefly mentioned above, this era corresponds to energies above 1TeV, or $t < 10^{-10}$ seconds.  As it has been shown in Table \ref{tab:timeline}, there are different phases in this category. Although the physics all processes relevant for these phases are very speculative and fascinating, since the main focus of the following thesis is on the Inflationary phase ($\simeq 10^{-34}$ seconds), we skip rest of the other parts and solely consider this phase in the following sections. \\
{\color{blue}{\textbf{ $\bigstar$ Bellow Collider Physics:}}} \\
The physical processes for this era can be divided into three main categories, `The first three minutes' after the Big-Bang where the universe is in the radiation dominated phase,  `From the Matter-Radiation equality to the CMB decoupling' and finally `the Large Scale Structure'. Among these three different phases is most relevant for our course study in the current thesis and thus after a very quick review of the physics at the first and last phases, we try to describe the second phase in more details. \\
As we mentioned above, from the $t \simeq 10^{-10}$ seconds toward today, the history of the universe can be well understood and observationally tested by the Particle Physics, Nuclear and Gravitational physics. Let us start with the time of the electroweak phase transition $t \simeq 10^{-10}s$, energies of order $100$ GeV. Above this energy, the electroweak symmetry is stored and all of the gauge bosons are massless. The interaction rate are high enough to keep the quark-lepton plasma in the thermal equilibrium. As time passes, the temperature of the universe drops and after $100$GeV, the symmetry between the electromagnetism and week interactions is broken and thus the $Z$ and $W^{\pm}$ bosons acquire their masses. At temperature 1MeV, $t \simeq 1$ second, neutrinos decouple from the rest of the matter. Shortly after this time, the temperature drops below the electron rest mass and thus the electron and positron annihilate and produce photons. However, the resulting photon-baryon fluid would be in the equilibrium. Then, about $T=0.1$ MeV, the strong interaction becomes important and thus the Neutrons and Protons are combined into the light elements( H, He, Li) during the Big-Bang Nucleosynthesis, $t \simeq 3$ mins.  After that, at energies of order 1eV, $t \simeq 10^{11}$ seconds, the matter and radiation densities becomes equal. At this phase the charged matter is tightly coupled with the photons. However, around 0.1eV, $t \simeq 380,000$ years, protons and electrons are combined to form the neutral Hydrogen atoms. Shortly after that, photons decouple and form the free-streaming cosmic microwave background radiation. As we will see in more details below, there are some anisotropies associated with the CMB temperature which provide the evidence for the primordial fluctuations in the matter density, being produced during the Inflationary phase. These small perturbations will grow during the gravitational instabilities and will form the large-scale structures in the late time universe. Small scales becomes non-linear and form the gravitationally bound objects which decouple from the expansion of the universe. Around $t \simeq 10^{8}$ years, associated with redshift $z \simeq 25$, high energy photons from the first stars start to ionize the Hydrogen inside the inter-galactic medium. This process of the `reionization' would be completed at $z \simeq 6$. Finally, at $t \simeq 10^{9}$ years, corresponds with the $z \simeq 0.3$, Dark Energy comes to dominate the expansion of the universe. As a result the background space-time enters to an accelerated expansion. 
 
\begin{table}[h!]
	\caption{Major Events in the History of the Universe, \cite{Baumann:2009ds}}
	\label{tab:timeline}
	\begin{center}
		\begin{tabular}{||l || r || r || r||}
			\hline 
			\hline 
			{\small {\color{blue}{Epoch}}} 
			& {\small Time} & {\small Energy} &  {\small Redshift}\\
			\hline 
			\hline
			{\small {\color{blue}{Planck Epoch?} }} & {\small $< 10^{-43}$ s} & {\small $10^{18}$ GeV} &
		    {\small $> 10^{46}$}	\\
		    {\small {\color{blue}{String Scale?}}} & {\small $\gtrsim 10^{-43}$ s} & {\small $\lesssim 10^{18}$ GeV} &  {\small $> 10^{46}$}\\
		    {\small {\color{blue}{Grand Unification?}}} & {\small $\sim 10^{-36}$ s} & {\small $10^{15}$ GeV} &  {\small $ < 10^{46}$} \\
			{\small  {\color{blue}{Inflation?}}} & {\small $\gtrsim 10^{-34}$ s} & {\small $\lesssim 10^{15}$ GeV} & {\small $ < 10^{46}$} \\
			 {\small {\color{blue}{SUSY Breaking?}}} & {\small $< 10^{-10}$ s} & {\small $> 1$ TeV} & {\small $ > 10^{20}$} \\
			{\small  {\color{blue}{Baryogenesis?}}} & {\small $< 10^{-10}$ s} & {\small $> 1$ TeV} & {\small $ > 10^{20}$} \\
			\hline
			\hline
			{\small  {\color{blue}{Electroweak Unification}}} &  {\small $10^{-10}$ s} & {\small 1 TeV}& {\small $10^{15}$ }  \\
			{\small  {\color{blue}{Quark-Hadron Transition}}} & {\small $10^{-4}$ s} & {\small $10^2$ MeV} &  {\small $10^{12}$ }\\
			{\small  {\color{blue}{Nucleon Freeze-Out}}} & {\small 0.01 s} & {\small 10 MeV} &  {\small $?$ }\\
			{\small  {\color{blue}{Neutrino Decoupling}}} & {\small 1 s} & {\small 1 MeV} &  {\small $6 \times 10^{9}$ }\\
			{\small {\color{blue}{BBN}}} & {\small 3 min} & {\small 0.1 MeV} & {\small $4 \times 10^{8}$ }\\
			\hline 
			\hline
			{\small  {\color{blue}{Matter-Radiation Equality}}} & {\small $10^4$ yrs} & {\small 1 eV} & {\small $3400$}\\
			{\small  {\color{blue}{Recombination}}} & {\small $10^5$ yrs} & {\small 0.1 eV} & {\small 1,100-1,400}\\
			\hline
			\hline
			{\small  {\color{blue}{Reionization}}} & {\small $10^8$ yrs} & {\small $2.6-7.0$ MeV} & {\small $25 - 6$}\\
			{\small  {\color{blue}{Galaxy Formation}}} & {\small $\sim 6 \times 10^8$ yrs} & & {\small $\sim 10$} \\
			{\small  {\color{blue}{Dark Energy}}} & {\small $\sim 10^9$ yrs} & {\small 0.33 meV} & {\small $\sim 2$} \\
			{\small  {\color{blue}{Solar System}}} & {\small $8 \times 10^9$ yrs} & & {\small 0.5} \\
			{\small  {\color{blue}{Today}}} & {\small $14 \times 10^{9}$ yrs} & {\small 0.24 meV} & {\small 0} \\
			\hline
	        \hline
		\end{tabular}
	\end{center}
\end{table}
Having presented a general picture about different phases during the cosmic expansion, in the following section, we only focus on what is most relevant to our case study. More precisely, we will pick up two specific pases during the cosmic history of the universe, say Inflation as well as the surface of the last scattering as a cosmological lab to test the primordial nature of the universe. 
\section{Inflationary Universe}
The goal of this section is to review the theory of the `Inflation' as a consistent solution to the several puzzles in the Standard Model of Cosmology. \\
We start with a short review of these puzzles. Then, we show how an early phase of accelerated expansion can solve these problems. We leave the observational signatures of this theory for the next chapters. 
\subsection{On the Big Bang Puzzles}
Before presenting the big bang puzzles, it is worth trying to answer the following question. \\
The question is that whether the initial conditions are supposed to be given by a physical theory or they can be fixed separately. \\
In order to answer this question, let us review what happens in classical mechanics. We recall that in order to determine the evolution of an object in classical mechanics, we need to provide the initial conditions additionally to the Newton's equation of motion. It is therefore not obvious whether a theory of cosmology is supposed to predict or explain the initial conditions of the universe or they can be fixed separately. The most natural expectation is that the current situation of the universe can be reached generically and without too much fine tuning in the initial conditions. However, as we see in the following, it turns out that the conventional Big Bang theory does require very fine tuned initial conditions to allow the universe to look like as what it is today. This poses the following puzzle in the Standard Model of Cosmology. It seems that our universe is the outcome of an `improbable accident' and thus is an unnatural phenomena, which if correct, makes cosmology a very disappointing subject! \\
However, as we will see in the following section, there is a brilliant idea, i.e. Inflationary theory, to by pass the above mentioned problem. \\
As we briefly discussed,  it seems that in the conventional Big bang theory, we require to have some fine tunings in the initial conditions of the universe. Here we try to illustrate what we mean by the fine tunings in the initial conditions in more details.
\subsubsection{Horizon Problem} 
\begin{figure}[t!]
	\centering
	\includegraphics[width=0.7\textwidth]{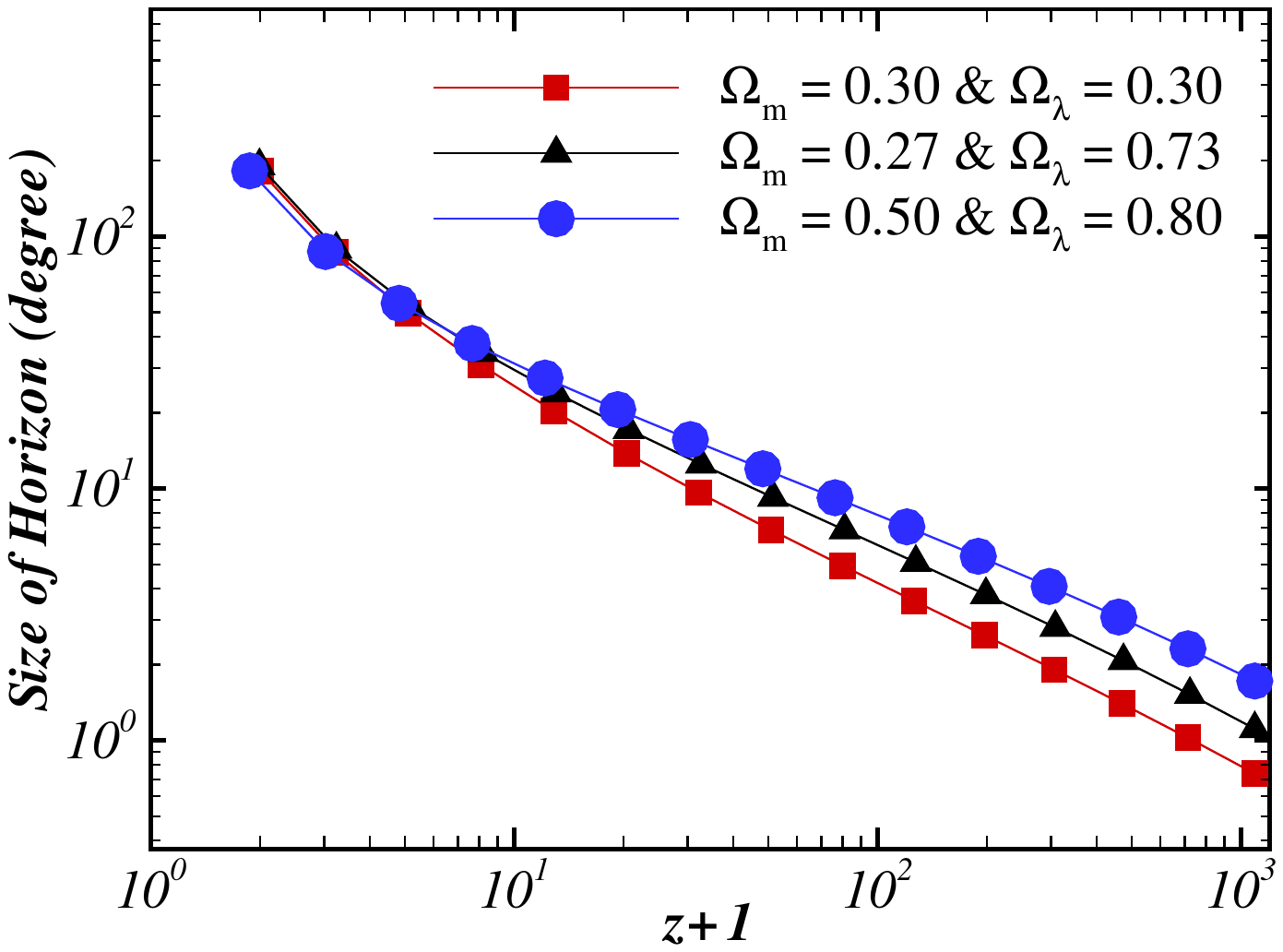}
	\caption{\label{fig:Behavior} The behavior of the particle horizon with time, \cite{Movahed}}
\end{figure}

\begin{figure}
	\centering
	\includegraphics[width=0.98\textwidth]{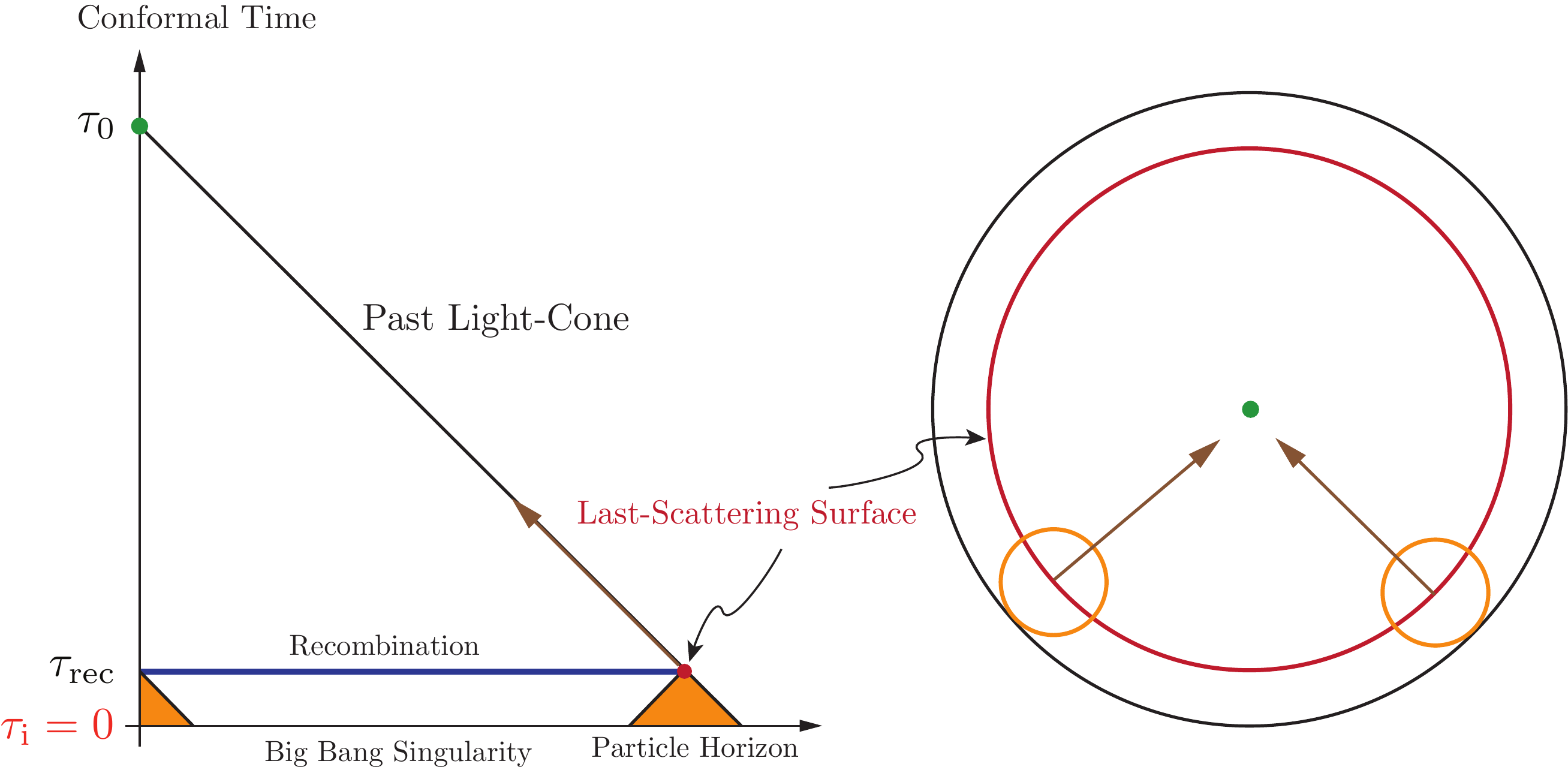}
	\caption{The CMB at last-scattering surface contains of $10^5$ causally disconnected regions, \cite{Baumann:2009ds}}
	\label{fig:Conformal1b}
\end{figure}

The key point in this problem is the behavior of the comoving particle horizon, $\tau$, in the conventional cosmology. We recall from Eq. (\ref{particleH}) that the particle horizon is defined in the following way,
\begin{equation}
\label{particleH2}
\tau \equiv \int_{0}^{t} \frac{dt'}{a(t')} = \int_{0}^{a} \frac{da}{H a^2} = \int_{0}^{a} d\ln{a} \left(\frac{1}{aH}\right),
\end{equation}
which is the maximum comoving distance that a light ray can travel between times 0 and t. \\
Here in order to express the Horizon problem in the best way, we represented it as an integral over the comoving Hubble radius, $\left(a H\right)^{-1}$. \\
It is easy to show that for a universe which is dominated by a fluid with equation of state $w$,  we would have, 
\begin{equation}
\label{comHubble}
(aH)^{-1} = H_0^{-1} a^{\frac{1}{2}(1+3w)}
\end{equation}
We should notice that the comoving Hubble radius depends on the combination $\left(1+ 3w\right)$. Plugging this back into Eq. (\ref{particleH2}), we obtain, 
\begin{equation}
\label{particleH3}
\tau \propto a^{\frac{1}{2}(1+ 3 w)}
\end{equation}
As we see, the qualitative behavior of the comoving horizon therefore depends on the combination $\left(1+ 3w\right)$. This leads to the well known horizon problem. The reason is that, in a decelerating universe, the comoving horizon grows with time which means that the fraction of the universe which are in the causal connection with us are growing with time. More explicitly, for a universe dominated by either the radiation or dust, the comoving particle horizon is given by, 
\begin{equation}
\tau = \int_0^a \frac{d a}{Ha^2} \propto
\left\{ 
\begin{array}{ll}
	a \quad \quad , \rm{RD}\\
	a^{1/2} \quad , \rm{MD}
\end{array} \right. ~,
\end{equation}
which shows that the comoving horizon grows with time. This means that the comoving scales which enter the horizon today where not in the causal contact at the surface of the last scattering. On the other hand, the near-homogeneity of CMB tells us that the universe 
must be extremely homogeneous at the last scattering surface. However, the comoving horizon on the surface of the last scattering was of order $2$ degrees, see for example Fig. \ref{fig:Behavior}.
This means that within the conventional cosmology, the universe had not have enough time to be in causal contact, and thus being homogeneous, on scales larger than $2$ degrees. So the question is that how the universe is extremely homogeneous and isotropic on the scales which contain many causally disconnected regions. \\
Having presented a summary of the horizon problem, it is worth trying to put it into some numbers. For this purpose, let us recall that our current universe is homogeneous and isotropic in the scales of order of the current horizon size, which is of order $ct_0 \simeq 10^{28}$cm. We can then calculate the size of the homogeneous and isotropic region back in the radiation dominated universe as, 
\begin{equation}
\label{homogeneous region}
d_{h}= ct_{0}\frac{a_{i}}{a_{0}}
\end{equation}
We can then compare this size with the physical horizon at this time, $d_{i} = c t_{i}$,
\begin{eqnarray}
\label{compare horizons}
\frac{d_{h}}{d_{i}}&=& \frac{ct_{0}}{ct_{i}}\frac{a_{i}}{a_{0}}  \nonumber\\
& =& \frac{t_{0}}{t_{i}}\frac{a_{i}}{a_{0}} 
\end{eqnarray}
Now in order to obtain an estimate about the above ratio, we recall that in the conventional cosmology the universe started with RD at the Planck time. In addition, since the temperature is proportional to the inverse of the scale factor, by comparing the temperature at the Planck time with that of the current universe, as well as comparing their time scales, we can easily find that the radius of the 
homogeneous region compared with the horizon radius is, 
\begin{equation}
\label{numeric compare horizons}
\frac{d_{h}}{d_{i}} \sim 10^{28}
\end{equation}
This means that at the Planck time, the number of the causally disconnected regions is at the order $10^{84}$. A similar calculation shows that this number is about $10^5$ at the surface of last scattering. This problem has been shown schematically in Fig. \ref{fig:Conformal1b}.
It is worth to present the ratio of the homogeneous to the horizon region in more transparent way as, 
\begin{equation}
\label{analytic compare horizons}
\frac{d_{h}}{d_{i}}\sim \frac{\dot{a}_{i}}{\dot{a}_{0}}
\end{equation}
This means that in a decelerating universe, the homogeneous radius is always larger than the horizon radius. We thus call the horizon problem the homogeneity problem. 
\subsubsection{Flatness Problem} 
As we mentioned before, the current observations confirm that our universe can be approximated with a nearly flat universe. The question is that why our universe is so close to the flat Euclidean space and moreover what does it say about the curvature of the universe in the past? \\
In order to quantify the problem, let us consider the Friedmann equation, (\ref{Friedmann}),
\begin{equation}
H^2 = \frac{1}{3} \rho(a) - \frac{k}{a^2}\, ,
\end{equation}
We can then try to express it with respect to the critical density at every redshift as, 
\begin{equation}
\label{equ:omega}
1-\Omega(a) = \frac{-k}{(aH)^2}
\end{equation}
where
\begin{equation}
\Omega(a) \equiv \frac{\rho(a)}{\rho_{\rm crit}(a)}\, , \qquad \rho_{\rm crit}(a) \equiv 3 H(a)^2
\end{equation}
We notice that here we have generalized the definition of the $\Omega$ to be time dependent. \\
Looking at Eq. (\ref{equ:omega}), we can easily see that, since in the conventional cosmology the comoving Hubble radius is always growing with time, the quantity $|\Omega(a)-1|$ also grows with time. This means that $\Omega=1$ is not a stable fixed point. Therefore our current near flat universe requires to have an extremely fine-tuned value of $\Omega$, to be very much close to 1, in the early universe. More explicitly, we can easily find that in order to explain the nearly flat universe today, the deviation form the flatness during the BBN, GUT and Planck epoch has to satisfy the following constraint conditions, 
\begin{eqnarray}
|\Omega(a_{\rm BBN})-1| &\le& {\cal O}(10^{-16})\, ,\\
|\Omega(a_{\rm GUT})-1| &\le& {\cal O}(10^{-55})\, ,\\
|\Omega(a_{\rm pl})-1| &\le& {\cal O}(10^{-61})\, .
\end{eqnarray}
Putting it in another way, we can also present the flatness problem in the following way, 
\begin{equation}
\label{density freidman}
(\Omega_{i} - 1) =(\Omega_{0} - 1)  \left( \frac{\dot{a_{0}}}{\dot{a_{i}}} \right)^2
\end{equation}
Where $i$ means any time before today. \\
As in the case of the horizon problem, we see that if in the past the universe was always decelerating, we always need a fine tuning in the initial value of the $\Omega$ to end up into our current observed universe.
\subsection{Inflation as a solution for the Big Bang puzzles}
As we discussed above, the key point in both of the horizon and flatness problems is the dynamical behavior of the comoving Hubble radius, $(aH)^{-1}$. More precisely, they arise because the comoving Hubble is a growing function in the conventional cosmology. So we may guess that, these puzzles may be solved if we insert a phase in the thermal history of the universe where the comoving Hubble horizon is a decreasing function and thus relax the necessary fine-tuning to reach the current universe. \\
This phase is called \textbf{\textit{inflation}} which is the main focus of the current thesis. \\
So it is worth to first of all, give an introduction to the theory of inflation and then show how this simple and intelligent idea can solve both of the Big Bang puzzles easily at the background level. Moreover, as we will see in the next chapter, at the perturbation level, inflation also make some predictions that can be checked observationally. We leave more details for the next chapter. \\
\subsubsection{The Main Idea of Inflation}
Let us start with the general definition of cosmic inflation, \\
\textit{Inflation is a stage of accelerated expansion where the strong energy condition is violated, i.e. $(1 + 3w) <0$, and thus the gravity acts as a repulsive force.} \\
In order to see how the violation of the strong energy condition leads to the accelerating expansion and thus make the gravity to behave as a repulsive force, we use the Einstein equation, Eq. \ref{Friedmann}, 
\begin{equation}
\label{strong-condition}
\frac{\ddot{a}}{a} = - \frac{1}{6M_{Pl}^2} \left(1+ 3w\right) \rho
\end{equation}
As we see from the above equation, in order to have an accelerating expansion, we have to break the strong energy condition. We should emphasize here that the ordinary matter can not violate the strong energy condition and thus may not be able to produce the accelerating expansion of the universe. The simplest form of energy that can give rise to such a behavior is the cosmological constant. However, as we will see later on, a pure cosmological constant can not gracefully exit from inflation and thus we need the field theory as a way to provide the necessary conditions for the accelerating expansion of the universe. 
\subsubsection{Solution of the Big Bang Puzzles within Inflation}
\begin{figure}[h]
	\centering
	\includegraphics[width=4.8cm]{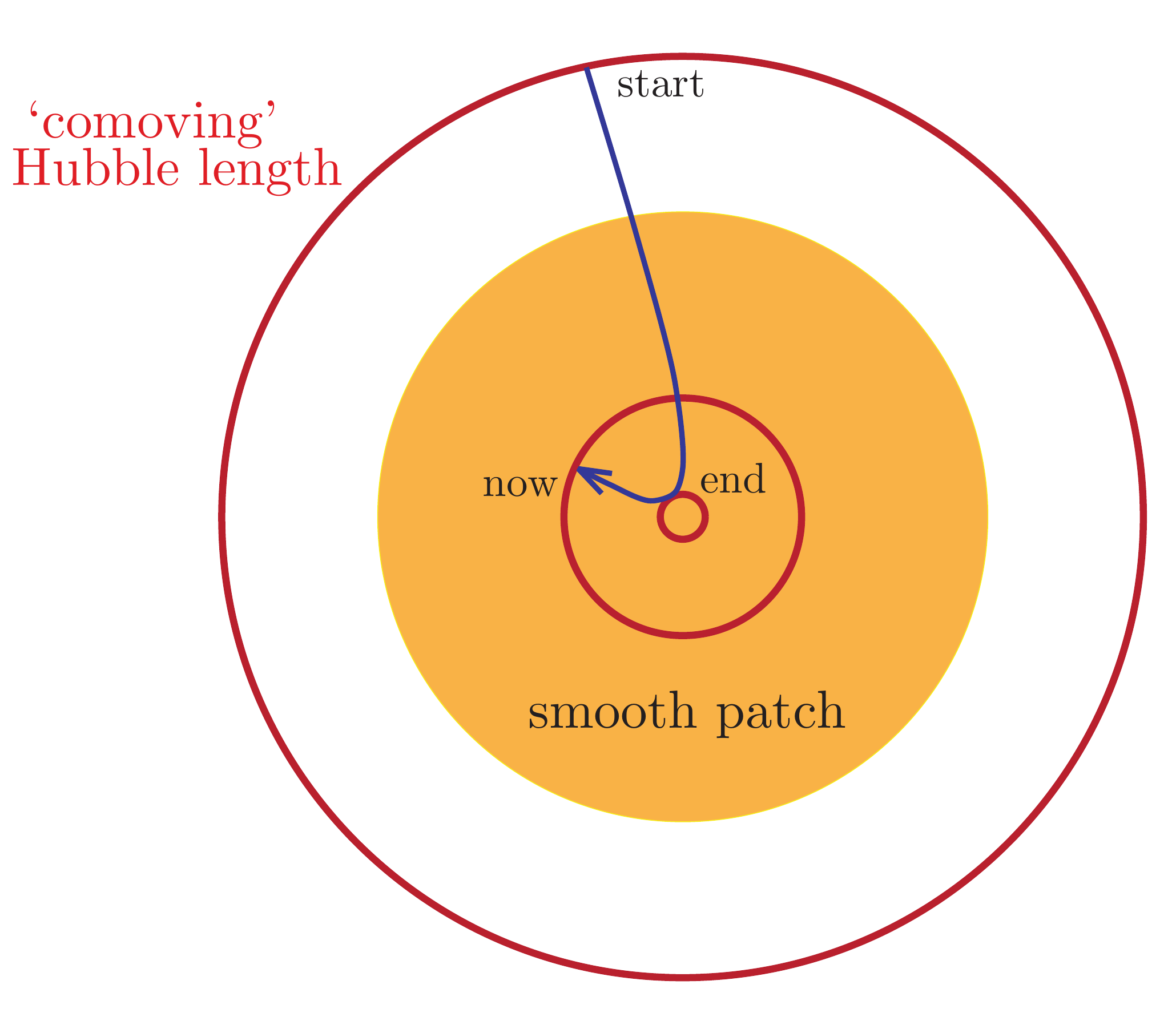}
	\hspace{0.6cm} \includegraphics[width=6.2cm]{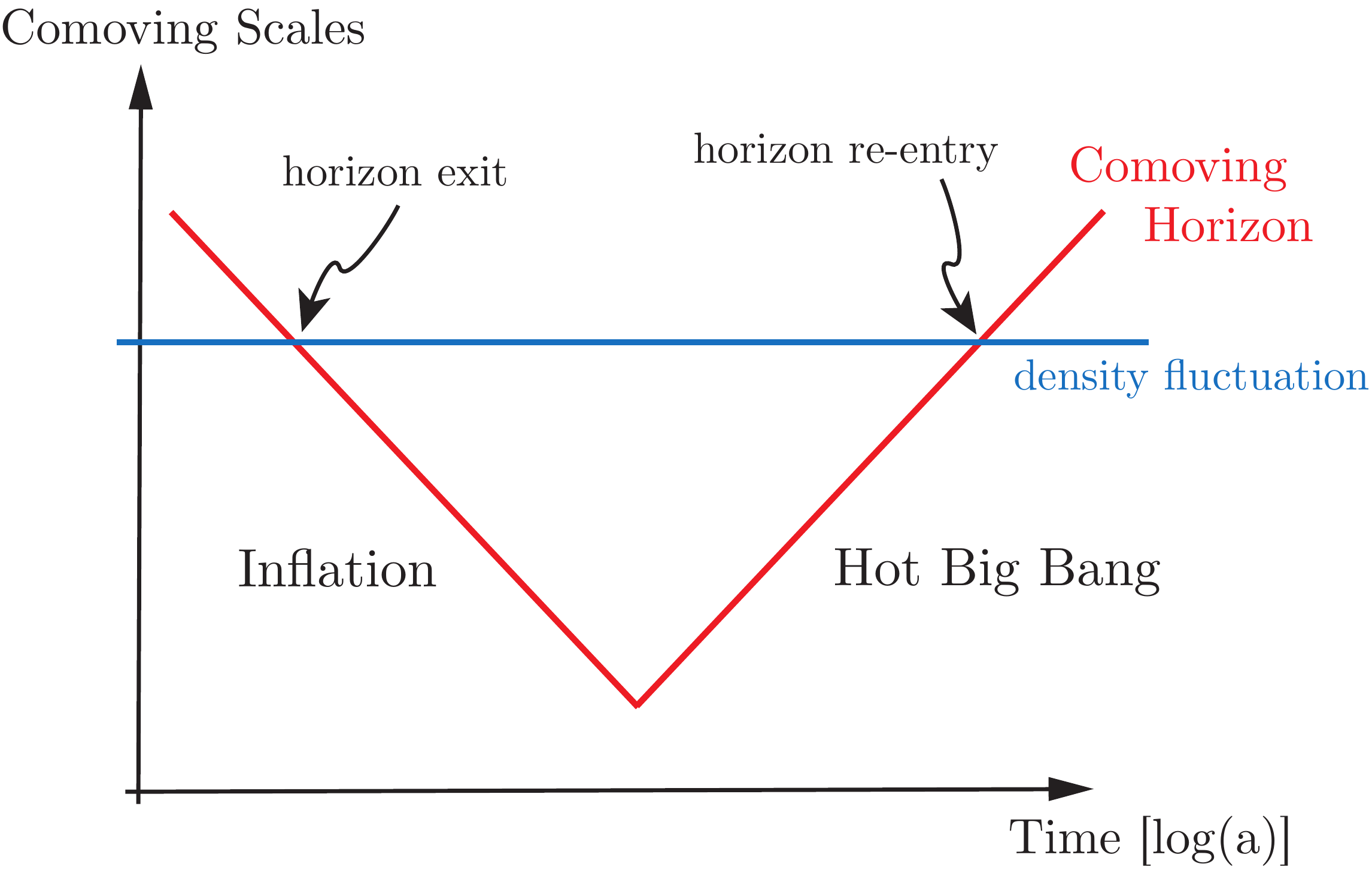}
	\caption{\label{fig:evo-horizon}{\it Left Panel:} 
		the evolution of the comoving Hubble horizon, $(aH)^{-1}$, in an inflationary universe. The comoving Hubble sphere shrinks during inflation and expands again after that. 
		{\it Right Panel:} Schematic solution to the horizon problem. At sufficiently early times, all of the relevant scales to cosmological observations today were inside the horizon. Then during inflation, they have been stretched outside the horizon. Finally, very recently, they have reentered the horizon again and initiated the homogeneous universe, \cite{Baumann:2009ds}.}
\end{figure}

Having presented the main idea of inflation, here we show how an accelerating expansion during the early universe can solve both of the Big Bang puzzles consistently. \\
{\color{blue}{\textbf{ $\blacklozenge$ Horizon Problem Resolved}}} \\
\textit{Qualitative Solution:} During inflation, the comoving Horizon is decreasing.
Therefore starting from a big enough value of the comoving horizon at the beginning of Inflation, 
all interesting scales were inside the horizon. They would then stretch during inflation and go outside the horizon. Finally, after inflation ends, the comoving horizon again increases and thus the scales which have left the horizon enter back inside the horizon. However, since they were in the casual connection initially, they make the CMB radiation to be extremely homogeneous and isotropic. The evolution of the comoving Hubble radius is given in Fig. \ref{fig:evo-horizon}. \\
\textit{Quantitative Solution:} Here we give a more detailed analysis under which conditions inflation can solve the horizon problem. More explicitly, we set the duration of inflation in such a way that at the last scattering surface we only have one causal region in the universe. \\
In order to do this, first of all, we calculate the `past light cone' and then demand that it is smaller than the `forward light cone'.\\
The past light cone is given by, 
\begin{equation}
\label{past light cone}
l_{p}(t_{dec}) = \int^{t_{0}}_{t_{dec}} \frac{dt}{a(t)}\sim 3 t_{0}\left[1-\left(\frac{t_{dec}}{t_{0}} \right) ^{1/3} \right]. 
\end{equation}
Since $t_{dec} \ll t_{0}$, the past light cone would be approximately equal to,
\begin{equation}
\label{appro-past light cone}
l_{p}(t_{dec}) \sim t_{0}= \frac{1}{H_{0}}.
\end{equation}
As a result, in order to solve the horizon problem, the particle horizon at the end of inflation must be larger than the comoving Hubble radius at the present time, 
\begin{equation}
\label{solve the horizon1}
\int^{t_{f}}_{t_{i}} \frac{dt}{a(t)} > \frac{1}{H_{0}}.
\end{equation}
Now taking into account the exponential growth of the scale factor during inflation, we obtain,
\begin{equation}
\label{solve the horizon2}
\int^{t_{f}}_{t_{i}} \frac{dt}{a(t)} \sim \frac{1}{H a(t_{i})} \left[1-e^{-H\left( t-t_{i}\right) }\right] \sim \frac{1}{H a(t_{i})}.
\end{equation}
Combining Eqs. (\ref{solve the horizon1}) and (\ref{solve the horizon2}), we yield
\begin{eqnarray}
\label{final-h}
\frac{1}{H a_{i}} &>& \frac{1}{H_{0} a_{0}} \nonumber\\
\frac{a_{0} a_{f}}{a_{i} a_{f}}&>& \frac{H}{H_{0}}.
\end{eqnarray}
Using the Friedmann equation and neglecting the curvature contribution,
\begin{equation}
\label{Friedman horizon}
\frac{H^2}{H_{0}^2} \sim \frac{\rho_{r}}{\rho_{m}} \left[\frac{a_{0}}{a_{f}}\right]^4
\end{equation}
and therefore we obtain, 
\begin{equation}
\label{number of h}
\frac{a_{f}}{a_{i}} \sim 10^{-2} \frac{a_{0}}{a_{f}}.
\end{equation}
Now if we assume that inflation happened at the GUT scale, we obtain, 
\begin{equation}
\label{e-number for horizon}
\frac{a_{f}}{a_{i}} >10^{26} \sim e^{60}.
\end{equation}
So we see that in order to solve the horizon problem, we need about $60$ e-folds of inflation. \\
{\color{blue}{\textbf{ $\blacklozenge$ Flatness Problem Resolved}}} \\
\textit{Qualitative Solution:} Since during inflation the comoving horizon is decreasing the  $\Omega =1$ is an attractor solution and the universe becomes flatter and flatter at the end of inflation. This solves the flatness problem. \\
\textit{Quantitative Solution:} Here we show how inflation solves the flatness problem. In order to show this, we rewrite Eq. (\ref{density freidman}) in the following way, 
\begin{eqnarray}
\label{flat quantive1}
\left( \Omega_{tot}^{0} -1\right)  &=& \frac{a^2_{i} }{a^2_{0}} \frac{H^2_{i}}{H^2_{0}} \left( \Omega_{tot}^{i}-1\right)  \nonumber\\
&=& \frac{a^2_{i}}{a^2_{f}} \frac{a^2_{f}}{a^2_{0}} \frac{H^2}{H^2_{i}} \left( \Omega_{tot}^{i} -1\right) 
\end{eqnarray}
We then use the current observational data for $\Omega_{tot}^0$ as, $(\Omega_{tot}^{0}-1)\sim 10^{-2}$ and demand that initially there was not any fine tuning on the value of the $\Omega_{tot}^i$, say $(\Omega_{tot}^{i}-1)\sim O(1)$.\\
Plugging these back into Eq. (\ref{flat quantive1}), yields,
\begin{equation}
\label{flat quantive2}
\frac{a_{i}}{a_{f}}\simeq 10^{-2} \frac{H_{0}}{H} \frac{a_{0}}{a_{f}}.
\end{equation}
Now following a procedure similar to what we have done for the horizon problem, we would get, 
\begin{eqnarray}
\label{flat quantive3}
\frac{a_{f}}{a_{i}}&\simeq &\frac{a_{0}}{a_{f}} \nonumber\\
&\simeq & 10^{28}  \nonumber\\
&\simeq &  e^{64} > e^{60}
\end{eqnarray}
This means that in order to solve the flatness problem, we would need to have at least 60 e-folds of inflation. 
\subsubsection{Scalar Driven Slow-Roll Inflation}
\begin{figure}
	\centering
	\includegraphics[width=0.7\textwidth]{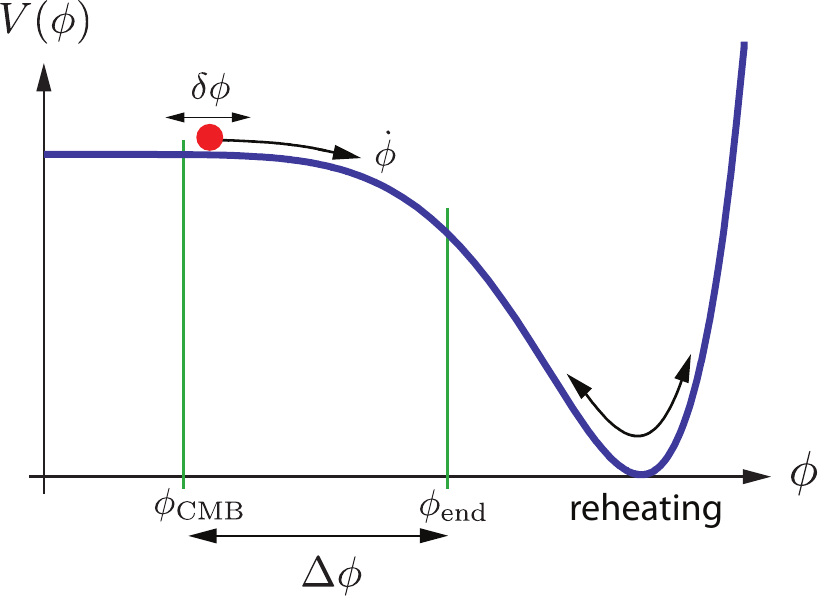}
	\caption{ The slow roll evolution of inflaton field during inflationary phase,\cite{Baumann:2009ds}.}
	\label{fig:smallfield}
\end{figure}

The simplest possible model for the inflation involves one scalar field, known as the \textit{inflaton} field, $\phi$. Here we skip the physical realization of this scalar field and just try to use it as a toy model to drive inflation. In order to make it as simple as possible, we assume that the inflaton field is coupled minimally to gravity. So the action describing this system is given by, 
\begin{equation}
\label{scalar field inflation}
S = \int d^4 x \sqrt{-g} \left[ \frac{M_{P}^2}{2}R + \frac{1}{2}g^{\mu\nu} \partial_{\mu} \varphi \partial_{\nu} \varphi - V(\varphi) \right] = S_{EH} + S_{\varphi}
\end{equation} 
The above action is the sum of the gravitational Einstein-Hilbert action as well as that of a scalar field with canonical kinetic term. \\
The energy-momentum tensor for the scalar field is given by, 
\begin{equation}
\label{energy momenton of scalar}
T_{\mu\nu}^{\varphi} \equiv - \frac{2}{\sqrt{-g}} \frac{\delta S_{\varphi}}{\delta g^{\mu\nu}} =  \partial_{\mu} \varphi \partial_{\nu} \varphi - g_{\mu\nu}\left(  \frac{1}{2}\partial^{\sigma} \varphi  \partial_{\sigma} \varphi + V(\varphi)  \right).
\end{equation} 
Assuming the Friedmann Robertson Walker as the background metric and turning off the spatial curvature, the components of the energy-momentum tensor take the following form, 
\begin{eqnarray}
\rho_\phi &=& {1\over 2}\dot\phi^2+V(\phi)\, , \\
p_\phi &=& {1\over 2}\dot\phi^2-V(\phi)\, .
\end{eqnarray}
Now by using the above expression for the energy density as well as the pressure, we can calculate the equation of state as, 
\begin{equation}
w_\phi  \equiv \frac{p_\phi}{\rho_\phi} = \frac{\frac{1}{2} \dot{\phi}^2 -V}{\frac{1}{2} \dot{\phi}^2 +V}.
\end{equation}
This shows that if the potential energy of a scalar field dominates over its kinetic term, then we can have a negative pressure, corresponding to a negative equation of state. \\
Now using the Klein Gordon equation along with first Friedman equation , we find the equation of motion for the inflaton field as, 
\begin{equation}
\ddot\phi+3H\dot \phi+V_{,\phi}=0 \qquad {,} \qquad H^2={1\over 3}\left({1\over 2}\dot\phi^2+V(\phi)\right)
\end{equation}
Where  $V_{, \phi} = \frac{dV}{d\phi}$. \\
\textit{Slow-Roll Parameters:} It turns out that during inflation, we define two useful dimensionless parameters, called slow-roll parameters, as 
\begin{equation}
\label{hubble slow-roll}
\epsilon \equiv -\frac{\dot{H}}{H^2} \qquad {,} \qquad \eta \equiv \frac{\dot{\epsilon}}{H\epsilon}.
\end{equation}
The first application of these parameters manifest itself in writing down the accelerated equation for the universe, 
\begin{eqnarray}
\frac{\ddot{a}}{a} &=& -\frac{1}{6M^{P}}\left( \rho_{\varphi} + 3 p_{\varphi} \right) = H^2 \left(1-\epsilon\right) \label{acceleration universe app1} \\
\epsilon &\equiv &\frac{3}{2} \left(\omega_{\varphi}+1 \right) = \frac{1}{2} \frac{\dot{\varphi}^2}{H^2} \label{acceleration universe app2}.
\end{eqnarray}
So as we discussed before, in order to get the accelerated expansion of the universe, the slow-roll parameter must be less than unity which can only happen when the kinetic term is smaller than the potential energy of the inflaton field, that is why we call it the slow-roll parameter. One example of such a model is shown in Fig. \ref{fig:smallfield}. The de sitter limit corresponds to $\epsilon \rightarrow 0 $. \\
In addition, the accelerated expansion is only possible for a long period of time if the second time derivative of the inflaton field is also small enough, 
\begin{equation}
|\ddot{\phi}| \ll |3H\dot{\phi}|, |V_{,\phi}|
\end{equation}
This can be translated to the smallness of the second slow-roll parameter, 
\begin{equation}
\eta = - \frac{\ddot{\phi}}{H \dot{\phi}}
\end{equation}
So in order to have the accelerated expansion for a long period of time, we need to have $\eta < 1$. \\
It is interesting to notice that the slow-roll parameters can also be expressed as specific conditions for the shape of the inflationary potential, 
\begin{eqnarray}
\epsilon_{v} &\equiv& \frac{M_{Pl}^2}{2} \left(\frac{V_{,\phi}}{V}\right)^2 \\
\eta_{v} &\equiv& M_{Pl}^2 \frac{V_{,\phi \phi}}{V}
\end{eqnarray}
The above slow-roll parameters are called the \textit{potential slow-roll} parameters while the ones defined in Eq. (\ref{hubble slow-roll}) are called the \textit{Hubble slow-roll} parameters. They can be related to each other via, 
\begin{equation}
\epsilon \approx \epsilon_{v} ~~~~,~~~~\eta \approx \eta_{v} - \epsilon_{v}
\end{equation}
In the slow-roll regime, $\epsilon_v, \eta_v \ll 1$, the background equations of motion are given by,
\begin{eqnarray}
H^2 &\approx& {1\over 3} V(\varphi) \approx \text{const}  \label{equ:SRH} \\
\dot{\varphi} &\approx& -\frac{V_{,\varphi}}{3H} \label{equ:SRphi}\
\end{eqnarray}
And the spacetime looks like de sitter, 
\begin{equation}
a(t) \sim e^{Ht}
\end{equation}
Inflation ends when, 
\begin{equation}
\epsilon \equiv 1.
\end{equation}
It is easy to calculate the number of e-folds before inflation ends as, 
\begin{eqnarray}
\label{equ:Nphi}
N(\varphi) &\equiv & \ln \frac{a_{end}}{a} \nonumber\\
&=&\int _t^{t_{end}} H d t = \int _\varphi^{\varphi _{end}} \frac{H}{\dot \varphi} d \varphi
\approx \int _{\varphi_{end}}^\varphi \frac{V}{V_{,\varphi}} d \varphi
\end{eqnarray}
As we saw before, in order to solve the horizon and flatness problems, we require about 60 number of e-folds, 
\begin{equation}
N_{tot} \equiv \ln{\left(\frac{a_{end}}{a_{start}}\right)} \gtrsim 60
\end{equation}
It is worth mentioning that the total required number of the e-folds depends on the energy scale of the inflation as well as on some details of the post-inflationary models. \\
We should also notice that the observed fluctuations in the CMB have been created around $50-60$ e-folds before the end of inflation. 
\section{Cosmic Microwave Background}
The goal of this section to give an introduction about the cosmic microwave background(CMB) radiation.\\ 
In this chapter, we only focus on the 
isotropic part of the cosmic microwave background. We start with a discussion about the history of the CMB discovery and then will review some features of the CMB.
\subsection{History of discovery of the CMB}
Since the universe is expanding, we expect that matter was hotter and denser in the past. This means that, if we go backward in cosmic time, there was an era in the thermal history of the universe when the electrons were too hot to be bound into atoms and due to the very rapid collisions between the free electrons and photons in this era the radiation was in the thermal equilibrium with the hot matter. As time passed, electrons and protons combine to form neutral matter and the radiation began a free expansion over the sky called CMB. \\
George Gamow and his collaborators recognized in the late 1940, for the first time, that the universe should now be filled with the black-body radiation. In 1950, Ralph Alpher and Robert Herman estimated the present temperature of this radiation to be about 5K. This work was already forgotten in the subsequent decades until in 1956 when a group at Princeton started searching for the cosmic microwave background. Their rough idea of the temperature, based on a nucleosynthesis  calculation of the P.J.E. Peeble, was about 10K. However, before they could complete their experiment, the radiation was discovered by Arno Penzias and Robert Wilson in their study of the noise background and in a radio telescope. Originally, Penzias and Wilson reported the temperature of the antenna to be about $(3.5\pm 1.0)$K at the wavelength 7.5 cm. \\
In addition,  their report of the temperature was in agreement with the black-body spectrum, 
\begin{equation} 
\label{black-body}
n_{T}(\nu) d \nu = \frac{ 8\pi \nu^2 d\nu}{exp{(h\nu/(k_B T))}-1},
\end{equation} 
where $n_{T}(\nu)$ is the number density of photons in the equilibrium with the matter at temperature T. 
However, this was not a prove that the radiation is necessarily a black-body spectrum. \\
In 1978 Penzias and Wilson were awarded the Nobel Prize in physics in honor of their finding. \\
After that, a larger number of measurements were made by many radio astronomers. However, they could not go to very small wavelengths, less than 0.3cm. So their formula was consistent with he `Rayleigh-Jeans' formula of the classical statistical mechanics, 
\begin{equation} 
\label{Rayleigh-Jeans}
h\nu n_{T}(\nu) \simeq  8\pi \nu^2k_B T
\end{equation} 
Finally, the Planck spectrum of the CMB was confirmed in 1990 by observations with the FIRAS radio matter which carried out by the Cosmic Background Explorer Satellite(COBE) launched in 1989 by NASA. \\
COBE proved that the backgound radiation has a very close black-body spectrum in the wavelength range 0.05cm to 0.5cm and after around 6 years of analysis, they reported a temperature $(2.725 \pm 0.00002)$K, \cite{Weinberg:2008zzc}. \\
In 2006, John C Mather and George F Smoot have been awarded the Nobel Prize in physics for their discovery of the blackbody and anisotropy of the CMB. \\
John C Mather coordinated the entire process of analyzing the COBE data as well as the primary responsibility for the experiment which revealed the black-body form of CMB. \\
George Smoot was responsible for measuring the Anisotropies in the temperature of the radiation. \\
NASA's second generation of space mission was Wilkinson Microwave Anisotropy Probe (WMAP) which was launched in 2001 to study the small anisotropies in the temperature in more details. 
The reported temperature of the CMB by WMAP is $T = 2.7255 \pm 0.00060$. \\
Finally, ESA's, European Space Agency, Planck was launched in 2009 to study the CMB in greater details than were done before. It covered the universe between 0.3mm to 11.1mm (corresponding to the frequencies between 27GHZ to 1THZ). The goal was to make a much more accurate separation between all of the different components.
A comparison of the resolution of the COBE, WMAP and Planck measurements are given in Fig. \ref{fig:compareCWP}.
\begin{figure}[t!]
	\centering
	\includegraphics[width=0.7\textwidth]{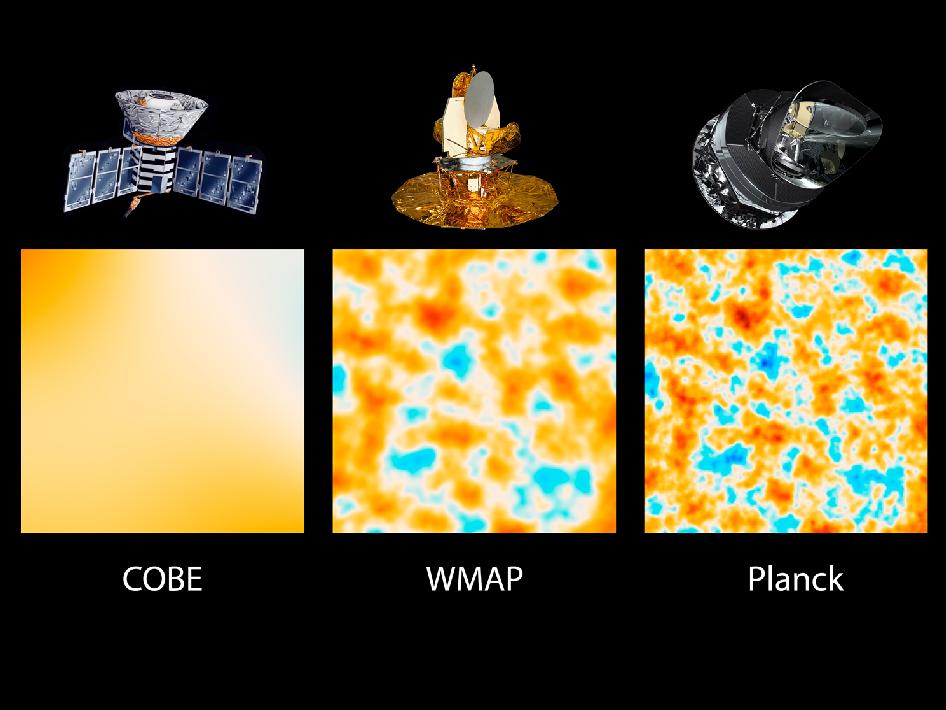}
	\label{fig:compareCWP}
	\caption{\label{fig:compareCWP} Comparison between the resolution of COBE, WMAP and PLANCK,\cite{NaSa}}
\end{figure}
\subsection{Features and evolution of the CMB}
As we already discussed, it has been shown by COBE that the CMB has a black-body spectrum. However, since this radiation is coming from far in the past when the temperature was of order 3,000K, we may wonder whether the
form of the spectrum is invariant under the expansion of the universe. \\
The simplest possible way to show this is assuming that there was a sharp transition, happening at $T_{L}$ which $L$ denotes the last scattering surface, between the time when the photon was in the thermal equilibrium with the matter and the time of free propagation phase. As a result, a photon with the frequency $\nu$ at $t>t_L$, had frequency $\nu a(t)/ a(t_L) $ at the time of the last scattering. So we can find the number density of the photons at time $t$ and with frequency in the range $\nu$ and $\nu + d\nu$ as, 
\begin{equation} 
\label{Rayleigh-Jeans}
n(\nu,t) d \nu(t) = \left(\frac{a(t_L)}{a(t)}\right)^3 n(\nu a(t)/ a(t_L), t_L) d(\nu a(t)/ a(t_L)),
\end{equation} 
where the dilution factor $\left(\frac{a(t_L)}{a(t)}\right)^3$ is due to the expansion of the universe.\\
We can simplify the above formula by using the black-body formula for the spectrum of the CMB, say Eq. (\ref{black-body}). It is easy to see the that the dependence on scale factor is canceled everywhere except in the exponential. This means that the expectrum remains the same but with a red-shifted temperature as, 
\begin{equation} 
\label{Rayleigh-Jeans}
T(t) = T(t_L) a(t_L)/a(t).
\end{equation}  
This confirms that the photon spectrum does not change over the expansion of the universe. \\
It is worth to mention that the above conclusion remains the same if we take into account the thickness of the last scattering while restrict to the elastic scattering between the photons and electrons which does not change the frequency of the photons \cite{Weinberg:2008zzc}.
\section{Summary}
In this chapter, we reviewed the standard model of cosmology briefly. We started with the kinematics and Dynamics of the universe and presented a brief review of the thermal history of the universe. Then, we showed that there are some problems in the standard model of cosmology and introduced the inflationary cosmology to solve these problems.  The simplest modes of inflation are based on a scalar field coupled minimally to gravity. The scalar field potential is flat enough to allow for about 60 e-foldings or so to solve the flatness and the horizon problems of the standard big-bang cosmology. Finally, we reviewed the cosmic microwave background with a brief history of its discovery as well as some specific features of the CMB.


\chapter{Cosmological Perturbation Theory And Inflationary Universe } 

\label{Chapter2} 

\lhead{Chapter 2. \emph{Cosmological Perturbation Theory}} 

\vspace{0.5cm}
\hrule \vspace{0.3cm}

\begin{quote}
\textbf{Abstract:} The goal of this chapter is to review the cosmological perturbation theory. After presenting this theory, we use it to understand the behavior of the perturbations during the epoch of inflation. We then calculate the power spectra of both scalar
and tensor fluctuations during inflation. Finally, we relate these results to observations of the cosmic
microwave background (CMB) radiation. By making this correspondence explicit, we can then put some constraints on inflationary predictions. 
\end{quote}

\vspace{0.1cm}  \hrule
\vspace{0.5cm}

\section{Review of The Cosmological Perturbation Theory}
\textbf{\textit{Overview:}} One of the central goals of the modern Cosmology is to explain the origin of primordial perturbations which are thought to be the seeds for the current structures in the universe. Before proposing the Inflationary Cosmology, the initial perturbations were taken for granted and their spectrum was fixed to be consistent with the observations. So the aim of the old fashioned cosmology was \textit{describing} the current universe by arranging the appropriate initial conditions. \\
Inflationary cosmology, on the other hand, tries to truly \textit{explain} the origin of the primordial perturbations and \textit{predict} precisely the spectrum. This means that the contemporary cosmology is falsifiable since we can test this theory by looking for its predictions and compare them with observations. \\
According to cosmic inflation, primordial perturbations originated from the quantum fluctuations. Then, during inflation, they are stretched to super horizon scales. Thus, inflation links the large scale structures of the universe to its microphysics origin. In addition, the resulting spectrum of these inhomogeneities seems not to be so much sensitive to the details of any particular model and has a nearly scale invariant shape. As a results, we would have a concrete predictions for the spectrum of the cosmic Microwave Background anisotropy.  \\
In the following, we will review the cosmological perturbation theory in details. We mainly follow the notation of \cite{Mukhanov, Weinberg:2008zzc, Baumann:2009ds, Gorbunov:2011zzc, Gorbunov:2011zz, Ruth Durrer}.
\subsection{Classification of Perturbations}
\textit{Linear Perturbation Theory:} According to the observations of the cosmic microwave background, the primordial universe was not completely homogeneous and it contained small inhomogeneities. However, since they were quite small, $\frac{\delta \rho}{\rho}\simeq 10^{-5}$, it is a good approximation to analyze them up to linear order in perturbations. This can be done by splitting every quantity, $Q(t, \mathbf{x})$, into a homogeneous part,$Q(t)$, which depends only on the cosmic time and a perturbation part, $\delta Q(t, \mathbf{x})$, which depends on both time and space, 
\begin{equation}
\label{linear-perturbation}
Q(t, \mathbf{x}) \equiv Q(t) + \delta Q(t, \mathbf{x})
\end{equation}
By using this strategy, we can also calculate the linearized Einstein equations as we will see in the following. \\
According to the Einstein equations, inhomogeneities in the matter distribution induce the metric perturbations and vice versa. These perturbations can be decomposed into the Scalar, Vector and Tensor parts and in the linear approximation, different types of perturbations evolve independently and can be analyzed separately. Here we classify the perturbations. \\ 
\textbf{\textit{Metric Perturbations:}} The metric of a flat homogeneous and isotropic FRW universe with tiny perturbations can be presented as, 
\begin{eqnarray}
\label{perturbed metric}  
ds^2 &=& g_{\mu \nu} dx^{\mu} dx^{\nu} \nonumber\\
&=& -(1+2\phi)dt^2 +2a B_{i} dx^i dt + a^2 \bigg{[} (1-2\psi)\delta_{ij} + E_{ij}\bigg{]} dx^i dx^j
\end{eqnarray}
where $\frac{\partial}{\partial x_i}$ denotes the partial spatial derivative and according to the scalar-vector-tensor(SVT) decomposition, we can represent $B_i$ and $E_{ij}$ as, 
\begin{equation}
B_i \equiv \partial_{i}B - S_{i}, ~~~~ \partial^i S_i = 0
\end{equation}
and, 
\begin{equation}
E_{ij} \equiv 2\partial_{ij}E +2 \partial_{(i}F_{j)} + h_{ij}, ~~~ \partial^i F_i = 0,~~~h^{i}_{i} = \partial^i h_{ij} =0
\end{equation}
Here $\phi$, $\psi$, $B_i$ and $E_{ij}$ are called the \textit{lapse} function, \textit{spatial curvature}, \textit{shift} vector and \textit{shear} tensor respectively. \\
In this way, we have decomposed the whole metric into scalar, vector and tensor parts. \\
\textbf{\textit{Matter Perturbations:}} 
Energy-Momentum tensor of a perfect fluid can be described by the following parameters, $\rho$, $p$, $u^{\mu}$ and $\Sigma^{\mu \nu}$ which refer to the energy density, pressure, the four velocity and the anisotropic stress, respectively. \\
In the following, we present the linear perturbations in each of the above quantities. \\
Let us start with the fluctuations in energy density and pressure. They can be defined in the following way, 
\begin{equation}
\label{delta rho-p}
\delta \rho(t, x^i) \equiv \rho(t,x^i) - \bar \rho(t) ~~~~,~~~~ \delta p(t, x^i) \equiv p(t, x^i) - \bar p(t)
\end{equation}
where the quantities with bar denote the background values. \\
Next, we consider the fluctuations in the four velocity. In this case, taking into account the fact that  $g_{\mu \nu} u^{\mu} u^{\nu} = -1$, the number of the independent components in $u^{\mu}$ is reduced by 1. So we have,
\begin{equation}
\label{definition of u}
u_\mu \equiv (-1-\phi, v_i) ~~~~,~~~~u^\mu \equiv (1 -\phi, v^i + B^i),
\end{equation}
where we have defined $u_{0}$ in such a way that the above constraint can be satisfied. \\
Finally, we are left with analyzing the anisotropic stress tensor. In this case, there are two crucial points that must be taken into account,\\
\textit{First}, since in the homogeneous and isotropic FRW universe the anisotropic stress tensor is zero, for this metric it is a perturbative quantity by its own. \\
\textit{Second} There are some features in $\Sigma^{\mu \nu}$ which can be used to 
reduce the number of independent components of this tensor. \\
They are given by, \\
$\bullet$ \textit{Transvelity with respect to $u^{\mu}$:}
\begin{equation}
\label{constraint1}
\Sigma^{\mu \nu} u_\nu = 0
\end{equation}
This means that $\Sigma^{00} = \Sigma^{0i} =0$. \\
$\bullet$ \textit{Traceless condition:}
\begin{equation}
\label{constraint2}
\Sigma^{\mu}_{\mu} = 0
\end{equation}
Combining Eqs. (\ref{constraint1}) and (\ref{constraint2}) leads to $\Sigma^i_i =0$.\\
So at the end of the day, the anisotropic stress is reduced to a symmetric, traceless three-tensor. \\
Considering the whole of the above features, we can finally present the energy-momentum tensor as,
\begin{eqnarray}
\label{perturbed energy-momentum}
T^0_0 &=& - (\bar \rho + \delta \rho) \nonumber\\
T^0_i &=& (\bar \rho +\bar p) v_i \nonumber\\
T^i_0 &=& -(\bar \rho + \bar p) (v^i + B^i) \nonumber\\
T^i_j &=& \delta^i_j (\bar p + \delta p) + \Sigma^i_j
\end{eqnarray}
Sa far, we have only considered one fluid. we can easily generalized the above picture to the case where we would have different types of fluids. In this case the total energy-momentum tensor is the sum of different components, 
\begin{equation}
\label{total energy}
T_{\mu \nu} = \sum_I T^I_{\mu \nu}.
\end{equation}
Translating this in terms of the different components of the energy-momentum tensor, we would have, 
\begin{eqnarray}
\label{perturbed adding}
\delta \rho &=& \sum_I \delta \rho_I \nonumber\\
\delta p &=& \sum_I \delta p_I \nonumber\\
(\bar \rho + \bar p) v^i &=& \sum_I (\bar \rho_I + \bar p_I) v_I^i \nonumber\\
\Sigma^{ij} &=& \sum_I \Sigma_I^{ij}
\end{eqnarray}
Form the above equation, we see that although the total energy density, pressure and the anisotropic stress tensor is the sum of different components, it is not the case for the four velocity. So it is convenient to define the three momentum density as, 
\begin{equation}
\label{momentum density}
\delta q^i \equiv (\bar \rho + \bar p) v^i,
\end{equation}
so in terms of the three momentum we would have, 
\begin{equation}
\label{adding moment}
\delta q^i = \sum_I \delta q^i_I
\end{equation}
\subsection{Gauge Transformations}
Let us consider the following coordinate transformation, 
\begin{equation}
\label{gaugetransformation}
x^{\alpha} \longrightarrow \tilde{x}^{\alpha} = x^{\alpha} + \xi^{\alpha}
\end{equation}
where $\xi^{\alpha}$ are infinitely small functions of the time and space. In the following, we split every functions into a time dependent background as well as the perturbation part and would read off the corresponding transformation law for different components. \\
$\bullet$ Scalar Perturbation: Consider a 4-scalar as,
\begin{equation}
q(x^{\alpha}) =  q^{(0)}(x^{\alpha}) + \delta q,
\end{equation}
where $q^{(0)}(x^{\alpha})$ refers to the background value of $q$, while $\delta q$ denotes the perturbation. We can easily verify that under the above gauge transformation, the perturbation $\delta q$ transforms as, 
\begin{equation}
\delta q \longrightarrow \delta \tilde{q} = \delta q - q^{(0)}_{,\alpha} \xi^{\alpha}
\end{equation}
$\bullet$ Vector Perturbation: Next, let us consider a 4-vector $V^{\alpha}$ as, 
\begin{equation}
V_{\mu}(x^{\alpha}) =  V_{\mu(0)}(x^{\alpha}) + \delta V_{\mu}
\end{equation}
we can show that under the above gauge transformation, $\delta V_{\mu}$ transforms as, 
\begin{equation}
\delta V_{\mu} \longrightarrow \delta \tilde{V}_{\mu} = \delta V_{\mu} - V^{(0)}_{\mu, \nu} \xi^{\nu} - V^{(0)}_{\nu} \xi^{\nu}_{, \mu}
\end{equation}
$\bullet$ Tensor Perturbation: Finally, let us consider the tensor, $g_{\mu \nu}$ field, as, 
\begin{equation}
g_{\mu \nu} = g^{(0)}_{\mu \nu} + \delta g_{\mu \nu}
\end{equation}
it is straightforward to show that under the gauge transformations, the tensor field transforms as, 
\begin{equation}
\label{gaugetensor}
\delta g_{\mu \nu} \longrightarrow \delta \tilde{g}_{\mu \nu}= \delta g_{\mu \nu} - g^{(0)}_{\mu \nu, \alpha} \xi^{\alpha} - g_{\alpha \nu}^{(0)} \xi^{\alpha}_{, \mu} - g_{\alpha \mu}^{(0)} \xi^{\alpha}_{,\nu}
\end{equation}
Just for the later references, it is convenient to rewrite $\xi^{\alpha}$ as, 
\begin{equation}
\label{decomposition xi}
\xi^{i} = \xi^{i}_{\perp} + \zeta^{,i}
\end{equation}
where $\xi^{i}_{\perp} $ is a divergence free 3-vector while $\zeta$ is a scalar function. \\
\subsection{Gauge Invariant/Physical Observables}
So far we have seen that under the general coordinate transformation, or gauge transformations, both perturbations change. As a result, in order to avoid the pitfall of fictitious gauge modes, it is extremely important to introduce the gauge invariant combinations of the metric and matter perturbations. Having this said, in the following, first of all, we find the transformation rules for the metric and matter components and then try to figure out which combinations of these fields give rise to a gauge invariant quantity. \\
Since at the linear order Scalar, Vector and Tensor perturbations are decoupled from each other, it is convenient to consider each type separately.\\
{\color{blue}{\textbf{\textit{Scalar Perturbations:}}}} \\
There are two different sources for the scalar perturbations. One comes from the scalar components of the metric, while the other come from the scalar components of the matter field perturbations. \\
Let us start with the \textit{metric} components. \\
The scalar part of the metric perturbation are defined as in Eq. (\ref{perturbed metric}).\\
Then, under the following gauge transformations, 
\begin{equation}
\label{gauge transformation}
t \longrightarrow t + \delta t ~~~,~~~
x^{i} \longrightarrow x^{i}+\delta^{ij} \partial_{j} \delta x
\end{equation}
we have, 
\begin{eqnarray}
\label{scalar transformation}
\phi &\longrightarrow & \phi - \dot{\delta}t \nonumber\\
B &\longrightarrow & B+a^{-1}\delta t - a\dot{\delta}x \nonumber\\
E &\longrightarrow & E-\delta x \nonumber\\
\psi &\longrightarrow & \psi+ H\delta t 
\end{eqnarray}
Combining the above equations, we can easily get different gauge invariant quantities as,  
\begin{equation}
\label{gauge in-phi}  
\Phi \equiv \phi - \frac{d}{dt} \left[ a^2\left(\dot{E}- B/a\right)  \right] 
\end{equation}
\begin{equation}
\label{gauge in-si}  
\Psi \equiv \psi +  a^2 H \left(\dot{E}- B/a\right) 
\end{equation}
We can easily check that under the above gauge transformations both of $\Phi$ and $\Psi$ are invariant. This means that if in one frame either  $\Phi$ or $\Psi$ are equal to zero, they will remain zero at any other physical frame. \\
Next, let us consider the gauge transformations of the \textit{matter} fields. The most important matter field, according to the Einstein equations, are the Energy-Momentum tensor. So it worth to consider it in what follows. Considering Eq. (\ref{perturbed energy-momentum}) and by using Eqs. (\ref{gaugetransformation}) and (\ref{gaugetensor}), we can get the transformation law for different components of this tensor as, 
\begin{eqnarray}
\label{energy-momentum transformation}
\delta\rho &\longrightarrow & \delta\rho - \dot{\bar{\rho}}\delta t \nonumber\\
\delta p &\longrightarrow & \delta p - \dot{\bar{p}}\delta t \nonumber\\
\delta q &\longrightarrow & \delta q + (\dot{\bar{\rho}} + \dot{\bar{p}})\delta t 
\end{eqnarray}
However, note that the anisotropic stress is gauge invariant. \\
Combining the above transformations, we can find the following gauge invariant quantities, \\
{\color{red}{$\bullet$}} \textit{Curvature Perturbation on Uniform-Density Hypersurfaces:} Geometrically, $\zeta$ measures the spatial curvature of the constant-density hypersurfaces. It has the following definition, 
\begin{equation}
\label{perturbation curvature} 
-\zeta \equiv \psi + \frac{H}{\dot{\rho}} \delta \rho
\end{equation}
{\color{red}{$\bullet$}} \textit{Entropic Perturbation:} This is given by the non-adiabatic part of the pressure and has the following definition, 
\begin{eqnarray}
\label{Entropy pressure} 
\delta p_{nad} &=& \delta p - \delta p_{ad} \nonumber\\
&=&  \delta p - \frac{\dot{p}}{\dot{\rho}} \delta \rho
\end{eqnarray} 
By using the above transformations, we can easily show that this is a gauge invariant quantity. \\
One very important properties of $\zeta$ is that for the adiabatic matters, where $\delta p_{nad} =0$,  it remains constant outside the horizon.\\
{\color{red}{$\bullet$}} \textit{Comoving Density:} This is given by the difference between the energy density and the scalar part of the 3-momentum, $q$, as,
\begin{equation}
\label{density Comoving} 
\delta \rho_{m} = \delta \rho - 3H \delta q
\end{equation}
{\color{red}{$\bullet$}} \textit{Comoving Curvature Perturbations:} This is given by the following expression, 
\begin{equation}
\label{comoving curvature} 
\cal{R} \equiv \psi - \frac{H}{\rho+p} \delta q
\end{equation}
Geometrically, this is the curvature perturbations transverse to the comoving world lines. \\
{\color{red}{$\bullet$}} \textit{Mukhanov-Sasaki Variable:} This is the field perturbation in the flat gauge, i.e. $\psi =0$, and has the following definition, 
\begin{equation}
\label{mukhanov sasaki} 
\delta \varphi_{\psi}\equiv \delta \varphi + \frac{\dot{\varphi}}{H} \psi
\end{equation} 
so the SM variable is given by $a\delta \varphi_{\psi}$
Having presented the gauge invariant quantities, in order to get some intuition about the meaning of some of the above quantities, it is convenient to consider the single field inflation and calculate the Comoving Curvature perturbation in this model. \\
In this case, the 3-momentum is given by $\delta q = -\dot{\varphi} \delta \varphi$. So we obtain, 
\begin{equation}
\label{R of single field} 
R \equiv \psi + \frac{H}{\dot{\varphi}} \delta \varphi
\end{equation}
In addition, in this model we have, 
\begin{equation}
\label{rho and phi} 
\delta \rho / \dot{\rho} \simeq \delta \varphi  / \dot{\varphi}
\end{equation}
This means that in this model, $\zeta$ and $R$ are equal to each other. \\
In addition, for the single field inflation, we can easily see that the Mukhanov-Sasaki variable is indeed a re-parametrization of the Comoving Curvature perturbation. \\
\begin{eqnarray}
\label{rewriting} 
\delta \varphi_{\psi} &\equiv & \delta \varphi + \frac{\dot{\varphi}}{H} \psi  \nonumber\\
& = &\frac{\dot{\varphi}}{H} \left( \psi - \frac{H}{\rho + p} \delta q \right) \nonumber\\ 
& = &\frac{\dot{\varphi}}{H} R    
\end{eqnarray}
So we see that the matter perturbation in one gauge looks like the metric perturbation in another gauge and vice versa. \\
{\color{blue}{\textbf{\textit{Vector Perturbations:}}}} \\
In general there are two different sources for the vector perturbations. However, since in this chapter, we are only considering the perturbations around a FRW background, we can safely assume that there is not any pure vectors in the matter sector and thus we are only left with the vector perturbations in the metric sector. We relax this assumption in the next chapters when we consider the anisotropic universe. \\
With this simplification in mind, here we present how the vectorial part of the metric is transformed under the gauge transformations. \\
The vectorial part of the metric perturbation is given by, 
\begin{equation}
\label{equ:Vmetric}
ds^2 = - d t^2 - 2 a(t) S_{i} d x^i d t +  a^2(t) [\delta_{ij}+2 F_{(i, j)} ] d x^i d x^j
\end{equation}
where $S_i$ and $F_i$ are divergence free vectors. \\
Then under the following gauge transformation,
\begin{equation}
\label{vector transformation} 
x^{i} \longrightarrow x^{i}+ \beta^{i} ~~~,~~~ \beta_{i,i}=0
\end{equation}
we have,
\begin{eqnarray}
\label{vector metric1 per} 
S_{i} \longrightarrow  S_{i} + a \dot{\beta_{i}} \\
F_{i} \longrightarrow  F_{i} - \beta_{i} , \label{vector metric2 per} 
\end{eqnarray}
Therefore the gauge invariant combination of these functions is, $\dot{F_{i}} +S_{i}/a$. This means that, among 4 degrees of freedom for these two vectors, there are only two independent components.  \\
{\color{blue}{\textbf{\textit{Tensor Perturbations:}}}} \\
The tensorial part of the metric is given by, 
\begin{equation}
\label{equ:Tmetric}
d s^2 = - d t^2 + a^2(t)[\delta_{ij} +  h_{ ij}] d x^i d x^j,
\end{equation}
where $h_{ij}$ are transverse, i.e. $h_{ij, i}=0$, and traceless, i.e. $h^i_{i}=0$. Tensor perturbations are gauge invariant. 
\subsection{Equations for Cosmological Perturbations}
Having presented the gauge transformations for both of the matter and metric perturbations, here we try to connect these two different pieces dynamically. This can be done by using the Einstein field equations. \\
We treat scalar, vector and tensor perturbations separately. 
\subsubsection{Scalar Perturbations}
As we already discussed, by using the Einstein field equations, we can correlate the metric perturbations with the matter field perturbations. Doing this, we have the following equations for the scalar perturbations, 
\begin{eqnarray}
\label{density} 
3H\left(\dot{\psi}+H\phi \right)+ \frac{k^2}{a^2}\left[ \psi + H\left( a^2 \dot{E} - aB \right] \right) &=& -4\pi G \delta \rho \\
\label{momentum} 
\dot{\psi}+H\phi &=& -4\pi G \delta q
\end{eqnarray}
Combining Eqs. (\ref{density}) and(\ref{momentum}), we would have, 
\begin{equation}
\label{poisson} 
\frac{k^2}{a^2}\Psi = -4\pi G \delta \rho_{m},
\end{equation}
which is a gauge invariant generalization of the Poisson equation. \\
In addition, the spatial parts of the Einstein Field equations yields the following equations, 
\begin{eqnarray}
\label{i and i} 
\ddot{\psi} + 3H \dot{\psi} + H \dot{\phi} + \left( 3H^2 + 2 \dot{H} \right) \phi &=& 4\pi G \left( \delta p -\frac{2}{3}k^2 \delta \Sigma \right) \\
\label{i and j} 
\left(\dot{E} - B/a \right) ^{.} +3H \left(\dot{E} - B/a \right) + \frac{\psi - \phi}{a^2} &=& 8\pi G \delta \Sigma
\end{eqnarray}
where the anisotropic stress is given by, 
\begin{equation}
\delta \Sigma _{ij} = \left[  \partial _{i} \partial_{j} + (k^2/3) \delta_{ij} \right] \Sigma.
\end{equation}
It is useful to write down Eq. (\ref{i and j}) in the \textit{longitudinal gauge}, i.e. $E = B = 0$, 
\begin{equation}
\label{i and j- long} 
\Psi - \Phi = 8\pi G a^2 \delta \Sigma.
\end{equation}
As a result, we see that in the absence of the anisotropic stress, we have, $\Psi = \Phi$. \\
In addition, by using the covariant conservation of the Energy-Momentum tensor, i.e. $\bigtriangledown_{\mu} T^{\mu\nu} =0$, we can further connect different perturbations as, 
\begin{eqnarray}
\label{continuity} 
\delta \dot{\rho} + 3H \left( \delta\rho + \delta p \right) &=& \frac{k^2}{a^2}\delta q + \left(\rho +p \right) \left[ 3 \dot{\psi} + k^2 \left(\dot{E} + B/a \right)  \right] \\
\label{Euler}
\delta \dot{q} + 3H \delta q &=& -\delta p + \frac{2}{3} k^2 \delta \Sigma - \left(\rho + p \right) \phi
\end{eqnarray}
This gives us the continuity and Euler equations respectively. \\
We can now rewrite Eq. (\ref{continuity}) in terms of the curvature perturbation, $\zeta$, as
\begin{equation}
\label{continuity 2} 
\dot{\zeta}= -H \frac{\delta p_{nad}}{\rho + p}-\Pi,
\end{equation}
where, as we mentioned before, $\delta p_{nad}$ is the non-adiabatic pressure and $\Pi$ is the scalar shear along the comoving world-lines, \cite{Lyth:2003im}, and can be measured with respect to the Hubble rate as,
\begin{eqnarray}
\label{sigma reletive H} 
\frac{\Pi}{H}&=& -\frac{k^2}{3H}\left\lbrace \dot{E} - (B/a)+ \frac{\delta q}{a^2(\rho + p)}\right\rbrace  \nonumber\\ &=& -\frac{k^2}{3 a^2 H^2}\zeta - -\frac{k^2 \Psi}{3 a^2H^2}\left[ 1-\frac{2\rho}{9(\rho + p)}\frac{k^2}{a^2H^2}\right].  
\end{eqnarray}
So far we have not considered any specific realization of the scalar field. Here we present a scalar field Lagrangian and would try to simplify the above equations for this case. \\
So let us consider the following scalar field Lagrangian. 
\begin{equation}
{L} = -\frac{1}{2} g^{\mu\nu}\varphi_{,\mu}\varphi_{,\nu} -V(\varphi)
\label{lag}
\end{equation}
Using the above Lagrangian, we can easily find the energy density, pressure and the momentum density as, 
\begin{eqnarray}
\delta\rho &=& \left[
\dot\varphi \left( \dot{\delta\varphi} -\dot\varphi \phi \right)
+ V_{\varphi}\delta\varphi \right] 
\label{eq:densityphi} \\
\delta p &=& \left[
\dot\varphi \left( \dot{\delta\varphi}-\dot\varphi \phi \right)
- V_{\varphi}\delta\varphi \right] 
\label{eq:pressurephi} \\
\delta q_{,i} &=& - \sum_I \dot{\varphi}
\delta\varphi_{,i}
\label{eq:mtmphi}
\end{eqnarray}
Where we have defined, $V_{\varphi} \equiv \partial V/\partial \varphi$. \\
Using the above equations, we can calculate the gauge invariant comoving energy density as, 
\begin{eqnarray}
\label{def:rhom}
\delta\rho_m = \left[ \dot\varphi \left(
\dot{\delta\varphi} -
\dot\varphi \phi \right) - \ddot\varphi \delta\varphi.
\right]
\end{eqnarray}
In addition, we can show that for the single field inflation the non-adiabatic perturbation is proportional to the comoving energy density, 
\begin{equation}
\delta p_{nad} = - \frac{2V_{,\varphi}}{3H\dot\varphi}
\delta\rho_m.
\end{equation}
On the other hand, according to Eq. (\ref{poisson}), since for the finite value of the $\Psi$, the comoving energy density is zero on the super-horizon scales, we can conclude that the perturbations are adiabatic in this limit. \\
Putting all of these different pieces together, we can find the equation of motion for the scalar perturbations as,  
\begin{eqnarray}
\label{scalarperturbation}
\ddot{\delta\varphi} + 3H\dot{\delta\varphi}
+ \frac{k^2}{a^2} \delta\varphi + V_{\varphi\varphi}
\delta\varphi =
-2V_{\varphi} \phi+ \dot\varphi \left[ \dot{\phi} + 3\dot{\psi} +
\frac{k^2}{a^2} (a^2\dot{E}-aB) \right]. \label{eq2:perturbation}
\end{eqnarray}
\subsubsection{Vector Perturbations}
Next, let us consider the vector perturbations. The vectorial part of the anisotropic stress is given by, 
\begin{equation}
\delta \Sigma_{ij} = \partial_{(i} \Sigma_{j)}~~~,~~~ \Sigma_{i,i} =0
\end{equation}
In this case, there are only two Einstein equations, 
\begin{eqnarray}
\label{vector a}
\dot{\delta q_i} + 3 H \delta q_i &=& k^2 \delta \Sigma_i   \\
\label{vector b}
k^2 (\dot {F_i} + S_i/a) & =& 16 \pi G \, \delta q_i
\end{eqnarray}
So in the absence of the anisotropic stress tensor, $\delta q_i$ is diluted by the expansion of the universe. This means that the guage invariant combination, $\dot {F_i} + S_i/a$, is also diluted and there would not remain any dominant contribution in the vector part. 
\subsubsection{Tensor Perturbations}
In this case, the Anisotropic stress plays the role of the source. however, in order to simplify the analysis, in the following, we neglect this part. So the equation of motion for the tensor part would be given by,   
\begin{equation}
\label{tensor3}
\ddot h + 3H \dot h + \frac{k^2}{a^2} h = 0.
\end{equation}
Eq. (\ref{tensor3}) is the only equations which governs the evolution of the tensor components and describes the evolution of the tensor mode in an expanding universe. 
\subsection{Primordial Spectrum During Inflation}
Having presented the perturbations, here we define the power spectrum of this perturbations, which as we will see in the following, can be fixed by the observations. \\
The power spectrum for the scalar perturbations is given by, 
\begin{equation}
\label{eq:defPR}
{\cal P}_R \equiv \frac{4\pi k^3}{(2\pi)^3} |R|^2
\end{equation}
In the same way, the power spectrum for the tensor modes is also given by, 
\begin{equation}
\label{eq:defPT}
{\cal P}_{ T} \equiv 2
\frac{4\pi k^3}{(2\pi)^3} |h|^2,
\end{equation}
where the additional factor 2 is due to the polarization of the gravitational waves. \\
\subsubsection{Scale-Dependence:}
The scale-dependence of the spectrum comes from the time-dependence of the Hubble parameter and is measured by the spectral indices defined via, 
\begin{equation}
\label{eq:defnR}
n_{R} - 1 \equiv \frac{{d}\ln {\cal P}_{R}}
{{d}\ln k} \biggr|_{k=aH}
\end{equation}
If $n_{R} =1$, the spectrum is called to be scale-invariant. \\
In a similar way, we can also define the spectral indices for the tensor mode as, 
\begin{equation}
\label{eq:defnT}
n_T \equiv \frac{{d}\ln {\cal P}_T}
{{d}\ln k} \biggr|_{k=aH}
\end{equation}
In this case, the spectrum is scale-invariant when $n_{T} = 0$. 
\section{The Evolution of the Perturbations during Inflation}
In this section, we focus on the evolution of the perturbations during inflation. We consider the scalar and tensor perturbations respectively. \\ 
{\color{blue}{\textbf{\textit{Scalar Perturbations:}}}} \\
Let us start with the scalar perturbations. 
In this case, we write down the equations which govern the evolution of the scalar fields and then by solving them figure out their behavior. However, in order to get some intuition about what is going on here, it is extremely useful if we start with the qualitative behavior of the perturbations and then try to proceed by considering the details of the analysis. 
\subsection{Qualitative Behavior of the Perturbations}
Here we present the qualitative behavior of the perturbations. Let us start with sub-horizon perturbations. In this case, the perturbations have quantum origin. So in order to make a very rough estimation for the amplitude of these quantum fluctuations on the physical scales $L$, $\delta \phi_L$, we can consider a finite volume, $L^3$, and assume that the field is nearly homogeneous within this region. For simplicity we can also neglect the mass of the field. So we would have the following expression for the action of this system, 
\begin{equation}
\label{action}
S \simeq \frac{1}{2} \int \left( \dot{X}^2 + ... \right) dt ~~~~~~,~~~~~~ X \equiv \delta \phi_L L^{3/2}
\end{equation}
where a dot denotes the derivative with respect to the physical time. Clearly, $X$ plays the role of the canonical quantization variable. So its corresponding conjugated momentum, $P$, is defined as, $P \equiv \dot{X} \simeq X/L$. \\
In addition, $X$ and $P$ also satisfy uncertainty relation, $\Delta X \Delta P \simeq 1$. Combining this with the above relation between the $X$ and $P$, we get $X \simeq \sqrt{L}$ or $\delta \phi_L \simeq L^{-1}$. This means that the amplitude of the massless scalar field is inversely proportional to the physical scales. \\
On the other hand, using the relation between the real space and the Fourier space quantum fluctuation, $\delta \phi_L \simeq |\delta \phi_k| k^{3/2}$, we can get the amplitude of the quantum fluctuations in the Fourier space as, 
\begin{eqnarray}
|\delta \phi_k|&\simeq& L^{-1} k^{-3/2} \\
&=& k^{-1/2}/a,
\end{eqnarray}  
where we have taken into account that the comoving wavenumber is given by, $k = a/L$. \\
So the typical size of the perturbation is given by, 
\begin{equation}
\delta_{\phi} \equiv |\delta \phi_k| k^{3/2} = k/a.
\end{equation}
This means that on scales less than the curvature scale, $\simeq H^{-1}$, the quantum fluctuations do not feel the curvature effect and they just oscillate and their amplitude decreases proportional to the inverse of the scale factor. However, as soon as the scale factor becomes of order $a_k \simeq k/H$, i.e. the mode exits the horizon, the curvature effects will be important and after that, as we will show in the following, the modes will be frozen. In this way, inflationary universe stretches the quantum fluctuations from a very short wavelength inside the Hubble patch into the cosmological scales. We should emphasize here that this can not happen for the decelerating universe, taking into account that for that case, the growth of the curvature scale is faster than the growth of the physical mode and as a result it is not possible that a typical mode cross the horizon and becomes super-horizon. So in this case, the amplitude of the perturbations decay continuously and eventually become negligible. A qualitative behavior of the perturbations is shown in Fig. \ref{fig:evolution}
\begin{figure}[t!]
	\centering
	\includegraphics[width=0.8\textwidth]{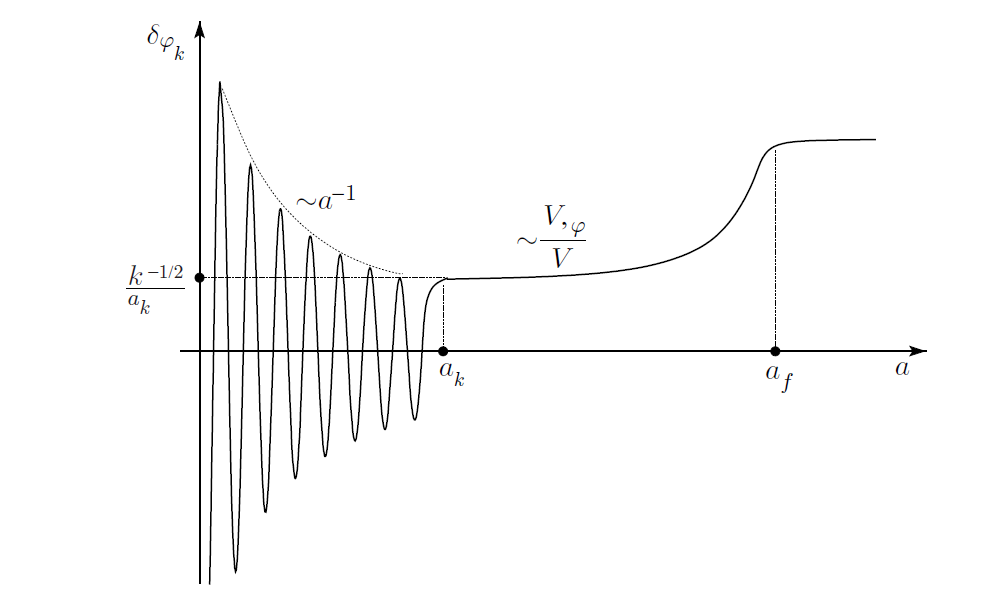}
	\label{fig:evolution}
	\caption{\label{fig:evolution} The evolution of the perturbations. While they are inside the horizon, they oscillate and decay proportional to the inverse of the scale factor. Right after horizon exit, their amplitude is frozen until the time of the horizon re-entry. 
	\cite{Mukhanov}}
\end{figure}
\subsection{Quantitative Behavior of the perturbations}
Having presented the qualitative behavior of the scalar perturbations during inflation, here we try to find the quantitative behavior of these perturbations. \\
Let us start with the equations of motion for the scalar perturbations in the flat gauge, $\psi =0$. In this gauge, Eq. (\ref{scalarperturbation}) can be written as, 
\begin{equation}
\label{eq:singlescalareom}
\ddot{\delta\varphi}_\psi + 3H\dot{\delta\varphi}_\psi
+ \left[ \frac{k^2}{a^2} + V_{\varphi \varphi}
- \frac{8\pi G}{a^3} \frac{{d}}{dt}
\left( \frac{a^3\dot\varphi^2}{H} \right)
\right] \delta\varphi_\psi = 0.
\end{equation}
Now in order to further simplify the above equation, we can define the following variables,
\begin{eqnarray}
v \equiv a{\delta\varphi}_\psi ~~~~~~~~~,~~~~~~~~~
z \equiv a\dot\varphi/H
\end{eqnarray}
Using the above variables, we can rewrite Eq. (\ref{eq:singlescalareom}) as, 
\begin{equation}
\label{veq}
v''+\left(k^2-\frac{z''}{z}\right)v=0,
\end{equation}
where a prime denotes the derivative with respect to conformal time. \\
We can also try to rewrite the mass term in terms of the slow-roll variables as,
\begin{equation}
\frac{z''}{z}=
(aH)^2 \left[
2+ 5\epsilon - 3\eta +9\epsilon^2-7\epsilon\eta +\eta^2 +\xi^2
\right], 
\label{gra}
\end{equation}
where we recall the definition of the slow-roll parameters as,
\begin{eqnarray}
\epsilon \equiv -\frac{\dot{H}}{H^2}\,,~~
\eta \equiv 2 \epsilon-\frac{\dot{\epsilon}}{2H\epsilon} \,,~~
\xi^2 \equiv \left(2\epsilon-\frac{\dot\eta}{H\eta}\right)\eta. 
\label{ep}
\end{eqnarray}
Now in order to proceed, we can neglect the time dependence of the slow-roll parameters. With this in mind, we would have,  
\begin{equation}
\tau \simeq -\frac{1}{(1-\epsilon)aH},
\end{equation}
as well as, 
\begin{equation}
\label{gra2}
\frac{z''}{z} = \frac{\nu_R^2-(1/4)}{\tau^2} ,~~~~~~
\nu_R \simeq \frac32 + 3\epsilon - \eta.
\end{equation}
Plugging these expressions back into Eq. (\ref{veq}), we can easily find the solution of this differential equation as, 
\begin{eqnarray}
\label{han}
v \simeq \frac{\sqrt{\pi|\tau|}}{2} e^{i(1+2\nu_R)\pi/4}
\left[ c_1 H_{\nu_R}^{(1)}(k|\tau|) +c_2 H_{\nu_R}^{(2)}(k|\tau|)
\right].
\end{eqnarray}
There are still two free parameters that must be fixed in the above solution, i.e. $c_1$ and $c_2$. They can be fixed by assuming a preferred ansatz for the initial condition which we choose it to be the Baunch-Davis vacuum in the far past, $k\tau\to-\infty$, 
\begin{eqnarray}
\label{vacnorm}
v \to \frac{e^{-ik\tau}}{\sqrt{2k}}
\end{eqnarray}
This leads to $c_1 =1$ and $c_2 =0$.\\
So the power spectrum on the short scales, $k\gg aH$, would be, 
\begin{eqnarray}
{\cal P}_{\delta\varphi} &\equiv& \frac{4\pi k^3}{(2\pi)^3} \left|
\frac{v}{a} \right|^2 \nonumber\\
&=& \left( \frac{k}{2\pi a} \right)^2
\end{eqnarray}
In addition, for the large scales, $k\ll aH$, we would have, 
\begin{equation}
{\cal P}_{\delta\varphi} \simeq
\left( (1-\epsilon)
\frac{\Gamma(\nu_R)}{\Gamma(3/2)} \frac{H}{2\pi} \right)^2
\left( \frac{|k\tau|}{2} \right)^{3-2\nu_R},
\label{latetimes}
\end{equation}
where we have used the following asymptotic behavior of the Hankel functions in the limit $k\tau\to0$, 
\begin{equation}
H_{\nu}^{(1)} (k|\tau|) \to -(i/\pi) 
\Gamma(\nu) (k|\tau|/2)^{-\nu}.
\end{equation}
As a result, for a massless scalar field, we would get the following power spectrum in the De Sitter universe, 
\begin{equation}
{\cal P}_{\delta\varphi} \to
\left( \frac{H}{2\pi} \right)^2 ~~~,~~~\frac{k}{aH} \to 0 
\end{equation}
So far we were looking for the solution of Eq. (\ref{veq}) at the slow-roll approximation. This approximation works very well at the beginning of inflation. The reason is that at this moment, and for the sub-Hubble scales, $k^2 \gg \frac{z''}{z}$, and as a result the exact shape of the $\frac{z''}{z}$ is not important. So we expect that Eq. (\ref{latetimes}) would be valid up to some times after the horizon exit. However, at later times, we need to use a large-scale limit which can be easily derived in terms of the comoving curvature perturbation,$\mathcal{R}$, 
\begin{eqnarray}
\label{Req}
\frac{1}{a^3\epsilon} \frac{{d}}{dt}
\left(a^3 \epsilon \dot{\mathcal{R}}\right)
+\frac{k^2}{a^2}{\mathcal{R}}=0
\end{eqnarray}
So on large scale limit, we would have, 
\begin{eqnarray}
\label{Rsolu}
\mathcal{R} = C_{1}+C_{2} \int \frac{dt}{a^3\epsilon}
\end{eqnarray}
where $C_1$ and $C_2$ are the constants of the integration. \\
Since after the horizon exit the second term in Eq. (\ref{Rsolu}) would decay very fast, we are left with the constant term, $C_1$. By matching the power spectrum at the horizon crossing, we can easily find $C_1$ as, 
\begin{equation}
C_1 = \frac{H^2}{\sqrt{2k^3}\dot{\varphi}}
\end{equation}
So the power spectrum of the comoving curvature perturbation is, 
\begin{equation}
{\cal P}_\mathcal{R} = \left( \frac{H}{\dot\varphi} \right)^2 {\cal
	P}_{\delta\varphi}\simeq
\left( \frac{H^2}{2\pi\dot\varphi} \right)^2_{k=aH}
\end{equation}
We can rewrite the above equation in terms of the potential as well as its first derivative as, 
\begin{equation}
{\cal P}_\mathcal{R}
= \left( \frac{128\pi}{3M_{P}^6}\frac{V^3}{V_\varphi^2}
\right)_{k=aH}.
\label{AS2}
\end{equation}
We recall that the amplitude of the comoving curvature perturbation is constant outside the horizon on very large scale. This means that its amplitude at the first and second horizon crossing which happen during the inflation and very late time, either during the radiation or matter epoch, respectively would be the same. This feature is shown in Fig. \ref{fig:scales}.
\begin{figure}[t!]
	\centering
	\includegraphics[width=0.8\textwidth]{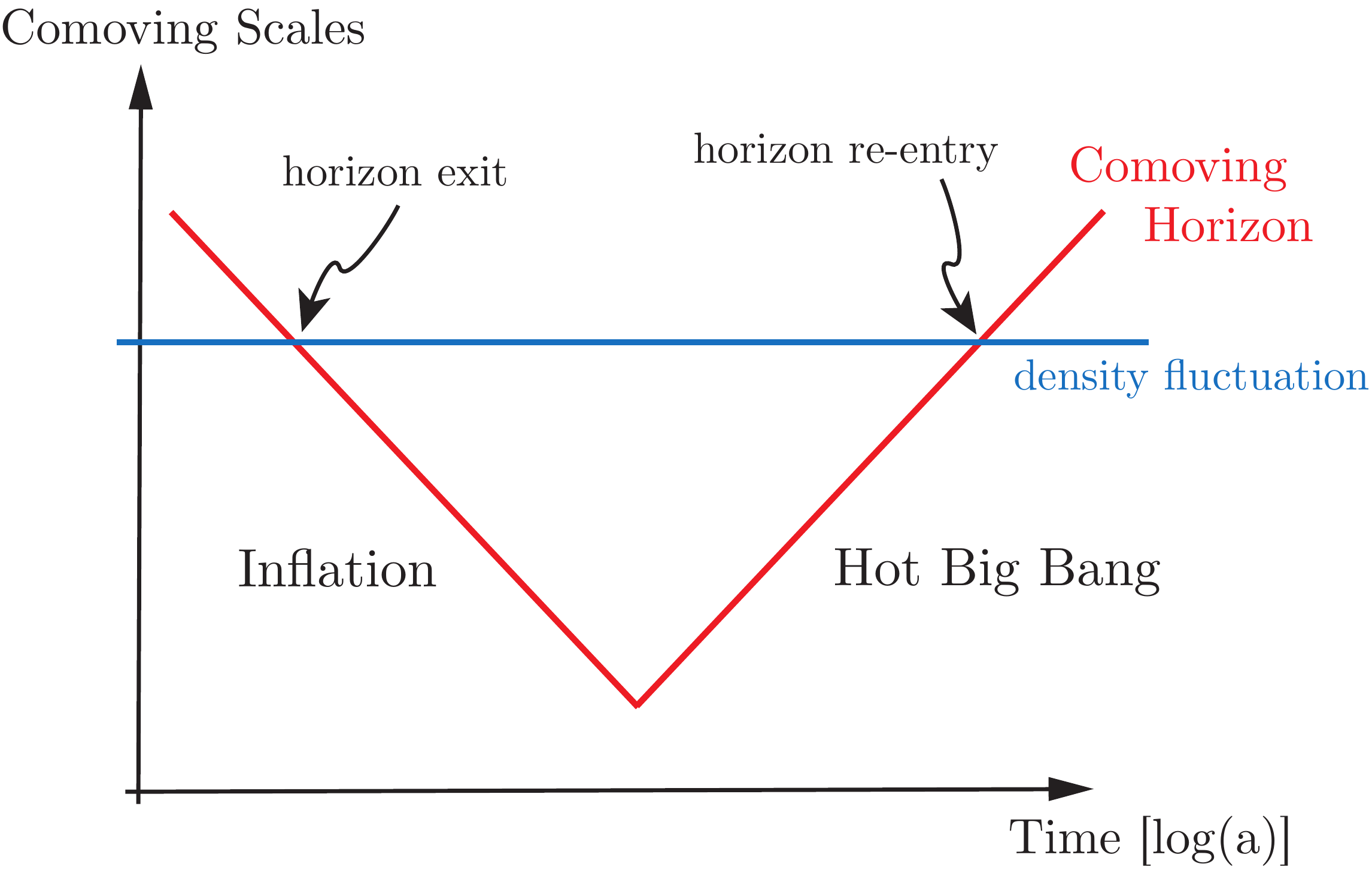}
	\label{fig:scales}
	\caption{\label{fig:scales} The creation and evolution of the perturbations during inflation. Fluctuations are created inside the horizon and due to shrinking the comoving Hubble would exit the horizon and would be frozen outside the horizon. Finally at some later time, either radiation or the matter dominance, they would re-enter the horizon again.
	\cite{Baumann:2009ds}}
\end{figure}
Having presented the form of the comoving curvature perturbation, we can now calculate the spectral indices for this model as,
\begin{eqnarray}
\label{index}
n_s-1 &=& 3-2\nu_R \nonumber\\
&=& -6\epsilon+2\eta
\end{eqnarray}
Since both of $\epsilon$ and $\eta$ are quite small during  inflation, we can conclude that the generated scalar perturbations during inflation are nearly scale invariant, $n_{s} \simeq 1$.\\
In addition, we can also define the running of the spectral index as,
\begin{eqnarray}
\alpha_{s} \equiv \frac{d n_s}{d\ln k}.
\biggr|_{k=aH}
\end{eqnarray}
Then by using the following expression, 
\begin{equation}
\frac{dx}{d\ln k}\biggr|_{k=aH}
=\left(\frac{dx}{dt}\right)
\left(\frac{dt}{d\ln a}\right)
\left(\frac{d\ln a}{d
	\ln k}\right)\biggr|_{k=aH}
= \frac{\dot{x}}{H} \biggr|_{k=aH}
\label{xderi}
\end{equation}
We can write $\alpha_{s}$ in terms of the slow-roll parameters as, 
\begin{eqnarray}
\alpha_s=16\epsilon \eta- 24\epsilon^2-2\xi^2.
\label{slow}
\end{eqnarray}
According to Eq. (\ref{slow}), since $\alpha_s$ is second order in terms of the slow-roll parameters, we expect that this would be quite small in the single field inflation. \\
{\color{blue}{\textbf{\textit{Tensor Perturbations:}}}} \\
Since the qualitative behavior of the tensor modes are very similar to the massless scalar fields, we skip the details and go directly to the main analysis for the tensor modes. 
We start from the main equation for the tensor mode, Eq. (\ref{tensor3}), and rewrite it in terms of the new variable, $u \equiv ah/2\sqrt{8\pi G}$, as
 \begin{equation}
 \label{tensor4}
 u'' + \left( k^2 - \frac{a''}{a} \right) u = 0.
 \end{equation}
 Comparing this equation with that for the scalar field, we see that they are very similar except that $z''/z$ has been replaced with $a''/a$,
 \begin{equation}
 \frac{a''}{a} = (aH)^2 (2-\epsilon).
 \end{equation}
Using the slow-roll approximation, we can simplify $a''/a$ as, 
\begin{eqnarray}
\frac{a''}{a} \simeq \frac{\nu_{\rm T}^2-(1/4)}
{\tau^2}~~~~,~~~~
\nu_T \simeq \frac32 + \epsilon
\end{eqnarray} 
We can now find the power-spectrum of the tensor mode on very large scale,$k\ll aH$, as
\begin{equation}
{\cal P}_{T}\simeq
\frac{64 \pi }{M_{P}^2} \left( (1-\epsilon)
\frac{\Gamma(\nu_T)}{\Gamma(3/2)} \frac{H}{2\pi} \right)^2
\left( \frac{|k\tau|}{2} \right)^{3-2\nu_T}
\label{PT}
\end{equation}
Quite similar to the scalar perturbations, we can also try to solve Eq. (\ref{tensor4}) on large scales, 
\begin{eqnarray}
h = D_{1} + D_{2} \int \frac{dt}{a^3}. 
\label{hsolu}
\end{eqnarray}
So again one of the terms decay very quickly and we are left with the constant solution. \\
As a result the power spectrum for the tensor mode would be, 
\begin{eqnarray}
\label{tenamplitude}
{\cal P}_{T} \simeq
\frac{64\pi}{m_{P}^2}
\left(\frac{H}{2\pi}\right)^2_{k=aH} \simeq
\frac{128}{3} \left( \frac{V}{m_{P}^4}
\right)_{k=aH}.
\end{eqnarray}
Having presented the evolution of the tensor mode, we can proceed with calculating the spectral index as, 
\begin{eqnarray}
n_{T}=-2\epsilon.
\label{nT}
\end{eqnarray}
In addition, the running of the spectral index is also given by, 
\begin{eqnarray}
\alpha_{T}=-4\epsilon(2\epsilon-\eta).
\label{alT}
\end{eqnarray}
Finally it is very useful to compare the power spectrum of the comoving curvature perturbation with that of the tensor mode as, 
\begin{eqnarray}
r \equiv \frac{{\cal P}_{T}}
{{\cal P}_\mathcal{R}} \simeq 16\epsilon.
\label{ratio}
\end{eqnarray}
Since $\epsilon\ll 1$, we can conclude that the amplitude of tensor modes are too smaller than the scalar perturbation. \\
in addition, combining Eqs. (\ref{nT}) and (\ref{ratio}), we would have, 
\begin{eqnarray}
r=-8n_{T}
\label{consistency}
\end{eqnarray}
\section{Observational Signatures of the Inflation }
In the previous sections, we have computed the power-spectrum of the comoving curvature perturbation as well as the tensor perturbations during inflation. Here, we relate these results to observations of the cosmic
microwave background (CMB) radiation. By making this correspondence explicit, we can then put some constraints on inflationary predictions. 
\subsection{Observational Constraints on $n_s$ and $r$}
Using the recent data from the Planck satellite, we can find the value of the
spectral index, $n_s$, running of the spectral index, $\alpha_s$, and the tensor to scalar ratio, $r$. The result is given in Figs. \ref{fig:ns-r} and \ref{fig:run}. \\
According to the Planck result, we would have,
\begin{eqnarray}
n_s &=& 0.9655 \pm 0.0062 \\
r_{0.002} &<& 0.10.
\end{eqnarray}
In addition, the joint analysis of the Planck and BKP(BICEP2/Keck Array + Planck) give us the following upper limit on the value of the tensor-to-scalar ratio as at $2\sigma$,  
\begin{eqnarray}
r_{0.002} &<& 0.08 ~~Planck~TT + lowP+ BKP \\
r_{0.002} &<& 0.09 ~~Planck~TT + lowP+lensing + ext+ BKP 
\end{eqnarray}
where we have used $k \simeq 0.002$ as the pivot scale for this measurement.\\
\begin{figure}[t!]
	\centering
	 \includegraphics[width=6.8cm]{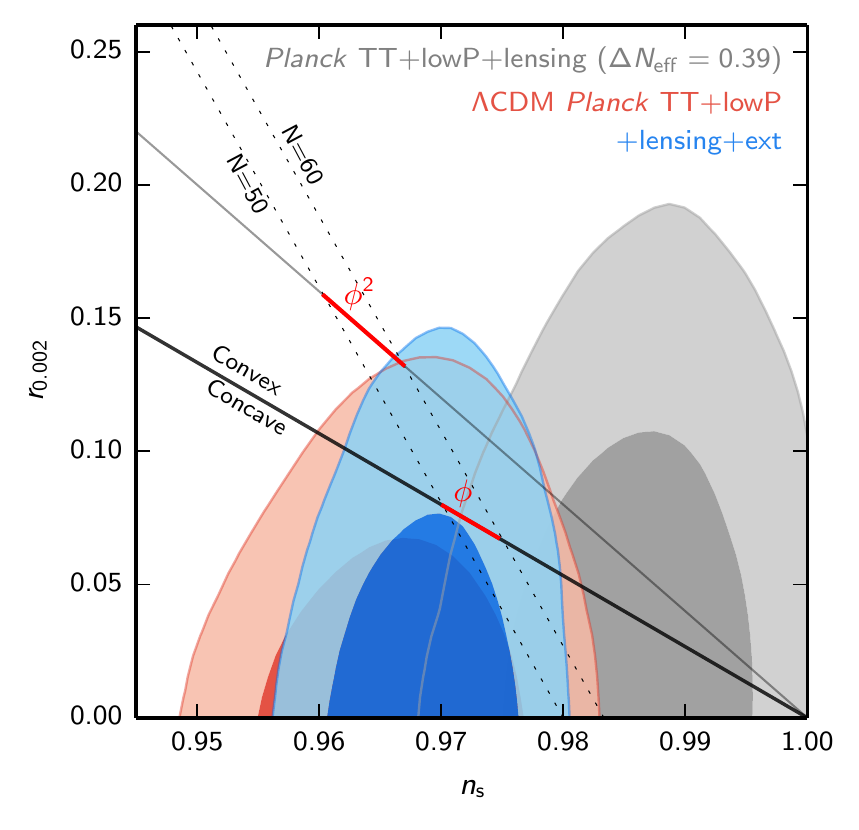}
	\hspace{0.6cm}
		\includegraphics[width=6.8cm]{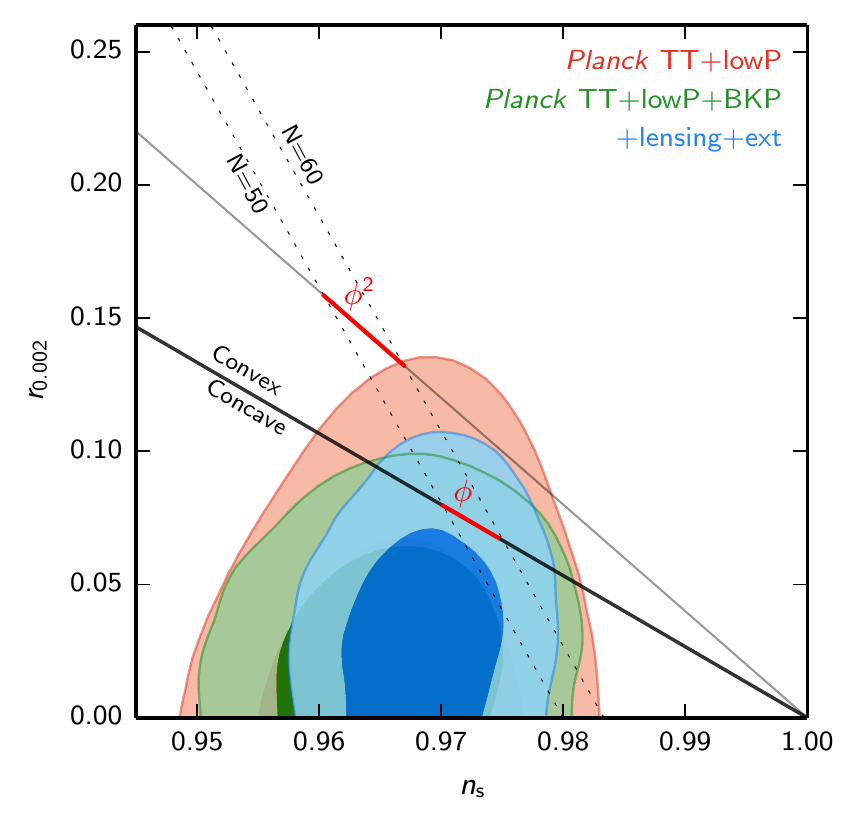}
	\label{fig:ns-r}
	\caption{\label{fig:ns-r} (\textit{Left Panel}) The constraints on the tensor-to-scalar ratio and spectral index in the $\Lambda CDM$ model by assuming negligible running. These results seem to be model dependent. For example, the gray contour shows how the results would be changed when we add the additional relativistic particles. The red lines show the approximate allowed range for $50<N<60$. (\textit{Right Panel}) Equivalent constraints for the $\Lambda CDM$ model, when we add also the B-mode polarization results. It seems that the quadratic potential is disfavored when we add the polarization data, \cite{Planck:2015xua}}
\end{figure}
According to the Planck result, we also have, 
\begin{eqnarray}
\frac{dn_s}{d\ln{k}} &=& -0.0126^{+ 0.0098}_{-0.0087} ,~~Planck~TT + lowP \\
\frac{dn_s}{d\ln{k}} &=& -0.0085 \pm 0.0076 ,~~Planck~TT,TE,EE + lowP \\
\frac{dn_s}{d\ln{k}} &=& -0.0065 \pm 0.0076 ,~~Planck~TT+ lowP + lensing \nonumber\\
&&~~~~~~~~~~~~~~~~~~~~~~~~~~ + ext + BKP.
\end{eqnarray} 
\begin{figure}[t!]
	\centering
	\includegraphics[width=8cm]{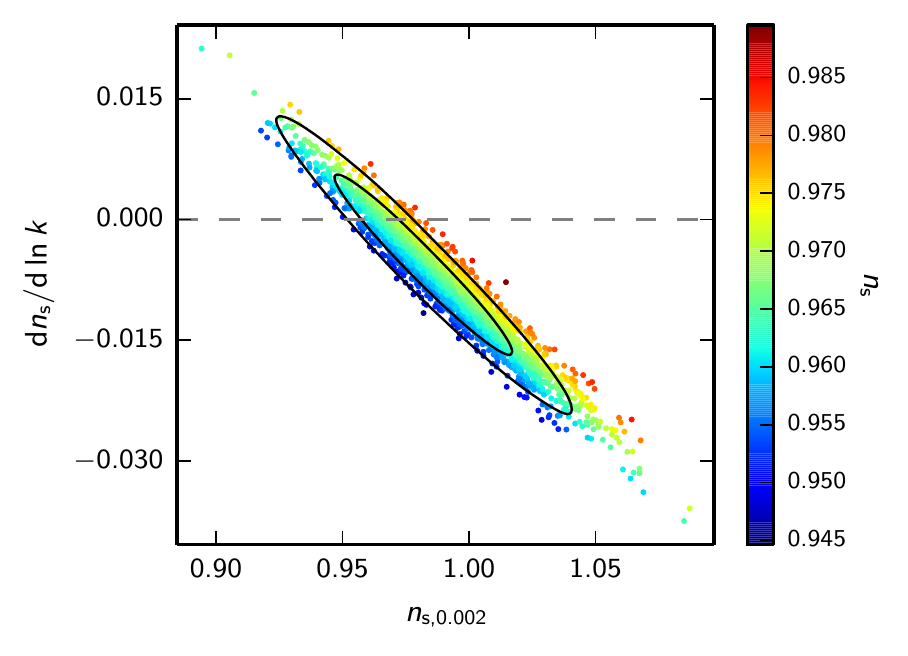}
	\label{fig:run}
	\caption{\label{fig:run} Constrains on the running of the spectral index in the model of $\Lambda CDM$. Although the Planck results are consistent with zero, it also allows a negative running.  \cite{Planck:2015xua}}
\end{figure}
\subsection{Gaussianity}
From the statistical physics, we recall that if a variable, e.g. $\mathcal{R}$, is Gaussian, then all of the odd correlation functions are zero and we are only left with the even correlation functions. In addition, all of the non-zero correlations can be calculated by using the power spectrum, the two point function. However, if $\mathcal{R}$ is not Gaussian, then the bispectrum is non-zero. There are many different ways for $\mathcal{R}$ to be non-Gaussian. So considering them in the most general case is rather involved. So in the following, we only consider a very simple family of the many different parameterizations for the non-gaussianity, \textit{local shape}, with the following form, 
\ba
\mathcal{R}(x) = \mathcal{R}_g(x)+ \frac{3}{5} f_{NL}\mathcal{R}^2_{g}(x). 
\ea
Here $\mathcal{R}_{g}$ refers to the Gaussian part of $\mathcal{R}$. In this simple case, the whole complication has been reduced into a single number $f_{NL}$. \\
Using the Planck data, the constraints on the value of  $f_{NL}$ is given by, 
\begin{equation}
f_{NL}^{local} = 2.5 \pm 5.7.
\end{equation}
Since this number is very small, it seems that the single field inflation is highly preferred by nature.
\subsection{CMB Anisotropies}
Having calculated the primordial curvature perturbation as well as the gravitational wave, one may wonder how these parameters are correlated with what an observer would see on the CMB.\\
\begin{figure}[htbp]
	\centering
	\includegraphics[width=0.75\textwidth]{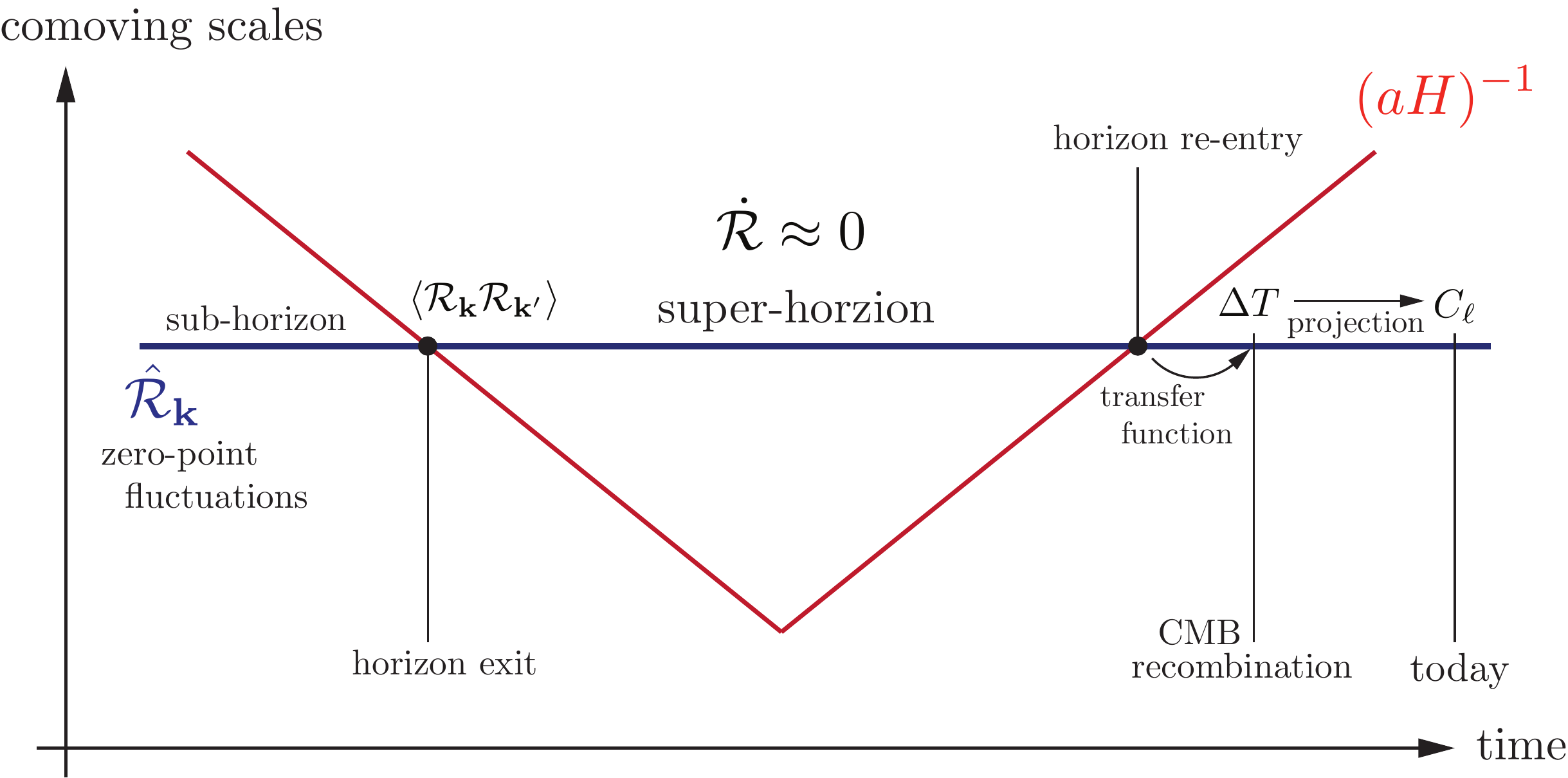}
	\caption{Generation and evolution of the perturbations during inflationary cosmology.  Fluctuations are created on subhorizon scales. Then due to the shrinking the Hubble parameters, they would eventually exit the horizon and freeze until horizon re-entry at late times. After the horizon re-entry, the fluctuations evolve and convert into anisotropies in the CMB, \cite{Baumann:2009ds}.}
	\label{fig:scales3}
\end{figure}
In oder to answer this question, some steps seems necessary to be taken. First of all, we need to figure out how $\zeta$ and $h$ would be related to the observables. Then, we would also need to take into account the possible evolution of these parameters as soon as they have entered the horizon, the schematic behavior of the perturbations is shown in Fig. \ref{fig:scales3}.\\
Generically we can write, 
\begin{equation}
Q_k(\tau) = T_Q(k, \tau, \tau_*) \zeta_k(\tau_*),
\end{equation} 
where $\tau_*$ denotes the moment of the horizon crossing and $T_{Q}$ also refers to the transfer function between the time of horizon re-entry till the time of the observation, $\tau$. The parameter $Q$ may be the temperature of the CMB and the goal is to use the CMB data to put constrains on the primordial curvature power spectrum. \\
For this purpose, it is convenient to use the harmonic expansion of the CMB map as, 
\begin{equation}
\label{equ:Tharm}
\Theta(\hat n) \equiv\frac{\Delta T(\hat n)}{T_0}= \sum_{\ell m} a_{\ell m} Y_{\ell m}(\hat n)\, ,
\end{equation}
where $\hat n$ denotes the direction in the sky and $a_{\ell m}$ is given by,
\begin{equation}
a_{\ell m} = \int d \Omega\, Y^*_{\ell m}(\hat n) \Theta(\hat n).
\end{equation}

Here, $Y_{\ell m}(\hat n)$ are the standard spherical harmonics and 
the magnetic quantum numbers satisfy $m= - \ell, \dots , + \ell$. \\
We can then combine the multipole moments $a_{\ell m}$ and calculate the
rotationally-invariant angular power spectrum as,
\begin{equation}
C_\ell^{TT} = \frac{1}{2\ell+1} \sum_m  \langle a^*_{\ell m} a_{\ell m} \rangle \, , \qquad {\rm or} \qquad  \langle a^*_{\ell m} a_{\ell' m'} \rangle  = C_\ell^{TT} \delta_{\ell \ell'} \delta_{m m'}.
\label{equ:Cl}
\end{equation}
\begin{figure}[t!]
	\centering
	\includegraphics[width=11cm]{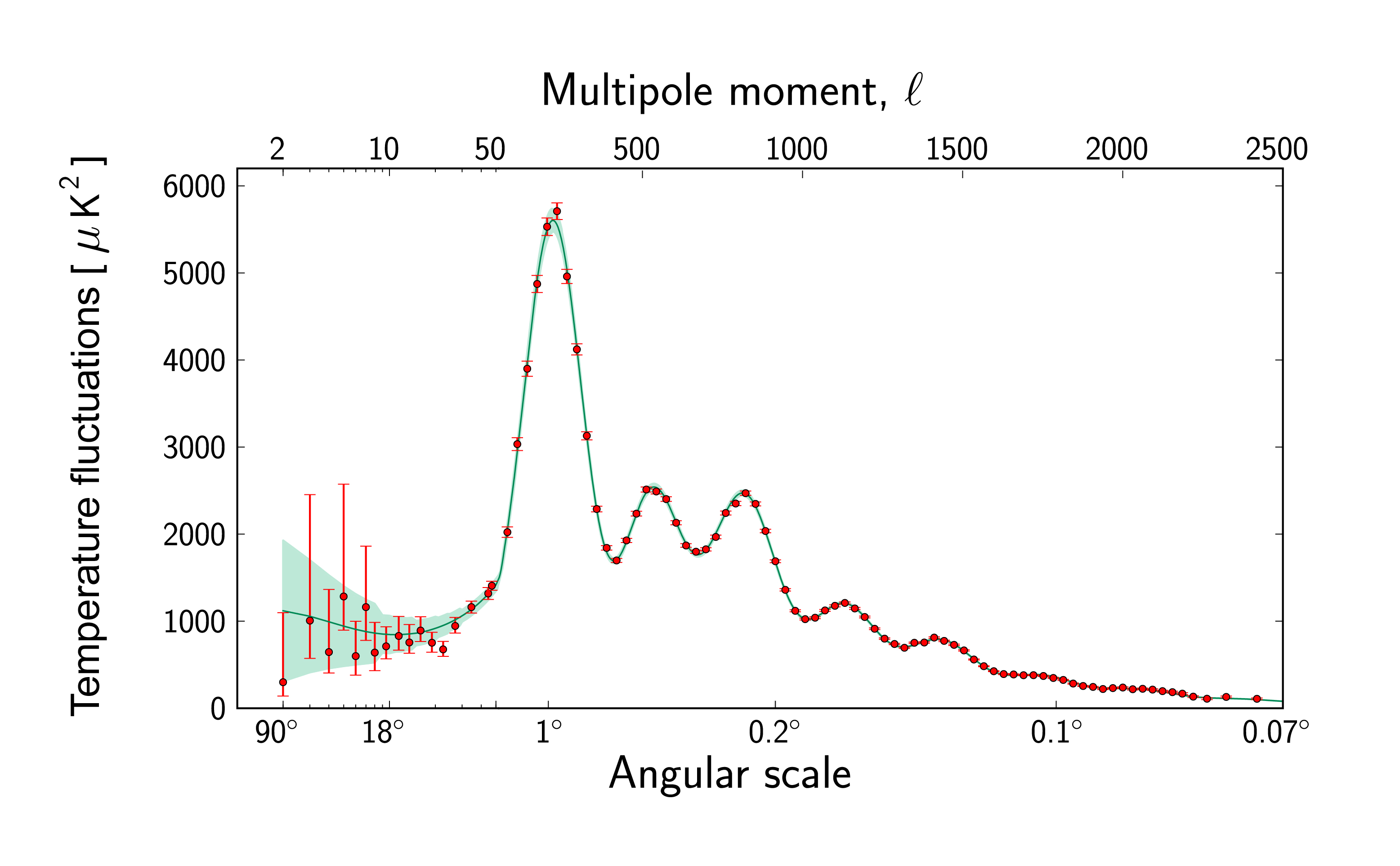}
	\label{fig:planck}
	\caption{\label{fig:planck} Angular power spectrum of CMB temperature fluctuations. \cite{plancksite}}
\end{figure}
We emphasize here that the 
angular power spectrum is indeed a very important tool in the statistical analysis of CMB. Fig. \ref{fig:planck} shows the Planck result for the angular power spectrum. The plot also contains a fit of the theoretical prediction of the CMB spectrum to the data. The theoretical curve depends on the background cosmological parameters as well as on the spectrum of initial fluctuations which is given by the Inflationary model. It is hence possible to use CMB for extracting information about both of them. \\
CMB temperature fluctuations are dominated by the curvature perturbation, $\zeta$.
The transfer function, $\Delta_{T \ell}(k)$, correlates the curvature perturbation to that of the  temperature fluctuation, $\Delta T$ through an integral over the $k$-space as,
\begin{equation}
\label{equ:alm} 
a_{\ell m} = 4\pi (-i)^\ell \int \frac{d^3 k}{(2\pi)^3} \, \Delta_{T \ell}(k)\, \zeta_{\bf k} \, Y_{\ell m}(\hat {\bf k}).
\end{equation}
Plugging Eq. (\ref{equ:alm}) back into Eq. (\ref{equ:Cl}) and using the following identity, 
\begin{equation}
\sum_{m=-\ell}^\ell Y_{\ell m}(\hat {\bf k}) Y_{\ell m}(\hat {\bf k}') = \frac{2\ell+1}{4\pi} P_\ell(\hat {\bf k} \cdot \hat {\bf k}')
\end{equation}
we obtain,
\begin{equation}
\label{equ:Cl2}
C_\ell^{TT} = \frac{2}{\pi} \int k^2 d k P_{\zeta}(k) \Delta_{T \ell}(k) \Delta_{T \ell}(k).
\end{equation}
For a generic value of $\ell$, the transfer functions must be obtained using CMBFAST or CAMB and the results depend on the background cosmology. So for a fixed background, the shape of the power-spectrum give us information about the initial spectrum, coming from the inflationary universe, \cite{Planck:2015xua},
\begin{equation}
P_{\zeta} (k_p) = 2.4 \times 10^{-9}
\end{equation} 
where according to the Planck, $k_p \simeq 0.05 Mpc^{-1}$. 
\section{Summary}
In this chapter, we presented the linear cosmological perturbation theory. As we have seen, at the linear order in perturbation, different components of perturbations are decoupled and we can thus consider them separately. We then used this mechanism to investigate the behavior of the perturbations during inflation. A very specific feature of inflation is that during this era the comoving Hubble horizon decreases. This means that after a while the comoving modes exit the horizon and will be frozen until the time of the horizon re-entry. We calculated the power spectrum of the scalar and tensor modes and also parameterized their scale-dependences. Finally, we tried to connect the inflationary parameters with the cosmic microwave background observations. We justified that the CMB observations can put some constraints on inflationary predictions.


\chapter{Anisotropic Inflation from Charged Scalar Fields } 

\label{Chapter3} 

\lhead{Chapter 3. \emph{Anisotropic Inflation from Charged Scalar Fields}} 

\vspace{0.5cm}
\hrule \vspace{0.3cm}

\begin{quote}
	\textbf{Abstract:} The goal of this chapter is to give an introduction to  anisotropic inflation, where the three dimensional rotational symmetry has been reduced to a planar symmetry. We consider different models of inflation including a $U(1)$ gauge field and charged scalar fields including symmetry breaking potential, chaotic inflation and hybrid inflation. We show that there exist attractor solutions where the anisotropies produced during inflation 
	becomes comparable to the slow-roll parameters.  
	In the models where the inflaton field is a charged scalar field  the gauge field becomes highly oscillatory at the end of inflation ending inflation quickly. Furthermore, in charged hybrid inflation the onset of waterfall phase transition at the end of inflation is affected significantly by the evolution of the background gauge field. 
\end{quote}

\vspace{0.1cm}  \hrule
\vspace{0.5cm}

\section{Introduction}

As we already discussed in last chapters, cosmic inflation proved to be a successful theory of early universe and the mechanism of structure formation \cite{Komatsu:2010fb}. The simplest modes of inflation are based on a scalar field coupled minimally to gravity.\\ 
There have been considerable interests on primordial anisotropies both observationally and theoretically. Observationally, there may be some indications of the statistical anisotropy of the comic microwave background (CMB),\cite{Eriksen:2003db} although the statistical significances of these findings are under debate \cite{Komatsu:2010fb, Hanson:2009gu, Hanson:2010gu}. On the theoretical sides there have been many attempts to construct models of inflation with vector fields or gauge fields
which can create sizable amount of anisotropy on curvature perturbations 
\cite{Maleknejad:2012fw, Ford:1989me, Kaloper:1991rw,  Kawai:1998bn,  Barrow:2005qv, Barrow:2009gx, Campanelli:2009tk, Golovnev:2008cf, Kanno:2008gn, Pitrou:2008gk, Moniz:2010cm, Boehmer:2007ut, Koivisto:2008xf}. These mechanisms can provide a seed of anisotropies at the order of few percent which may be detectable on CMB \cite{Ackerman:2007nb}, \cite{Yokoyama:2008xw}, \cite{Dimopoulos:2009vu, Dimastrogiovanni:2010sm, ValenzuelaToledo:2009af}.

There have been different approaches to implements vector field in models of inflation where the vector field breaks the gauge symmetry explicitly. One fundamental problem in these models, as demonstrated in \cite{Himmetoglu:2008zp}, is the appearance of ghost which render the system unstable and physically unacceptable. Therefore, it is crucial that the vector field is protected by a gauge symmetry so the longitudinal mode of the vector field excitations is not physical. On the other hand, because of the conformal invariance of models with gauge fields, any excitation of gauge field during inflation is diluted and can not seed the desired anisotropies. Therefore, it is essential that one breaks the conformal invariance while keeping the gauge symmetry explicit. This approach was employed in different contexts in 
\cite{Martin:2007ue, Watanabe:2009ct, Kanno:2009ei, Emami:2009vd, Gumrukcuoglu:2010yc, Watanabe:2010fh, Dimopoulos:2010xq, Dulaney:2010sq}.

In this chapter we would like to study different inflationary models where there is a non-zero background  $U(1)$ gauge field, $A_\mu$, coupled to a complex scalar field. The charged scalar field can be either the inflaton field or the waterfall field of hybrid inflation. Furthermore, in order to break the conformal invariance and produce large enough anisotropies, as explained above, we assume that the gauge field has a time-dependent gauge kinetic coupling with the kinetic energy in the form of $\frac{-f(\phi)}{4 } F_{\mu \nu} F^{\mu \nu}$. To be specific, we consider three following models: \\
{\color{red}{$\bullet$}} Symmetry breaking hilltop inflation: In this model  inflaton is a charged complex scalar field with the symmetry breaking (Mexican hat) potential.\\
{\color{red}{$\bullet$}} Charged hybrid inflation: This is the standard hybrid inflation \cite{Linde:1993cn} where the inflaton field $\phi$ is real but now the waterfall field $\psi$ is charged under $U(1)$ gauge field. \\
{\color{red}{$\bullet$}} Chaotic inflation where the inflaton field $\phi$ is charged under the $U(1)$ field. \\
The results of this chapter are based on \cite{Emami:2010rm}

\section{Background Equations}
\label{background}
Here we present the action for the cases of symmetry breaking potential where the 
inflaton field $\phi$ is charged under the $U(1)$ gauge field with a $\phi$-dependent  gauge kinetic coupling $f^{2}(\phi)$. The action and the background equations for the other cases can be obtained accordingly.  

As in \cite{Emami:2009vd} the action is
\ba
\label{action4} S= \int
d^4 x  \sqrt{-g} \left [ \frac{M_P^2}{2} R - \frac{1}{2} D_\mu \phi
\,  D^\mu \bar \phi -   \frac{f^{2}(\phi)}{4} F_{\mu \nu} F^{\mu
	\nu}  - V(\phi, \bar \phi) \right]
\ea
where $M_P^{-2} = 8 \pi G$,
for $G$ being the Newton constant and the overline represents the
complex conjugation. The covariant derivative is given by
\ba
D_\mu
\phi = \partial_\mu  \phi + i \e \,  \phi  \, A_\mu
\ea
where $\e$ is
the dimensionless gauge coupling of $A_\mu$ to $\phi$. As usual, the
gauge field strength is given by
\ba F_{\mu \nu} = \nabla_\mu A_\nu
- \nabla_\nu A_\mu  = \partial_\mu A_\nu - \partial_\nu A_\mu \, .
\ea
We work with potentials which have axial symmetry where $V$ and
$f(\phi)$ are only functions of $\phi \bar \phi=  |\phi |^2$. It is
more instructive to decompose the inflaton field into the radial and
angular parts \ba \phi(x) = \rho(x) \,  e^{i \theta(x)}\, , \ea so
$V=V(\rho)$ and $f^2(\phi)=f^2(\rho)$. As usual, the action
(\ref{action4}) is invariant under local gauge transformation \ba
\label{transformation} A_\mu \rightarrow A_\mu - \frac{1}{e}
\partial_\mu \epsilon(x) \quad , \quad \theta \rightarrow \theta +
\epsilon(x) \, . \ea
With this decomposition, the action  (\ref{action4}) is transformed into
\ba
\label{action2} S&=& \int d^4 x \sqrt{-g} \bigg{[} \frac{M_P^2}{2}
R -  \frac{1}{2} \partial_\mu \rho
\partial^\mu \rho-
\frac{\rho^2}{2}  \left( \partial_\mu \theta + \e A_\mu  \right)
\left( \partial^\mu \theta + \e A^\mu  \right) \nonumber\\
&&- \frac{f^2(\rho)}{4}
F_{\mu \nu} F^{\mu \nu}  - V(\rho) \bigg{]} 
\ea
The corresponding
Klein-Gordon equations of motion are
\ba \label{theta-Eq}
\partial_\mu\,  J^\mu =0
\\
\label{rho-Eq}
\partial_\mu \left[  \sqrt{-g} \partial^\mu \rho \right] -
\frac{ J_\mu J^\mu}{\rho^3 \sqrt{-g} }   -\sqrt{-g}\,(
V_\rho+\frac{f(\rho)f_\rho(\rho)}{2}F_{\mu \nu} F^{\mu \nu})=0 \, ,
\ea
accompanied by with the Maxwell's equation
\ba \label{Maxwell}
\partial_\mu \left(  \sqrt{-g}\, f^2(\rho)\, F^{\mu \nu} \right) = \e J^\nu \, ,
\ea
where the current $J^\nu$ is defined by
\ba J^\nu \equiv  \rho^2
\sqrt{-g} \left( \partial^\nu \theta + \e A^\nu  \right)  \, .
\ea
The conservation of $J^\mu$ from Eq. (\ref{theta-Eq}) is a
manifestation of the axial symmetry imposed on $V$. Interestingly,
Eq. (\ref{theta-Eq}) is not independent from Maxwell's equation,
where  $F^{\mu \nu}$ being anti-symmetric leads to $\partial_\mu
\partial_\nu F^{\mu \nu} = \partial_{\mu} J^{\mu}=0$.

Finally, the stress energy momentum tensor, $T_{\alpha \beta}$, for
the Einstein equation, $G_{\alpha \beta} = 8 \pi G \, T_{\alpha \beta}$, has the following form,

\ba \label{energy-momentum} 
T_{\alpha \beta} &=&
\frac{-f^2(\rho)}{4} g_{\alpha \beta} F_{\mu \nu} F^{\mu \nu}
+f^2(\rho) F_{\alpha \mu} F_\beta^{\, \mu} +  \partial_\alpha \rho
\partial_\beta \rho +  \frac{J_\alpha J_\beta}{\rho^2 | g|} \nonumber\\
&&-g_{\alpha \beta} \left[ \frac{1}{2} \partial_\mu \rho
\partial^\mu \rho +  \frac{J_\mu J^\mu}{2 \rho^2 | g|}  +V
\right]\, . \ea

We are interested in the effects of a non-zero background gauge field on the evolution of system.
To fix the gauge we use the Coulomb-radiation gauge $A_0= \partial_i A^i=0$. Since our background is only time-dependent, from the constraint $J^0=0$ one concludes that $\dot \theta=0$ at the level of background. The inclusion of a non-zero background gauge field breaks the Lorentz invariance explicitly since a preferred direction is singled out in the background space-time.  We take our background gauge field to have the form  $A_{\mu}=(0,A(t),0,0)$.

As in \cite{Watanabe:2009ct} our background metric has the following form
\ba
\label{metric4} ds^2 = - dt^2 + e^{2\alpha(t)}(e^{-4\sigma(t)}d x^2
+e^{2\sigma(t)}(d y^2 +d z^2)) \, .
\ea
Here $\alpha(t)$ measures the background number of e-foldings, $\dot \alpha$
represents the background isotropic Hubble expansion rate while $\dot \sigma(t)$
measure the anisotropic expansion rate. For a universe with small anisotropies, we require that
$|\dot \sigma/\dot \alpha| \ll 1$.

Assuming that the fields $\rho, A, \alpha$ and $\sigma$ are only function of
$t$ the background equations of motion are
\ba
\label{back-A-eq4}
\partial_t{\left(  f^2(\rho) e^{\alpha + 4 \sigma} \dot A        \right)}& =& - \e^2 \rho^2 e^{\alpha + 4 \sigma}  A \\
\label{back-rho-eq4}
\ddot\rho+3\dot \alpha\dot \rho+ V_\rho+ \left(
-f(\rho)f_\rho(\rho)\dot A^2 +\e^2 \rho A^2   \right) e^{-2\alpha+4\sigma}&=&0  \\
\label{Ein1-eq4}
\frac{1}{2}\dot
\rho^2+V(\rho)+ \left(   \frac{1}{2}f^2(\rho)\dot
A^2 +\frac{\e^2\rho^2}{2}A^2 \right) e^{-2\alpha+4\sigma}
&=&
3 M_P^2 \left(   \dot \alpha^2-\dot \sigma^2 \right)  \\
\label{Ein2-eq4}
V(\rho)+  \left(  \frac{1}{6}f^2(\rho)\dot
A^2+\frac{\e^2\rho^2}{3}A^2  \right)e^{-2\alpha+4\sigma}
&=& M_P^2 \left( \ddot \alpha    + 3 \dot \alpha^2 \right)  \\
\label{anisotropy-eq4}
\left(\frac{1}{3}f^2(\rho)\dot A^2  -\frac{\e^2\rho^2}{3}A^2    \right) e^{-2\alpha+4\sigma}
&=& M_P^2\left( 3\dot \alpha \dot \sigma+ \ddot \sigma      \right)\, .
\ea
In the limit where $\e=0$ these equations reduce to those of \cite{Watanabe:2009ct}. One can also check that not all equations above are independent. For example,
Eq. (\ref{Ein1-eq4}) can be obtained from the remaining four equations.

From Eq. (\ref{back-rho-eq4}), the total energy density, ${\cal E}$, governing the dynamics  of the inflaton field is given by
\ba
\label{Veff4}
{\cal E}= \frac{\dot \rho^2}{2}+V+  e^{-2\alpha+4\sigma} \left( \frac{1}{2}f^2(\rho)     \dot
A^2+\frac{\e^2\rho^2}{2} A^2 \right) \, .
\ea
Since the second and the third terms above
come from the gauge field, we refer to them respectively as the kinetic energy and potential energy associated with the gauge field.  Using Eqs.(15) and (16), the equation for acceleration of the
universe is given by
\ba
\label{accel4}
\ddot\alpha +\dot\alpha^2 =
-2\dot\sigma^2-\frac{1}{3M_{P}^2}\dot
\rho^2+\frac{1}{3M_{P}^2}\left[V-\frac{1}{2}f^2(\rho)\dot
A^2e^{-2\alpha+4\sigma} \right] \, .
\ea
Interestingly, the term in effective potential proportional to $\e$ cancels out in this expression.
One also observes that for inflation to take place, corresponding to $\ddot\alpha +\dot\alpha^2 >0$, one requires that the background potential $V(\rho)$ dominates over the contribution from the gauge field kinetic energy.

In the following we are interested in configuration where inflation take places with small anisotropies such that
\ba
\label{delta}
\delta \equiv
|\frac{\dot \sigma} {\dot \alpha}| \ll 1.
\ea
Small amount of anisotropies in background inflationary dynamics may be acceptable assuming that they do not impose too much anisotropies on CMB temperature power spectrum.  In order for anisotropies to be small, the contribution of gauge field to the total energy density
should be small compared to the background potential. To parametrize this, we define the ratios
$R_1$ and $R_2$ via
\ba
\label{R12}
R_1 \equiv \frac{\dot A^2 f(\rho)^2 e^{-2 \alpha}}{2 V} \quad , \quad
R_2 \equiv \frac{ \e^2 \rho^2\, A^2  e^{-2 \alpha}}{2 V} \, .
\ea
For the contribution of the gauge field energy density into the total energy density to be small
we require $R_{1}, R_2 \ll 1$.  In this limit where the anisotropy is smaller than the slow-roll parameters (defined below), from Eq. (\ref{anisotropy-eq4}) we obtains
\ba
\label{delta-R124}
\delta \simeq \frac{2}{3} (R_1 - R_2)\, .
\ea

One of our goal in this work is to see the behavior of $R_{1,2}$ during inflation. As we shall see, during early stage  inflation both $R_{1}$  and $R_{2}$ are
very small and inflation is basically driven by the background potential $V$ and one can treat
the system as isotropic inflation. As in \cite{Watanabe:2009ct}, sometime during inflation, $R_1$ rises quickly such that its contribution to the Klein-Gordon equation governing the scalar field dynamics can not be neglected. This is an attractor mechanism and as we shall see
below $R_1$ scales like the slow-roll parameters once the system is in the attractor regime.
Interestingly, the Hubble expansion rate is still predominantly given by the background potential $V$ but one should check that
the anisotropy given by Eq. (\ref{anisotropy-eq4}) is under control. One new effect in our model is that sometime at the end of inflation $R_2$ becomes comparable to $R_1$ 
and the  gauge field oscillates very rapidly. Because of  the interaction term $\e^2 \rho^2 A_x^2$, the rapid oscillations of the gauge field induce rapid changes in inflaton effective mass, violating the slow-roll conditions and ending inflation abruptly. Here we would like to study these three distinct inflationary phases in some details. For convenience, we refer to these three stages of inflation as phase one, two and three, respectively. However, note hat the third inflationary stage is very short compared to other two inflationary stages.

As a measure of slow-roll parameters and 
phase change we define the dimensionless quantity $\epsilon $  and $\eta$ given by
\ba
\label{epsilon-eta4}
\epsilon \equiv  -\frac{\ddot \alpha}{\dot \alpha^2} \quad , \quad 
\eta \equiv \frac{\ddot \rho}{3 \dot \alpha \dot \rho} \, .
\ea
When the slow-roll approximation holds $\epsilon , \eta \ll 1$ and  inflation ends when $\epsilon , \eta \simeq 1$.  In our anisotropic inflationary models, $\eta$ has sudden jumps which represents the onsets of phase changes. In
{\bf Fig. \ref{eta-fig}} we have plotted $\eta $ for some parameters value which clearly indicates the jumps in $\eta$.


\section{Symmetry Breaking Hilltop Inflation}
\label{symmetry}

We start with an almost isotropic configuration with negligible anisotropies such that the gauge field contributions in expansion rate Eq. (\ref{Ein1-eq4}) and the Klein-Gordon equation (\ref{back-rho-eq4}) are negligible, corresponding to $R_{1}, R_{2} \ll \epsilon , \eta$.  However, we would like to allow the gauge field kinetic energy to increase such that $\delta$ increases towards the  allowed observational bounds.

To be specific, we work with the symmetry breaking (Mexican hat) potential which is theoretically well motivated in a  model with an Abelian gauge field
\ba
\label{pot}
V= \frac{\lambda}{4} \left( | \phi |^2 - \frac{M^2}{\lambda}  \right)^2 
\equiv  \frac{\lambda \mu^4}{4 } \left ( \hat \rho^2 -1 \right)^2 \, .
\ea
Here $\lambda$ is a dimensionless coupling and for the later convenience we have defined the dimensionless variable $ \hat \rho \equiv \rho/\mu$ where $\mu \equiv   M/\sqrt{\lambda}$. The potential has global minima at $ \rho = \mu $ or $\hat \rho=1$.
In this picture, we assume inflaton starts at the top of the potential
and proceeds towards the global minima. As is well  known cosmic strings are produced at the end of inflation which can have interesting observational effects.  

In \cite{Watanabe:2009ct} the authors studied the simple chaotic inflationary potential $V= m^2 \phi^2/2$ for a real scalar field. As we shall see, their conclusion about the existence of the
attractor mechanism during the second phase and the behavior of $R_1$ and $\delta$ will also hold in our case. However, the contribution of 
gauge coupling $\e$ via $R_2$ opens up new inflationary phase and the dynamics of the system at the end of inflation is quite different than what studied in \cite{Watanabe:2009ct}.

During the inflation, which happens mostly in the hilltop regions of the potential, one may approximate potential (\ref{pot}) as
\ba
V\simeq \frac{M^4}{4 \lambda} - \frac{M^2}{2}\rho^2 \, .
\ea
For this approximation to be valid, one has to satisfy $\hat \rho \ll 1$ during inflation.

Since we are interested in small anisotropies, $R_1, R_2, \delta \ll 1$,  the background expansion is given as in standard slow-roll inflationary models with
\ba
\label{Hubble4}
{\dot \alpha}^2 \simeq \frac{V}{3 M_P^2}  \simeq \frac{M^4}{12 \lambda M_P^2} \, .
\ea

\begin{figure}[t]
	\centering
	\includegraphics[width=9cm]{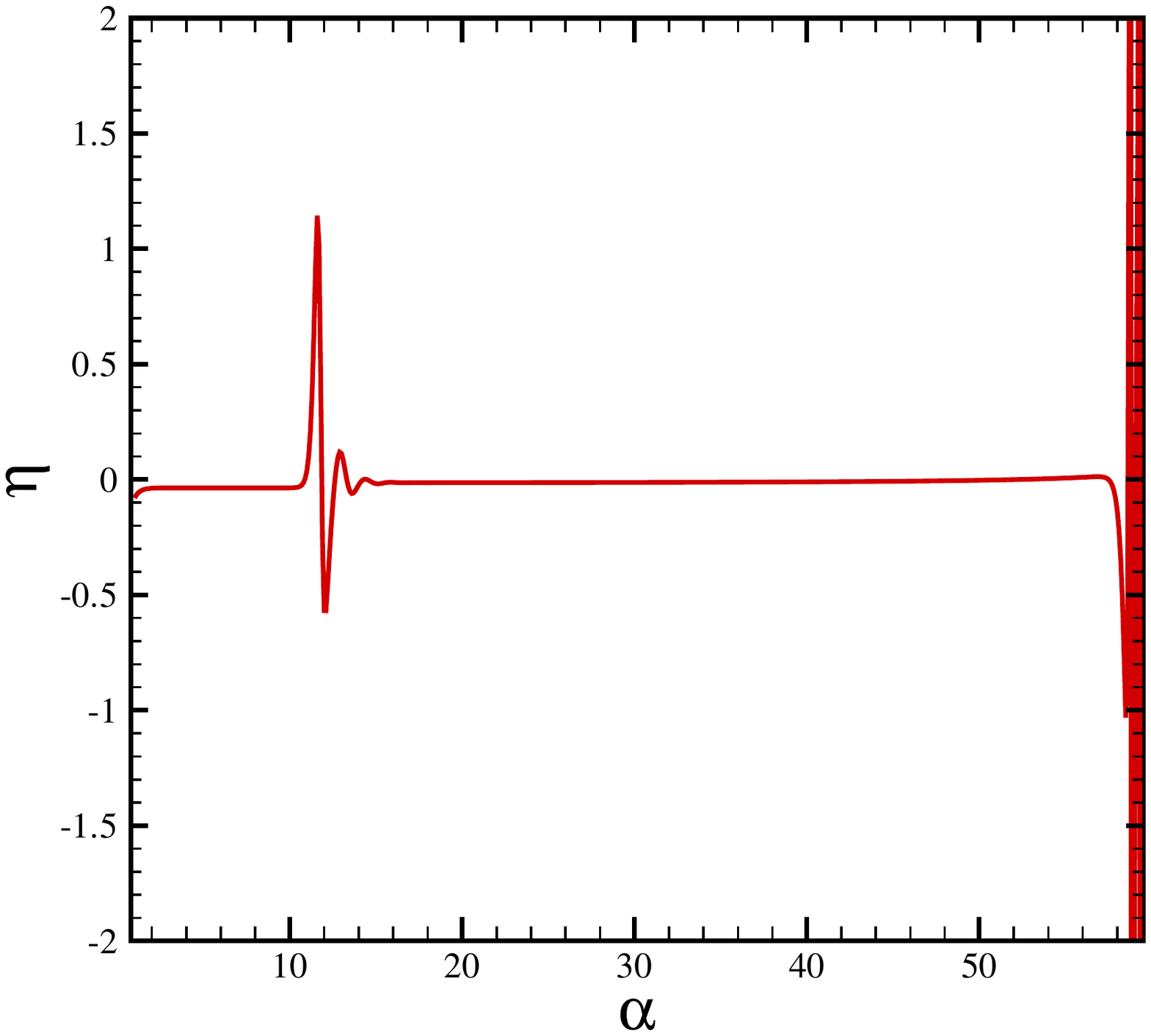}
	\caption{Here we have plotted $\eta$ defined in Eq. (\ref{epsilon-eta4}), with $\lambda= 2.5 \times 10^{-13}$, $M= 3\times 10^{-6} M_P$, $p=50$, $\rho_{in} = M_p/5 $ and $\e=1$.
		The first phase change, at e-folding $\alpha\simeq 10$, happens when $R_1 $ becomes comparable to $\epsilon$. The second phase change happens very close to the end of inflation (in about one e-folding towards the end of inflation) when
		$R_2 $ also becomes comparable to $\epsilon$.}
	\vspace{0.5cm}
	\label{eta-fig}
\end{figure}

Our numerical investigations show that the phase changes happen when there are sudden changes in $\dot \rho$ in a short period. This can be seen in the plot of $\eta$ in
{\bf Fig. \ref{eta-fig} } where there are two sudden jumps. The first jump  corresponds to  the transition from phase one to phase two where the $R_1$ contribution in Eq. (\ref{back-rho-eq4}) becomes important. The second jump, very close to the end of inflation,
represents the transition from the second phase to the third phase. This happens when the right hand side of Eq. (\ref{back-A-eq4}) can not be neglected and it eventually affects the evolution of $\rho$ in Eq. (\ref{back-rho-eq4}).  As we shall see the third period is very short
and inflation ends abruptly once the gauge field starts to oscillate during the third phase.

\subsection{Phases One and Two}

At the early stage of inflation, the contribution of gauge field in total energy density
and the scalar field equation is completely negligible corresponding to $R_{1}, R_{2} \ll \epsilon, \eta$.  Inflation proceeds as in standard slow-roll hilltop inflation and in order for the slow-roll condition to be satisfied one requires that $M_P^2 (V_\rho/V)^2$ and
$M_P^2 V_{\rho\rho}/V$ both to be much smaller than one. These in turn yields
$p_c \gg 1$ where 
\ba
\label{pc}
p_c \equiv \frac{M^2}{2 \lambda M_P^2} \, .
\ea
The Hubble expansion rate is given
by Eq. (\ref{Hubble4}) while the scalar field equation in the slow-roll approximation is
\ba
\label{rho-phase1}
\rho' \simeq  4 \frac{ \lambda M_P^2}{M^2} \rho \quad   \rightarrow  \quad
\rho \simeq \rho_{in} e^{2 \alpha/p_c} \, ,
\ea
where $\rho_{in}$ represents the initial value of the inflaton field. Also here and below, prime denotes derivative with respect to $\alpha$, the number of e-foldings. We use the convention such that at the start of inflation $\alpha=0$ and the total number of e-foldings measured at the end of inflation is $\alpha= \alpha_f \simeq 60$ to solve the flatness and the horizon problem. From Eq. (\ref{rho-phase1}) the number of e-folds as a function of $\rho$  during the first phase is 
\ba 
\label{back-alpha0}
\alpha(\rho)  \simeq  \frac{p_c}{2} \ln \left(  \frac{ \rho}{ \rho_{in}}      \right)   \, .
\ea

So far we have not specified the form of $f(\rho)$, the time-dependent  gauge kinetic coupling. In order for the perturbative gauge theory to be under control we demand that the effective gauge kinetic
coupling $g_A(\alpha) = f(\rho)^{-1}$ to be small during inflation and approaches unity at the end of inflation for some yet unknown dynamical mechanism, that is $g_A(\alpha_f) =1$. This indicates that $f(\rho)$ is a decreasing function during inflation.   As mentioned before, we would like the gauge field contribution to the energy density to be subdominant but big enough to play some roles in anisotropy and scalar field equations. To determine the form of $f(\rho)$ we note that  during the first two phases, $R_2 \ll R_1$  so the terms proportional to
$\e$ in background equations (\ref{back-A-eq4})-(\ref{anisotropy-eq4}) can be neglected.
From Eq. ({\ref{back-A-eq4}}) one obtains $\dot A  \propto f(\rho)^{-2}e^{-\alpha} $ so
$R_1 $ scales like $R_1 \propto  f(\rho)^{-2}  e^{-4 \alpha} \sim   f(\rho)^{-2} \rho^{-2 p_c}$. Therefore, for the critical coupling $f_c \equiv ( \mu/\rho)^{p_c}$ the gauge field kinetic energy remains fixed during the first two phases.
As we started with negligible $R_1$ in phase one, then it remains negligible afterwards, i.e. $R_1 \ll \epsilon$. In order to increase $R_1$ during the second phase  we consider the gauge kinetic coupling
\ba
\label{f}
f(\rho) = \left(\frac{\mu}{\rho} \right)^{p} = \hat \rho^{- p} \, ,
\ea
with $p> p_c$
such that the gauge field kinetic energy becomes important during the second and third stages of inflation.

As explained above, during the first two phases the right hand side of Eq. ({\ref{back-A-eq4}}) can be neglected and using Eq. (\ref{f})
one obtains 
\ba
\label{A-prime}
\hat A' = \left( \xi \hat \rho^2 \right)^p  e^{-\alpha} \, ,
\ea
where the dimensionless gauge field is defined via $\hat A \equiv  A/\mu$ and 
$\xi$ is a constant of integration. Note that we defined the constant of integration in this way so the subsequent analysis becomes more simplified. Physically, $\xi$ is measured by the initial value of $R_1$ at the start of inflation, $\alpha=0$.  Plugging Eq. (\ref{A-prime}) into 
Eq. (\ref{R12}) during the first two phases one obtains 
\ba
\label{R1-12}
R_1 \simeq \frac{p_c}{3} e^{-4 \alpha} \left( \xi \hat \rho \right)^{2p} \, ,
\ea
and therefore the initial value of $R_1$ is 
\ba
\label{R1in}
R_{1 \, {in}} \simeq \frac{ p_c}{3}   \left( \xi \hat \rho_{in} \right)^{2 p} \, .
\ea

Since we demand that $R_{1,2} \ll 1$, the Friedmann equation is still given by Eq. (\ref{Hubble4}). Combined with Eq. (\ref{f}), the inflaton field equation in the slow-roll limit is cast into
\ba
\label{chi-eq}
(\hat \rho^2)' - \frac{4\,  \hat \rho^2}{p_c}  + \frac{2 p\,  \xi^{2p}}{3} e^{- 4 \alpha} \left(  \hat \rho^2  \right)^p = 0\, .
\ea                 

As can be seen from Eq. (\ref{R1-12}), $\left( \xi \hat \rho  \right)^{2p}$ is very small during the first phase of inflation so one can neglect the last term in Eq. (\ref{chi-eq}) and the solution is given by Eq. (\ref{rho-phase1}).  
For this to take place, we need to make sure that at the start of inflation, $\alpha=0$, the third term in  Eq. (\ref{chi-eq}) is indeed much smaller than the second term. Using the expression
of $R_{1\, in}$ given in Eq. (\ref{R1in}) this condition is transformed into
\ba
\label{cond1}
R_{1\, in} \ll \frac{2}{p} \hat \rho_{in} \, .
\ea

\begin{figure}[t]
    \centering
	\includegraphics[width=7cm]{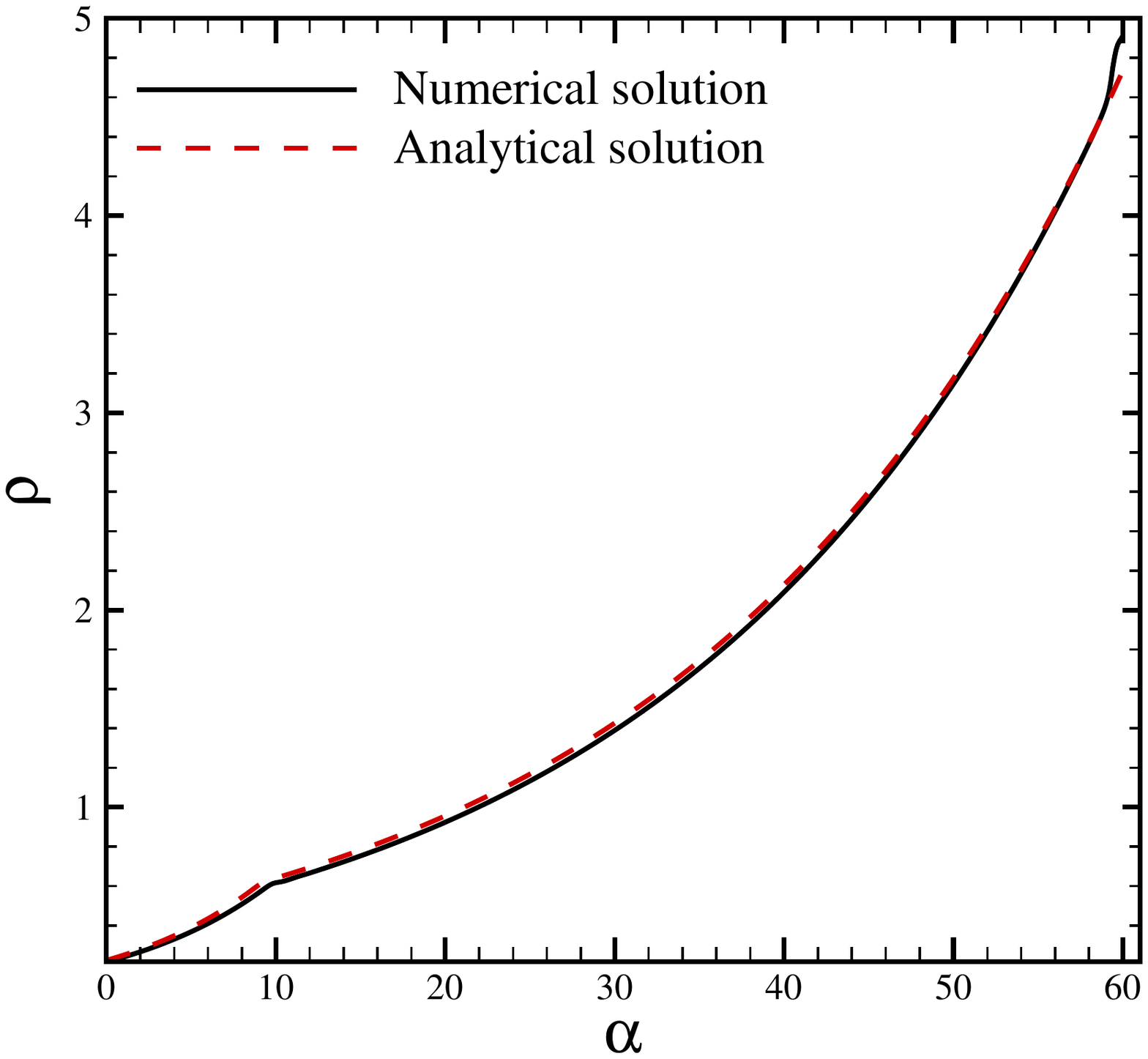} 
	\includegraphics[width=7cm]{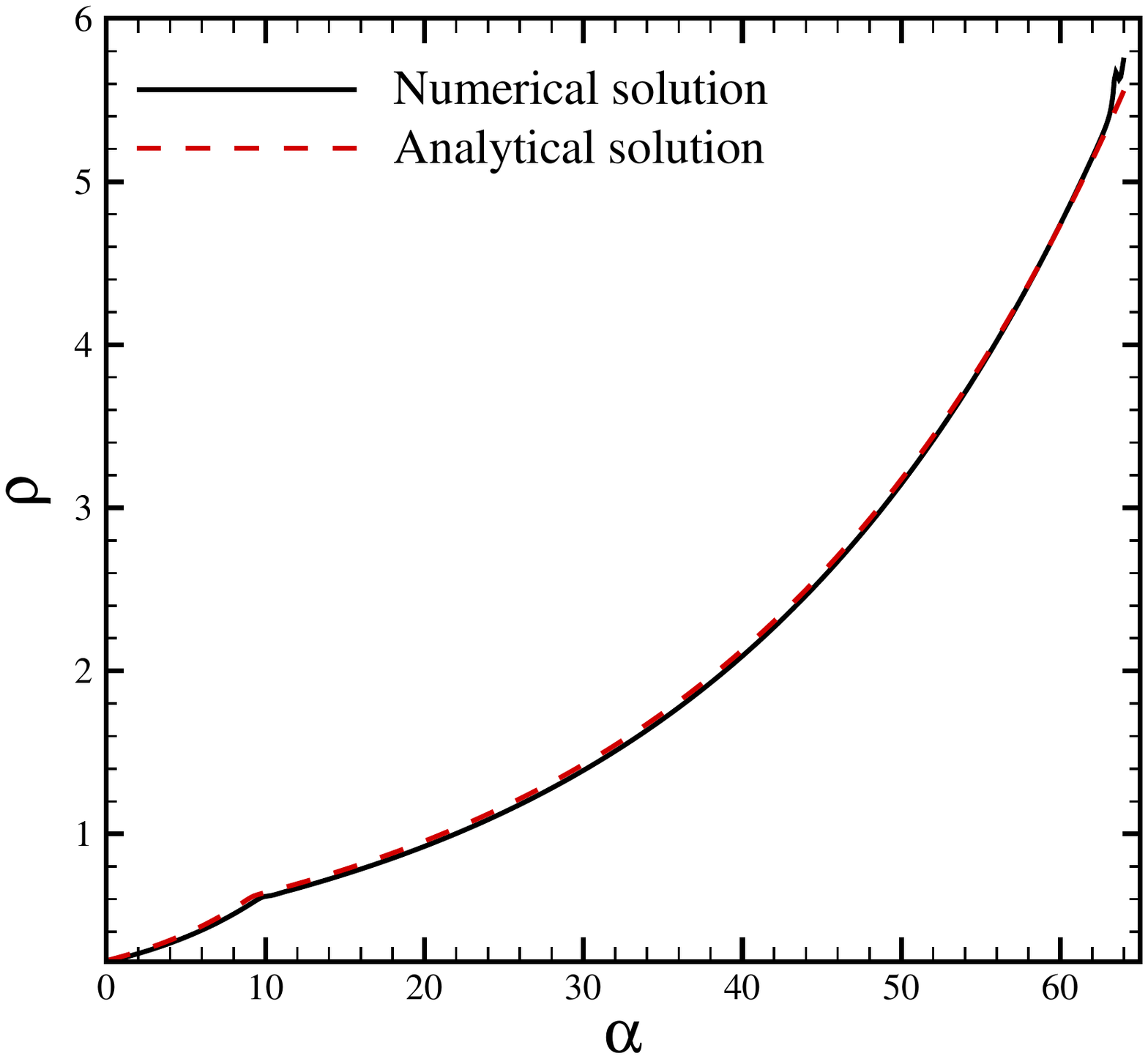}
	\caption{Here we plot our analytical solution for $\rho(\alpha)$, Eq. (\ref{hat-rho1}),
		shown by the red dashed curve, and compare it to the full numerical solution denoted by the solid black curve. The agreement between them is very good. The left figure corresponds to $\e=1$ whereas for the right figure $\e=10^{-4}$. As argued, the time of first phase change, which here is at $\alpha_1 \simeq 10$, is independent of the value of $\e$ and is well approximated by our analytical formula Eq. (\ref{alpha1}). All other parameters are as in {\bf Fig.} \ref{eta-fig}. }
	\vspace{0.5cm}
	\label{r-fig-sym}
\end{figure}

As inflation proceeds and $\hat \rho$ increases the last term in Eq. (\ref{rho-phase1})
catches up with the second term and one should take the effect of this term into account. This is exactly when the gauge field contribution into the inflaton equation, Eq. (\ref{back-rho-eq4}), becomes important as promised. Eq. (\ref{rho-phase1}) can be solved with the answer
\ba
\label{hat-rho1}
\hat \rho  \simeq \frac{ \hat \rho_{in} e^{\frac{2 \alpha}{p_c}  } }{ \left[ 1 + \frac{p^2 p_c}{6 (p - p_c) } \left(  \xi \hat \rho_{in} \right)^{2p} e^{\frac{4 (p- p_c) \alpha}{p_c} }  \right]^{1/2p} }  \, .
\ea
During the first inflationary phase, the second term in the denominator is much smaller than unity and the solution to the  above equation reduces to our previous result, Eq. (\ref{rho-phase1}). The transition from the first phase to the second phase happens when the two terms in the denominator above become comparable. Defining the first phase transition to take place at $\alpha= \alpha_1$, one obtains
\ba
\label{alpha1}
\alpha_1 &\simeq& \frac{p_c}{4 (p-p_c)} \ln \left[  \frac{6 (p-p_c)}{p^2 p_c  \left( \xi \hat \rho_{in} \right)^{2p}}\right] \nonumber\\
&\simeq& \frac{ p_c}{4 (p-p_c)} \ln \left[  \frac{2 (p-p_c)}{p^2 R_{1\, in}} 
\right] \, ,
\ea
where to get the final answer Eq. (\ref{R1in}) has been used. This is an interesting result because the onset of the first phase transition is controlled by the anisotropy at the start of inflation, $R_{1\, in}$, and the parameter $p$.  As can be seen from Eq. (\ref{alpha1}), the smaller is the value of initial anisotropy  $R_{1\, in}$, the longer it takes for the system to enter 
the second inflationary regime. 
We have checked that Eq. (\ref{alpha1}) gives a good estimate of $\alpha_1$ compared to the full numerical results. 

In {\bf Fig.} \ref{r-fig-sym} we have plotted our analytical solution for $\rho(\alpha)$ compared to the full numerical analysis. As can be seen they are in very good  agreement. Since $R_2 \ll R_1$ in this period, the value of $\alpha_1$ is independent of the gauge coupling $\e$ as can be seen from the plots. The left figures correspond to $\e=1$ whereas for the right figure, $\e=10^{-4}$.

In order for the phase transition to take place during the 60 observable e-folds, we require $\alpha_1 < 60$. This in turn impose the following lower bound on $R_{1\, in}$
\ba
\label{R1-lower}
R_{1\, in} > \frac{2(p-p_c)}{p^2} \exp \left[  -\frac{240(p-p_c)}{p_c}  \right] \, .
\ea
This is reasonable, because the smaller is the initial value of anisotropy, the longer it takes for the gauge field kinetic energy to become significant to affect the dynamics of the inflaton field. For values of $p_c$ comparable to $p$  Eq. (\ref{R1-lower}) can easily be satisfied and the first phase transition takes place during the physically relevant  window of inflation.

During the second phase, $\alpha > \alpha_1$, the solution (\ref{hat-rho1}) quickly approaches to its attractor solution and
\ba
\label{hat-rho2}
\left( \xi \hat \rho \right)^{2p} e^{-4 \alpha} \simeq \frac{6 (p-p_c)}{p^2 p_c} \, .
\ea
If we plug back this equation into the inflaton equation (\ref{chi-eq}), we find that the last term in   (\ref{chi-eq}) behaves as a constant source term with the magnitude $4 (p-p_c)/p p_c$ turned on at the time of  phase change. This explains the kink in $\rho$ behavior seen in {\bf Fig.}
\ref{r-fig-sym}.

\begin{figure}[t]
	\includegraphics[width=7cm]{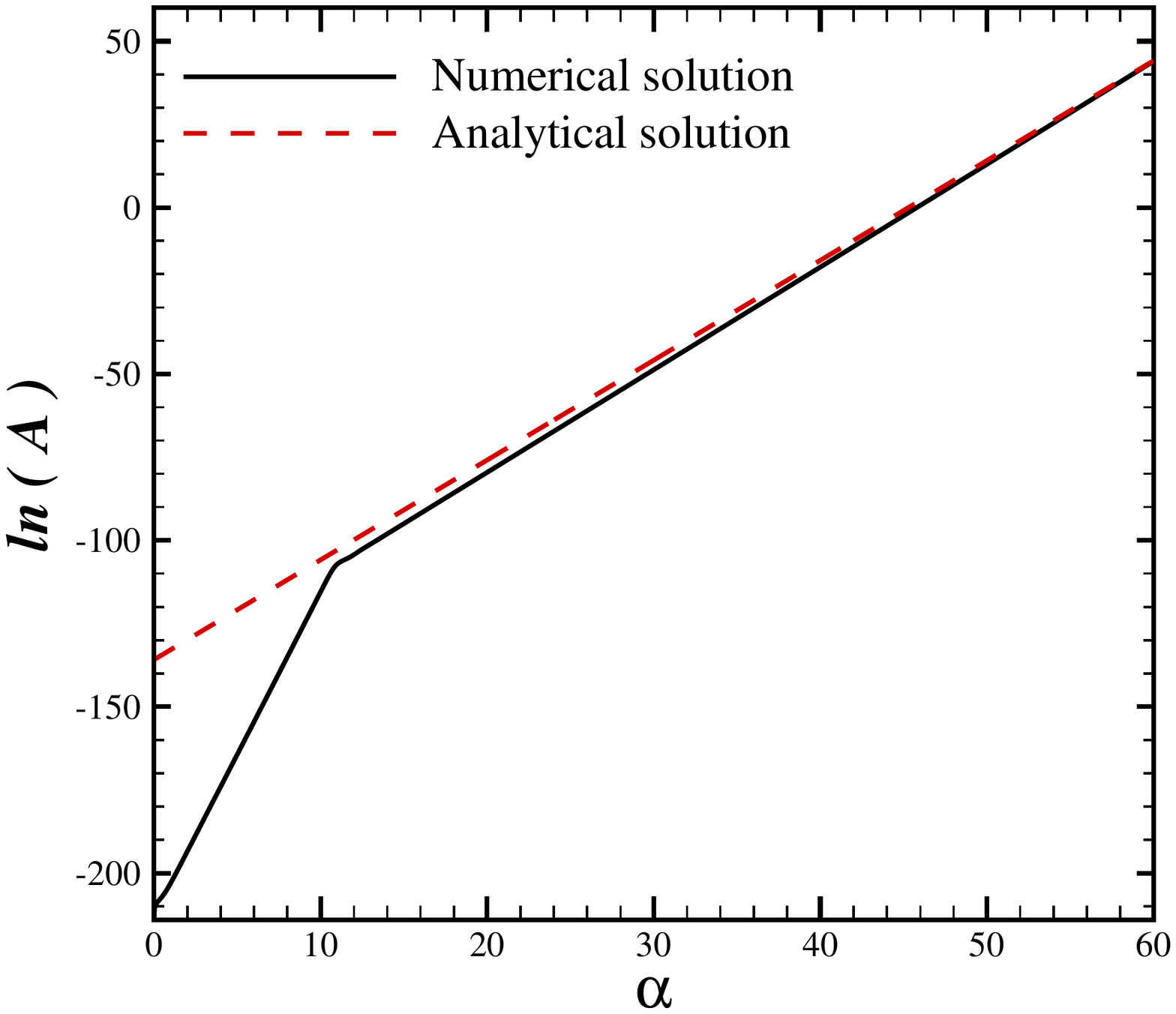} \hspace{0.5cm}
	\includegraphics[width=7cm]{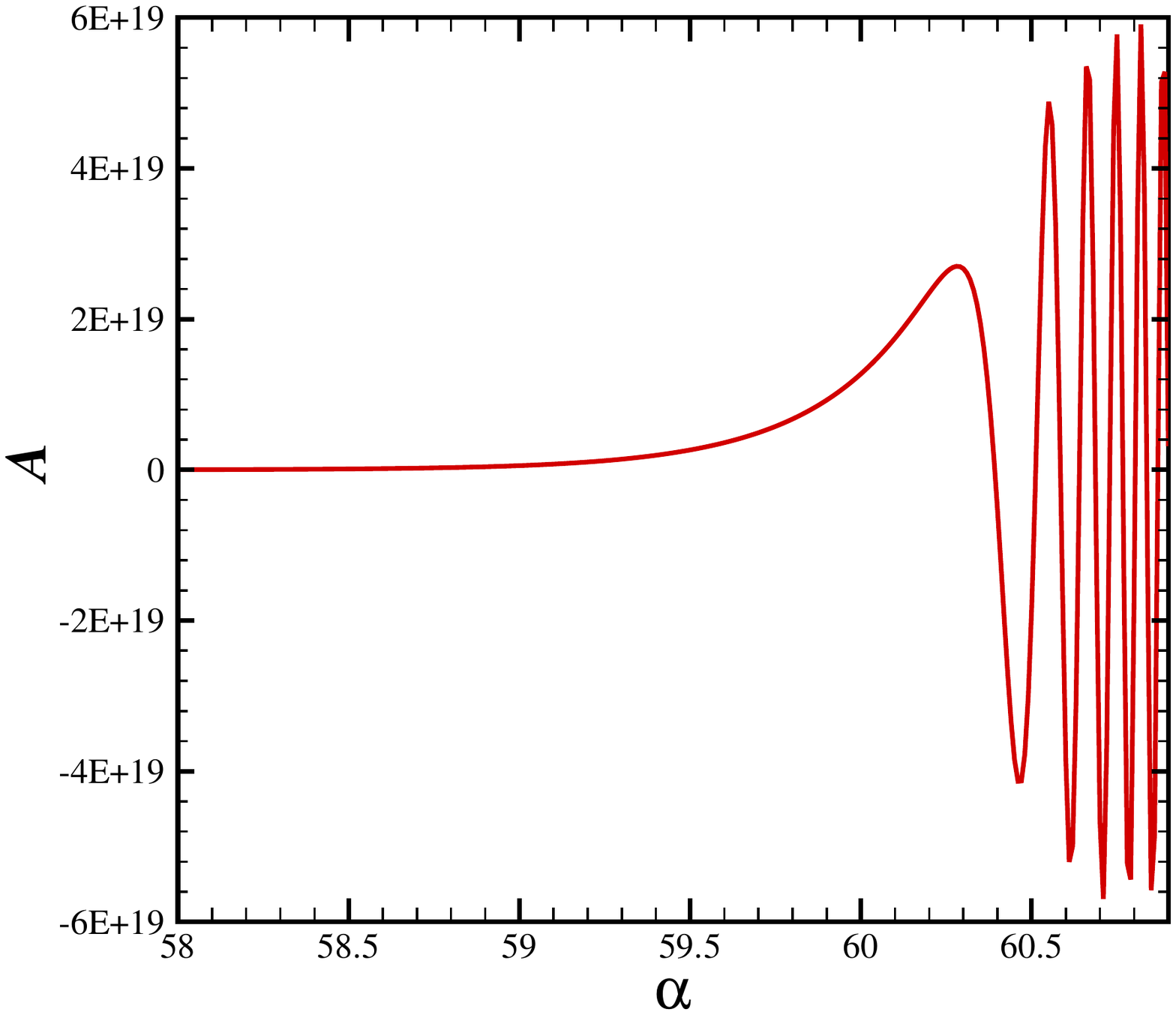}
	\caption{Here we plot  the evolution of the gauge field. The left graph represents $\ln A$ where
		the red dashed-dotted curve is our analytical solution Eq. (\ref{A3-Bes}) whereas the solid black curve is the full numerical solution. The agreement between our analytical solution Eq. (\ref{A3-Bes}) valid for the second and third phase and the full numerical result is good. Also the change of the slope of $\ln A$ form the first phase to the second phase is in good agreement with our other analytical result, Eq. (\ref{A-prime2}), valid for the first two phases. The right graph represents $A$ during the last few e-foldings. The oscillatory behavior suggested by Eq. (\ref{A3-Bes}) is clearly seen. The start of the third phase, corresponding to the first peak is well approximated by our analytical estimation Eq. (\ref{alpha2}). All parameters here are as in {\bf Fig.} \ref{eta-fig}.
		\vspace{0.5cm}
		\label{sym-A-fig} }
\end{figure}

It is also instructive to look into the gauge field evolution in this stage. Plugging Eq. (\ref{hat-rho1}) into Eq. (\ref{A-prime}) one can find an analytic expression for $\hat A'$ valid for both phase one and two with the asymptotic behavior 
\ba
\label{A-prime2}
\hat A' \simeq  \left\{
\begin{array}{c}
	(\xi \hat \rho_{in}^2 )^{p} \,  \exp\left[{\frac{(4 p- p_c) \alpha}{p_c} }\right]   \quad  \quad \quad \alpha< \alpha_1
	\\
	\frac{6 (p-p_c) }{p^2 p_c \xi^p} \, e^{3 \alpha} \, ~~~~~~~~~~~~~~ \quad \quad \quad \alpha> \alpha_1 
\end{array} \right. 
\ea
This indicates that the gauge field increases like $e^{3 \alpha}$ during the second phase
whereas it was increasing with a slightly higher rate, $ \exp\left[{\frac{(4 p- p_c) \alpha}{p_c} }\right]$, during the first inflationary stage. The change in the slope of the evolution of $\ln A$
clearly can be seen in {\bf Fig.} \ref{sym-A-fig}.

Finally we are in a position to find the form of $R_1$ and $\delta$. During the first inflationary stage, plugging Eqs. (\ref{A-prime2}) and (\ref{back-alpha0}) into  Eq. (\ref{R12}), yields
\ba
\label{delta-first}
\delta \simeq \frac{2}{3} R_1  \simeq \frac{2}{3} R_{1\, in} 
\exp\left[ \frac{4 (p-p_c) \alpha}{p_c} \right]  \, .
\ea
As expected, $\delta$ increases exponentially during the first inflationary stages with the initial amplitudes set by $R_{1\, in} $. During the second stage it reaches its  attractor value. Plugging Eq.  (\ref{hat-rho2}) into Eq. (\ref{R12}), during the attractor regime we have
\ba
\label{R1-hilltop}
\delta \simeq \frac{2R_1}{3} \simeq \frac{4(p-p_c)}{3p^2}.
\ea
This attractor value is fairly independent of  the initial conditions.  

In chaotic model studies in \cite{Watanabe:2009ct}  it was shown that $R_1$ follows the slow-roll parameters $R_1 \sim \epsilon$. Here we show this conclusion also holds for our case. To see this, note that in the slow-roll limit
$
\epsilon \simeq 2 R_1 + \frac{3\dot \rho^2}{2 V}.
$
Using Eq. (\ref{hat-rho2}) in Eq. (\ref{chi-eq}) one can approximately find that   
$\epsilon \simeq 2 R_1 + 4 \hat \rho^2 p_c/p^2$. Since  $\hat \rho <1$ during the second phase 
one concludes that 
\ba
\label{epsilon-R1}
\epsilon \gtrsim 2 R_1 \, .
\ea

\begin{figure}[t]
    \centering
	\includegraphics[width=7cm]{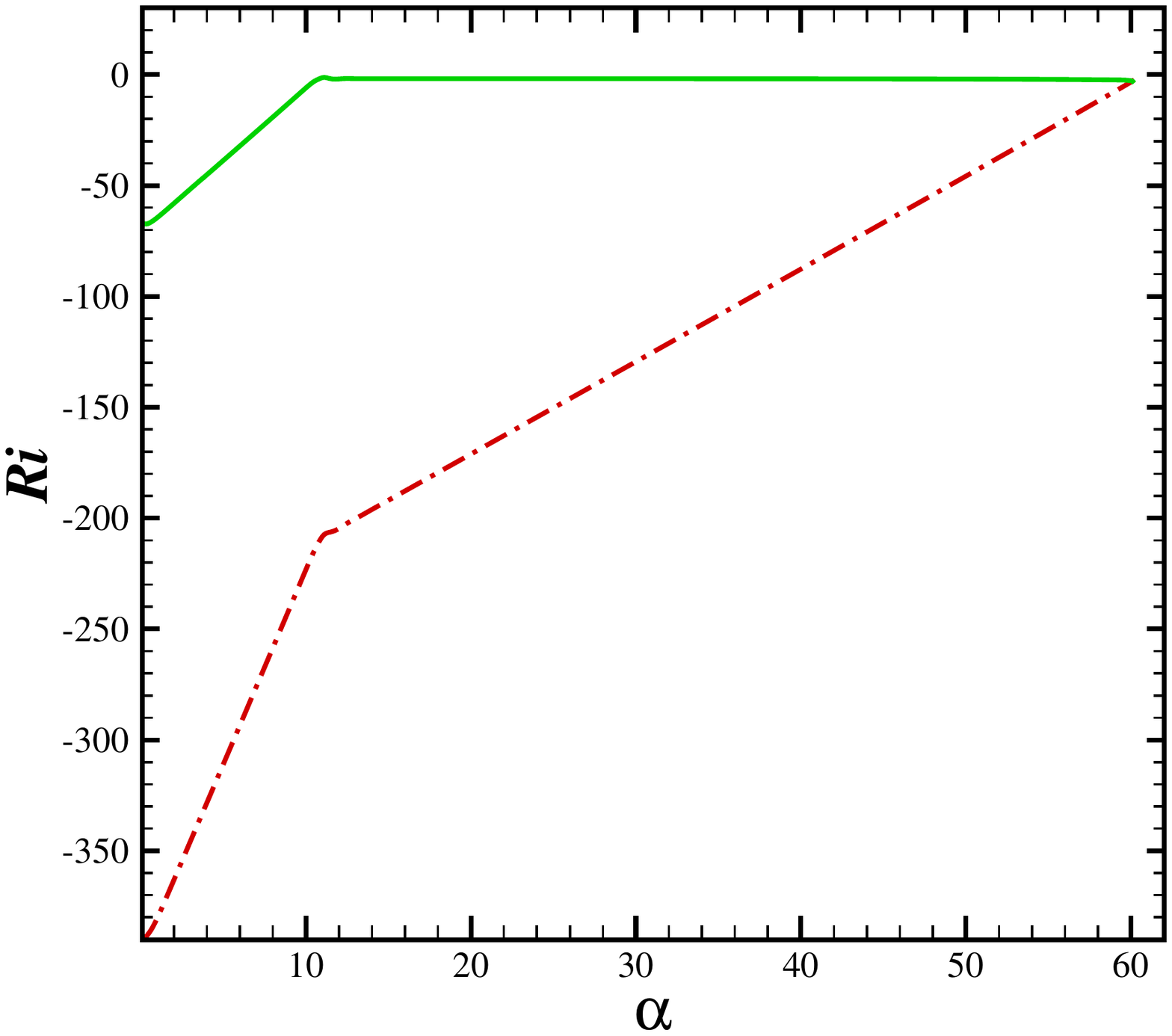} \hspace{0cm}
	\includegraphics[width=7cm]{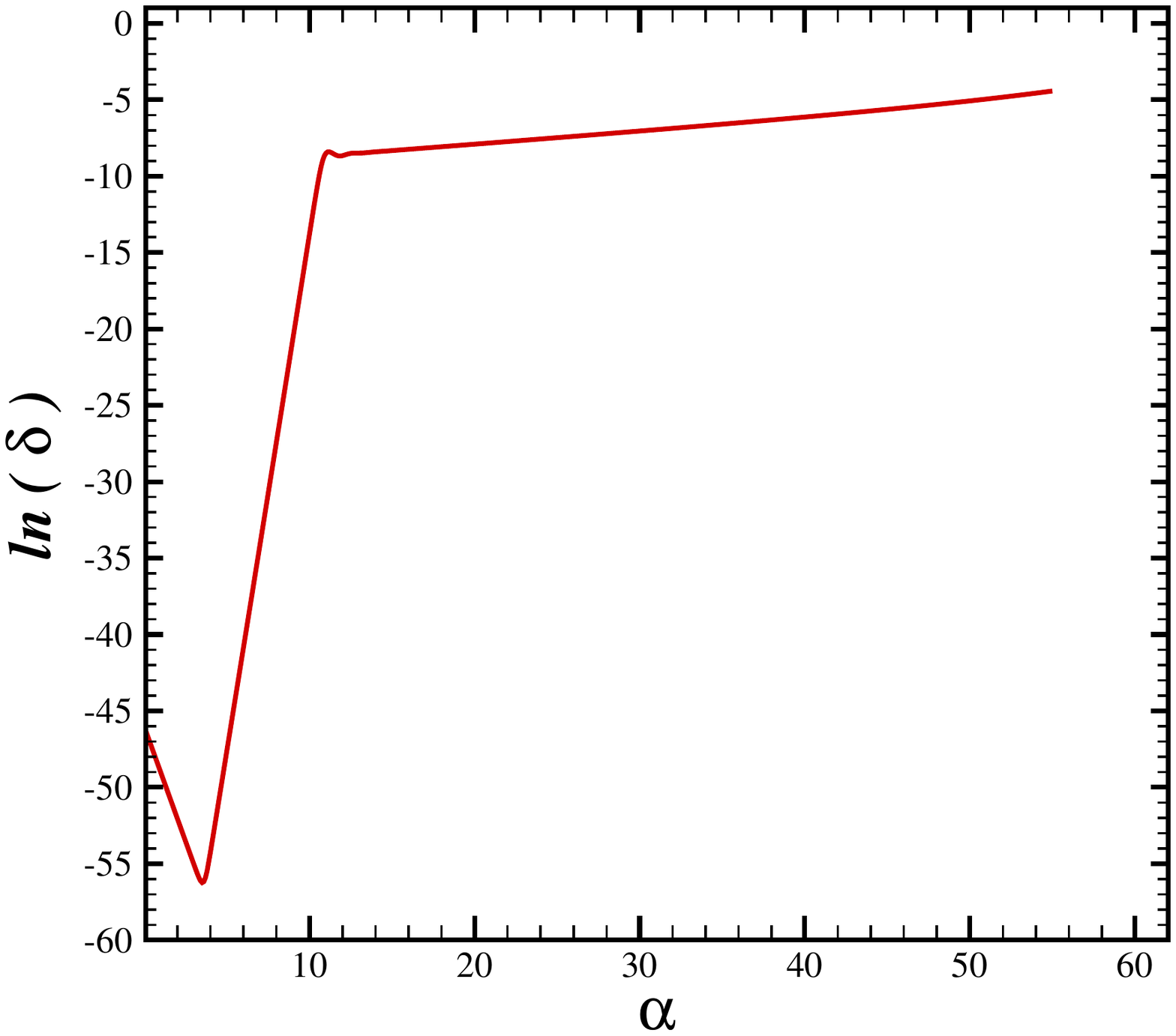} \hspace{1cm}
	\caption{In left figure we plot  $\ln (R_1/\epsilon) $ ( upper solid green curve) and $\ln( R_2/\epsilon)$ (lower dashed-dotted red curve). The phase change takes place at $\alpha_1 \simeq 10$ followed by the attractor regime denoted by the almost horizontal line where $ R_1 \propto \epsilon $.
		As explained in the text, $R_2$ is very small compared to $R_1$ until the very end of inflation when they become comparable and inflation ends shortly after that. Right: $\ln \delta$ is presented. The attractor behavior during the second inflationary stage is clear.
		All parameters here are as in {\bf Fig.} \ref{eta-fig}. }
	\vspace{0.5cm}
	\label{Ri-delta-fig}
\end{figure}
In {\bf Fig.} \ref{Ri-delta-fig} we have plotted the ratio $R_1/\epsilon$ and the anisotropy $\delta$. As can be seen, our analytical formulae Eqs. (\ref{epsilon-R1})  and (\ref{R1-hilltop}) are in good agreement with the full numerical results. As explained before, the solution during the second phase quickly reaches the attractor regime where $R_1\simeq \epsilon \simeq \delta$ and the fraction of gauge field energy density to the total energy density  in the Friedmann equation is at the level of slow-roll parameter.  The attractor phase can clearly be seen from the behavior of $R_1$ and $\delta$ in {\bf Fig.} \ref{Ri-delta-fig}.

As the gauge field $A$ increases exponentially the effective potential for the inflaton field increases as $\e^2 A^2 \rho^2 e^{-2 \alpha} \propto \e^2  \rho^2 e^{4 \alpha}$ and  the slow-roll condition quickly terminates at the final stage of inflation. This is the third stage where $R_2 $ becomes comparable to $R_1$. Below we study this phase in some details.

\subsection{Final stage of inflation}

Now we consider the final stage of inflation when the right hand side of Eq. (\ref{back-A-eq4}) can not be neglected. Using Eq. (\ref{hat-rho2}) in Eq. (\ref{back-A-eq4}), the equation of motion for the gauge field can be approximated to
\ba
\label{A-eq1}
\hat A''  - 3  \hat A' +  \beta  e^{4 \alpha} \hat A=0
\ea
where the dimensionless parameter $ \beta$ is defined via
\ba
\label{beta}
\beta \equiv \frac{36 \e^2 \, (p-p_c)}{\lambda \, p^2 p_c^2 \, \xi^{2p}} \, .
\ea
The solution to this equation is in the form of Bessel functions 
\ba
\label{A3-Bes}
\hat A = e^{\frac{3 \alpha}{2}} \left[ a_1 J_{3/4} \left(  \frac{ \sqrt { \beta}}{2} e^{2 \alpha}  \right) + a_2 Y_{3/4} \left(  \frac{ \sqrt { \beta}}{2} e^{2 \alpha}   \right) \right] \, ,
\ea
where $a_1$ and $a_2$ are the constants of integration. 

From the form of Eq. (\ref{A-eq1}) it is seen that the third inflationary phase starts when the last term in Eq. (\ref{A-eq1}) is comparable to the second term. This means that 
$\sqrt \beta e^{2\alpha_2} \simeq 1$ where $\alpha=\alpha_2$ is the start of the third inflationary stage. This gives 
\ba
\label{alpha2}
\alpha_2 \simeq  \frac{1}{4} \ln \left[ \frac{\lambda \, p^2 p_c^2\,  \xi^{2p}}{36\,  \e^2 \, (p-p_c)} \right] \, .
\ea
We have checked numerically that this expression gives a very good estimate for $\alpha_2$, the onset of transition from the second inflationary stage to the third inflationary stage. Shortly after $\alpha > \alpha_2$, the argument of the Bessel function  exponentially increases and  the gauge field starts to oscillate. This in turn triggers 
a sharp increase in the slow-roll parameters $\epsilon$ and $\eta$ and inflation ends abruptly. 
This can be seen  in the plot of $\eta$ shown in {\bf Fig.} \ref{eta-fig}.

For the consistency of our setup we require that $\alpha_1 < \alpha_2$, i.e. the third inflationary stage takes place after the second inflationary stages. Comparing Eqs. (\ref{alpha2}) and (\ref{alpha1}), and assuming $(p-p_c)/p_c \sim 1$, one requires
\ba
\e^2 \ll \lambda p^4 \xi^{2p} R_{1\, in} \, .
\ea
Because $\hat \rho_{in} \ll 1$ and $p\gg 1$, this condition can easily be met.

The condition $\sqrt \beta e^{2\alpha_2} \simeq 1$  indicates that during the final stage of inflation the arguments of the Bessel functions in Eq. (\ref{A3-Bes}) are bigger than unity while during the first two inflationary stage the arguments of the Bessel functions are small. Using the small argument limit of the Bessel function,  $J_{3/4} (x)  \sim  x^{3/4}$ and $Y_{3/4}(x) \sim x^{-3/4}$ for $x\ll 1$ one concludes that the term containing $Y_{3/4}$ decays quickly as inflation proceeds and the term containing $J_{3/4}$ survives in Eq. (\ref{A3-Bes}). 
More specifically, $J_{3/4} \left( \sqrt \beta\,  e^{2\alpha}/2 \right) \simeq 
(\sqrt \beta/2)^{3/4}  e^{3 \alpha/2}$ and comparing this with Eq. (\ref{A-prime2}) during the second inflationary stage  one can fix  the coefficient $a_1$ to obtain
\ba
\label{A3-Bes}
\hat A \simeq  \frac{2^{5/4} (p-p_c)}{ p^2 p_c \, \xi^p \, \beta^{3/8}} \, 
e^{3 \alpha/2}  J_{3/4} \left(  \frac{ \sqrt { \beta}}{2} e^{2 \alpha}  \right) \, .
\ea
Note that this expression works for the second and the third inflationary stages whereas the formula Eq. (\ref{A-prime2}) works for the first two inflationary stages. We have checked that 
Eq. (\ref{A3-Bes}) is in good qualitative agreement with the full numerical analysis. In {\bf Fig}
\ref{sym-A-fig} we have compared Eq. (\ref{A3-Bes}) with the full numerical result and the agreement between them is good. Also in {\bf Fig}
\ref{sym-A-fig} we have plotted the behavior of $A$ for the last few e-foldings. The start of the third phase is when the argument of the Bessel function in Eq. (\ref{A3-Bes}) becomes at the order of unity given by Eq. (\ref{alpha2}). This corresponds to the first peak in the plot of $A$
in the right figure of {\bf Fig} \ref{sym-A-fig}.

It is also instructive to compare $R_2$ with $R_1$ during the final stage of inflation. 
As explained before, $R_1$ and $R_2$ measure respectively the gauge field kinetic energy and potential energy compared to the background inflationary potential. Physically, we expect that during the final stage of inflation $R_2$ rises quickly and becomes comparable to $R_1$ and $\epsilon$. 
In {\bf Fig} \ref{Ri-delta-fig} we have compared $R_2$ with $R_1$ and $\delta$. Initially 
$R_2$ is very small, but during the final stage of inflation $R_2$ rises quickly and becomes comparable to $R_1$. Physically this means that the inflaton mass receives a time-dependent contribution of the form $\e^2 A^2 \rho^2 e^{-2 \alpha}$ and the slow-roll conditions are violated soon after the gauge field starts to oscillate.
This  conclusion is supported in both {\bf Fig} \ref{sym-A-fig} and {\bf Fig} \ref{eta-fig}.

Finally it is also instructive to look into the behavior of the inflaton field as a function of the the strength of the gauge coupling $\e$. As argued before, during the first two inflationary stages 
$\e$ does not play important roles and the evolution of the inflation proceeds as in $\e=0$. In particular, the position of the first kink, $\alpha_1$, is quite insensitive to the value of $\e$.
On the other hand, $\e$ controls the end of the second inflationary  phase, $\alpha_2$, but only logarithmically. In the right plot of {\bf Fig.} \ref{r-fig-sym} we have changed $\e$ by four orders of magnitudes. Correspondingly, $\alpha_2$, and the total number of e-foldings 
changed by 4. This is consistent with our analytical formula Eq. ({\ref{alpha2}})
which for $\e \rightarrow 10^{-4} \e$  predicts an increase of e-foldings of 
$( \ln 10^{8})/4 \simeq 4.6 $.

\section{Charged Hybrid Inflation}
\label{hybrid}

In the previous example the inflaton field was a complex field and was responsible for the symmetry breaking. Now we consider the case where the inflaton field is real, while the symmetry breaking 
is controlled by another complex scalar field, the waterfall field. The action is
\ba
\label{action3} S=\int d^4 x  \sqrt{-g} \left [ \frac{M_P^2}{2} R - \frac{1}{2} \partial_\mu \phi
\,  ^\mu\phi- \frac{1}{2} D_\mu \psi
\,  D^\mu\bar\psi - \frac{f^{2}(\phi)}{4} F_{\mu \nu} F^{\mu
	\nu}- V(\phi, \psi, \bar \psi) \right]  \, .
\ea
As explained above, $\phi$ is the inflaton field while $\psi$ is the complex waterfall field. 

The potential is as in standard hybrid inflation \cite{Linde:1993cn}
\ba
\label{pot3} V(\phi, \psi, \bar \psi)=\frac{\lambda}{4 }  \left(  |\psi|^2 - \frac{M^2}{\lambda} \right)^2 + \frac{g^2}{2} \phi^2 |\psi|^2 + \frac{m^2}{2} \phi^2 \, .
\ea   
We are interested in the configuration where the potential is axially symmetric and $V(\psi, \bar \psi , \phi)= V(\chi, \phi)$ where $ \psi(x) = \chi(x) \,  e^{i \theta(x)}$. 
Following the same metric ansatz as in Eq. (\ref{metric4}) and taking $A_{\mu}=(0,A(t),0,0)$ the equations of motion are
\ba
\label{back-A-eq34}
\partial_t{\left(  f^2(\phi) e^{\alpha + 4 \sigma} \dot A \right)}& =& - \e^2 \chi^2 e^{\alpha + 4 \sigma}  A \\
\label{back-phi-eq34}
\ddot\phi+3\dot \alpha\dot \phi+ \phi(m^2+g^2\chi^2) -f^2(\phi)_{,\phi}\dot A^2 e^{-2\alpha+4\sigma}&=&0  \\
\label{back-chi-eq34}
\ddot\chi+3\dot \alpha\dot \chi+ \left(\frac{\lambda}{4}(\chi^2-\frac{M^2}{\lambda})+g^2\phi^2
+e^2 A^2 e^{-2\alpha+4\sigma}\right) \chi&=&0  \\
\label{Ein1-eq34}
\dot
\phi^2+\dot \chi^2+2V(\phi,\chi)+ \left(f^2(\phi)\dot
A^2 +\e^2\chi^2A^2 \right) e^{-2\alpha+4\sigma}
&=&
6 M_P^2 \left(\dot \alpha^2-\dot \sigma^2 \right)  \\
\label{Ein2-eq34}
V(\phi,\chi)+  \left(  \frac{1}{6}f^2(\phi)\dot
A^2+\frac{\e^2\chi^2}{3}A^2  \right)e^{-2\alpha+4\sigma}
&=& M_P^2 \left( \ddot \alpha    + 3 \dot \alpha^2 \right)  \\
\label{anisotropy-eq34}
\left(\frac{1}{3}f^2(\phi)\dot A^2  -\frac{\e^2\chi^2}{3}A^2 \right) e^{-2\alpha+4\sigma}
&=& M_P^2\left( 3\dot \alpha \dot \sigma+ \ddot \sigma\right)\, .
\ea 

From Eq. (\ref{Ein1-eq34}), the total energy density determining  the expansion rate of the universe is given by
\ba
\label{energy24}
{\cal E}=V(\phi,\chi) + e^{-2\alpha+4\sigma} \left( \frac{1}{2}f^2(\phi)     \dot
A^2+\frac{\e^2\chi^2}{2} A^2 \right) \, .
\ea

The interesting new effect is that the gauge coupling $\e$ induces a new time-dependent mass term for the waterfall field in the form $\e^2 e^{-2 \alpha} A^2 \,  \chi^2$. This can be seen both
in total energy density and also in equation governing the dynamics of the waterfall field, Eq. (\ref{back-chi-eq34}).  As in standard hybrid inflation we work in the vacuum dominated regime where the waterfall field is very heavy during inflation so $\chi$ quickly settles down to its instantaneous minimum $\chi=0$ during inflation. In standard hybrid inflation  models, inflation ends when  inflaton field reaches a critical value,
$\phi=\phi_c \equiv \frac{M}{g}$, where the waterfall field becomes tachyonic and rolls down very quickly to its global minimum $\psi=\mu\equiv M/\sqrt{\lambda}, \phi=0$ ending inflation very efficiently. In our model to find the moment when the waterfall field becomes tachyonic, let us calculate the $\chi$ field  effective mass
\ba
\label{chi-mass4}
\frac{\partial^2 V}{\partial \chi^2}\large|_{\chi=0}  = g^2 (\phi^2 - \phi_c^2) + \e^2 e^{- 2 \alpha} A^2 \, .
\ea
In the absence of the gauge field, the onset of waterfall field instability is when $\phi= \phi_c$. However, in the presence of the gauge field the time when the  tachyonic instability is triggered 
is modified. Indeed, if either of $\e$  or the background gauge field $A$ are very large, then the onset of tachyonic instability can be significantly altered and inflation will end before $\phi$ reaches $\phi_c$. This can have profound effects on the dynamics of waterfall phase transition and symmetry breaking \cite{Felder:2000hj, Dufaux:2010cf}. Furthermore, one needs to revisit the question of tachyonic preheating in this case.

The condition of waterfall phase transition, Eq. (\ref{chi-mass4}), can be rewritten as
\ba
\label{transition4}
\hat \phi^2 + \frac{\e^2}{g^2} \hat A^2 e^{-2\alpha} -1=0
\ea
where we have defined the dimensionless fields $\hat \phi \equiv \phi/\phi_c$ and 
$\hat A \equiv A/\phi_c$. In this notation, in the absence of gauge field the waterfall phase transition happens at $\hat \phi=1$. If the gauge field is expected to play important roles in determining the dynamic of waterfall phase transition then one requires the second term in Eq. (\ref{transition4}) to become at the order of unity at the time of transition. Below we will study under what conditions on model parameters this condition can be met.

As mentioned above, we assume the waterfall  field is very heavy during inflation and the potential driving inflation is 
\ba
V\simeq \frac{M^4}{4\lambda}+\frac{1}{2}m^2\phi^2  \, .
\ea
In order for the inflaton field to be light during inflation so the slow-roll conditions are met we need $p_c\gg1$ where now $p_c$ is defined via 
\ba
p_c \equiv \frac{M^4}{2 \lambda m^2 M_P^2} \, .
\ea
Furthermore, the assumption that the waterfall field is very heavy during inflation requires
$\lambda M_P^2 /M^2 \gg1$. Finally, the condition of vacuum domination during inflation 
is met if $\lambda/g^2 \ll M^2/m^2$.

As in our previous symmetry breaking example we assume the anisotropies are negligible corresponding to $\delta \lesssim \epsilon$ throughout inflation 
so the background expansion is still given by Eq. (\ref{Hubble4}). 
As in symmetry breaking example, inflation starts with the isotropic limit where $R_1 \ll \epsilon$ at the early stage of inflation. As inflation proceeds, $R_1$ rises quickly and we enter the second phase of inflation where the gauge field dynamics affect the evolution of the inflaton field in Eq. (\ref{back-phi-eq34}). However, unlike the previous example, the final stage of inflation will be very different where now inflation ends violently once the waterfall field becomes tachyonic.

During the first phase of inflation, the inflaton dynamics is
\ba
\label{phase1-hybrid4}
\phi' + \frac{2 \phi}{p_c} =0 \rightarrow  \phi \simeq \phi_{in} e^{-2 \alpha/p_c}  \, ,
\ea
where $\phi_{in}$ is the initial value of the inflaton field.  The number of e-foldings is 
\ba
\alpha = -\frac{p_c}{2} \ln \left( \frac{\phi}{\phi_{in}} \right)  \, .
\ea

Now we need to determine the form of the gauge kinetic coupling, $f(\phi)$, such that 
$R_1$ rises quickly during the second phase of inflation. Since during inflation $\chi=0$, then 
the gauge field equation (\ref{back-A-eq34}) can easily be solved with $A' \sim e^{-\alpha} f(\phi)^{-2}$. Consequently, $R_1$ scales like $R_1 \sim f(\phi)^{-2} e^{-4 \alpha} \sim 
f(\phi)^{-2} \phi^{2 p_c}$. Therefore, for the critical coupling $f_c \sim \phi^{p_c}$,  $R_1$
remains fixed and the fraction of the gauge field kinetic energy to the background energy density remains fixed. As in previous example, we allow for the following coupling
\ba
\label{f24}
f(\phi) =\left(  \frac{\phi}{\phi_c}\right)^p = \hat \phi^{p} \, ,
\ea
with $p> p_c$ so the energy density of the gauge field increases as inflation proceeds. As in the symmetry breaking case, we expect to enter the attractor regime where $R_1 \simeq \epsilon$ till inflation ends via tachyonic instability. With gauge kinetic coupling given by Eq. (\ref{f24}), the gauge field equation can be solved easily and
\ba
\label{A-prime-hybrid4}
\hat A' = e^{-\alpha} \left( \xi \hat \phi^2 \right)^{-p} \, ,
\ea
where $\xi$ is a constant of integration.  Plugging  the gauge field solution into the inflaton field equation yields
\ba
(\hat \phi^2)' + \frac{4 \hat \phi^2}{p_c}  - \frac{2p \,  \xi^{-2p }}{3} ( \hat \phi^2)^{-p} e^{-4\alpha} =0 \, .
\ea
In comparison to symmetry breaking analysis, it is interesting to note that 
one can reproduce the previous results with the replacements $p \rightarrow -p$ and $p_c \rightarrow -p_c $.  Like in symmetry breaking example, this equation can be solved easily with the solution during the first stage of inflation given by Eq. (\ref{phase1-hybrid4}) while during the second stage of inflation, for $\alpha > \alpha_1$, one obtains the attractor solution
\ba
\label{hat-phi-eq4}
\left( \xi \hat \phi \right)^{2p} e^{4 \alpha} \simeq \frac{p^2 p_c}{6 (p-p_c)} \, .
\ea
Furthermore,  plugging Eq. (\ref{hat-phi-eq4}) into gauge field equation (\ref{A-prime-hybrid4}) yields
\ba
\label{A-hybrid4}
\hat A \simeq \frac{2 (p-p_c) \xi^{p}}{p^2 p_c} e^{3\alpha} \, .
\ea
As expected, the gauge field increases exponentially.

We can also calculate the level of anisotropy in this scenario. For the first inflationary stage, $\delta$ is given as in Eq. (\ref{delta-first}) whereas during the second stage it reaches the attractor value
\ba
\label{R1-hybrid4}
\delta \simeq \frac{2}{3} R_1 \simeq 
\frac{4(p-p_c)}{3 p^2} \frac{\lambda m^2}{g^2 M^2} \, .
\ea
As expected, $R_1$ reaches the scaling solution during the second inflationary stage.  Like in symmetry breaking case one obtains $\epsilon \gtrsim 2 R_1$.

Now we have all the tools to answer our original question that under what conditions the gauge field can play a role in triggering the water field phase transition. As explained below Eq. (\ref{transition4}), this can happen when the combination $(\e^2/g^2) \hat A^2 e^{-2\alpha_f}$
is comparable to unity at the end of inflation when $\alpha= \alpha_f \simeq 60$ and $\hat \phi=1$. Using Eqs. (\ref{hat-phi-eq4}) and (\ref{A-hybrid4}),  and noting that by definition 
$\hat \phi=1$ when $\alpha= \alpha_f$, one obtains the interesting results 
\ba
\frac{\e^2}{g^2} \hat A^2 e^{-2\alpha}\large|_{\alpha_f} \simeq  \frac{2 \e^2 (p-p_c)}{3g^2 p^2 p_c} \sim \frac{\e^2}{p^2 g^2} \, .
\ea
This indicates that for $\e \ll p \, g$ the gauge field does not play important role in triggering the waterfall field tachyonic instability and inflation ends as in standard hybrid inflation. However, for $\e \gg p \, g$, then the gauge field shuts off inflation before $\phi =\phi_c$
and the dynamics of the waterfall phase transition, symmetry breaking and tachyonic preheating would be drastically different than what happens in standard hybrid inflation. Because of the inhomogeneous end of inflation  large non-Gaussianities can be produced at the end of inflation in the light of \cite{Yokoyama:2008xw}. 
\section{Chaotic Inflation}
\label{chaotic}

As our final example, here we briefly study the case of chaotic inflation. Many of our previous results also apply here. For this purpose, we will be brief here only emphasizing our main results  and compare them to the results of \cite{Watanabe:2009ct} where they assumed $\e=0$.  
To be specific, we concentrate on the quadratic potential 
\ba
V= \frac{m^2}{2} |\phi|^2 \, .
\ea 
The equation of motions are the same as in symmetry breaking example.  During the first inflationary stage, the relevant equations are
\ba
\label{back-chaotic}
\dot \alpha^2 \simeq \frac{m^2 \rho^2}{6 M_p^2} \quad , \quad
3 \dot \alpha \dot \rho + m^2 \rho \simeq 0 \, ,
\ea 
where $\rho \equiv | \phi|$. Note that in our slow-roll limit the Friedmann equation above also holds throughout the inflationary period. These equations can easily be solved to give
\ba
\label{rho-chaotic1}
\rho^2 \simeq \rho_{in}^2 - 4 M_P^2 \alpha \, .
\ea

Now we determine the form of the desired gauge kinetic coupling. As in previous examples 
we start with an almost isotropic configuration with negligible anisotropies such that the gauge field contributions in expansion rate Eq. (\ref{Ein1-eq4}) and the Klein-Gordon equation (\ref{back-rho-eq4}) are negligible. However, we would like to allow the gauge field kinetic energy to increase such that $\delta$ is within the observational bounds. As  can also be seen from the solution of Eq. (\ref{back-A-eq4}), in order for the gauge field energy density to remain constant during the first inflationary stage  the gauge kinetic coupling should have the critical form $f_c \sim e^{2 \alpha}$. For the power law inflationary potentials with $V \propto \rho^n$ the critical gauge kinetic coupling  is given by $f_c \sim e^{ \rho^2 /n M_P^2}$. However, as in \cite{Watanabe:2009ct},  for the energy density of the gauge field to increases during the course of inflation we consider $f = e^{c \rho^2/n M_P^2}$ with $c>1$. For our specific example with $n=2$ we consider the gauge coupling 
\ba 
f = e^{c \rho^2/2M_P^2} \, .
\ea

During the first two inflationary stage one can neglect the right hand side of Eq. (\ref{back-A-eq4}) and the gauge field evolution is given by
\ba
\label{Adot-chaotic}
\dot A = p_A \exp\left[  - \alpha -  \frac{c\rho^2}{M_p^2} \right] \, ,
\ea
where $p_A$ is a constant of integration defined in \cite{Watanabe:2009ct}. Plugging this into the inflaton equation results in
\ba
\label{rho-chaotic-eq2}
( \rho^2)' + 4 M_P^2  - \frac{4 c \, p_A^2}{m^2}  \exp\left[  -4 \alpha -  \frac{c\rho^2}{M_p^2} \right] = 0\, .
\ea 
\begin{figure}[t]
	\includegraphics[width=7cm]{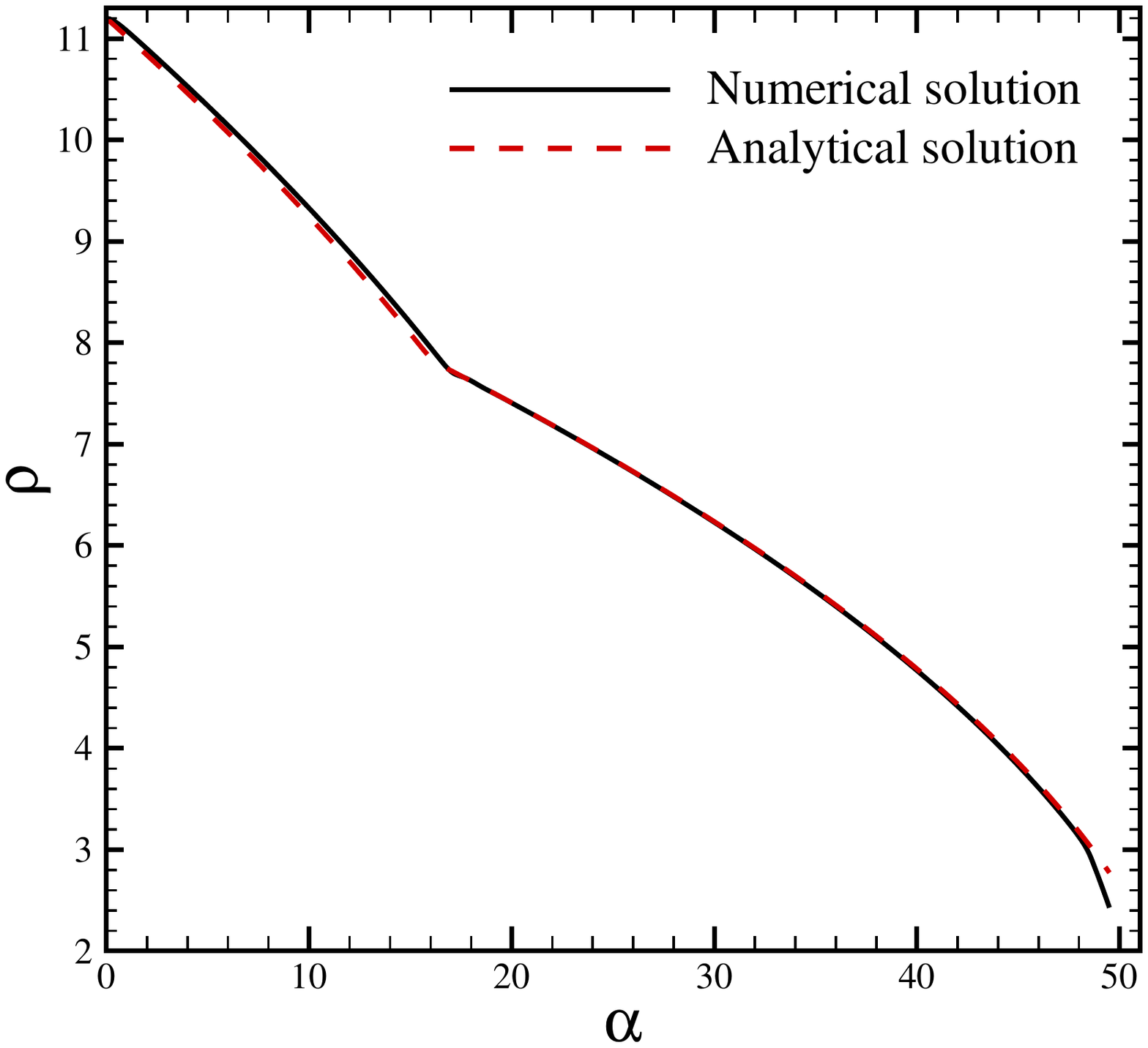} \hspace{1cm}
	\includegraphics[width=7cm]{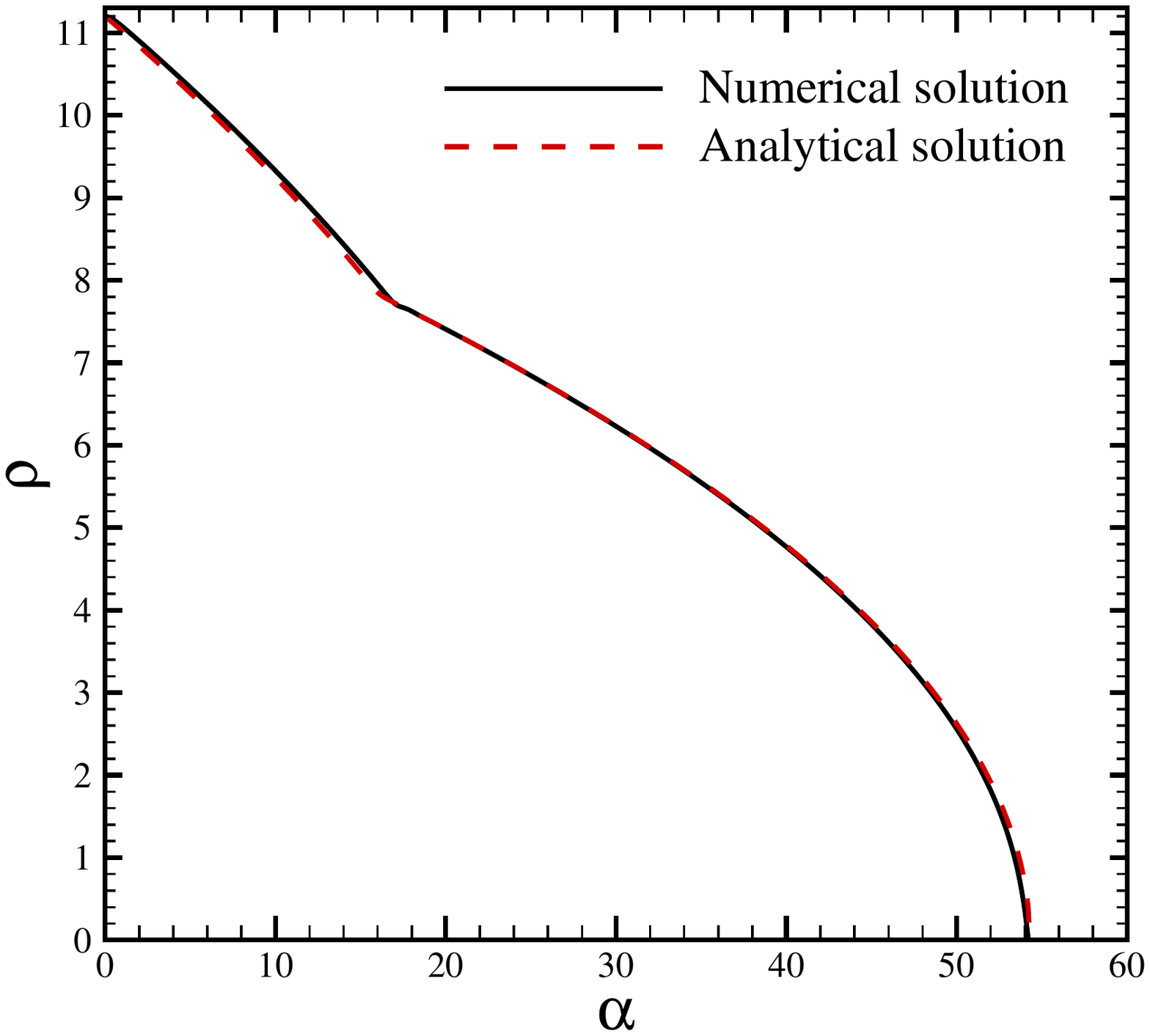}
	\caption{Here we plot the evolution of $\rho(\alpha)$ in chaotic inflation. The dashed-dotted red curve is the analytical solution whereas the solid black curve is the full numerical result.
		As can be seen, the agreement between them is very good. The left figure corresponds to $\e=0.1$ whereas for the right figure $\e=0$.  As explained below, the position of the first kink is independent of the value of $\e$ and is well approximated by Eq. (\ref{alpha1-chaotic}). 
		However, the total number of e-foldings  depends logarithmically  on $\e$. Other parameters are $m=10^{-6} M_P$, $\rho_{in} = 11.2 M_P$ and $c=2.5$. }
	\vspace{0.5cm}
	\label{r-fig-chaotic}
\end{figure}

During the first inflationary stage, when the third term in Eq. (\ref{rho-chaotic-eq2}) is negligible compared to the second term,  the solution is given  by Eq. (\ref{rho-chaotic1}). The phase change happens when the last term above becomes comparable to the second term. Because of the attractor mechanism the solution rapidly converges to
\ba
\label{rho-chaotic2}
\exp\left(  -4 \alpha -  \frac{c\rho^2}{M_p^2} \right) \simeq \frac{m^2 M_P^2 (c-1)}{c^2 p_A^2} \, .
\ea
Note that this solution is analogous of Eq. (\ref{hat-rho2}) for the symmetry breaking example.
Now plugging Eq. (\ref{rho-chaotic2}) into Eq. (\ref{rho-chaotic-eq2}) yields
\ba
3 \dot \alpha \dot \rho = -\frac{m^2}{c} \rho \, .
\ea
This should be compared to the $\rho$ equation during the first inflationary phase where now $m^2 \rightarrow m^2/c$. This sudden change of mass induces a jump in $\ddot \rho$ which clearly can be seen in {\bf Fig.\ref{r-fig-chaotic} }.

Plugging solutions (\ref{rho-chaotic2}) and (\ref{rho-chaotic1}) into Eq. (\ref{Adot-chaotic})
one finds that $\dot A$ scales like $e^{  (4 c-1) \alpha}$ and $e^{3 \alpha}$
during the first and second inflationary stage respectively. The behavior of the gauge field during the first and second inflationary stages 
has a profile very similar to the plot on the left hand side of {\bf Fig.} \ref{sym-A-fig}. 
It is also instructive to look into the anisotropy parameter. During the first inflationary stage
one has
\ba
\delta \simeq \frac{2}{3} R_{1\, in} \exp \left[ 4 (c-1) \alpha 
\right] \, ,
\ea
whereas during the second inflationary stage it reaches the attractor value
\ba
\delta \simeq \frac{2 M_P^2}{3 \rho^2} \frac{c-1}{c^2} \simeq \frac{c-1}{3c} \epsilon \, .
\ea
As expected, $\delta \simeq \epsilon$ during the attractor regime.

Note that Eq. (\ref{rho-chaotic-eq2}) can be solved with the general answer 
\ba
\label{rho-chaotic-full}
\rho^2 \simeq -\frac{4 M_P^2 \alpha}{c} - \frac{M_P^2}{c} \ln \left[ \frac{4 (c-1)}{ K - e^{- 4 (c-1) \alpha} \left( K- 4 (c-1) e^{c \rho_{in}^2/M_P^2} \right)} \right] \, ,
\ea
where $K\equiv 4 c^2 p_A^2/m^2 M_P^2$. This has Eqs. (\ref{rho-chaotic1}) and (\ref{rho-chaotic2})
as the two limiting solutions. In {\bf Fig.} \ref{r-fig-chaotic} we have compared this analytical solution with the full numerical results and the agreement between them is very good. The time of the first phase change, $\alpha_1$, is when the second term in the denominator above becomes comparable to the first term which results in  
\ba
\label{alpha1-chaotic}
\alpha_1 \simeq  \frac{c \rho_{in}^2}{4(c-1)} + \frac{1}{4(c-1)} \ln \left[ \frac{m^2 M_P^2 (c-1)}{c^2 p_A^2} \right]
\simeq \frac{1}{4(c-1)} \ln \left[ \frac{c-1}{c^2} \frac{M_P^2}{\rho_{in}^2 R_{1\, in}}
\right] \, .
\ea
This indicates that the smaller is the value of the initial anisotropy $R_{1\, in}$, the longer it takes for the system to enter the attractor regime.  We have checked that this analytical expression gives a good estimate of $\alpha_1$.

\begin{figure}[t]
	\includegraphics[width=7cm]{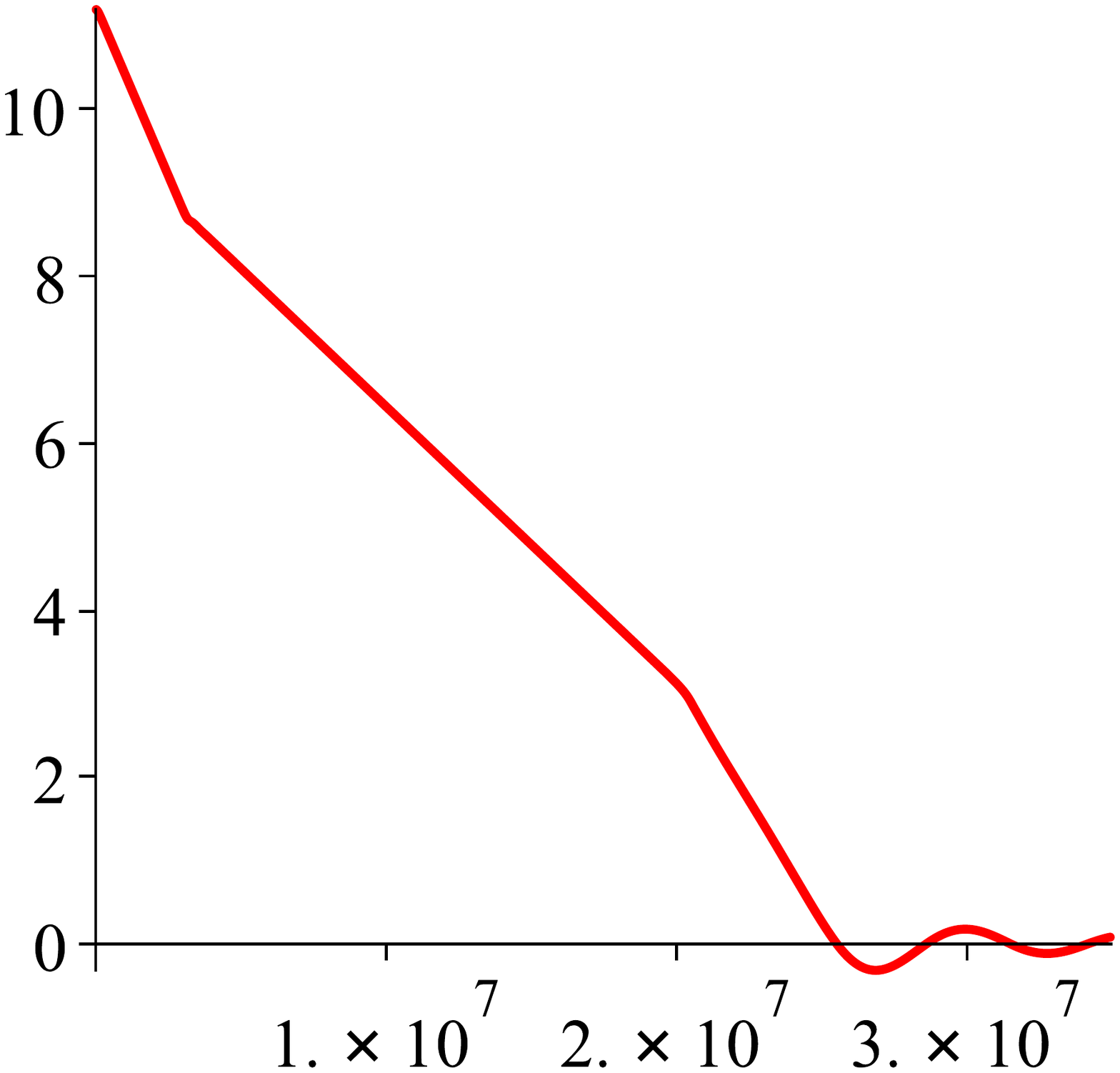} \hspace{0.5cm}
	\includegraphics[width=7cm]{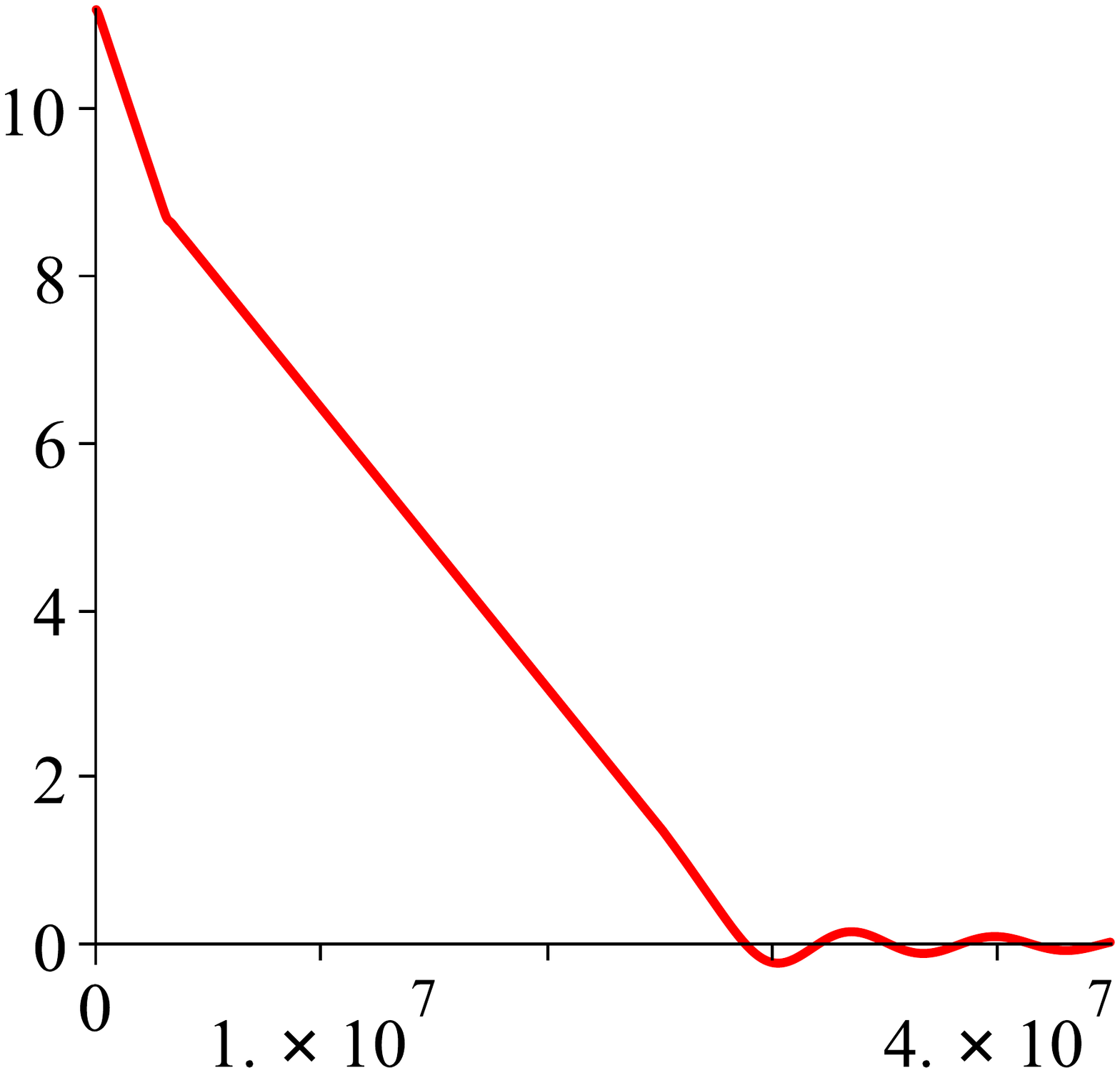} 
	\caption{ Here we present  the evolution of the inflaton field with respect to the time coordinate in chaotic inflation.  The left figure corresponds to $\e=0.1$ whereas for the right figure $\e=0$.  In this plot, the existence of the second kink and the duration of the third inflationary phase can be seen clearly when we turn on the gauge coupling $\e$. }
	\vspace{0.5cm}
	\label{rho-time}
\end{figure}

Like in our previous examples, as the gauge field increases exponentially during the second inflationary stage the right hand side of Eq. (\ref{back-A-eq4}) becomes important and one enters the final inflationary stage. The gauge field equation (in the slow-roll limit) has the same form as 
Eq. (\ref{A-eq1}) where now the parameter $\beta$ is given by
\ba
\beta \equiv \frac{6 \e^2 M_P^4 (c-1)}{c^2 p_A^2}\, .
\ea
As in symmetry breaking case the solution is given by the Bessel function Eq. (\ref{A3-Bes}).
The start of the third inflationary stage, $\alpha=\alpha_2$, is when the argument of the Bessel function becomes comparable to unity so one obtains  $\alpha_2 \simeq\frac{-1}{4} \ln \beta $.
Our numerical analysis shows that this expression gives a good estimate of $\alpha_2$. As the argument of the Bessel function increases exponentially, the gauge field starts to oscillate rapidly. This in turn produces an oscillating effective mass for the inflaton in the form of
$\e^2 \rho^2 A^2 e^{-2 \alpha}$ and the slow roll conditions are terminated quickly, ending inflation abruptly.  Our numerical analysis shows that usually inflation ends when the gauge field makes one or two oscillations in less than one e-fold.  The behavior of the gauge field during the final inflationary stage is similar to the plot on the right hand side of {\bf Fig.}
\ref{sym-A-fig} for the symmetry breaking potential. Also the behavior of the inflaton field as a function of time is presented in {\bf Fig.} \ref{rho-time}. The start of the third inflationary stage and the existence of the second kink can be seen clearly when we turn on
$\e$. However, as mentioned above, the third inflationary stage is very short, less than an e-fold. This can also be seen from {\bf Fig.} \ref{r-fig-chaotic}  where the evolution of the inflaton field is presented as a function of e-foldings.
\section{Summary}
\label{conclusion}

In this chapter, we studied anisotropic inflation in models with charge scalar fields. We have shown that, the system reaches the attractor solutions sometime during inflation where the ratios $\delta/\epsilon$ and $R_1/\epsilon$, measuring the level of anisotropies, become at the order of unity. This attractor mechanism is fairly independent of the initial conditions. One can tune the model parameters such that the time of phase changes, denoted here by $\alpha_1$ and $\alpha_2$, take place within the first few e-foldings relevant for the CMB. 

The new interesting effect in our model is the effect of the gauge coupling $\e$ on the dynamics of the inflaton field and the gauge field. At the final stage of inflation, the term on the right hand side of Eq. (\ref{back-A-eq4}) becomes important and gauge field become highly oscillatory.  Because of the interaction $\e^2 \rho^2 A^2$, the oscillations of the gauge field induce an effective time-dependent mass term for the inflaton field and inflation ends shortly after the gauge field starts to oscillate.  We have studied these effects in  examples of symmetry breaking and chaotic inflation  models. Both of our main results here, that is the existence of the attractor solutions and the oscillatory behavior of the gauge field at the end of inflation, show up  similarly in these two models. \\
We also studied the effect of the gauge field on the dynamics of the  waterfall field in charged hybrid inflation model.  Because of the coupling $\e^2 \rho^2 A^2$ the onset of waterfall phase transition can be significantly different than in standard hybrid inflation. Furthermore, the highly oscillatory behavior of the gauge field and its coupling to the inflaton field can play important roles in the studies of tachyonic preheating and reheating. Also, as noticed in \cite{Yokoyama:2008xw}, the inhomogeneous end of inflation can have interesting effects for non-Gaussianities in this model.\\
In this chapter we considered only the background dynamics. We study the cosmological perturbations of this model in the next chapters.


\chapter{Curvature Perturbations in Anisotropic Inflation with Symmetry Breaking} 

\label{Chapter4} 

\lhead{Chapter 4. \emph{Anisotropic Inflation from Charged Scalar Fields}} 

\vspace{0.5cm}
\hrule \vspace{0.3cm}

\begin{quote}
\textbf{Abstract:} In this chapter, we study curvature perturbations in the anisotropic inflationary model  with a complex scalar field charged under a $U(1)$ gauge field in Bianchi I universe. Due to Abelian Higgs mechanism, the gauge field receives an additional longitudinal mode. We verify that the dominant contributions into statistical anisotropies come from matter field perturbations and one can neglect the contributions from the metric perturbations. It is shown that the contribution of longitudinal mode into the statistical anisotropic power spectrum, though exponentially small,
has an opposite sign compared to the corresponding contribution from the transverse mode.
We obtain an upper bound on gauge coupling in order to satisfy the observational constraints
on curvature perturbations anisotropy.
\end{quote}

\vspace{0.1cm}  \hrule
\vspace{0.5cm}
\section{Introduction}

As we already discussed in the previous chapter, there may be indications of statistical anisotropies on CMB \cite{Hanson:2009gu, Hanson:2010gu} which can not be generated in simple  inflationary models which are based on only scalar fields. This opens up the interesting possibilities that primordial seeds in generating curvature perturbations during inflation may not be statistically isotropic.\\
One can parameterize the statistical anisotropy via \cite{Ackerman:2007nb}
\be
P_{\zeta} (\vec k) = P_0(k) \left(  1+ g_* \cos^2 \theta  \right)
\ee
in which $P_{\zeta} (\vec k)$ represents the curvature perturbations and $\theta$ is the angle between the preferred direction in the sky which breaks the rotational invariance and the momentum vector $\vec k$.  Constraints from CMB and large scale structure indicate that $| g_*| \lesssim 0.01$
\cite{Groeneboom:2009cb, Pullen:2010zy}. \\
Motivated by these observations, in the last chapter, we considered an inflationary model in Bianchi type I universe in which the inflaton field was a complex scalar field charged under the $U(1)$ gauge field with the electric charge coupling $\e$. We saw that, under the Abelian Higgs mechanism, the gauge symmetry is spontaneously broken and the  gauge field acquires a dynamical mass in the form $\e^2 \rho^2 A_\mu A^\mu$ in which $\rho$ is the radial component of the complex inflaton field $\phi$.
In addition, in order to break the conformal invariance, we also introduced a gauge kinetic coupling where the $U(1)$ gauge kinetic coupling is a function of the radial part of the inflaton field, $\rho$, with the action $\Delta {\cal L} = \frac{-f(\rho)^2}{4} F_{\mu \nu} F^{\mu \nu}$.
Then by choosing $f(\rho)$ such that $f(\phi) \propto a^{-2}$, the electric field would be constant at the background level and the gauge field quantum fluctuations remain scale invariant. As shown in \cite{Watanabe:2009ct} the inflationary system admits an attractor solution in which the anisotropy reaches a small but cosmologically detectable level.  Cosmological perturbation for this model in which inflaton field is a real scalar field with no charge coupling to the gauge field $A_\mu$ is studied in great details in
\cite{ Gumrukcuoglu:2010yc,  Watanabe:2010fh, Dulaney:2010sq, Yamamoto:2012sq, Bartolo:2012sd, Funakoshi:2012ym}. \\
Here, we perform the cosmological perturbation theory for our charged model in details. As we shall see this has interesting implications for the cosmological perturbations and in generating statistical anisotropies. Namely, as in usual Abelian Higgs mechanism, one scalar degrees of freedom is eaten by the gauge field and  the longitudinal mode of the gauge field is excited. As a result, along with the two transverse modes of the gauge field, the longitudinal excitations will also contribute into anisotropy analysis.\\
This chapter is based on \cite{Emami:2013bk}.
\section{Anisotropic Inflation from Charged Scalar field }
\label{model}
Here we present our anisotropic setup. It contains a charged inflaton field 
$\phi$ which is charged under the $U(1)$ gauge field $A_\mu$ with the electric charge $\e$. Since we considered its background evolution in full details in Chapter \ref{Chapter3}, we skip writing down the action and reference the reader to Eqs. (\ref{action4}) through (\ref{accel4}) for some details about the form of the action as well as the equations of motion. \\
In addition, as we have seen in the last chapter, it turns out that $R_2$ is negligible until very near to the end of inflation. So we can just skip it and replace $R_1$ with $R$ as, 
\ba
\label{R-def-5}
R \equiv \frac{\dot A_x^2 f(\rho)^2 e^{-2 \alpha}}{2 V} \, .
\ea
Then, in order
for the anisotropy to be small we require that $R \ll 1$. 

\subsection{The Attractor Solution}

It is more convenient to express the background metric (\ref{metric4}) in the following form
\ba
\label{Bianchi-metric5}
ds^2 = a(\eta)^2 (d \eta^2 + d x^2) + b(\eta)^2 ( d y^2 + d z^2)
\ea
in which $a= e^{\alpha - 2 \sigma}$ and $b=e^{\alpha + \sigma}$.
Here we have defined the conformal time $d\eta $ via $dt = a(\eta) d \eta$.
Let us define the slow-roll parameters
\ba
\epsilon_H \equiv - \frac{\dot H}{H^2}  \quad , \quad
\eta_H \equiv \epsilon_H - \frac{\ddot H}{2 H \dot H}
\quad , \quad \dot \epsilon_H = 2 H \epsilon_H (2 \epsilon_H - \eta_H)
\ea
We are working in  the slow-roll limit in which $\epsilon_H , \eta_H \ll 1$. To leading order in slow-roll parameter and anisotropy $a \simeq b \simeq -1/H \eta$.

Although the anisotropy is very small, $R\ll 1$, so the Hubble expansion rate in modified Friedmann equation (\ref{Ein1-eq4}) is mainly dominated by the isotropic potential term, but the back-reactions of the gauge field on the inflaton field induce an effective mass for the inflaton  as given by the last two terms in Eq. (\ref{back-rho-eq4}). This in turn will affect the dynamics of the inflaton field. As shown in \cite{Watanabe:2009ct} the system reaches an attractor solution in which $R \propto \epsilon_H$. For this to happen we need $f(\rho) \propto a^{n}$
with $n\simeq -2$. Indeed, the background expansion is given by
\ba
\label{a-scale5}
a \propto \exp \left[ - \int d \rho \frac{V}{M_P^2 V_\rho} \right] \, .
\ea
So if one chooses
\ba
\label{f-scale5}
f \propto \exp \left[ -n  \int d \rho \frac{V}{M_P^2 V_\rho} \right]
\ea
this yields $f \propto a^{n}$.

The exact form of $f$ therefore depends on $V(\rho)$. For the chaotic potential used in \cite{Watanabe:2009ct}  we have
\ba
V= \frac{1}{2} m^2 \rho^2  \quad \quad  \rightarrow \quad \quad
f(\rho) = \exp {\left( \frac{c\rho^2}{2 M_P^2}  \right)}
\ea
with $c$ a constant very close to unity. As shown in \cite{Watanabe:2009ct} during the attractor phase the effective inflaton mass is reduced by the factor $1/c$ such that
during the attractor phase the inflaton evolution is given by
$d \rho/d \alpha \simeq - M_P^2V_{,\rho}/c V$. The cosmological perturbations for this background was studied in details in  \cite{Gumrukcuoglu:2010yc,  Watanabe:2010fh, Dulaney:2010sq, Yamamoto:2012sq, Bartolo:2012sd, Funakoshi:2012ym} with the conclusion that in order not to produce too much anisotropy one needs $c-1 \sim 10^{-5}$.

For our model, following \cite{Emami:2010rm}, we consider the symmetry breaking potential which is physically well-motivated for the charged scalar field in the light of Abelian Higgs mechanism. The potential is
\ba
\label{sym-V5}
V= \frac{\lambda}{4} \left( |\phi|^2 - \frac{\cM^2}{\lambda}  \right)^2
\ea
in which $\lambda$ is a dimensionless coupling. The potential has global minima at
$\mu = \pm \cM/\sqrt \lambda$. The inflaton field rolls near the top of the potential so in the slow-roll limit, the potential can be approximated by
\ba
\label{sym-V25}
V\simeq \frac{ \cM^4}{4\lambda} - \frac{\cM^2}{2} \rho^2 \, .
\ea
From Eq. (\ref{a-scale5}) we have
\ba
a \propto \rho^{-p_c/2} \quad , \quad  p_c \equiv \frac{\cM^2}{2 \lambda M_P^2} \, .
\ea
To have a long enough period of slow-roll inflation we require $p_c \gg 1$.

Motivated by this, from Eq. (\ref{f-scale5}) we see that to find an attractor solution with a near scale invariant gauge field power spectrum (i.e. a scale invariant electric field power spectrum) we take \cite{Emami:2010rm}
$ f(\rho) \propto \rho^{-p} $ with $p$ very close to $p_c$.  Noting that $\rho \propto a^{-2/p_c} \propto (-\eta)^{-2/p_c}$ this yields
\ba
\label{f-eq5}
f= \left( \frac{\eta}{\eta_e}\right)^{2 c}  \quad , \quad c \equiv \frac{p}{p_c} \, ,
\ea
in which $\eta_e$ indicates the time of end of inflation. We assume that at the end of inflation
$f$ reaches its canonical value $f(\tau_e) =1$ and the isotropic FRW universes emerges at the end of inflation.
As we shall see, the strength of anisotropy is measured by the parameter $I$ given by
\ba
I \equiv \frac{c-1}{c} = \frac{p-p_c}{p}\, .
\ea
During the attractor phase \cite{Watanabe:2009ct, Emami:2010rm}
\ba
R \simeq \frac{I \epsilon_H}{2}  \quad \quad , \quad \quad
\frac{\dot \sigma}{H} \simeq \frac{2R }{3}
\simeq \frac{\epsilon_H}{3}.
\ea
This indicates that the anisotropy is at the order of slow-roll parameter during the attractor phase.

At the background level there is no restriction on the value of $c$ or $I$, only one requires 
$c \ge 1$ to reach the attractor solution. However, as we shall see from the perturbation theory in next Sections, in order not to produce too much anisotropies one requires $c\rightarrow 1$ and $I \ll 1$. 

In this picture inflation ends when the back-reaction of the gauge field on the inflaton field
via the interaction $\e^2 \rho^2 A_\mu A^\mu$ induces a large mass for the inflaton.
Comparing this with the inflaton mass $\cM$, inflation ends when
$\e^2 e^{-2 \alpha_e}A_x^2 (\eta_e) \sim \cM^2$ in which $\alpha_e$ indicates the number of e-folds at the end of inflation. As shown in \cite{ Emami:2010rm} the end of inflation depends logarithmically on $\e$. More specifically, noting that during the attractor phase  \cite{ Emami:2010rm} $A_x \propto e^{3 \alpha}$, we obtain
\ba
\alpha_e \sim -\frac{\ln \e}{2} + ...
\ea
where dots indicate the dependence on other parameters such as $p_c$ and the initial value of the gauge field. As one  expects, the larger is the gauge coupling $\e$, the shorter is the period of inflation. This is easily understood from the induced mass term $\e^2 A_\mu A^\mu \rho^2$ for the inflaton field due to Higgs mechanism.

\subsection{Perturbations}

Now we look at the perturbations of the background metric (\ref{Bianchi-metric5}). Because the gauge field has a component along the $x$-direction, the three-dimensional rotation invariance is broken into a subset of two-dimensional rotation invariance in $y-z$ plane. Therefore, to classify our perturbations, we can look at the transformation properties of the physical fields under the rotation in $y-z$ plane.
As mentioned in \cite{Dulaney:2010sq, Gumrukcuoglu:2010yc,  Watanabe:2010fh} the metric and matter perturbations are divided into scalar and vector perturbations for a general rotation in $y-z$ plane. It is also important to note that there are no tensor excitations in two dimensions.
The most general form of metric perturbations is
\ba
\label{deltag5}
\delta g_{\alpha \beta} &= &  \left(
\begin{array}{c}
- 2 a^2 A~~~~~~~  a^2 \partial_x \beta~~~~~~~~~~~~~~~ a\, b \left( \partial_i B + B_i \right)
\\\\
~~~~~~~~~~~~~~~~~~~~      - 2 a^2 \bar \psi   ~~~~~~~~~~~~~       a b\,  \partial_x \left( \partial_i \gamma + \Gamma_i \right)\\\\
~~~~~~~~~~~~~~~~~~~~~~~~~~~~~~~~~~~~~~~~~~~~~~~~~~~~~~~~~~     b^2 \left( - 2 \psi \delta_{ij} + 2 E_{, ij} + E_{(i,j)} \right)
\end{array}
\right ) \nonumber\\
&&
\ea
Here $A, \beta, B, \bar \psi, \gamma, \psi $ and $E$ are scalar perturbations and $B_i, \Gamma_i$ and $E_i$ are vector perturbations subject to transverse conditions
\ba
\label{transverse5}
\partial_i E_i = \partial_ i B_i = \partial_i \Gamma_i =0 \, .
\ea
where we have defined, $ E_{(i,j)} \equiv E_{i,j} + E_{j,i}$.\\
In Appendix \ref{appendix1} we have presented the properties of metric perturbations under a general coordinate transformation. In our analysis below we chose the following gauge
\ba
\label{gauge05}
\psi= \bar \psi = E = E_i =0 \, ,
\ea
which from Appendix \ref{appendix1}  one can check that it  is a consistent gauge.
Note that the gauge (\ref{gauge05}) is similar to the flat gauge in standard FRW background.

As for the matter sector we choose the unitary gauge $\theta=0$, so $\phi$ is real. Also, exploiting the  two-dimensional rotation symmetry, in Fourier space  we choose
\ba
\label{ab5}
\overrightarrow{k}= (k_{x} , k_{y} , 0) ~~~,~~~ k_{x} = k \cos{\theta} ~~~,~~~k_{y} =\frac{b}{a} k \sin{\theta}~~.
\ea
Therefore the scalar and vector perturbations of the matter sector, $\delta A_\mu^{(S)}$  and $\delta A_\mu^{(V)}$, are
\ba
\delta A_\mu^{(S)} = (\delta A_0, \delta A_x, \partial_y M, 0) \quad \quad , \quad \quad
\delta A_\mu^{(V)} =(0, 0, 0, D) \, .
\ea
With these decompositions of the metric and matter fields into the scalar and vector sectors, one can check that these modes do not mix with each other and one can look at their excitations and propagation separately. In this work we concentrate on the anisotropies generated from scalar excitations which are more dominant compared to the anisotropies generated by vector excitations. Therefore, for the rest of analysis we set $D= \Gamma_i= B_i=0$.
\subsection{Slow-roll Approximations}

In next Section we need to calculate the second order action in the slow-roll approximations.
Here we present some useful equations in the slow-roll approximation which will be employed in next section.  Including the first slow-roll and anisotropy corrections into the background expansion one can check that
\ba
\label{ab-coorections5}
a \simeq  H^{-1} (-\eta)^{-1 - \epsilon_H} \quad , \quad
b \simeq  H^{-1}(-\eta)^{-1 - \epsilon_H - I \epsilon_H} \, .
\ea
Our convention is such that at the start of inflation $a_{in}=1$ with number of e-folds $N_{in}=0$. The total number of e-folds at the end of inflation is $N_e$ with $N_e\simeq 60$
to solve the flatness and the horizon problem. Furthermore, at the end of inflation
$\eta =\eta_e \rightarrow 0$. With this convention, for the CMB scale modes $k_{CMB}$, we have
$N_e = -\ln (- k_{CMB} \, \eta_e)$.
In our discussions below, we concentrate on CMB scale modes so to simplify the notation we denote $k_{CMB}$  by $k$.

From the above formula, and using Eq. (\ref{f-eq5}) for the function $f(\eta)$,
one can obtain the following expressions which would be useful later on
\begin{align}
\label{slow roll5}
&&\frac{a^{'}}{a} &= (-\eta)^{-1}(1 + \epsilon_{H} )   &&\frac{b^{'}}{b} = (-\eta)^{-1}(1 + \epsilon_{H} + I\epsilon_{H})\nonumber\\
&&\frac{a^{''}}{a} &= (-\eta)^{-2}(2 + 3\epsilon_{H})  &&\frac{b^{''}}{b} = (-\eta)^{-2}(2 + 3\epsilon_{H} + 3 I\epsilon_{H})\nonumber\\
&&\frac{k^{'}}{k} &= (-\eta)^{-1}(-\sin^2{\theta}I\epsilon_{H})   &&  \frac{k^{''}}{k} = (-\eta)^{-2}(-\sin^2{\theta}I\epsilon_{H}) \nonumber\\
&& \frac{f'}{f} &= (-\eta)^{-1} ( -2 - 2 \epsilon_H - \eta_H + 2 I \epsilon_H )   &&  \frac{f^{''}}{f} = (-\eta)^{-2}( 2 + 9 \epsilon_H - 3 \eta_H + 6 I \epsilon_H   ) \, ,
\end{align}
in which a prime indicates derivative with respect to conformal time.

For the future reference, the following equations are helpful
\ba
\label{epsilon-eq5}
\epsilon_H \simeq \frac{8 \lambda^2 M_P^2 \rho^2}{\cM^4} \quad , \quad
H \simeq \frac{\cM^2}{\sqrt{12 \lambda} M_P} \, .
\ea

\section{Second Order Action}
\label{sec-ac}

Here we present the second order action for the scalar  perturbations. Our goal is to find the second order action both for the free fields and for the interactions. As we shall see the fields $\delta A_0, A, \beta$ and $B$ are non-dynamical in the sense that they have no time-derivatives in the action. As a result, their equations of motion give constraints which can be used to eliminate them in terms of the remaining dynamical fields
$\delta \rho, \delta A_1, M$  and $ \gamma$.

The second order action for the scalar perturbations is
\ba
\label{S2-scalar5}
S_2^{(S)} &=& \int d \eta d^3 x \left[ 2 b b' A_{,x} \beta_{, x} + a b (\frac{a'}{a} +\frac{b'}{b} )
A_{,y} B{, y} +  a b \gamma_{, xy} A_{, xy} - a^2 b^2 V(\rho_0) A^2 
\right. \nonumber \\ &&\left.
- \frac{\e^2}{2} b^2 \rho_0^2 A_x^2 A^2 - \frac{ab}{2} \beta_{, xy} B_{, xy}
+ a'b \gamma_{, xy} \beta_{, xy}+ \frac{ab}{2} \gamma_{, xy} \beta'_{, xy}
+ \frac{\e^2}{2} b^2 \rho_0^2 A_x^2 \beta_{x}^2 
\right. \nonumber \\ &&\left.
+ \frac{a^2}{4} \beta_{,x y}^2  -\frac{b^2}{2}  B_{,xy}\gamma'_{, xy} 
+ \frac{b^2}{2} (\frac{b'}{b} - \frac{a'}{a})\gamma_{, xy} B_{, xy}
+ \frac{b^2}{4} B_{, xy}^2 
- \frac{\e^2}{2} b^2 A_x^2 \rho_0^2 \gamma_{, xy}^2 
\right. \nonumber \\ &&\left.
+ \frac{b^2}{4}  (\gamma'_{, xy})^2  + \frac{b^2 f^2}{2 a^2} (A_x')^2\gamma_{, xy}^2 - \frac{b^2}{4} (\frac{b''}{b} - \frac{a''}{a}) \gamma_{, xy}^2
+ \frac{b^2}{2} \delta \rho'^2
- b^2 \rho_0' A \delta \rho'
\right. \nonumber \\ &&\left.
- b^2 \rho_0' \beta_{,x} \delta \rho_{,x}  - a b \rho_0' B_{, y} \delta \rho_{, y}
 - \frac{b^2}{2} \delta \rho_{,x}^2 - \frac{a^2}{2} \delta \rho_{,y}^2
- \e^2 b^2 \rho_0^2 A_x \beta_{, x} \delta A_0  
 \right. \nonumber \\ &&\left.
+ \frac{\e^2}{2} b^2 \rho_0^2 \delta A_0^2 - \frac{\e^2}{2} b^2 \rho_0^2 \delta A_1^2 - \frac{\e^2}{2} b^2 A_x^2 \delta \rho^2 - 2\e^2 b^2 \rho_0 A_x \delta \rho \delta A_1
- \frac{\e^2}{2} a^2 \rho_0^2 M_{, y}^2
\right. \nonumber \\ &&\left.
+ \e^2 a b \rho_0^2 A_x \gamma_{, xy} M_{, y}
 - \e^2 b^2 \rho_0^2 A_x A \delta A_1
-\e^2 b^2 \rho_0 A_x^2 A \delta \rho + \frac{f^2 b^2}{2 a^2} \delta A_1'^2
+\frac{f^2 b^2}{2 a^2} \delta A_{0 , x}^2  \right. \nonumber \\ &&\left.
- \frac{f^2 b^2}{ a^2} \delta A_1' \delta A_{0, x} -  \frac{f^2 b^2}{ a^2} A_x' A \delta A_1'
+  \frac{f^2 b^2}{ a^2} A_x' A \delta A_{0, x}  - \frac{f^2 b}{ a} A_x' \gamma_{, xy} M_{,y}'
\right. \nonumber \\ &&\left.
+\frac{f^2 b}{ a}  A_x'  \gamma_{, xy} \delta A_{0, y} - \frac{f^2 b}{ a}  A_x' B_{, y} \delta A_{1, y} 
+ \frac{f^2 b}{ a} A_x' B_{, y} M_{, xy} + \frac{f^2}{2} M_{y}'^2 + \frac{f^2}{2} \delta A_{0, y}^2 
\right. \nonumber \\ &&\left.
- f^2 M_{, y}' \delta A_{0, y} - \frac{f^2}{2} \delta A_{1, y}^2 -  \frac{f^2}{2}
M_{, xy}^2 + f^2 \delta A_{1, y} M_{, xy} + 2 \frac{f f_{,\rho} b^2}{a^2} A_x'\delta A_1' \delta \rho
\right. \nonumber \\ &&\left.
- 2 \frac{f f_{,\rho} b^2}{a^2} A_x' \delta A_{0, x} \delta \rho - \frac{f f_{,\rho} b^2}{a^2} A \delta \rho +  \frac{ f_{,\rho}^2 b^2}{2a^2}A_x'^2 \delta \rho^2 + \frac{f f_{,\rho \rho} b^2}{2a^2} A_x'^2 \delta \rho^2 
\right. \nonumber \\ &&\left.
- \frac{a^2 b^2}{2} V_{, \rho \rho} \delta \rho^2 - a^2 b^2 V_{, \rho} A \delta \rho ~~~
\right]
\ea
in which a prime indicates derivative with respect to conformal time.

As mentioned above, the excitations $\delta A_0, A, \beta$ and $B$ have no time-derivatives so they are non-dynamical. The details of eliminating the non-dynamical excitations in terms of
dynamical perturbations are given in Appendix \ref{reduced}.

The final second order action is a complicated function of $\delta \rho, \delta A_1, M$  and $ \gamma$. Specifically, integrating out  $\delta A_0, A, \beta$ and $B$ one encounters the functions $\lambda_i$ and
$\bar \lambda_i$ as defined in Eqs. (\ref{lambdak1}) - (\ref{lambda17}). At this level it seems hopeless to get any insight into the form of the action and the prospects for analytical analysis. Happily, the analysis becomes considerably simple if one notice the following effects. Looking at the formulae for $\lambda_i$ and $\bar \lambda_i$ it is evident that  $\bar \lambda_1$ is the  key parameter which controls the form of other $\lambda_i$ and $\bar \lambda_i$.  Now let us look at the function $\bar\lambda_1$
\ba
\label{b-lambda15}
\bar \lambda_1 \equiv \frac{b^2}{2 a^2} k^2 f^2 + \frac{\e^2 }{2} b^2 \rho^2 \, .
\ea
Following the procedures of integrating out the non-dynamical fields in Appendix \ref{reduced} one can check that $\bar \lambda_1$ comes from integrating out $\delta A_0$.
Neglecting the anisotropy for the moment, the ratio of the second term in $\bar \lambda_1$
compared to the first term scales like $\e^2 a^2/ f^2 \sim \e^2 H^2  \eta_e^4/\eta^6 $. Therefore, during the early stages of inflation in which $-\eta \gg -\eta_e$, the second term in $\bar \lambda_1$ is completely negligible compared to the first term. In this limits all $\lambda_i$
and $\bar \lambda_i$ collapse to simple forms and we will be in the limit somewhat similar to
\cite{Watanabe:2010fh}. In this limit the effect of gauge coupling $\e$ is sub-dominant in the action and the leading interaction comes from the gauge kinetic coupling $f^2(\phi) F^2$.
On the other hand, as inflation proceeds the second term in
$\bar \lambda_1$ eventually dominates and we enter the second phase in which all
$\lambda_i$ and $\bar \lambda_i$ are proportional to $\e^2$. In this limit, the interaction induced from the symmetry braking, $\e^2 \rho^2 A_\mu A^\mu$, becomes as important as the interaction from the gauge kinetic coupling.   We will elaborate more on this issue later on when we present the dominant interactions for the transverse and longitudinal modes.

Having this said, one may wonder why $\bar \lambda_1$ plays such a prominent role.
The answer to this question is provided in Section \ref{matter-leading} in which we demonstrate that the leading interactions come from the matter sector. So it is not surprising that only $\bar \lambda_1$, which originates from integrating out $\delta A_0$, will have a prominent effect
while the other parameters $\lambda_i$ and $\bar \lambda_i$, which have their origins in integrating out metric fields $A, \beta$ and $B$, are negligible.

The time when the two terms in $\bar \lambda_1$ become comparable, denoted by $\eta_c$, is given by
\ba
\label{etac5}
-\eta_c =  \left(\frac{\e \rho}{-H k }\right)^{1/3} (-\eta_{e})^{2/3}=
\left( \frac{3 \e M_{P} \sqrt{2\epsilon_H}}{\cM^2} \right)^{1/3} (-\eta_e)^{2/3} \, .
\ea
In order to obtain the last equality, we considered  the CMB scale modes in which
$k= a_{in} H = H$.  Eq. (\ref{etac5}) indicates a $k$-dependence in $\eta_c$. However, as we will see explicitly below, the leading contributions from the inflaton field and the transverse mode are blind to
this $k$-dependence.

It is also instructive to look at $N_c$, the number of e-folds when $\eta=\eta_c$. Using $\eta \simeq -1/a H$ and Eq. (\ref{etac5}) we have
\ba
\label{Nc5}
N_c \simeq \frac{2 N_e}{3} - \frac{1}{3} \ln \left( \e \sqrt{\frac{3 \epsilon_H}{2 \lambda}}
\right) \simeq \frac{2 N_e}{3} \, .
\ea
The last approximation is valid for typical parameter values such that the logarithmic correction
in Eq. (\ref{Nc5}) is at the order of unity. Our convention is such that at the start of inflation
$N_{in}=0$ and  the total number of e-folds at the end of inflation is $N_e$.  With $N_e\simeq 60$ to solve the flatness and the horizon problem we obtain  $N_c \sim 40$.

\subsection{Second Order Action in the Slow-roll Approximation}
After integrating out the non-dynamical fields, the remaining dynamical fields are $\delta \rho,  \delta A_1, M$ and $\gamma$. However, for the gauge field excitations, the physically relevant fields are the transverse mode $D_1$ and the longitudinal mode $D_2$ which are related to $\delta A_1$ and $M$ via
\ba
\label{D125}
D_{1} &\equiv& \delta A_{1} -ik \cos{\theta}M \\
D_{2}&\equiv& \cos{\theta} \delta A_{1} + ik \sin^2{\theta}M \, .
\ea
Here we present the second order action in the slow-roll limit for the dynamical variables
$\delta \rho,  D_1$ and  $D_2$.  The action is presented separately for $\eta< \eta_c$ and $\eta_c < \eta < \eta_e$. 

In this work we are interested in anisotropy generated in curvature perturbation power spectrum. Note that,  as discussed in Appendix \ref{appendix1}, the scalar perturbations 
$\gamma$ will furnish one polarization of tensor perturbations in isotropic universe after inflation. Therefore the interactions $L_{\delta \rho \gamma}, L_{D_1 \gamma}$ and $L_{D_2 \gamma}$ will not contribute into curvature perturbation anisotropy and we do not present them in this section. However, they are presented in the Appendices when we present the whole second order action for the scalar perturbation.

\subsection{ $\eta  < \eta_c $ }
\label{eta-less}
First we consider the period in which $\eta < \eta_c$ so the term containing $\e$ in
$\lambda_i$ and $\bar \lambda_i$ are negligible and the first term in $\bar \lambda_1$ in Eq. (\ref{b-lambda15}) dominates.

As we shall see in next section, in order not to produce too much anisotropy, one requires $I \ll 1$ (i.e. c $\rightarrow 1$) which we will assume in all our analysis below.  Considering the leading corrections from the slow-roll and anisotropy expansion yields (for details see Appendix \ref{second slow})
\begin{align}
\label{Total scalar action5}
S_{2}^{(1)} = \int d\eta \,  d^3k \bigg{(} L_{\rho \rho}  + L_{D_{1}D_{1}} + L_{D_{2}D_{2}} +    L_{\rho D_{1}} +  L_{\rho D_{2}} +     L_{ D_{1} D_{2}} \bigg{)}~,
\end{align}
in which the free fields Lagrangians are
\ba
\label{rhorho action15}
L_{\rho \rho} &=& \frac{1}{2}|\overline{\delta \rho}'|^2 + \frac{1}{2}\bigg{[} -k^2 + (-\eta)^{-2}\bigg{(}2+9\epsilon_{H}-6\frac{\eta_{H}}{1-I}
-12\frac{I}{1-I} \nonumber\\
&&
(1-2\sin^2{\theta})\bigg{)}\bigg{]}|\overline{\delta \rho}|^2
\\
\label{D1D1  action15}
L_{D_{1} D_{1}} &=& \frac{1}{2}|\overline{D_{1}}^{'}|^2 + \frac{1}{2}\bigg{[} -k^2 + (-\eta)^{-2}\bigg{(}2+9\epsilon_{H}-3\frac{\eta_{H}}{1-I}\bigg{)}\bigg{]}|\overline{D_{1}}|^2
\\
\label{D2D2  action15}
L_{D_{2} D_{2}} &=& \frac{1}{2}|\overline{D_{2}}^{'}|^2 + \frac{1}{2}\bigg{[} -k^2 + (-\eta)^{-2}\bigg{(}2+3\epsilon_{H} + I\epsilon_{H}\bigg{)}\bigg{]}|\overline{D_{2}}|^2 \, .
\ea
Here we have defined the canonically normalized fields via
\begin{align}
\label{canonical variables15}
\overline{\delta \rho}_{k} &\equiv b \delta \rho_{k}\equiv u_{k} \\
\label{D1bar5}
\overline{D_{1k}}&\equiv \frac{b}{a}f \sin{\theta} D_{1k}\equiv \frac{b}{a}f \sin{\theta} v_{k}\\
\label{D2bar5}
\overline{D_{2k}}&\equiv \frac{\e \cM^2}{2\sqrt{2}\lambda k M_{P}}\sqrt{\frac{\epsilon_{H}}{1-I}} bD_{2k} \equiv \frac{\e \cM^2}{2\sqrt{2}\lambda k M_{P}}\sqrt{\frac{\epsilon_{H}}{1-I}} b w_{k} 
\end{align}

The interaction Lagrangians relevant for curvature perturbations anisotropy are
\ba
\label{rhoD1new5}
L_{\rho D_{1}} &=& \left(\frac{1}{\eta}\right)\frac{b^2}{a} \sqrt{6I} \sin^2{\theta} f \Big{(} \delta \rho^{*} D_{1}^{'} + c.c.\Big{)} - \left( \frac{a^2}{f\eta} \right) \e^2 \sqrt{\frac{I \epsilon_{H}^2}{\lambda}}M_{P} \sin^2{\theta}\times \nonumber\\
&&\Big{(} \delta \rho^{*} D_{1} + c.c.\Big{)} 
\\
\label{rhoD2new5}
L_{\rho D_{2}} &=& \left(\frac{1}{\eta}\right) \left(\frac{ab^2}{8}\frac{\e^2 \cM^4}{\lambda^2 k^2 f}\right)\sqrt{6I} \epsilon_{H} \cos^3{\theta}\Big{(} \delta \rho^{*} D_{2}^{'} + c.c.\Big{)} - \left( \frac{a^2}{f\eta} \right) \e^2 \sqrt{\frac{I \epsilon_{H}^2}{\lambda}}\times \nonumber\\
&&M_{P} \cos{\theta}\Big{(} \delta \rho^{*} D_{2} + c.c.\Big{)}
\ea

As mentioned  before, Eqs. (\ref{rhorho action15})-(\ref{D2D2  action15}) represent the free-field actions for  $\bar \delta \rho,  \bar D_1$ and $\bar D_2$. As expected, during this phase in which the effect of symmetry breaking term $\e^2 \rho^2 A_\mu A^\mu$ is sub-leading, similar to \cite{Watanabe:2010fh}, Eqs. (\ref{rhorho action15})-(\ref{D2D2  action15}) represent nearly massless fields with almost scale-invariant power spectrum.
The interaction terms are given by Eqs. (\ref{rhoD1new5}) and (\ref{rhoD2new5}). For technical reasons the interaction terms are presented in terms of the original non-canonical fields. 

To calculate the induced anisotropy in curvature perturbation power spectrum, we are interested in interactions between the gauge field and the inflaton field given by $L_{\rho D_1}$ and $L_{\rho D_2}$ in Eqs. (\ref{rhoD1new5}) and (\ref{rhoD2new5}). First let us look at the interaction between the transverse mode and the inflaton field, $L_{\rho D_15}$.  From Eq. (\ref{rhoD1new5}) we see that
$L_{\rho D_1}$ has two contributions. The first term in  $L_{\rho D_1}$ comes from the gauge kinetic coupling $f^2(\phi) F^2$ which is similar to models such as
\cite{Watanabe:2010fh} with a real inflaton field. However, the second term in $L_{\rho D_1}$
comes from the interaction $\e^2 \rho^2 A_\mu A^\mu$ which originates  from the symmetry breaking effects. This interaction does not exist in models where $\phi$ is a real field. One  can easily check that for $\eta < \eta_c$  the first term in $L_{\rho D_1}$ dominates over the second term. The two interactions in $L_{\rho D_1}$
become comparable near $\eta = \eta_c$. This is understandable, since during the period $\eta< \eta_c$, the effects of symmetry breaking are small and the system proceeds as in \cite{Watanabe:2010fh}.

Now let us look at  $L_{\rho D_2}$,  the interaction between the longitudinal mode and the inflaton field.  As expected the longitudinal mode becomes physical because of the symmetry breaking effect $\e^2 \rho^2 A_\mu A^\mu$ so both terms in Eq. (\ref{rhoD2new5}) are proportional to $\e^2$. The last term in $L_{\rho D_2}$ comes directly from the interaction
$\e^2 \rho^2 A_\mu A^\mu$. However, the first term in  $L_{\rho D_2}$ is somewhat non-trivial. As we shall see in Section \ref{matter-leading}, after integrating out $\delta A_0$ a coupling in the form  $\delta \rho^*D_2'+ c.c.$ appears which
cancels the corresponding term coming from  $f^2 F^2$ interaction during the phase $\eta < \eta_c$. As a result, the derivative coupling $\delta \rho^*D_2'+ c.c.$ during the first phase comes from sub-leading interactions so it contains $\e^2$. Finally, comparing the two terms  in Eq. (\ref{rhoD2new5})
one can check that during the phase $\eta< \eta_c$ the second term in Eq. (\ref{rhoD2new5})
is smaller than the first term by a factor $1/p_c \ll 1$.

It is also instructive to compare $L_{\rho D_1}$ and $L_{\rho D_2}$ during this phase. Relating
$D_1$ and $D_2$  to the normalized field $\bar D_1$ and $\bar D_2$ as given in Eqs. (\ref{D1bar5}) and (\ref{D2bar5}) and assuming that $\bar D_1$ and $\bar D_2$ have similar amplitudes one can check that
\ba
\label{L-ratio5}
\frac{L_{\rho D_1}}{L_{\rho D_2}} \sim \frac{k\,  f}{ \e \rho a} \gg1
\ea
in which Eq. (\ref{epsilon-eq5}) have been used to eliminate $\epsilon_H$. The conclusion that
$L_{\rho D_1} \gg L_{\rho D_2}$ is understandable since during the first phase the effects of the
coupling $\e$ is negligible.

To summarize, the leading interaction during the phase $\eta < \eta_c$ is given by the first term in Eq.(\ref{rhoD1new5}) from the transverse mode interaction $L_{\rho D_1}$. As mentioned, this interaction originates from the gauge kinetic coupling interaction $f(\rho)^2 F^2$. As a result, the induced anisotropy originated from this phase is similar to models with a real inflaton field such as in  \cite{Watanabe:2010fh}.

\subsection{ $\eta_c< \eta  < \eta_e $ }
\label{eta-big}
As we mentioned below Eq. (\ref{b-lambda15}) during the period
$\eta_c< \eta  < \eta_e  $ the effect of the gauge coupling $\e$ becomes important.  During this phase the dominant contributions  in $\lambda_i$ and $\bar \lambda_i$ in Eqs. (\ref{lambdak1}) - (\ref{lambda17}) come from the terms containing $\e$. Expanding to leading order in terms of the slow-roll parameters and $I$ and concentrating on CMB-scale modes which are expected to be super-horizon by the time $\eta=\eta_c$,
the second order action is
\begin{align}
\label{Total scalar action25}
S_{2}^{(2)} = \int d\eta d^3k \bigg{(} L_{\rho \rho} +  L_{D_{1}D_{1}} + L_{D_{2}D_{2}} +   L_{\rho D_{1}} +  L_{\rho D_{2}} \bigg{)}~,
\end{align}
where,
\ba
\label{rhorho action5}
L_{\rho \rho} &=& \frac{1}{2}|\overline{\delta \rho}'|^2 + \bigg{[} \frac{1}{\eta^{2}} -\left(\frac{\e^2 I \epsilon_{H} \lambda }{\cM^4}\right) \left(\frac{1}{f^{2}\eta^{2}}\right)\bigg{]}|\overline{\delta \rho} |^2
\\
\label{D1D1  action5}
L_{D_{1} D_{1}} &=& \frac{1}{2}| \overline{D_{1}}'|^2 + \bigg{[} \frac{1}{\eta^{2}} -\left(\frac{3\e^2 \epsilon_{H}}{4\lambda}\right) \left(\frac{1}{f^{2}\eta^{2}}\right)\bigg{]}| \overline{D_{1}} |^2
\\
\label{D2D2  action5}
L_{D_{2} D_{2}} &=& \frac{1}{2}| \overline{D_{2}}'|^2 + \bigg{[} \frac{1}{\eta^{2}} -\left(\frac{3\e^2 \epsilon_{H}}{4\lambda}\right) \left(\frac{1}{f^{2}\eta^{2}}\right)\bigg{]}| \overline{D_{2}} |^2
\ea
\ba
\label{rhoD1 action205}
L_{\rho D_{1}} &=& \left(\frac{\sin^2{\theta}}{\eta}\right)\left( \frac{b^2}{a} \sqrt{6I} f \Big{(} \delta \rho^{*} D_{1}^{'} + c.c.\Big{)} - \left( \frac{a^2}{f}\right) \e^2 \sqrt{\frac{I \epsilon_{H}^2}{\lambda}}M_{P} \Big{(} \delta \rho^{*} D_{1} + c.c.\Big{)}
\right) \nonumber\\
\\
\label{rhoD2 action205}
L_{\rho D_{2}} &=& \left(\frac{\cos{\theta}}{\eta}\right)
\left(\frac{b^2}{a} \sqrt{6I} f \Big{(} \delta \rho^{*} D_{2}^{'} + c.c.\Big{)} - \left( \frac{a^2}{f} \right) \e^2 \sqrt{\frac{I \epsilon_{H}^2}{\lambda}}M_{P} \Big{(} \delta \rho^{*} D_{2} + c.c.\Big{)}\right) \nonumber\\
\ea

During the second phase  the canonical variables $\overline{\delta \rho}_{k}$ and $\overline{ D_1}_{k}$ are the same as defined in Eqs. (\ref{canonical variables15}) and (\ref{D1bar5}) while the canonical normalized field $\overline{D_2}_k$ is
\ba
\label{canonical variables2}
\overline{D_{2k}}&\equiv& \frac{b}{a} f D_{2k}\equiv \frac{b}{a} f w_{k} \, .
\ea

As in the first phase, for the purpose of  calculating the curvature perturbations power spectrum, we look into interactions  between $\delta \rho$ and other fields. As before, the interaction  $L_{\rho \gamma}$ does not have any directional dependence so we have not considered it in above action. Therefore we are left with $L_{\rho D_{1}}$ and $L_{\rho D_{2}}$.

The crucial difference compared to the first phase is that once the second term in Eq. (\ref{b-lambda15}) dominates over the first term, the effects of gauge coupling $\e$ from the interaction $\e^2 A_\mu A^\mu$ become important. To see this, let us look at the interactions $L_{\rho D_1}$ and $L_{\rho D_2}$ given in Eqs. (\ref{rhoD1 action205}) and (\ref{rhoD2 action205}). One can easily check that in both Eqs. (\ref{rhoD1 action205}) and (\ref{rhoD2 action205}), the terms containing $\e^2$ are much larger than the first terms containing $D_1'$ and $D_2'$ which
come from the gauge kinetic coupling $f^2 F^2$. In this view, during $\eta_c < \eta< \eta_e $ the dominant interaction in the system is $\e^2 A_\mu A^\mu$ and not $f^2 F^2$. This is in contrast to the first phase in which, as we saw in the previous subsection,  the interaction $f^2 F^2$ was the dominant one and the effects of symmetry breaking were not important.

It is also instructive to compare the forms of $L_{\rho D_{1}}$ and $L_{\rho D_{2}}$
for these two phases. From Eq. (\ref{rhoD1 action205}) and (\ref{rhoD1new5}) we see that
$L_{\rho D_{1}}$ has the same functional form in both phases. However, $L_{\rho D_{2}} $
has different functional forms in two phases. The last terms in Eq. (\ref{rhoD2 action205})
and (\ref{rhoD2new5}) are the same. This is reasonable since this term directly originates from
the interaction $\e^2 A_\mu A^\mu$. However, the first terms in Eq. (\ref{rhoD2 action205})
and (\ref{rhoD2new5}),  containing the derivative coupling of $\delta \rho^*D_2'+ c.c.$,  have different forms in these two phases. Intuitively, this is somewhat non-trivial. However, as we shall show explicitly in Section  \ref{matter-leading}, this difference originates from integrating out $\delta A_0$.
After integrating out $\delta A_0$, a coupling in the form  $\delta \rho^*D_2'+ c.c.$ appears which
cancels the corresponding term coming from  $f^2 F^2$ interaction in the first phase. As a result, the derivative coupling $\delta \rho^*D_2'+ c.c.$ during the first phase comes from sub-leading interactions so it contains $\e^2$. However, during the second phase, the leading terms
in derivative coupling $\delta \rho^*D_2'+ c.c.$ survives and as a result the first term in
Eq. (\ref{rhoD2 action205}) gets the usual form similar to derivative coupling in Eq. (\ref{rhoD1 action205}).

Comparing Eq. (\ref{canonical variables2}) with Eq. (\ref{D2bar5}) we see that
$\overline {\delta \rho}$ and $\overline D_1$ have the same forms in both phases but
$\overline D_2$ have different forms in two phases. Also Eq. (\ref{rhoD1 action205}) is proportional to $\sin^2{\theta}$ while Eq. (\ref{rhoD2 action205}) is proportional to $\cos{\theta}$. As a result we can guess that the contributions of the longitudinal mode in $g_*$ has a  different sign than the corresponding contributions from the transverse mode.  So the question arises whether or not we can produce a positive $g_*$ factor from the longitudinal mode (from \cite{Watanabe:2010fh} we know that $g_{*}$ is negative for the transverse modes). We will come back to this question when we calculate the power spectrum of curvature perturbations.

Having obtained the quadratic action we also need to know the wave function solution for
$\delta \rho, D_1$ and $D_2$. For the first phase the answer is simple: since all modes are nearly massless, the mode functions of $\overline{\delta \rho}, \overline D_1$ and $\overline D_2$ are simply the mode function of the massless scalar fields with the Bunch-Davies initial condition. More specifically
\begin{align}
\label{mode function0}
M_{j\textbf{k}} &= m_{j\textbf{k}}a_{j\textbf{k}} + m^{*}_{j(-\textbf{k})}a^{\dag}_{j(-\textbf{k})} ~~~,~~~ j = (\overline{\delta \rho} , \overline{\gamma}  , \overline{D_{1}} , \overline{D_{2}})  \nonumber \\
m_{j\textbf{k}} &\equiv \frac{1}{\sqrt{2k}}e^{-ik\eta}(1-\frac{i}{k\eta}) \, .
\end{align}
The profile of the outgoing solution for $\eta_c <\eta < \eta_e$ is given in details in
Appendix \ref{outgoing modes}. Here
we demonstrate that during the second phase the inflaton excitations and the gauge field excitations remain nearly massless so one can still use the free wave function given in Eq. (\ref{mode function0}). To verify that the perturbations remain nearly massless during the second phase  it is instructive to look at the times when the arguments of the Hankel functions  Eqs. (\ref{mode function-u}),  (\ref{mode function-v}) and (\ref{mode function-w}) becomes the order unity. This can be interpret as the times when the modes become massive so it oscillates towards the end of inflation.  Defining $\eta_u$ as the time when the inflaton field fluctuations $u_k$ become massive we have $\eta_u \simeq \Omega^{1/4}$. As a result, the number of e-folds towards the end of inflation when $u_k$ is massive, $\Delta N_u \equiv \ln (\eta_u/\eta_e)$, is given by
\ba
\label{Nu}
\Delta N_u \simeq \frac{1}{4} \ln \left( \frac{\e^2 I \epsilon_H \lambda M_P^4}{\cM^4}
\right) \simeq   \frac{1}{4}  \ln \left( \frac{\e^2 I \epsilon_H }{\lambda p_c^2} \right)
\simeq \frac{1}{4}\ln(10^3 \e^2) \, ,
\ea
in which in the last approximation we assumed the typical model parameters of symmetry breaking inflation $\lambda \sim 10^{-13},  \epsilon_H \sim 10^{-2}$, $p_c \sim N_e$, and as we shall see below, $I \sim 10^{-5}$. Therefore, if $\e \lesssim 1$ which is a natural choice, we see that $\Delta N_u \sim 2$. As a result, for $\e$ not exponentially large, the inflaton field excitations remain nearly massless almost during entire period of  inflation. As a result, in our analysis of power spectrum in next section we can treat $u_k$ as nearly massless field excitations.

Also one can  check that $\Delta /\Omega \sim p_c^2/I$. As a result, the time
$\eta_v \simeq  \Delta^{1/4}$  when the gauge field excitations become massive,
and the corresponding number of e-foldings $\Delta N_v \equiv \ln (\eta_v/\eta_e)$, is given by
\ba
\Delta N_v \simeq \frac{1}{4} \ln \left(  \frac{\e^2 \epsilon_H}{\lambda} \right)
\simeq \frac{1}{4} \ln( 10^{10} \e^2) \simeq 6 + \frac{1}{2} \ln \e \, .
\ea
This indicates that   for typical model parameters $\Delta N_v - \Delta N_u \simeq 4$ so
$\Delta N_v \sim 6$. Therefore, we can also safely conclude that the gauge field excitations are
nearly massless during most of the period of inflation. Finally, one can also easily check that
$\eta_c \gg \eta_u, \eta_v$, so at the time $\eta =\eta_c$, all fields excitations are nearly massless to very good approximations.

\section{Power Spectrum of  Curvature Perturbations}
\label{power-spec}

We are ready to calculate the curvature perturbation power spectrum.  We are interested in anisotropies generated in curvature perturbation power spectrum. The anisotropies are generated by interactions $L_{\rho D_1}$ and $L_{\rho D_2}$ from the coupling of the transverse and longitudinal modes to $\delta \rho$. The corresponding Feynman diagrams are
given in Fig. 1.

\begin{figure}
	\includegraphics[ width=0.9\linewidth]{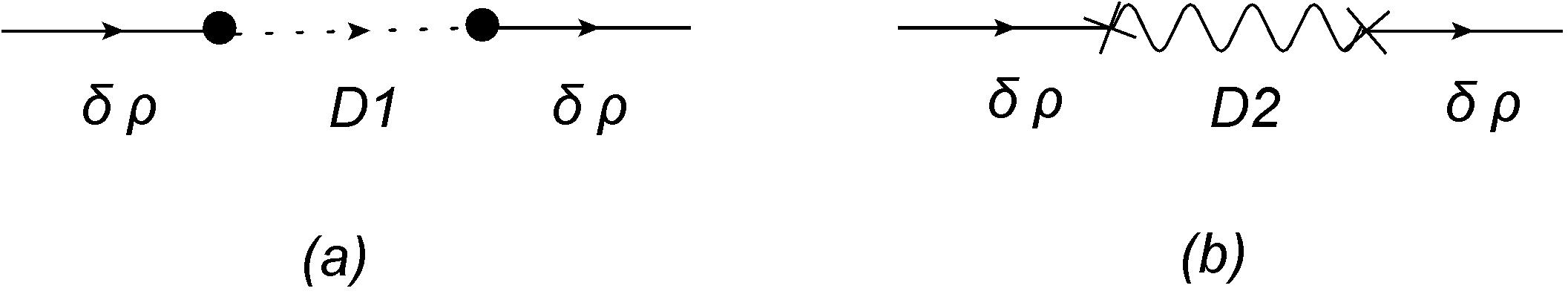}
	\caption{ The transfer vertices for the interactions of the inflaton field $\delta \rho$ with the gauge field excitations $D_1$ and $D_2$. The left figure represents $L_{\rho D_1}$ as given by Eq. (\ref{rhoD1 action205})
		while the right figure represents $L_{\rho D_2}$ given by Eq. (\ref{rhoD2 action205}). }
	\label{stop}
\end{figure}
Using the standard In-In formalism for the curvature perturbation power spectrum
\cite{Weinberg:2005vy, Chen:2010xka, Chen:2009zp} we have
\begin{align}
\label{n point function}
<\overline{\delta \rho}^2(\eta_{e})> = \bigg{<} \bigg{|}\bigg{[} \overline{T} \exp{\left(i\int_{\eta_{0}}^{\eta_{e}}H_{I}(\eta')d\eta'\right)}\bigg{]} \overline{\delta \rho}^2(\eta)\bigg{[} T \exp{\left(-i\int_{\eta_{0}}^{\eta_{e}}H_{I}(\eta')d\eta'\right)}\bigg{]}  \bigg{|}\bigg{>} \, ,
\end{align}
where $T$ and $\overline{T}$ respectively denote the time-ordered and anti-time-ordered products and $H_{I}$ refers to the interaction part of the Hamiltonian in the interaction picture.
As for $\eta_0$ we can take $\eta_0 \rightarrow -\infty$ so the modes of interests were originally deep inside the horizon.\\
To leading order the contribution of anisotropy in inflaton power spectrum, $\Delta <\overline{\delta \rho}^2(\eta_{e})>$, is
\begin{align}
\label{commutator}
\Delta <\overline{\delta \rho}^2(\eta_{e})> = - \int_{\eta_{0}}^{\eta_{e}} d\eta_{1} \int_{\eta_{0}}^{\eta_{1}}d\eta_{2} \bigg{[} H_{I}(\eta_{2}) , \bigg{[} H_{I}(\eta_{1}) , \overline{\delta \rho}^2(\eta)\bigg{]}\bigg{]} \, .
\end{align}
As discussed in details in previous Section the derivative interactions of the  longitudinal mode, terms containing $D_2'$,  have different forms in  phases  $\eta < \eta_c$ and $\eta > \eta_c$. To take this into account, we  can write the interaction Hamiltonian as follows
\ba
\label{Hamiltonian}
H_{I}(\eta) &=&  -\left(\frac{1}{\eta}\right)\left(\frac{b^2f}{a}\sqrt{6I}\sin^2{\theta}\right)
\left(\delta \rho ^{*} D_{1}'+c.c. \right) + \left( \frac{a^2}{f\eta} \right) \e^2 \sqrt{\frac{I \epsilon_{H}^2}{\lambda}}M_{P} \sin^2{\theta} \nonumber\\
&&
\Big{(} \delta \rho^{*} D_{1} + c.c.\Big{)} +  \left( \frac{a^2}{f\eta} \right) \e^2 \sqrt{\frac{I \epsilon_{H}^2}{\lambda}}M_{P} \cos{\theta}\Big{(} \delta \rho^{*} D_{2} + c.c.\Big{)}\nonumber\\
&&
-\left(\frac{1}{\eta}\right)\left(\frac{b^2f}{a}\sqrt{6I}\cos{\theta}\right)\left(\delta \rho ^{*} D_{2}'+c.c. \right)\theta(\eta-\eta_{c}) \nonumber\\
&& \equiv H_{1} + H_{2} + H_{3} + H_{4} \,  ,
\ea
in which the form of interactions $H_i, i=1,..4$, is read off in order from the above equation.
Here we used the step function $\theta (\eta - \eta_c)$ to take into account the change in the form of interaction after $\eta > \eta_c$ for the longitudinal mode.

Plugging back Eq. (\ref{Hamiltonian}) into the Eq. (\ref{commutator})  the non-zero terms are
\begin{align}
\label{Correction Powerspectrum}
\frac{\Delta <\overline{\delta \rho}^2(\eta_{e})>}{<\overline{\delta \rho}^2(\eta_{e})>}
&= \frac{\Delta <\overline{\delta \rho}^2(\eta_{e})>_{11}}{<\overline{\delta \rho}^2(\eta_{e})>} + \frac{\Delta <\overline{\delta \rho}^2(\eta_{e})>_{12}}{<\overline{\delta \rho}^2(\eta_{e})>} + \frac{\Delta <\overline{\delta \rho}^2(\eta_{e})>_{21}}{<\overline{\delta \rho}^2(\eta_{e})>} \nonumber \\
& ~~~ + \frac{\Delta <\overline{\delta \rho}^2(\eta_{e})>_{22}}{<\overline{\delta \rho}^2(\eta_{e})>} + \frac{\Delta <\overline{\delta \rho}^2(\eta_{e})>_{33}}{<\overline{\delta \rho}^2(\eta_{e})>} + \frac{\Delta <\overline{\delta \rho}^2(\eta_{e})>_{34}}{<\overline{\delta \rho}^2(\eta_{e})>} \nonumber \\
& ~~~ 
+ \frac{\Delta <\overline{\delta \rho}^2(\eta_{e})>_{43}}{<\overline{\delta \rho}^2(\eta_{e})>} + \frac{\Delta <\overline{\delta \rho}^2(\eta_{e})>_{44}}{<\overline{\delta \rho}^2(\eta_{e})>}
\end{align}
In this notation, $\Delta <\overline{\delta \rho}^2(\eta_{e})>_{ij}$ represents the contribution of the two interactions from $H_i$ and $H_j$ in Eq. (\ref{Hamiltonian}).

Now we calculate each term in Eq. (\ref{Correction Powerspectrum}) in turn.  The contributions from the transverse mode $D_1$ (and $D_1'$) are
\begin{align}
\label{Correction Powerspectrum11}
\frac{\Delta <\overline{\delta \rho}^2(\eta_{e})>_{11}}{<\overline{\delta \rho}^2(\eta_{e})>}
&= \frac{192 I}{|u^{(0)}(\eta_{e})|^2}
\int_{\eta_{0}}^{\eta_{e}} d\eta_{1} \int_{\eta_{0}}^{\eta_{1}}d\eta_{2} \left(\frac{\eta_{1}\eta_{2}}{\eta_{e}^4}\right)\bigg{(}\sin^4{\theta} Im \bigg{[}u(\eta_{1})u^{*}(\eta_{e})\bigg{]}\nonumber\\
&~~~ Im \bigg{[}u(\eta_{2})u^{*}(\eta_{e})v^{*'}(\eta_{1})v^{'}(\eta_{2})\bigg{]}\bigg{)} = 24 I \sin^2{\theta} N_{e}^2
\end{align}
\begin{align}
\label{Correction Powerspectrum12}
\frac{\Delta <\overline{\delta \rho}^2(\eta_{e})>_{12}}{<\overline{\delta \rho}^2(\eta_{e})>}
&= \frac{32 \e^2 I \epsilon_{H}\sqrt{6}}{|u^{(0)}(\eta_{e})|^2 \sqrt{\lambda}}\frac{M_{P}}{H}
\int_{\eta_{0}}^{\eta_{e}} d\eta_{1} \int_{\eta_{0}}^{\eta_{1}}d\eta_{2} \left(\frac{\eta_{1}}{\eta_{2}^4}\right)\bigg{(}\sin^4{\theta} Im \bigg{[}u(\eta_{1})u^{*}(\eta_{e})\bigg{]} \nonumber\\
& ~~~Im \bigg{[}u(\eta_{2})u^{*}(\eta_{e})v^{*'}(\eta_{1})v(\eta_{2})\bigg{]}\bigg{)} = -\frac{31}{245} \e^2 I \epsilon_{H} \sqrt{\frac{6}{\lambda}} \frac{M_{P}}{H} \sin^2{\theta}
\end{align}
\begin{align}
\label{Correction Powerspectrum21}
\frac{\Delta <\overline{\delta \rho}^2(\eta_{e})>_{21}}{<\overline{\delta \rho}^2(\eta_{e})>}
&= \frac{32 \e^2 I \epsilon_{H}\sqrt{6}}{|u^{(0)}(\eta_{e})|^2 \sqrt{\lambda}}\frac{M_{P}}{H}
\int_{\eta_{0}}^{\eta_{e}} d\eta_{1} \int_{\eta_{0}}^{\eta_{1}}d\eta_{2} \left(\frac{\eta_{2}}{\eta_{1}^4}\right)\bigg{(}\sin^4{\theta} Im \bigg{[}u(\eta_{1})u^{*}(\eta_{e})\bigg{]} \nonumber\\
&~~~
Im \bigg{[}u(\eta_{2})u^{*}(\eta_{e})v^{*}(\eta_{1})v^{'}(\eta_{2})\bigg{]}\bigg{)}
= -\frac{2}{7} \e^2 I \epsilon_{H} \sqrt{\frac{6}{\lambda}} \frac{M_{P}}{H} \sin^2{\theta}N_{e}
\end{align}
\begin{align}
\label{Correction Powerspectrum22}
\frac{\Delta <\overline{\delta \rho}^2(\eta_{e})>_{22}}{<\overline{\delta \rho}^2(\eta_{e})>}
&= \frac{32 \e^4 I \epsilon_{H}^2}{|u^{(0)}(\eta_{e})|^2 \lambda}\frac{M_{P}^2}{H^2}
\int_{\eta_{0}}^{\eta_{e}} d\eta_{1} \int_{\eta_{0}}^{\eta_{1}}d\eta_{2} \left(\frac{\eta_{e}^4}{\eta_{1}^4\eta_{2}^4}\right)\bigg{(}\sin^4{\theta} Im \bigg{[}u(\eta_{1})u^{*}(\eta_{e})\bigg{]}
\nonumber\\
&~~~ Im \bigg{[}u(\eta_{2})u^{*}(\eta_{e})v^{*}(\eta_{1})v(\eta_{2})\bigg{]}\bigg{)} = \frac{9 }{1078} \frac{\e^4 I \epsilon_{H}^2}{\lambda} \frac{M_{P}^2}{H^2} \sin^2{\theta}
\end{align}
where $|u^{(0)}(\eta_{e})|^2 = \frac{1}{2k^3\eta_{e}^2}$ is the amplitude of the free inflaton field fluctuations. Note that, as we showed at the end of the previous Section,  both the inflaton field excitations and the gauge field excitations  remain nearly massless during most of the period of inflation, so we have used the massless mode function approximations for $u_k(\eta)$ and $v_k(\eta)$ given in Eq. (\ref{mode function0}).

The first term, Eq. (\ref{Correction Powerspectrum11}), is the same as in models of real inflaton field  \cite{Watanabe:2010fh}. However, the next three terms Eqs. (\ref{Correction Powerspectrum12}), (\ref{Correction Powerspectrum21}) and
(\ref{Correction Powerspectrum22}) are originated from the interaction $\e^2 \rho^2 A^2$
which does not exist in models with a real inflaton field.  Also note  the relative sign between  Eqs. (\ref{Correction Powerspectrum12}) and  (\ref{Correction Powerspectrum21}) compared to (\ref{Correction Powerspectrum22}).

The contributions of the longitudinal mode, $D_2$ (and $D_2'$) are
\begin{align}
\label{Correction Powerspectrum33}
\frac{\Delta <\overline{\delta \rho}^2(\eta_{e})>_{33}}{<\overline{\delta \rho}^2(\eta_{e})>}
&= \frac{32 \e^4 I \epsilon_{H}^2}{|u^{(0)}(\eta_{e})|^2 \lambda}\frac{M_{P}^2}{H^2}
\int_{\eta_{0}}^{\eta_{e}} d\eta_{1} \int_{\eta_{0}}^{\eta_{1}}d\eta_{2} \left(\frac{\eta_{e}^4}{\eta_{1}^4\eta_{2}^4}\right)\bigg{(}\cos^2{\theta} Im \bigg{[}u(\eta_{1})u^{*}(\eta_{e})\bigg{]}
\nonumber\\
& ~~~~~
Im \bigg{[}u(\eta_{2})u^{*}(\eta_{e})w^{*}(\eta_{1})w(\eta_{2})\bigg{]}\bigg{)} \nonumber\\
& ~~~~~ = \frac{18}{5} (\e^2 I \epsilon_{H}\lambda) \frac{M_{P}^4}{\cM^4} (k \eta_{e})^2 \cos^2{\theta}
\end{align}

\begin{align}
\label{Correction Powerspectrum34}
\frac{\Delta <\overline{\delta \rho}^2(\eta_{e})>_{34}}{<\overline{\delta \rho}^2(\eta_{e})>}
&= \frac{32 \sqrt{6} \e^2 I \epsilon_{H}}{|u^{(0)}(\eta_{e})|^2 \sqrt{\lambda}}\frac{M_{P}}{H}
\int_{\eta_{c}}^{\eta_{e}} d\eta_{1} \int_{\eta_{c}}^{\eta_{1}}d\eta_{2} \left(\frac{\eta_{2}}{\eta_{1}^4}\right)\bigg{(}\cos^2{\theta} Im \bigg{[}u(\eta_{1})u^{*}(\eta_{e})\bigg{]}
\nonumber\\
& ~~~~~
Im\bigg{[}u(\eta_{2})u^{*}(\eta_{e})w^{*}(\eta_{1})w^{'}(\eta_{2})\bigg{]}\bigg{)} \nonumber\\
& ~~~~~= 32 \sqrt{\frac{3\epsilon_{H}}{\lambda}} (\e I \lambda^2) \frac{H M_{P}^4}{\cM^6} (k^2 \eta_{e}) \cos^2{\theta}
\end{align}

\begin{align}
\label{Correction Powerspectrum43}
\frac{\Delta <\overline{\delta \rho}^2(\eta_{e})>_{43}}{<\overline{\delta \rho}^2(\eta_{e})>}
&= \frac{32 \sqrt{6} \e^2 I \epsilon_{H}}{|u^{(0)}(\eta_{e})|^2 \sqrt{\lambda}}\frac{M_{P}}{H}
\int_{\eta_{c}}^{\eta_{e}} d\eta_{1} \int_{\eta_{0}}^{\eta_{1}}d\eta_{2} \left(\frac{\eta_{1}}{\eta_{2}^4}\right)\bigg{(}\cos^2{\theta} Im \bigg{[}u(\eta_{1})u^{*}(\eta_{e})\bigg{]} 
\nonumber\\
& ~~~~~
Im \bigg{[}u(\eta_{2})u^{*}(\eta_{e})w^{'*}(\eta_{1})w(\eta_{2})\bigg{]}\bigg{)} \nonumber\\
& ~~~~~= -8 \sqrt{\frac{2}{3\lambda}} (I \lambda^2 k^2) \frac{H M_{P}^3}{\cM^4} \left(\frac{3 \e M_{P}\sqrt{2\epsilon_{H}}}{\cM^2} \right)^{2/3} \cos^2{\theta}\left(-\eta_{e}\right)^{4/3}
\end{align}

\begin{align}
\label{Correction Powerspectrum44}
\frac{\Delta <\overline{\delta \rho}^2(\eta_{e})>_{44}}{<\overline{\delta \rho}^2(\eta_{e})>}
&= \frac{192 I}{|u^{(0)}(\eta_{e})|^2} \int_{\eta_{c}}^{\eta_{e}} d\eta_{1} \int_{\eta_{c}}^{\eta_{1}}d\eta_{2} \left(\frac{\eta_{1}\eta_{2}}{\eta_{e}^4}\right)\cos^2{\theta} Im \bigg{[}u(\eta_{1})u^{*}(\eta_e)\bigg{]}
\nonumber\\
& ~~~~~
Im \bigg{[}u(\eta_{2})u^{*}(\eta_e)w^{*'}(\eta_{1})w^{'}(\eta_{2})\bigg{]} \nonumber\\
& ~~~~~= \frac{\pi^2}{27}\frac{\e M_{P} \cM\eta_{e}^2}{\Gamma(3/4)^4} \left(I(\frac{3I}{2})^{3/4}\sqrt{\epsilon_{H}}\right) (N_{e} - N_{c})^2 \cos^2{\theta}
\end{align}
Note that the contributions from the longitudinal mode are sourced by $\e$ and are scale-dependent. Furthermore, these terms all have positive powers of $\eta_e$ which are
exponentially small as expected.  As discussed in previous Section there is a cancelation in derivative couplings of the longitudinal mode between the terms coming from the $f^2 F^2$ interaction and a term coming from integrating out $\delta A_0$. As a result,
as shown in  Eq. (\ref{L-ratio5}),
the leading interaction from the longitudinal mode during most of the period of inflation ($0< N < N_c$) is much smaller than the leading interaction of the transverse mode.  This justifies why the anisotropy generated from the longitudinal mode is much smaller than the anisotropy generated from the transverse mode.

As a result the  fractional change in the curvature perturbations
power spectrum due to anisotropy, to leading order, is
\ba
\label{partd}
\frac{\Delta <\overline{\delta \rho}^2(\eta_{e})>}{<\overline{\delta \rho}^2(\eta_{e})>} &=& 24 I \sin^2{\theta} N_{e}^2 -\frac{31}{245} \e^2 I \epsilon_{H} \sqrt{\frac{6}{\lambda}} \frac{M_{P}}{H} \sin^2{\theta} -\frac{2}{7} \e^2 I \epsilon_{H} \sqrt{\frac{6}{\lambda}} \frac{M_{P}}{H} \sin^2{\theta}N_{e} \nonumber\\
&& + \frac{9}{1078} \frac{\e^4 I \epsilon_{H}^2}{\lambda} \frac{M_{P}^2}{H^2} \sin^2{\theta} + \frac{18}{5} (\e^2 I \epsilon_{H}\lambda) \frac{M_{P}^4}{\cM^4} (k \eta_{e})^2 \cos^2{\theta} \nonumber\\
&& + 32 \sqrt{\frac{3\epsilon_{H}}{\lambda}} (\e I \lambda^2) \frac{H M_{P}^4}{\cM^6} (k^2 \eta_{e}) \cos^2{\theta} -8 \sqrt{\frac{2}{3\lambda}} (I \lambda^2 k^2) \frac{H M_{P}^3}{\cM^4} \times\nonumber\\
&&\left(\frac{3\e M_{P}\sqrt{2\epsilon_{H}}}{\cM^2} \right)^{2/3} \cos^2{\theta}\left(-\eta_{e}\right)^{4/3} + \frac{\pi^2}{27}\frac{\e M_{P} \cM\eta_{e}^2}{\Gamma(3/4)^4} \times
\nonumber\\
&& \left(I(\frac{3I}{2})^{3/4}\sqrt{\epsilon_{H}}\right) (N_{e} - N_{c})^2 \cos^2{\theta} \, .
\ea
In this formula, $N_e$ stands for the total number of e-folds which we take to be $60$ and $N_c$ is the number of e-fold from the start of inflation till $\eta= \eta_c$ given by Eq. (\ref{Nc5}).

As mentioned before, the first four terms in Eq. (\ref{partd}) come from the transverse mode. The first term is similar to
\cite{Watanabe:2010fh} while the next three terms are due to the charge effects which do not exist in models with a real inflaton field.
However, the last four terms in Eq. (\ref{partd}) are due to longitudinal mode which also
do not exist in models with a real inflaton field. However,  since they are suppressed with the powers of $\eta_{e}$ we conclude that their contributions into $g_{*}$ is very small.  As a result, the dominant contribution in $g_*$ comes  from the transverse mode.
Since $\sin^2 \theta = 1- \cos^2 \theta$,
the leading correction to anisotropy power spectrum in  Eq. (\ref{partd})  is
\ba
\label{g-lead}
g_* \simeq - 24 I N_{e}^2 + \frac{2}{7} \e^2 I \epsilon_{H} \sqrt{\frac{6}{\lambda}} \frac{M_{P}}{H}  N_{e}  - \frac{9}{1078}  \frac{\e^4 I \epsilon_{H}^2}{\lambda} \frac{M_{P}^2}{H^2} \, .
\ea
The interesting thing is that the two contributions of the transverse mode in $g_*$, the last two terms in Eq. (\ref{g-lead}), have different signs. However, one can easily check that  the sign of $g_*$ is always negative, so the positive contribution from the term containing $\e^2$ is always offset by the negative term containing $\e^4$. This is intuitively understandable, since we expect that a total positive contribution in $g_*$ comes from the longitudinal mode which are exponentially suppressed in this model while we do not expect the net contribution from the transverse mode
to give a positive contribution in $g_*$. This is consistent with the results in \cite{Watanabe:2010fh}.

Demanding that $|g_*| < 0.01$ in order not to produce too much anisotropy,
we find that  $I\simeq 10^{-6}$ and \textbf{$ \e^2 \leq 10 \sqrt{\frac{\lambda}{I \epsilon_{H}^2}} \frac{H}{M_{P}}$}.
For typical model parameters in symmetry breaking inflation, this leads to $\e \lesssim 10^{-3}$.

As observed in \cite{Bartolo:2012sd}  the infra-red (IR) modes of the vector field perturbations remain frozen on super-horizon scales which accumulate to renormalize the background gauge field. As a result, this can lead to a large value of $g_*$ unless one takes
$N_e \sim 60$ as we have assumed here.

\section{The Origin of the Leading Interactions Terms}
\label{matter-leading}

Having calculated the anisotropic power spectrum through complicated procedure of integrating out
the non-dynamical fields and approximating $\bar \lambda_1$ and other $\lambda_i$ and
$\bar \lambda_i$, one may wonder what the origins of the leading interaction terms $L_{\rho D_1}$ and $L_{\rho D_2}$, or alternatively $L_{\rho A_{1} }$, $L_{\rho M}$  and $L_{A_{1} M}$, are. Are they coming from the metric perturbations or from the matter sector? 

The full second order action containing both the matter perturbations  and the metric perturbations contributions are given in Appendix \ref{reduced}. Subsequently, in Appendix \ref{second slow} we have presented the leading order actions in slow-roll approximation which were used in Section \ref{eta-less} and \ref{eta-big}. Here we show that these leading interactions actually come from the matter perturbations. In other words, below we show that the contributions of the  matter sector are actually the same leading terms which were used in
Section \ref{eta-less} and \ref{eta-big}.

To show this first we integrate out $\delta A_0$ and then read off the interaction terms containing the matter perturbations.  The leading terms in the matter sector coming from integrating out $\delta A_{0}$ are
\begin{align}
\label{lead A0}
&-\frac{k^2 \cos^2{\theta}}{\bar \lambda_{1}}\frac{b^4}{2a^3}\frac{\sqrt{6I}}{\eta}f^3(\delta \rho^* \delta A_{1}' + c.c.)-\frac{k^3\cos{\theta} \sin^2{\theta}} {\bar \lambda_{1}}\frac{b^4}{2a^3}\frac{\sqrt{6I}}{\eta}f^3(i\delta \rho^* M' + c.c.)\nonumber\\
&-\frac{b^4}{4a^4}\frac{k^3}{\bar \lambda_{1}} f^4 \sin^2{\theta}\cos{\theta}(i M' \delta A_{1}^{'*} + c.c.) \, .
\end{align}
On the other hand, the leading terms for the matter perturbations present in the original
action  (without  integrating out any fields) are
\begin{align}
\label{lead action}
\frac{b^2}{a}\frac{\sqrt{6I}}{\eta}f (\delta \rho^* \delta A_{1}' + c.c.)+ \frac{b^2}{2a^2}k^3 f^2 \sin^2{\theta}\cos{\theta}(iM \delta A_{1}^{*} + c.c.)
-\e^2b^2 \rho A_{x}\Big{(}\delta \rho \delta A_{1}^{*} + c.c. \Big{)}\, .
\end{align}
So by adding Eq. (\ref{lead action}) and (\ref{lead A0}) we can obtain all the leading interaction terms  for $L_{\rho A_{1} }$ , $L_{\rho M}$ and $L_{A_{1} M}$ as,
\begin{align}
\label{Leading Terms}
&\left(\frac{b^2}{a}\frac{\sqrt{6I}}{\eta}f-\frac{k^2 \cos^2{\theta}}{\bar \lambda_{1}}\frac{b^4}{2a^3}\frac{\sqrt{6I}}{\eta}f^3\right)\bigg{(}\delta \rho^* \delta A_{1}' + c.c. \bigg{)} - \frac{k^3\cos{\theta} \sin^2{\theta}} {\bar \lambda_{1}}\frac{b^4}{2a^3}\frac{\sqrt{6I}}{\eta}f^3 \times \nonumber\\
& \bigg{(}i\delta \rho^* M' + c.c.\bigg{)} + \left(\frac{b^2}{2a^2}k^3 f^2 \sin^2{\theta}\cos{\theta}\right)\bigg{(}iM \delta A_{1}^{*} + c.c.\bigg{)}-\left(\frac{b^4}{4a^4}\frac{k^3}{\bar \lambda_{1}} f^4 \sin^2{\theta}\cos{\theta}\right)
\nonumber\\
&
\bigg{(}iM' \delta A_{1}^{*'} + c.c.\bigg{)}-\e^2b^2 \rho A_{x}\Big{(}\delta \rho \delta A_{1}^{*} + c.c. \Big{)}\, .
\end{align}
Interestingly, this is the whole leading action which was used in previous sections
to calculate the anisotropic power spectrum.

As a result, the leading interaction terms for the first phase, $\eta < \eta_c$, are
\ba
\label{Leading Terms1}
L_{\mathrm{lead.}}&=& \left(\frac{1}{\eta}\right)\left(\frac{b^2f}{a}\sqrt{6I}\sin^2{\theta}\right)\left(\delta \rho ^{*} D_{1}'+c.c. \right) - \left( \frac{a^2}{f\eta} \right) \e^2 \sqrt{\frac{I \epsilon_{H}^2}{\lambda}}M_{P} \sin^2{\theta}\Big{(} \delta \rho^{*} D_{1} + c.c.\Big{)}\nonumber\\
&& - \left( \frac{a^2}{f\eta} \right) \e^2 \sqrt{\frac{I \epsilon_{H}^2}{\lambda}}M_{P} \cos{\theta}\Big{(} \delta \rho^{*} D_{2} + c.c.\Big{)}
\ea
Interestingly, this is exactly the leading term interaction as obtained in Eq. (\ref{rhoD1new5}). Similarly, for the second phase, $\eta > \eta_c$, Eq. (\ref{Leading Terms}) yields
\ba
\label{Leading Terms2}
L_{\mathrm{lead.}}&=& \left(\frac{1}{\eta}\right)\left(\frac{b^2f}{a}\sqrt{6I}\sin^2{\theta}\right)\left(\delta \rho ^{*} D_{1}'+c.c. \right) - \left( \frac{a^2}{f\eta} \right) \e^2 \sqrt{\frac{I \epsilon_{H}^2}{\lambda}}M_{P} \sin^2{\theta}
\nonumber\\
&& \Big{(} \delta \rho^{*} D_{1} + c.c.\Big{)}-  \left( \frac{a^2}{f\eta} \right) \e^2 \sqrt{\frac{I \epsilon_{H}^2}{\lambda}}M_{P} \cos{\theta}\Big{(} \delta \rho^{*} D_{2} + c.c.\Big{)}
+\left(\frac{1}{\eta}\right)\times
\nonumber\\&& 
\left(\frac{b^2f}{a}\sqrt{6I}\cos{\theta}\right)\left(\delta \rho ^{*} D_{2}'+c.c. \right)
\ea
As expected, this expression is the sum of the leading interaction terms Eqs. (\ref{rhoD1 action205}) and (\ref{rhoD2 action205}).

In summary we conclude that the leading interactions in generating anisotropies originate from the matter sector and one can neglect the metric perturbations in calculating the leading order corrections to the curvature perturbations power spectrum . Computationally, this is a very important result which considerably simplifies the perturbation analysis in similar models. This conclusion was also reached in  \cite{Bartolo:2012sd}.

This also explains why in the processes of integrating out the non-dynamical fields only  $\bar \lambda_1$ plays prominent roles. As mentioned below Eq. (\ref{b-lambda15}) $\bar \lambda_1$ originates from integrating out $\delta A_0$ which is the non-dynamical field in the matter sector. On the other hand,
other $\lambda_i$ and $\bar \lambda_i$ originate from integrating out the non-dynamical fields
$A, B$ and $\beta$ in the metric side which should not play prominent roles as expected from the above results.

\section{Summary}
\label{Summary}

In this chapter, we have studied anisotropy generated in an anisotropic inflationary scenario with a complex scalar field charged under the $U(1)$ gauge field. Because of the Abelian Higgs mechanism, the gauge field obtains the dynamical mass $\e^2 \rho^2 A_\mu A^\mu$. As a result, the angular excitations of the complex scalar field is eaten by the gauge field so the longitudinal component of $A_\mu$ becomes excited. \\
There are two types of interactions in the system. The first interaction originates from the gauge kinetic coupling $f(\rho)^2 F^2$ while the second interaction comes from the symmetry breaking effect $\e^2 \rho^2 A_\mu A^\mu$. These interactions induce exchange vertices  between $\delta \rho$ and the transverse and the longitudinal modes encoded in  the interactions $L_{\rho D_1}$ and $L_{\rho D_2}$. As discussed in details in Section \ref{sec-ac} the dominant interaction during the period $0< N< N_c$ is $L_{\rho D_1}$ originated from
$f(\rho)^2 F^2$ which is similar to models with a real inflaton field. As a result the leading exchange vertex is given by the derivative coupling of the transverse mode. However, during the phase $N_c \leq N \leq N_e$ the dominant interaction is given by $\e^2 \rho^2 A_\mu A^\mu$. Correspondingly, the dominant exchange vertices are the terms in $L_{\rho D_1}$ and $L_{\rho D_2}$ containing the coupling $\e^2$. \\
The leading contributions to anisotropic power spectrum are  given in Eq. (\ref{partd}) and
Eq.  (\ref{g-lead}). The first four terms in  Eq. (\ref{partd}) come from the interaction of $\delta \rho $ with the transverse mode, $L_{\rho D_1}$.
In terms of Feynman diagrams this interaction is represented  by the exchange vertex shown in Fig. 1 (a). This is similar to the result obtained  in  \cite{Watanabe:2010fh} plus the contributions in Eq.  (\ref{g-lead}) containing the effects of $\e$. As we showed, the sign of $g_*$ is always negative.  In order to satisfy the observational constraints on curvature perturbation power spectrum we obtain $I \lesssim 10^{-5}$ and $\e \lesssim 10^{-3}$.
In addition, unlike \cite{Watanabe:2010fh}, the longitudinal mode $D_2$ also contributes into the anisotropic power spectrum.  In terms of the Feynman diagrams this interaction is represented  by the exchange vertex shown in Fig. 1 (b). However,  the longitudinal mode contributes only towards the end of inflation and its contributions to the anisotropic power spectrum are hugely suppressed compared to the contribution from the transverse mode. \\
We also verified that the leading interactions in anisotropic power spectrum come from the matter sector perturbations. In other words, to calculate the leading order corrections into the power spectrum, one can neglect the metric perturbations. Computationally, this knowledge simplifies the analysis considerably.

In this chapter we have only calculated anisotropy in  curvature perturbation power spectrum. However, after inflation ends the Universe becomes isotropic. As a result, we restore the usual two degrees of freedom associated with the tensor perturbations. One can specifically check that the scalar perturbation $\gamma$ and the vector perturbations $\Gamma_i$ furnish two polarizations of tensor perturbations after inflation ends. Note that $\Gamma_i$, subject to 
$\partial _i \Gamma_i =0$ during anisotropic inflation, has only one degrees of freedom 
($\Gamma_3$ in our convention) so it can account only for one tensor polarization while the other polarization is given by $\gamma$ as expected. As shown in Appendices \ref{reduced}
and \ref{second slow} the interactions  $L_{\delta \rho \gamma}, L_{\gamma D_1 }$ and $L_{\gamma D_2}$ are generated in our system. As a result  there will be cross correlation between the tensor and scalar perturbations in the form of $\langle \delta \rho \gamma \rangle $ as studied in \cite{Watanabe:2010fh}. Following the in-in formalism analysis,  the cross correlation $\langle \delta \rho \gamma \rangle $ has contributions from the interaction  $L_{\delta \rho \gamma}$ and also contributions  from the second order action
$ L_{\delta \rho D_1} L_{\gamma D_1 } $. As in \cite{Watanabe:2010fh} we expect to have a contribution  like $- 24 I \sqrt{\epsilon_H} N^2 \sin^2 \theta $ in $ \langle \delta \rho \gamma \rangle$.  In addition, our analysis shows that we also obtain contributions proportional to $\e^2$ and $\e^4$ with the structure similar to the corresponding 
terms in $g_*$ in Eq. (\ref{g-lead}).  A complete analysis of the scalar and tensor perturbations cross-correlation  is an interesting question and we will come back to it in the next chapters.


\chapter{$ \delta N$  Formalism in Anisotropic Inflation and \\
Large Anisotropic Bispectrum and Trispectrum } 

\label{Chapter5} 

\lhead{Chapter 5. \emph{Cosmological Perturbation Theory}} 

\vspace{0.5cm}
\hrule \vspace{0.3cm}

\begin{quote}
\textbf{Abstract:} 
In this chapter, we present a consistent $\delta N$ formalism for calculating the curvature perturbations in anisotropic
cosmological backgrounds. We then employ our $\delta N$ formalism to calculate the power spectrum, the bispectrum and the trispectrum in models of anisotropic inflation with the background gauge fields in Bianchi I universe. Our results coincide exactly with the previous results obtained from in-in formalism. We study the Suyama-Yamaguchi inequality for the amplitudes of the bispectrum and the trispectrum in the presence of anisotropic shapes.
\end{quote}

\vspace{0.1cm}  \hrule
\vspace{0.5cm}
\section{Introduction}
As we have already seen in chapters \ref{Chapter3} and \ref{Chapter4}, recently there have been 
many interests in anisotropic inflation both from the observation as well as the theoretical point of view. From the theoretical point of view, some interesting model of anisotropic inflation were proposed in \cite{Watanabe:2009ct, Emami:2010rm} in which with proper choice for the gauge kinetic coupling, the inflationary system admits an attractor solution in which the gauge field energy density, i.e. the electric field energy density, and the metric anisotropy reaches a small but cosmologically observable level, see chapter \ref{Chapter3} for more details. \\
The cosmological perturbations of these models have been also done recently, see chapter \ref{Chapter4} for more details.  
These analysis are based on standard in-in formalism which proved technically difficult due to anisotropic background. On the other hand, experiences with
$\delta N$ formalism \cite{Sasaki:1995aw, Wands:2000dp, Lyth:2004gb, Lyth:2005fi, Naruko:2012fe, Naruko:2012um, Sugiyama:2012tj, Dias:2012qy}
in models of inflation with scalar fields showed that $\delta N$ analysis are technically
much easier to handle  when calculating the curvature perturbations and their correlations such
as power spectrum and bispectrum. Therefore it will be very helpful to extend the
standard $\delta N$ formalism 
to models of anisotropic backgrounds such as \cite{Watanabe:2009ct}. This is one of our main goal in this chapter. \\
There have been works in the literature employing the conventional
$\delta N$ formalism for the models  containing vector or gauge fields but the effects of anisotropic background were not taken into account, i.e. the gauge field is treated on the same footing as the scalar fields in an FRW background.\\
In this chapter, we present a consistent $\delta N$ formalism for anisotropic backgrounds such as in \cite{Watanabe:2009ct} in which the background metric is in the form of Bianchi I. After presenting our $\delta N$ formalism, we calculate the power spectrum and reproduce exactly the results in \cite{Watanabe:2010fh, Bartolo:2012sd}. We also calculate the bispectrum which coincides exactly with the results of \cite{Bartolo:2012sd}. \\
In addition, it is also worth to check whether the Suyama-Yamaguchi (SY) inequality \cite{Suyama:2007bg}, \cite{Sugiyama:2011jt}, \cite{Smith:2011if, Assassi:2012zq} between the amplitude of the Bispectrum in the squeezed limit, $f_{NL}$, and the amplitude of the trispectrum in the collapsed limit, $\tau_{NL}$, is hold when the primordial perturbations are not statistically isotropic. We will study this question in the context of anisotropic inflation. \\
This chapter is based on \cite{Abolhasani:2013zya}. 

\section{ $\delta N$ formalism for anisotropic backgrounds}
\label{deltaN-sec}

In this section we extend the $\delta N$ formalism \cite{Sasaki:1995aw, Wands:2000dp, Lyth:2004gb, Lyth:2005fi, Naruko:2012fe, Naruko:2012um, Sugiyama:2012tj, Dias:2012qy} to anisotropic backgrounds. 
First we present the background fields equations. After presenting the general metric perturbations,  we look into the fields equations using a gradient expansion method, which is an expansion in $\epsilon$ defined via
\ba
\label{epsilon-eq}
\epsilon \equiv \frac{k}{a H} \, ,
\ea
We should emphasize here that $\epsilon$ here is different than the slow-roll parameter, $\epsilon_H$. $k$ represents the wave number in Fourier space.  
We demonstrate that the separate universe picture works, that is, in the limit $\epsilon \ll 1$ the background fields equations are locally hold inside each homogenized patch.  This proof is valid to all order in perturbation theory.

\subsection{Background Equations}
Our background is  the Bianchi I metric  with the scale factors $a_1(t), a_2(t)$ and $a_3(t)$ 
\ba
\label{Bianchi-metric1}
ds^2 = -dt^2 + a_1(t)^2 d x^2 +a_2(t)^2 d y^2 +a_3(t)^2 d z^2 \, .
\ea
We adopt the notations used in \cite{Miedema:1993} in which 
\ba
H_i(t)= \dfrac{\dot{a_i}}{a_i} \qquad , \qquad H \equiv \dfrac{1}{3} \sum_{i=1}^3 H_i \, ,
\ea
in which $H_i$ is the Hubble expansion rate for the $i$-th spatial direction, $i=1,2,3$ and a dot indicates the derivative with respect to $t$.

The components of background Ricci tensor are
\ba
\label{Ricci-back}
R^0{}_0 &=& 3 \dot H + \sum_{k} H_k ^2
\\
R^0{}_i&=&0 
\\
R^i{}_j&=& \delta^i{}_j \left(\dot{H}_i + 3HH_i \right).
\ea
The background Ricci scalar is 
\ba
R =  6 \dot H + 18 H^2 - \sum_{k > k'} H_k H_{k'} \, 
\ea
To solve the Einstein fields equations we have to specify our energy momentum tensor.
The general  energy momentum tensor $T_{\mu \nu}$ for an imperfect fluid has the form \cite{ellis98}
\ba
\label{eq:stress6}
T_{\mu \nu} = (\rho + p)\,u_{\mu}\,u_{\nu}+ p\,g_{\mu \nu}+ q_{\mu}\,u_{\nu} + u_{\mu}\,q_{\nu} +  \pi_{\mu \nu}
\ea
supplemented with the following conditions 
\ba
q_{\mu}\,u^{\mu} = 0 \quad , \quad   \pi^{\mu}{}_{\mu} = 0  \quad , \quad  ~\pi_{\mu \nu} = \pi_{\nu \mu} \quad , \quad 
~\pi_{\mu \nu}\,u^{\nu} = 0 \ , \nonumber
\ea
where $u^\mu$ is the fluid's four-vector velocity,  $\rho$ is the relativistic energy density, $p $ is the isotropic pressure, $\pi_{\mu \nu} $ is the trace-free  anisotropic pressure (stress) and $q^\mu$ usually is referred  to as ``heat conduction'', which is also the energy flux relative to $u^\mu$.

The special case of a perfect fluid is identified with $\pi_{\mu \nu}= q^\mu=0$ so we recover the standard form of $T_{\mu \nu}$ for the perfect fluid 
\ba
\label{eq:pf}
T_{\mu \nu} &=& (\rho+p)\,u_{\mu}\,u_{\nu} + p\,g_{\mu \nu}
\quad \quad \mathrm{( perfect \, \,  \,  fluid)} \, .
\ea

For the comoving coordinate associated with the fluid we have 
\ba
u^{\mu} = (1, \vec{0})  \qquad , \qquad u_{\mu} = (-1, \vec{0}),
\ea
so the Einstein equations can be read as 
\ba
\label{00-back6}
3  {\cal H}^2&\equiv& \sum_{i > j} \bar H_j \bar H_{j} =  \frac{\bar \rho}{M_P^2}
\\
\bar T^0{}_i &=& \bar q_i =0
\\
\label{i=j-back6}
M_P^2 \dot{\bar H}_i&=&-3  M_P^2 \bar H \bar H_i + \dfrac{1}{2} (\bar \rho-\bar p) + \bar \pi^{i}{}_{i}
\ea
Here we have used the convention that $\bar H_i$ represents the background Hubble expansion rates while  $\bar \rho, \bar p$ and  so on represent the background fluid's properties. We also defined
${\cal H}$ as the effective Hubble expansion rate appearing in Friedmann equation, Eq. (\ref{00-back6}). Note, ${\cal H}$ here should not be confused with the Hubble expansion rate defined for conformal time usually used in literature.

Finally, the energy conservation equation $u_{\mu} \nabla_{\nu} T^{\mu \nu} = 0$ results in
\ba
\label{cont.-back6}
-u_{\mu} \nabla_{\nu} T^{\mu \nu} = \dot{\bar \rho} + 3 H (\bar \rho + \bar p) + \bar H_j \bar \pi
^{i}{}_{j} \delta^{j}_i=0 .
\ea
in which again we have $\bar H = \sum_{i} \bar H_i/3$. 

Note that, in this model, the Hubble parameter appearing in $(0,0)$ component of Einstein equation, ${\cal H}$,  and the Hubble parameter appearing in continuity equation, $\bar H$, are not equal. The  difference between them is given by
\ba
\dfrac{ {\bar H}^2-{\cal H}^2 }{\bar H^2}= \dfrac{1}{6} \, \dfrac{\sum (\bar H-\bar H_i)^2}{\bar H^2}  \, .
\ea
As a result $\bar H > {\cal H}$.

\subsubsection{  Example: $U(1)$ gauge fields in an expanding background}

As an example of non-perfect fluid with anisotropic pressure and heat conduction, consider 
the standard $U(1)$ gauge field theory in an expanding background. This theory will be the base of anisotropic inflation in next section. The action is $L_{em} = -F_{\mu \nu} F^{\mu \nu}/4$ in which $F_{\mu \nu}=\partial_\mu A_\nu - \partial_\nu A_\mu$ is the field strength associated with the $U(1)$ gauge field $A_\mu$.

The electric field, $E_\mu$, and the magnetic field, $H_\mu$,  are given by,
\begin{equation}
E_\mu=F_{\mu\nu}u^\nu
\end{equation}
\\ and
\begin{equation}
H_\rho=\frac 1{2}\eta _{\rho\mu \nu \sigma}u^\mu F^{\nu \sigma} \, .
\label{mf}
\end{equation}
The electromagnetic energy-momentum tensor, $T^{\mu \nu}_{em}$, is
\begin{equation}
T^{\mu \nu}_{em}= F^{\sigma \mu}F_\sigma^{\hspace{1mm}\nu}- \frac{1}{4}g^{\mu \nu}F_{\sigma \rho}F^{\sigma \rho} \, .
\label{Tem}
\end{equation}
For  an observer comoving with the fluid $T_{em}^{\mu \nu}$ can be written as \cite{barrow97} 
\begin{equation}
T_{em}^{\mu \nu}=\frac12\left( E^2+H^2\right) u^\mu u^\nu+\frac
16\left( E^2+H^2\right) h^{\mu \nu}+2u^{(\mu}\eta
^{\nu)\alpha \beta \gamma}u_\alpha E_\beta H_\gamma+\pi^{\mu \nu},  \label{Tem1}
\end{equation}
where $\eta^{\mu \nu \alpha \beta}$ is the four-dimensional totally antisymmetric volume element  ($\eta_{0123} = \sqrt{- \det g}$), $h_{\mu \nu} = g_{\mu \nu} + u_\mu u_\nu$ is the projection matrix,  $E^2=E_\mu E^\mu$ and $H^2=H_\mu H^\mu$, respectively, are the magnitudes of the  electric and the magnetic fields and $\pi_{\mu \nu}$ is a traceless and  space-like symmetric tensor given by
\begin{equation}
\pi^{\mu \nu}_{em}= \frac{1}{3}\left(E^2+H^2\right)h^{\mu \nu}- E^\mu E^\nu-H^\mu H^\nu .  \label{Mten1}
\end{equation}
Eq. (\ref{Tem1}) can be compared with the energy momentum tensor for a generic
imperfect fluid defined in Eq. (\ref{eq:stress6}) which yields 
\begin{eqnarray}
\rho _{em} &=&\frac 1{2}\left( E^2+H^2\right) ,  \label{muem}
\\
&&  \nonumber \\
p_{em} &=&\frac 16\left( E^2+H^2\right) ,  \label{pem} \\
&&  \nonumber \\
q_{em}^\mu &=&\eta ^{\mu \nu   \alpha \beta}u_\nu E_\alpha H_\beta,  \label{qem} \\
&&  \nonumber \\
\pi ^{\mu \nu } &=&\pi^{\mu \nu}_{em}.  \label{piem}
\end{eqnarray}

\subsection{Perturbations}

Let us now consider the fields equations with perturbations. In our $\delta N$ analysis we adopt the notation used in \cite{Sugiyama:2012tj}. The order of spatial derivative or the so-called gradient expansion is denoted by $\epsilon=k/aH$ while the order of smallness of perturbations are denoted by $\delta$. In principle, one has to consider different gradient expansion parameters $\epsilon_i$ for different directions $\epsilon_i = k/a_i H_i$. However, to simplify the analysis we assume $\epsilon_i \sim \epsilon$ so there is no hierarchy for gradient expansions along different directions. 

We use the standard ADM formalism for the metric decomposition as follows
\ba
\label{ADM}
ds^2 = -d{\cal N}^2 + \gamma_{ij} \left( dx^i + \beta^i dt \right) \left( dx^j + \beta^j dt\right) \, ,
\ea
in which ${\cal N}$ is the lapse function, $\beta_i$ are the shift vectors, and $\gamma_{ij}$ represent the spatial three-dimensional  metric. The spatial indices $i=1,2,3$ are raised or lowered by the spatial metric  $\gamma_{ij}$. Furthermore,  we decompose the spatial metric as follows
\ba
\label{gamma-ij}
\gamma_{ij} = a_i(t) a_j(t) e^{\psi_i (\mathbf{x},t)+\psi_j (\mathbf{x},t)} \tilde{\gamma}_{ij} \, ,
\ea
where $a_i(t)$ is the average scale factor for the $i$-th spatial direction and $\psi_i(\mathbf{x},t)$  are equivalent to curvature perturbation $\psi$ in the isotropic limit. In linear perturbation theory $ \beta^i, \psi_i $ and $ \tilde {\gamma}_{ij}$ are small perturbations at the order  ${\cal O}({\delta})$ with $\delta \ll1$.  But in our analysis below, we  do not use the assumption that
$\delta \ll 1$ so our analysis are valid to all order in perturbation theory. 

Note that in general Bianchi Type-I model we considered here  there is no spatial symmetry so all physical degrees of freedom are in the form of scalar perturbations, encoded  in ${\cal N}, \beta_i, \psi_i$ and $\tilde{\gamma}_{ij} , i \neq j$
and  there is no vector or tensor perturbations.


An important step in dealing with the gradient expansion ordering of Einstein equations 
is the order of the shift functions $\beta^i$. We note that at the background level 
$\beta^i=0$. As a result one expects that the background metric should be valid globally 
in the limit $\epsilon \rightarrow 0$ and, as employed in  \cite{Lyth:2004gb}, one can assume 
\ba
\label{beta-order5}
\beta^{i} = {\cal O} (\epsilon).
\ea
The ordering of $\beta^i $  in Eq. (\ref{beta-order}) was also obtained in \cite{Sugiyama:2012tj} with the assumption that the anisotropic pressure is first order in gradient expansion. 
We look into ordering of $\beta^i$ more rigorously in Appendix \ref{App-B}  and verify Eq. (\ref{beta-order5}).  
Furthermore, as we demonstrated in Appendix  \ref{off-Ein}, 
it can be shown that the non-diagonal spatial metric components, $\gamma_{ij}$, to all orders in perturbations  theory  are also at the first order of gradient expansion
\ba
\gamma_{i\neq j} = {\cal O} (\epsilon) \, .
\ea

Now we have all the necessary materials for performing the gradient expansion analysis for the Einstein equations. Here we emphasize that the following expansions are valid to the first order of gradient expansion $\epsilon$ but to all orders of perturbations $\delta$.

The $(0,0)$ component of perturbed Einstein tensor is
\ba
\label{00-pert}
G^0{}_0 =  \dfrac{-1}{{\cal N}^2}  \sum_{i > j} (\bar H_i+ \dot{\psi}_{i})(\bar H_j+ \dot{\psi}_{j}) +{\cal O} (\epsilon^2)
\ea
Combining Eq. (\ref{00-pert}) with the background $(0,0)$ component 
equation, Eq. (\ref{00-back6}), yields 
\be
\dfrac{-M_P^2}{{\cal N}^2}  \sum_{i > j} (\bar H_i+ \dot{\psi}_{i})(\bar H_j+ \dot{\psi}_{j})=  \rho (\mathbf{x},t)+O(\epsilon^2)
\ee
Locally, as a function of $(\mathbf{x},t)$, the above equation takes the form
\be
\label{Fried-pert}
3 M_P^2 {\cal H}^2(\mathbf{x},t) =  \rho (\mathbf{x},t) +O(\epsilon^2),
\ee
in which
\ba
\label{calH}
{\cal H}^2(\mathbf{x},t) \equiv \frac{1}{3}\sum_{i > j} H_i(\mathbf{x},t) H_{j} (\mathbf{x},t) \, ,
\ea
with the following generalization of local Hubble expansion parameter $H_i(\mathbf{x},t)$
\ba
\label{H-local}
H_i(\mathbf{x},t) \equiv \dfrac{ \bar H_i(t) + \dot{\psi}_i (\mathbf{x},t)}{\cal N} \, .
\ea
As a result one can readily associate the average local Hubble expansion rate $H (\mathbf{x},t) $ as 
\ba
\label{H-average6}
H (\mathbf{x},t) \equiv  \frac{1}{3} \sum_i H_i(\mathbf{x},t) = \dfrac{ \bar H(t) + \frac{1}{3}\sum_i\dot{\psi}_i (\mathbf{x},t)}{\cal N}
\ea
in which the background average Hubble expansion rate $\bar H$ is $\bar H= \sum_i \bar H_i/3$.

Now we look at the energy conservation equation in its contracted form 
$u_{\mu} \nabla_{\nu} T^{\mu \nu} = 0$.  At the background level the energy conservation
equation is given by Eq. (\ref{cont.-back6}). Defining the fluid's proper time $\tau$ via
$\frac{d}{d\tau} = u^{\mu} \nabla_{\mu} \simeq \frac{1}{\cal N} \frac{d}{dt} +{\cal O}(\epsilon^2)$, the perturbed
energy conservation equation is
\ba
\dfrac{d\rho(\mathbf{x},t)}{d \tau}+3 H(\mathbf{x},t) \left(\rho(\mathbf{x},t)+p(\mathbf{x},t)\right) +\left[-u_{\mu} \frac{d}{d\tau}q^{\mu} + \nabla_{\mu} q^{\mu}- u_{\mu} \nabla_{\nu} \pi^{\mu \nu}\right]=O(\epsilon^2)\,,
\label{cont}
\ea
in which $H(\mathbf{x},t)$ is the average local Hubble expansion rate defined in Eq. \eqref{H-average6}. By using Eq. \eqref{aniso-pres-con} and \eqref{heat-con} the above equation takes the following simple local form
\ba
\dfrac{d\rho(\mathbf{x},t)}{d \tau}+3 H(\mathbf{x},t) \left( \, \rho (\mathbf{x},t)+p(\mathbf{x},t) \, \right)+ \sum_i{\pi}^i{}_i(\mathbf{x},t)   H_i (\mathbf{x},t) ={\cal O}(\epsilon^2) \, .
\label{cont-sim}
\ea
in which ${\pi}^i{}_i(\mathbf{x},t) =  \bar {\pi}^i{}_i + \delta {\pi}^i{}_i(\mathbf{x},t)$  to all orders in perturbations.

So again we conclude that  our separate universe  recipe works and it is enough to replace any background function $f(t)$ by its local form $f(\mathbf{x},t)$ and also using new local directional Hubble parameters $H_i (\mathbf{x},t)$. Our prescription will be satisfactory if we can also check the $(i=j)$ components of Einstein equations which are identical to the dynamical equations  of $\pi^i_i$. The diagonal spatial components of Ricci tensor  can be read as 
\ba
R^i{}_i =\dfrac{dH_i(\mathbf{x},t)}{d \tau} + 3 H(\mathbf{x},t)H_i(\mathbf{x},t) + {\cal O} (\epsilon^2) \, ,
\ea
so the $(i=j)$ components of Einstein equation simply modifies the corresponding background equation, Eq. (\ref{i=j-back6}), as follows (for the off-diagonal components of Einstein equation see Appendix \ref{off-Ein})
\ba
\label{dH-pert}
M_P^2 \dfrac{dH_i(\mathbf{x},t)}{d \tau}&=&-3 M_P^2 H(\mathbf{x},t)H_i(\mathbf{x},t) + \dfrac{1}{2} \left(\rho(\mathbf{x},t)-p(\mathbf{x},t) \right)  + \pi^{i}{}_{i}(\mathbf{x},t)
\ea

Now we have a complete set of local fields equations, Eq. (\ref{Fried-pert}), Eq. (\ref{cont-sim}) and Eq. (\ref{dH-pert}),  mimicking the corresponding background equations,
Eq. (\ref{00-back6}),  Eq. (\ref{cont.-back6}) and Eq. (\ref{i=j-back6}),  
with the  local Hubble parameters $H_i(\mathbf{x},t) $ defined in Eq. (\ref{H-local}).  We emphasize again that this set of equations are valid to all order in perturbations $\delta$ but to the first order of gradient expansion $\epsilon$.

The separate Universe approach discussion is now complete. The $\delta N$ formalism is also at hand noting that from the  equations above one has
\ba
N_i(\mathbf{x},t_1,t_2) \equiv \int_{t_1}^{t_2} H_i (\mathbf{x},t) {\cal N} dt = \int_{t_1}^{t_2} \bar H_i  dt + \int_{t_1}^{t_2} \dot{\psi}_i  dt
\ea
So one readily finds
\ba
N_i(\mathbf{x},t_1,t_2) - \bar{N}_i(t) = \psi_i(t_2) -  \psi_i(t_1)
\ea
Now defining the average  expansion by 
\ba
N(\mathbf{x},t_1,t_2)= \frac{1}{3}\sum_i N_i(\mathbf{x},t_1,t_2) 
= \int_{t_1}^{t_2} H (\mathbf{x},t) {\cal N} dt
\ea
one obtains
\ba
\delta N(\mathbf{x},t_1,t_2)=      N(\mathbf{x},t_1,t_2) - \bar{N}(t) = \psi(t_2) -  \psi(t_1)
\ea
in which   $\psi (\mathbf{x},t)$ is defined as the average of $\psi$
\ba
\psi(\mathbf{x},t) \equiv \dfrac{1}{3}  \sum_{i} \psi_i(\mathbf{x},t) \, .
\ea

We are interested in curvature perturbation on surface of constant energy density.  The curvature perturbation $\zeta$ via
\ba
\label{zeta}
-\zeta  = \psi - \frac{H}{\dot \rho} \delta \rho \, ,
\ea
is gauge invariant.  But this definition just works to the first order in perturbations $\delta$. The definition of $\zeta$
to all orders of perturbation theory can be found in \cite{Lyth:2004gb}. However, as it is shown below, we calculate  $\delta N$ on the surface of uniform energy density so the definition of $\zeta$ to nonlinear orders is irrelevant for our purpose. 

The relation between $\zeta$ and $\delta N$ therefore is  
\ba
\label{zeta-N}
\zeta(\mathbf{x},t) = \delta N(\mathbf{x},t_i,t_f)  \, ,
\ea
in which the initial surface is a flat surface $\psi=0$ and the final surface should be a uniform energy density surface $\delta \rho= 0$.

Here a comment is in order. The  diagonal components of the anisotropic pressure,  $\delta \pi^{i}_i$ (no sum over $i$), are non-zero at the background level  so their perturbations are expected to play some roles in the curvature perturbation analysis.  However, the non-diagonal spatial components of anisotropic pressure and the heat conduction terms are absent at the background level so their perturbations will dilute quickly. The diagonal anisotropic pressure plays two different roles in the curvature perturbation analysis, a direct effect and an indirect effect. The direct effect can be seen from the continuity equation, Eq. \eqref{cont-sim}, in which $\delta \pi^{i}_i$ contributes to the Hubble expansion rate. This effect, by using Eq.\eqref{cont-sim}, can be quantified as follows
\ba
\label{N-formula}
N(\mathbf{x},t_i,t_f) =\int_{t_i}^{t_f} H(\mathbf{x},t)   d\tau=  -\dfrac{1}{3} \int_{t_i}^{t_f}  dt \dfrac{\dot{\rho}(\mathbf{x},t)}{\rho(\mathbf{x},t)+p(\mathbf{x},t)} - \dfrac{1}{3} \int_{t_i}^{t_f} dt Q(\mathbf{x},t)  \, ,
\ea
in which $Q(\mathbf{x},t) $ is defined as
\ba
\label{Q-eq}
Q(\mathbf{x},t) &=& \dfrac{\cal N}{\rho+p} \left[-u_{\mu} \frac{d}{d\tau}q^{\mu} + \nabla_{\mu} q^{\mu}- u_{\mu} \nabla_{\nu} \pi^{\mu \nu}\right] \nonumber\\
& =& \dfrac{{\cal N}(\mathbf{x},t)}{\rho(\mathbf{x},t)+p(\mathbf{x},t)} \sum_i H_i(\mathbf{x},t) \pi^i{}_i (\mathbf{x},t)  + {\cal O} (\epsilon^2).
\ea
The above equation shows that the diagonal anisotropic pressure components $\delta \pi^{i}{}_{i}$ contributes to $\delta N$ through their effect on continuity equation as captured by
the term containing $Q$ in Eq. (\ref{N-formula}).

The  indirect effect of anisotropic pressure is more subtle and sometimes can be more important than the contribution from the term containing $Q$ above. This effect can be understood as the back-reactions of fields responsible for anisotropic pressure on the dynamics of other background fields  such as the inflaton field. The $\delta N$ formalism automatically includes this indirect effect. We will see this effect in next section in application of our $\delta N$ formalism for models of  anisotropic inflation.

\section{Anisotropic Inflation}
\label{anisotropic inflation}

In this section, we present the model of anisotropic inflation with a $U(1)$ gauge field originally presented in \cite{Watanabe:2010fh} which provide a non-trivial setup to employ our
$\delta N$ formalism.

However, in order to make it easier to proceed, we only consider the real inflaton case which means that in the given action in Eq. (\ref{action4}) we put $\e =0$. \\
To employ the $\delta N$ formalism, as usual we need to have a good control of the background dynamics. In oder to make it easier to compare with the previous results, we assume that the gauge field has a non-zero classical value along the x-direction so $A_\mu =(0,A_{x}(t),0,0)$. As a result, the background space-time is in the form of  Bianchi I Universe with the metric given in Eq. (\ref{metric4}). Then, like before, $H \equiv \dot \alpha$ is the average Hubble expansion rate, $H_a \equiv \dot a/a$
and $H_b \equiv \dot b/b$ are the expansion rates along the spatial directions $x$ and $y$  
and $\dot \sigma/H \equiv (H_b - H_a)/ H$ is a measure of anisotropic expansion.
\subsection{The background dynamics}
The fields equations for this system can be easily driven by putting $\e=0$ in Eqs. (\ref{back-A-eq4})-(\ref{anisotropy-eq4}). Since in this case the inflaton field is real, the equation of motion for $A_{x}$ (the Maxwell equation), Eq. (\ref{back-A-eq4}), is easily solved as
\ba
\label{gaugefield}
\dot{A_{x}}= f(\phi)^{-2}e^{-\alpha(t)-4\sigma(t)}p_{A} \, ,
\ea
where $p_{A}$ is a constant of integration. \\
Now in order to compare our results with 
that of \cite{Watanabe:2009ct}, we consider the Chaotic potential,
\ba
\label{Chaoticpotential6}
V= \frac{1}{2} m^2 \phi^2  \quad \quad  \rightarrow \quad \quad
f(\phi) = \exp {\left( \frac{c\phi^2}{2 M_P^2}  \right)}
\ea
with $c$ a constant very close to unity and we have used the results in Eqs. (\ref{a-scale5}) and (\ref{f-scale5}). In our discussion below we take the form of $f$, in
terms of $a= e^{\alpha}$, to be
\ba
\label{f-form-6}
f= \left( \frac{a}{a_f} \right)^{-2 c}  \simeq \left( \frac{\eta}{\eta_e}\right)^{2c} \, ,
\ea
in which $a_e$ and $\eta_e$ represent the value of the scale factor and the conformal time at the end of inflation. \\
As shown in \cite{Watanabe:2009ct} the system reaches the attractor solution in which
$R$ is given by
\ba
\label{R-app-6}
R = \frac{c-1}{2c}\epsilon_{H} = \frac{1}{2}I\epsilon_{H} \, ,
\ea
where we have defined $I\equiv\frac{c-1}{c}$ and $\epsilon_H \equiv \dot H/H$ is the slow-roll parameter. Combined with the definition of $R$ in Eq. (\ref{R-def-5}) we obtain
\ba
\label{cons-eq-6}
\dot A^2 f^2 e^{-2 \alpha} = I \, \epsilon_H V \, .
\ea
As we shall see below, this equation will be the key equation to find $\delta N$ in terms of $\delta \phi$ and $\delta \dot A$.\\
During the attractor phase the inflaton evolution is given by
\ba
\label{klinGordon6}
M_P^{-2}\frac{d \phi}{d \alpha} \simeq -\frac{V_\phi}{ V} + \frac{c-1}{c}\frac{V_\phi}{ V} \, .
\ea
Interestingly, this means that the back-reactions of the gauge field on the inflation field change the effective mass of the inflaton field as given by the second term above.\\
Using Eq. (\ref{Chaoticpotential6}) in Eq. (\ref{klinGordon6}) results in the following equation
\ba
\label{klinN6}
\phi_{e}^2 - \phi^2 = 4 M_P^2 \alpha (1-I)
\ea
in which $\phi_e$ is the value of $\phi$ at the end of inflation. We choose the convention such that $\alpha_e=0$, so during inflation $\alpha<0$. Eq. (\ref{klinN6}) clearly shows the effect of the gauge field back-reactions on the evolution of the inflaton field. The fact that the
evolution of the inflaton field is affected by the gauge field, as given by the correction factor $(1- I)$ in Eq. (\ref{klinN6}) is the key to calculate $\delta N$ in the presence of gauge field. In passing, we comment that in the previous applications of $\delta N$ in the literature for models with the  gauge fields, this important effect is not taken into account. In other words, $\delta N$ in these papers  have been written with treating $\delta A_\mu$ in the same footing as $\delta \phi$ in an FRW background without taking into account the back-reactions of the gauge field
in the evolution of inflaton field and in the dynamics of the anisotropic background.\\
In connection with our discussion in previous section the energy density, pressure, momentum density   and  stress associated with the electro-magnetic field are given by
\begin{eqnarray}
\label{muem-eq}
\rho _{em} &=&\frac 1{2}\left( E^2+B^2\right) = \frac{3}{2}I \epsilon_{H}H^2 ,  \label{muA}\\
p_{em} &=&\frac 16\left( E^2+B^2\right) = \frac{1}{2}I \epsilon_{H}H^2  ,  \label{pA} \\
q_{em}^i &=&\eta ^{ijkq}u_jE_kB_q =0 ,  \label{qA} \\
\end{eqnarray}
and
\be
\pi_{\mu}^{\nu}=\left[ \begin{array}{cccc}
	0&0&0&0\\
	0&-2I\epsilon_{H}H^2&0&0\\
	0&0&I\epsilon_{H}H^2&0\\
	0&0&0&I\epsilon_{H}H^2
\end{array}\right ].
\ee
Plugging the value of $\rho_{em}$ into definition of $R$ and using the attractor value
Eq. (\ref{R-app-6}) we obtains $\rho_{em} \simeq R V$ as advertised before. Also Eq. (\ref{muem-eq}) indicates that  $E=\sqrt{3I\epsilon_{H}}H$ and $B=0$.


\subsection{$\delta N$ in anisotropic inflation}

Our goal here is to calculate the curvature perturbations in this model by employing our
$\delta N$ formalism. As we argued before the contribution of the gauge field into the Hubble expansion rate and total energy density is sub-dominant. This means that the surface of end of inflation is controlled only by the inflaton field. However, the gauge field plays an important role in Klein-Gordon equation and in the evolution of the inflaton field as can be seen in Eq.(\ref{klinN6}). 

Perturbing  Eq.(\ref{klinN6}) we have,
\ba
\label{pert-klin6}
2\phi \delta \phi = -4\delta N + 4N \delta I \, .
\ea
As a result
\ba
\label{delta N}
\delta N = -\frac{1}{2}\phi \delta \phi + N\delta I \, .
\ea
Note that, in order to connect to the standard notation we made the replacement $\alpha \rightarrow N$ so $\delta \alpha = \delta N$ from now on. Also note that $R$ is related to $I$ by Eq. (\ref{R-app-6}) so by $\delta I$, we actually mean $\delta (R/\epsilon)$.
Since it is easier to work with $\delta I$ than $\delta R$, we use $\delta I$ from now on.

The first term in Eq. (\ref{delta N}) is the contribution of the inflaton field, while the second term is due to the  back-reaction of the gauge field on the inflaton dynamics. Also by perturbing Eq. (\ref{cons-eq-6}) we have
\ba
\label{deltaA16}
\epsilon_{H} \delta I = -12 R \delta N + 4 R \frac{\delta \dot{A_x}}{\dot{A_x}}
\ea
Now combining Eq. (\ref{delta N}) and Eq. (\ref{deltaA16}) we have
\ba
\label{total delta N}
\left(1 + \frac{12 R N}{\epsilon_{H}}\right)\delta N = -\frac{\phi}{2 M_P^2} \delta\phi + \frac{4R N }{\epsilon_{H}}\frac{\delta \dot{A_x}}{\dot{A_x}}
\ea
Now using Eq. (\ref{R-app-6}) we have $R N/\epsilon_H \sim I N$. As we shall see, we require $N I  \ll 1$ in order not to produce too much anisotropy in power spectrum so we can neglect the second term in the left hand side of Eq. (\ref{total delta N}) and
\ba
\label{App delta N}
\delta N \simeq -\frac{\phi}{2 M_P^2} \delta\phi + 2 I N \frac{\delta \dot{A_x}}{\dot{A}} \, .
\ea
This is our result for $\delta N$ to linear order in terms of $\delta \phi$ and $\delta \dot A$. Interestingly, since in this model the leading contribution into the anisotropic power spectrum
comes from the electric field instead of the magnetic field, we see that in $\delta N$ only 
$\delta \dot{A}$ and not $\delta A$ appears. This should be compared with the conventional models of $\delta N$ involving scalar fields $\phi_I$ in which
$\delta \phi_I$ and not $\partial_t  \delta {\phi_I}$ appears. This is because for light scalar fields $\delta \dot \phi_I$ become negligible on super-horizon scales once the attractor solution has been reached. 

To calculate the power spectrum and the higher order correlations, we have to know the behavior of  $\frac{\delta \dot{A_x}}{\dot{A}}$ outside the horizon. For this purpose, we have to solve the mode function for $\delta A_i$ with the initial Bunch-Davies vacuum deep inside the horizon.  As shown in \cite{Bartolo:2012sd} the canonically normalized gauge field quantum fluctuations are given by 
\ba
\label{canonical gauge field}
\delta A_i =   \sum_{\lambda = \pm} \int \frac{d^3k}{(2\pi)^{3/2}}e^{i \overrightarrow{k}.\overrightarrow{x}}\vec{\epsilon}_{\lambda}(k) \frac{\widehat{V}_i}{f}  \, ,
\ea
in which
\ba
\widehat{V} = a_{\lambda}(\overrightarrow{k})V_{\lambda}(k) + a_{\lambda}^{\dagger}(-\overrightarrow{k})V_{\lambda}^{*}(k) \, .
\ea
Here  $a_{\lambda}(\overrightarrow{k})$ and $a_{\lambda}^{\dagger}(\overrightarrow{k})$
represent the annihilation and the creation operators and $\epsilon_\lambda$ for $\lambda =\pm$ represents the circular polarization with the properties
$\vec k \,  . \,   \vec \epsilon_\pm(\vec k) =0 \, , \,  \vec k \, \times \,  \vec \epsilon_\pm(\vec k) =
\mp i k \, \vec \epsilon_\pm(\vec k)  \, , \vec \epsilon_\lambda (-\vec k) = \vec \epsilon_\lambda\, (\vec k)^*$, normalized via
$\vec \epsilon_\lambda(\vec k)  \, . \,  \vec \epsilon_{\lambda'}(\vec k')= \delta_{\lambda \lambda'}$ and
\ba
\sum_{\lambda}\epsilon_{\lambda,i}(\vec{k})\epsilon^{*}_{\lambda,j}(\vec{k}) = \delta_{ij} - \hat{k}_{i}\hat{k}_{j} \, .
\ea

The mode functions satisfy the evolution equation
\begin{align}
\label{canonical gauge field equation}
V_{\lambda}(k)''+ \left( k^2 - \frac{f''}{f} \right)V_{\lambda}(k)=0 \, ,
\end{align}
where the prime denotes the derivative with respect to conformal time $d \eta= dt/a(t)$. For  $f$ given in 
Eq. (\ref{f-form-6}) the normalized gauge field mode function is the same as that of  a massless scalar field in dS space with
\begin{align}
\label{canonical gauge field}
V_{\lambda}(k) \simeq \frac{1+ i k \eta}{\sqrt{2}k^{3/2}\eta} e^{-ik\eta} \, .
\end{align}
Using this form of the wave function and the attractor solution Eq. (\ref{cons-eq-6})
one can easily show that on super-horizon scales
\ba
\label{mode-deltaA-6}
\frac{\delta \vec{\dot{A}}}{\dot{A}} = \sum_{\lambda}\vec{\epsilon}_{\lambda} \frac{\sqrt{3}H}{\sqrt{2I\epsilon_{H}k^3}}    \quad \quad (k < a H)
\ea
In particular, we see that on super-horizon scale, $ \delta A_x/\dot A_x$ is a constant.

Now we are in the position to calculate the total effect of the gauge field in curvature perturbation $\zeta$.  From Eq. (\ref{N-formula})  and Eq. (\ref{App delta N})
we see that there are two different terms that encode the contributions of the gauge field in $\zeta$.  Eq. (\ref{App delta N}) encodes the indirect effects of the gauge field on $\delta N$ originating from its back-reaction on inflaton field dynamics. However, the direct
contribution of the gauge field in $\delta N$ is encoded in term $Q$ in  Eq. (\ref{N-formula}).
As we shall prove below, the contribution from the $Q$ term in the curvature perturbation 
is negligible and the leading contribution of the gauge field in curvature perturbation is 
from its back-reaction effects in Eq. (\ref{App delta N}).

Calculating $Q$ from Eq. (\ref{Q-eq}) yields (note that in this model $q^{\mu}$ is proportional to the product of the electric and magnetic fields and since in this model the magnetic field is zero therefore there is no correction from $q^{\mu}$)
\begin{eqnarray}
\label{Q app term}
Q &=& \frac{1}{(\rho+p)}\left( H_{a}\pi^{1}_1 + 2 H_{b}\pi^{2}_2 \right)\nonumber\\
~&=& \frac{2H^2}{(\rho+p)}\left(H_{b} - H_{a} \right)I \epsilon_{H}\nonumber\\
~&=& I^2 H \epsilon_{H} \, ,
\end{eqnarray}
where we have used $(\rho+p) \simeq \dot{\phi}^2 = 2 H^2 \epsilon_{H}$ and $H_{b} - H_{a}= H I\epsilon_{H}$.

Perturbing Eq. (\ref{Q app term}), we have
\begin{eqnarray}
\label{Q pert term}
\delta Q &=&2 I H\epsilon_{H}\delta I \nonumber\\
~&=& 2 I H \left(-12 R \delta N + 4 R \frac{\delta \dot{A_x}}{\dot{A}}\right) \, .
\end{eqnarray}
Now integrating Eq. (\ref{Q pert term}) over $t$ we have,
\begin{eqnarray}
\label{int Q pert term}
\int_{t_{1}}^{t_{2}}\delta Q  dt&=& \int_{t_{1}}^{t_{2}}  2 H I\left(-12 R \delta N + 4 R \frac{\delta \dot{A_x}}{\dot{A}}\right) d t \nonumber\\
~&=&  2 I N\left(-12 R \delta N + 4 R \frac{\delta \dot{A_x}}{\dot{A}}\right)
\end{eqnarray}
To perform this integral, we have assumed hat $H$ and $R$ are nearly constant in the slow-roll approximation. Furthermore, $\delta \dot A_x/\dot A_x$ is also nearly constant as can be seen from  Eq. (\ref{mode-deltaA-6}).

Eq. (\ref{int Q pert term}) indicates that the contribution of $Q$ in $\delta N$ is at
the order of $I N R \sim N I^2$. However, as we shall see, in order not to produce too much
anisotropy we require $I N^2 < 1$ so $ N I^2 \ll 1$ and we can safely neglect the contribution of  $Q$ in $\delta N$. 

In conclusion, the only contribution of the gauge field in curvature perturbations comes in
$\delta N$ as given by the second term in  Eq. (\ref{App delta N}).  As a result  we have
\begin{eqnarray}
\label{psiA}
\zeta & = &  \delta N \nonumber\\
~&=& -\frac{\phi}{2 M_P^2} \delta\phi + 2IN\frac{\delta \dot{A_x}}{\dot{A}}.
\end{eqnarray}

We are interested in curvature perturbation power spectrum ${\cal P}_\zeta$ defined
via
\ba
\label{curvaturepower}
\langle \hat{\zeta}_{\vec{k}_{1}}\hat{\zeta}_{\vec{k}_{2}}\rangle = (2 \pi)^3 \delta^3(\vec{k}_{1}+\vec{k}_{2}) { P}_{\zeta}(\vec k_1)
\quad , \quad {\cal P}_{\zeta}(\vec k) = \frac{k_1^3}{2 \pi^2} P_\zeta (\vec k) \, . 
\ea
We decompose the power spectrum into the isotropic part, ${\cal P}_{0}$,
coming from the $\delta \phi$ contribution in Eq. (\ref{psiA}) and the anisotropic power spectrum, $\Delta {\cal P}$,  coming from $\delta \dot A_x$ in Eq. (\ref{psiA})
\ba
\label{power spectrum}
{\cal P}_{\zeta} \equiv {\cal P}_{0} + \Delta {\cal P} \, .
\ea
As usual the isotropic  power spectrum  is given by
\ba
\label{deltaA}
{\cal P}_{0} = \frac{H^2}{8\pi^2 M_P^2 \epsilon_{H}} \, .
\ea
To calculate the anisotropic power spectrum we note that $\delta \phi$ and $\delta \dot A_x$ are
mutually uncorrelated so  $\langle \delta \phi \delta \dot A_x \rangle |_* =0$. As a result
\ba
\label{gfactor}
\Delta {\cal P} &=& \frac{k_1^3}{2 \pi^2} 4 I^2 N^2 \left\langle  \frac{\delta \dot A_x(k_1)}{A_x}
\frac{\delta \dot A_x(k_2)}{A_x} \right\rangle  \nonumber\\
&=& \frac{k_{1}^3}{2\pi^2} \frac{6IH^2}{\epsilon_{H}k_{1}^3}N^2 \sin^2{\theta} \nonumber\\
&=& 24 \, I N^2 {\cal P}_{0} \sin^2{\theta}
\ea
in which the angle $\theta $ is defined via $\cos \theta = \hat n. \hat k$.
Now comparing this with the anisotropy factor $g_*$ defined via
\ba
{\cal P}_{\zeta}(\vec{k}) = {\cal P}_{0} \left( 1 + g_{*} (\hat k . \hat n)^2  \right) \, .
\ea
we obtaine
\ba
\label{g-star}
g_* = -24 I N^2
\ea
Very interestingly this is the result obtained in \cite{Watanabe:2010fh, Bartolo:2012sd} using
the standard in-in formalism.   The advantage of using $\delta N$ formalism is that we only needed to use the background attractor solutions with the information about $ \delta \dot A$ at the time of horizon crossing. This should be compared with the tedious analysis employed in 
\cite{Watanabe:2010fh, Bartolo:2012sd, Emami:2013bk} using in-in formalism to calculate $\Delta {\cal P} $. Physically, one expects that the $\delta N$ method to be applicable in this model. The reason is that all the dynamics between the inflaton field and the gauge field are in the form of local interactions and the dynamics of modes are trivial inside the horizon and at the time of horizon crossing. 

As mentioned in \cite{Watanabe:2010fh, Bartolo:2012sd}, in order to satisfy the observational bound $|g_*| <0.1$, and taking $N=60$ to solve the horizon and the flatness problems,  we require $I < 10^{-6}$. As a result $c$ is very close to unity and
$R\ll1$ as advertised before.

As emphasized in \cite{Bartolo:2012sd} if one allows $N$ to be too large then the 
accumulative anisotropies produced from the IR modes can become too large. Therefore, the total number of e-foldings should not be too large in this model.


\section{Bispectrum and Trispectrum}
\label{bispectrum}

In this section we calculate the bispectrum and the trispectrum in the anisotropic inflation model studied in previous section using our $\delta N$ method and compare the results with the corresponding results in \cite{Bartolo:2012sd} and \cite{Shiraishi2012} obtained from
the standard in-in formalism. As we shell see the results for bispectrum and the trispectrum
are in exact agreements.

As usual, in order to calculate the bispectrum and the trispectrum in $\delta N$ formalism, we have to expand $\delta N$ to higher orders in perturbations. The expansion of $\delta N$ to
linear order is given in previous section in Eq. (\ref{psiA}). Here we generalize it to second order. To this purpose, we perturb the attractor solution Eq. (\ref{cons-eq-6}) to second order in fields perturbations
\begin{align}
\label{attractor-per-6}
\frac{\delta I}{I}&=  \frac{2 f_{,\phi}}{f}\delta \phi +  \frac{2 \delta \dot{A_x}}{\dot{A}} + 
\left[ \bigg{(}\frac{f_{,\phi}}{f}\bigg{)}^2 + \frac{ f_{,\phi \phi}}{f}   \right] \delta \phi^2  + \bigg{(}\frac{\delta \dot{A}}{\dot{A}}\bigg{)}^2 +  \frac{4 f_{,\phi}}{f} \frac{\delta \dot{A_x}}{\dot{A}} \delta \phi  \nonumber\\
& -2 \delta N \bigg{(}1+  \frac{2 f_{,\phi}}{f}\delta \phi +  \frac{2 \delta \dot{A_x}}{\dot{A}}   \bigg{)} +2 \delta N^2
\, .
\end{align}
This formula gives a relation between $\delta I, \delta N$ and different powers of $ \delta \phi$ and  $\delta \dot A$. On the other hand, Eq. (\ref{delta N}) from the perturbation of the evolution of $\phi(\alpha)$  gives a relation between $\delta I, \delta N$ and $\delta \phi$ which is valid to all orders in $\delta I$ and $\delta N$.  Now plugging back Eq.(\ref{attractor-per-6}) into Eq.(\ref{delta N}) and  keeping the leading corrections from $I \ll 1$ we obtain
\ba
\label{delta N total-6}
\delta N = N_{\phi} \delta \phi + N_{\dot{A}} \delta \dot{A} + \frac{N_{\phi \phi}}{2} \delta \phi^2  +  \frac{N_{\dot{A} \dot{A}}}{2} \delta \dot{A}^2 + N_{\phi \dot{A}} \delta \phi \delta \dot{A_x}
\ea
in which to leading order in $I$
\ba
N_\phi \simeq  -\frac{\phi}{2 M_P^2}   \quad , \quad
N_{\phi \phi} \simeq \frac{2 f_{,\phi}^2}{f^2} + \frac{ 2f_{,\phi \phi}}{f} + \frac{\phi^2}{M_P^4}
+\frac{ 4  \phi}{M_P^2}  \frac{f_{,\phi}}{f}
\ea
and
\ba
N_{, \dot A} \simeq \frac{2 I N}{\dot A} \quad , \quad N_{, \dot A \dot A} \simeq \frac{2 I N}{\dot A^2}
\quad , \quad
N_{, \phi \dot A} \simeq \frac{4 I N}{\dot A} \frac{f_\phi}{f}
\ea
Having calculated $\delta N$ to second order in  Eq. (\ref{delta N total-6}), we can calculate the bispectrum  $B_{\zeta}(\vec k_{1}, \vec k_{2}, \vec k_{3})$ defined via
\ba
\label{bi- def}
\langle \zeta (\vec k_{1}) \zeta (\vec k_{2}) \zeta (\vec k_{3}) \rangle &\equiv& \left( 2 \pi \right)^3 \delta^3 \left( \vec k_{1} +  \vec k_{2} +  \vec k_{3}\right) B_{\zeta}(\vec k_{1}, \vec k_{2}, \vec k_{3}) 
\ea
There are three contributions into bispectrum; (a): $N_{, \phi \phi}N_{,\phi} N_{,\phi} \langle \delta \phi ^4 \rangle$, (b): $N_{, \dot{A}  \dot{A}}N_{, \dot{A}} N_{,\dot{A}} \langle
\delta \dot{A}^4 \rangle$ and (c): $N_{, \phi \dot{A}}N_{, \phi} N_{,\dot{A}}\langle
\delta \phi ^2\delta \dot{A}^2 \rangle$. The term (a) is the one expected from scalar field theory and is very small. The term (b) is purely from the gauge field while the term (c) is from the mixing of inflaton and the gauge field. One expects that the contribution of term (c) to be sub-leading as compared to the contribution of the term (b). Indeed, a direct analysis shows that the ratio of (b) to (c) is $N$ so for $N\sim 60$ one can safely neglect the contribution from the term (c).  In conclusion,  the leading contribution  to the  bispectrum comes from $\langle\delta  \dot A^4 \rangle $ and
\ba
\label{leading three0-6}
\langle \zeta (k_{1}) \zeta (k_{2}) \zeta (k_{3})  \rangle  &\simeq& \frac{1}{2}
N_{, \dot{A}\dot{A}}( k_{1}) N_{, \dot{A}}( k_{2}) N_{, \dot{A}} ( k_{3}) \int \frac{d^3p}{(2\pi)^3} \langle \delta \dot{A}_{x} (\vec k_{1})  \delta \dot{A}_{x} (\vec k_{2})
\delta \dot{A}_{i} (\vec p) \delta \dot{A}_{i} (\vec k_{3} - \vec p) \rangle \nonumber\\
&&+ 2 \mathrm{perm.} 
= 4 I^3 N(k_{1}) N(k_{2}) N(k_{3}) \int \frac{d^3p}{(2\pi)^3} \times \nonumber\\
&&~~~\langle \delta \dot{A}_{x} (\vec k_{1})  \delta \dot{A}_{x} (\vec k_{2})
\delta \dot{A}_{i} (\vec p) \delta \dot{A}_{i} (\vec k_{3} - \vec p)\rangle + 2 \mathrm{perm.}
\ea
in which $N(k_i)$ represents the time when the mode $k_i$ leaves the horizon.
Now in the Coulomb gauge $A_0=0$, the gauge field perturbations $\delta A_i (\vec k)$ are given by \cite{Bartolo:2012sd}
\ba
\frac{\delta \vec{\dot{A}}}{\dot{A}}|_{t_k} = \sum_{\lambda}\vec{\epsilon}_{\lambda} \frac{\sqrt{3}H}{\sqrt{2I\epsilon_{H}k^3}}
\ea

Plugging these back in Eq. (\ref{leading three0-6}) yields
\ba
\label{leading three-6}
\langle \zeta (k_{1}) \zeta (k_{2}) \zeta (k_{3})  \rangle  &\simeq&
288 I N(k_{1}) N(k_{2}) N(k_{3}) \left( 2 \pi \right)^3 \delta^3 \left( \vec k_{1} +  \vec k_{2} +  \vec k_{3}\right) \times \nonumber\\
&& \bigg{(} C(\vec k_{1}, \vec k_{2})P_0(k_{1})P_0(k_{2}) + 2 \mathrm{perm.} \bigg{)} \, ,
\ea
in which the momentum shape function $C(\vec k_1, \vec k_2) $ is defined via
\ba
C(\vec k_{1}, \vec k_{2})\equiv\bigg{(}1 -   (\widehat k_1.\widehat n)^2  -   (\widehat k_2.\widehat n)^2 + 
(\widehat k_1.\widehat n) \,  (\widehat k_2.\widehat n) \,  (\widehat k_1.\widehat k_2)  \bigg{)}
\ea
To obtain Eq. (\ref{leading three-6})
we have used  $P_0{(k_{1})} = \frac{H^2}{4k_{1}^3\epsilon_{H}M_{P}^2}$
for the isotropic power spectrum and
\ba
\label{modeA}
\langle \frac{\delta \dot{A}_{i}(\vec k_{1})}{ \dot{A}} \frac{\delta \dot{A}_{j}(\vec k_{2})}{ \dot{A}} \rangle = \frac{3H^2}{2I\epsilon_{H}k^3M^2_{P}}
\bigg{(} \delta_{ij} - \widehat{k}_{1i}\widehat{k}_{1j} \bigg{)}\left( 2 \pi \right)^3 \delta^3 \left( \vec k_{1} +  \vec k_{2}\right)
\ea
Using  Eq. (\ref{leading three-6}), one can calculate the bispectrum as
\ba
\label{fnl leading-6}
B_\zeta(\vec k_1, \vec k_2, \vec k_3)=  288 I N(k_1) N(k_2) N(k_3) \left( C(\vec k_{1}, \vec k_{2}) P_0(k_{1})P_0(k_{2})
+ 2 \mathrm{perm.} \right)
\ea
This completes our results for the bispectrum. As expected, the shape of the bispectrum is anisotropic. Very interestingly our formula Eq. (\ref{leading three-6}) and Eq. (\ref{fnl leading-6})
agree exactly with the result of \cite{Bartolo:2012sd} obtained using the standard in-in formalism.

To calculate $f_{NL}$ we go to the squeezed limit $k_{1} \ll k_{2} \simeq k_{3}$ in which
\ba
f_{NL} (\vec k_1, \vec k_2, \vec k_3) = \lim_{k_1 \rightarrow 0} \frac{5}{12}
\frac{B_\zeta(\vec k_1, \vec k_2, \vec k_3)}{P_\zeta(k_1) P_(k_2)} \, .
\ea
In this limit we get
\ba
\label{fnl squeezed}
f_{NL} &=& 240 I N(k_{1}) N(k_{2})^2 C(\vec k_{1}, \vec k_{2})   \quad \quad  \quad
( k_{1} \ll k_{2} \simeq k_{3} )  \\
&\simeq& 10 N\,  |g_*| \, C(\vec k_1, \vec k_2)
\ea
Taking $N \sim 60$ and $|g_* | \sim 0.1$ and neglecting the orientation-dependence  in $f_{NL}$ this leads  to large non-Gaussianity $f_{NL} \sim 60$.

Now we are in the position to calculate the trispectrum of our model. The trispectrum is defined via
\ba
\label{taunl}
\langle \zeta (k_{1}) \zeta (k_{2}) \zeta (k_{3}) \zeta (k_{4}) \rangle = (2 \pi)^3 \delta^3 \left( \vec k_{1} +  \vec k_{2} +  \vec k_{3} + \vec k_{4}\right) T_\zeta (\vec k_1, \vec k_2, \vec k_3, \vec k_4) \, .
\ea
In the collapsed limit $\vec k_1 + \vec k_3 = \vec k_2 + \vec k_4 =0$
we can calculate the parameter $\tau_{NL}$ via
\ba
\label{tau-NL-def}
\tau_{NK}(\vec k_i) = \lim_{\vec k_1 + \vec k_3\rightarrow 0} \frac{1}{4}
\frac{T_\zeta(\vec k_1, \vec k_2, \vec k_3, \vec k_4)}{P_\zeta( | k_1 + k_3| ) P_\zeta (k_1) P_\zeta (k_3)}
\ea
The trispectrum $\langle \zeta (\vec k_{1}) \zeta ( \vec  k_{2}) \zeta (\vec  k_{3}) \zeta (\vec k_{4}) \rangle$ has
6 contributions in the forms of
$$
N_{, \phi \phi}^2N_{,\phi}^2  \langle \delta\phi^6\rangle  \,\, , \,\,
N_{, \phi \dot{A}}^2N_{,\phi}^2  \langle \delta \phi ^4 \delta \dot{A}^2  \rangle  \,\, , \,\,
N_{, \phi \dot{A}}^2N_{,\dot{A}}^2 \langle \delta \phi ^2 \delta \dot{A}^4 \rangle
\,\, , \,\,
$$ 
$$
N_{, \phi \dot{A}}N_{, \dot{A} \dot{A}}N_{,\dot{A}} N_{, \phi} \langle \delta \phi^2
\delta \dot{A}^4 \rangle
\,\, , \,\,
N_{, \phi \dot{A}}N_{, \phi \phi}N_{,\dot{A}} N_{, \phi} \langle \delta \phi^4 \delta \dot{A}^2
\rangle
$$
and $N_{,\dot{A} \dot{A}}^2N_{,\dot{A}}^2 \langle \delta \dot{A}^6\rangle$.
As in the case of bispectrum, one can easily check that the last term has the dominant contribution in trispectrum. Therefore  to leading order we have
\ba
\label{leading four}
\langle \zeta (\vec k_{1}) \zeta (\vec k_{2}) \zeta (\vec k_{3}) \zeta (\vec k_{4}) \rangle &\simeq& N_{, \dot{A}\dot{A}}(k_{1})N_{, \dot{A}\dot{A}}(k_{2}) N_{, \dot{A}}(k_{3})N_{, \dot{A}}(k_{4})\int \frac{d^3p}{(2\pi)^3}  \int \frac{d^3q}{(2\pi)^3}
\nonumber\\
&&\left\langle \delta \dot{A}_{x} (\vec k_{3}) \delta \dot{A}_{x} (\vec k_{4})
\delta \dot{A}_{i} (\vec q)\delta \dot{A}_{i} (\vec k_{1}-\vec q) \delta \dot{A}_{j} (\vec p)\delta \dot{A}_{j} (\vec k_{2}-\vec p)
\right\rangle + 5 \mathrm{perm.} \nonumber\\
&=& 3456 I N(k_{1}) N(k_{2}) N(k_{3}) N(k_{4}) \bigg{(}  D(\vec k_{3}, \vec k_{4}, \vec k_{1}+\vec k_{3}  )P(k_{3}) P(k_{4}) \nonumber\\
&&~~
P(|\vec k_{1}+\vec k_{3}|) +  \mathrm{11 perm.}  \bigg{)}\left( 2 \pi \right)^3 \delta^3 \left( \vec k_{1} +  \vec k_{2} +  \vec k_{3} + \vec k_{4}\right) \, ,
\ea
in which $ D(\vec k_{3}, \vec k_{4}, \vec k_{1}+\vec k_{3}  )$ refers to the trispectrum's shape function and is given by
\ba
\label{four shape}
D(\vec k_{3}, \vec k_{4}, \vec k_{1}+\vec k_{3}) &=& 1 -  (\widehat k_4.\widehat n)^2 
-  (\widehat k_3.\widehat n)^2  - (\widehat{ k_1 + k_3}. \widehat n)^2 +
(\widehat k_3. \widehat n)  (\widehat k_4. \hat n)  (\widehat k_3. \widehat k_4) 
\nonumber\\
&& + (\widehat k_4. \widehat n)   (\widehat{ k_1 + k_3}. \widehat n)  (\widehat{ k_1 + k_3}. \widehat k_4)
 (\widehat k_3. \widehat n)   (\widehat{ k_1 + k_3}. \widehat n)  (\widehat{ k_1 + k_3}. \widehat k_3) 
 \nonumber\\
&& - (\widehat k_3. \widehat n)    (\widehat k_4. \widehat n)    (\widehat{ k_1 + k_3}. \widehat k_3)  (\widehat{ k_1 + k_3}. \widehat k_4)  \, .
\ea
Comparing Eq. (\ref{four shape}) with the definition of trispectrum we obtain
\ba
\label{T-zeta}
T_\zeta(\vec k_1, \vec k_2, \vec k_3, \vec k_4) &=&  3456 I N(k_{1}) N(k_{2}) N(k_{3}) N(k_{4}) \bigg{(}  D(\vec k_{3}, \vec k_{4}, \vec k_{1}+\vec k_{3}  )P_\zeta(k_{3}) P_\zeta(k_{4})P_\zeta(|\vec k_{1}+\vec k_{3}|) \nonumber\\
&& + \mathrm{11 perm.}  \bigg{)} \, .
\ea
Now going to the collapsed limit $\vec k_1 + \vec k_3 = \vec k_2 + \vec k_4 =0$
and using the definition of $\tau_{NL}$ given in Eq. (\ref{tau-NL-def}) we obtain
\begin{align}
\label{tauNL collapsed}
\tau_{NL}(k_{1}, k_{2}, k_{3}, k_{4}) \simeq 3456 I N(k_{3})^2 N(k_{4})^2  D(\vec k_{3}, \vec k_{4}, \vec k_{1}+\vec k_{3}  ) \, .
\end{align}
As in the case of bispectrum the trispectrum is anisotropic so $\tau_{NL}$ has 
direction-dependence. Comparing our trispcetrum with the results of \cite{Shiraishi2012} obtained from in-in formalism,  we find the exact agreement between these two results.

Now we are in the position to check the SY inequality between $f_{NL}$ and $\tau_{NL}$ which states 
\ba
\label{SY}
\tau_{NL} \ge \left( \frac{6}{5} f_{NL} \right)^2 \, .
\ea
The importance of SY inequality as a tool to rule out inflationary scenarios was studied
in \cite{Komatsu:2010hc, Sugiyama:2011jt}.  The SY inequality presented in the form of 
Eq. (\ref{SY}) is for the models in which $f_{NL}$ and $\tau_{NL}$ are either scale-invariant or have the same
scale-dependence. In our case, see also \cite{Rodriguez:2013cj, BeltranAlmeida:2011db},  we have complicated shape-dependent for $f_{NL}$ and $\tau_{NL}$ so the original SY inequality as given in Eq. (\ref{SY}) is not applicable.
Instead, in \cite{Assassi:2012zq} a general integral representation of SY is proved which states
\ba
\int d^3 q_1 d^3 q_2 \tau_{NL} (\vec q_1, \vec k - \vec q_1, \vec q_2, -\vec q_2 - \vec k)
P_\zeta(q_1) P_\zeta(q_2) \ge
\left( \int d^3 q \frac{6}{5} f_{NL} (\vec q, -\vec q -\vec k, \vec k) P_\zeta(q)
\right)^2\nonumber\\
\ea
in which $k\rightarrow 0$. As demonstrated in \cite{Shiraishi2012} this integral form of
SY inequality does hold in our model if we assume $g_* <1$. Indeed, the condition $g_*<1$ 
is necessary from the observational constraints on the power spectrum and the consistency of our starting assumption that the anisotropic power spectrum is sub-leading so  $\Delta { \cal P} < {\cal P}_0$.

To see qualitatively that the SY inequality in its simple form, Eq. (\ref{SY}), does hold in our model, let us neglect the direction-dependence in $f_{NL}$ and $\tau_{NL}$ coming from $D(\vec k_{3}, \vec k_{4}, \vec k_{1}+\vec k_{3}  )$ and $C(\vec k_1, \vec k_2) $. As a result
\begin{align}
\label{ration}
\frac{\tau_{NL} (k_{1}, k_{2}, k_{3}, k_{4})}{(\frac{6}{5})^2f_{NL}(k_{3})f_{NL}(k_{4})} \simeq   \frac{1}{g_*} \,     \quad \quad \quad \quad (g_* < 1)
\end{align}
in which we have taken $N(k_i) = N$  and $g_* \simeq -24 I N^2$ from Eq. (\ref{g-star}).
Demanding that $g_* < 1$ from the cosmological observations and also from the consistency of our analysis we conclude that   $\tau_{NL} > \frac{6}{5} ( f_{NL})^2$ so the SY inequality does hold.


\section{Summary}
\label{discussions}

In this chapter, we have presented the consistent $\delta N$ formalism in anisotropic backgrounds.
We have demonstrated that the separate universe approach works.  In each homogenized patch 
the local continuity equation and the local Friedmann equation hold which have the same form as the corresponding background equations. We note that the Hubble expansion rate appearing in continuity equation, $H( \mathbf{x}, t)$, and the Hubble expansion rate appearing in Friedmann equation, ${\cal H}( \mathbf{x}, t)$, are different.  

The anisotropic pressure has two different effects in $\delta N$ analysis. One is the direct effect encoded by the term containing $Q$  in Eq. (\ref{Q-eq}). The second effect is indirect and comes from the back-reactions of the source of anisotropic pressure on the dynamics of other fields, such as the inflaton field. We have calculated these two effects in model of anisotropic inflation containing a $U(1)$ gauge field. We have shown that the second effect, i.e. the back-reaction effect, is much larger than the direct effect coming from the $Q$
term in $\delta N$ formula. In previous works in the literature, this back-reaction effect during inflation was not taken into account. The gauge  field contribution $\delta A_i$ is added trivially as a non-interacting field during inflation.

Taking into account the back-reaction of gauge field on inflaton dynamics, we have calculated $\delta N$ to linear and to second order in perturbations. We have demonstrated that our $\delta N$ formalism exactly reproduces the power spectrum and the bispectrum results obtained in previous works using standard in-in formalism. This is a non-trivial verification of the validity of our $\delta N$ analysis. The advantage in using $\delta N$ formalism is that all we need to know to calculate the power spectrum and higher order correlations is the background dynamics and the profile of gauge field fluctuations on super-horizon scales. This method seems to be considerably simpler than the standard in-in formalism.\\
We also calculated the bispectrum and the trispectrum in anisotropic inflation model. The bispectrum and the trispectrum are both orientation-dependent and scale-dependent. As a result the SY inequality in its simple form is not applicable. However,  a generalization of the SY inequality in its integrated form indeed holds.


\chapter{Primordial Anisotropies in Gauged Hybrid Inflation} 

\label{Chapter6} 

\lhead{Chapter 6. \emph{Cosmological Perturbation Theory}} 

\vspace{0.5cm}
\hrule \vspace{0.3cm}

\begin{quote}
\textbf{Abstract:} In this chapter, we 
study primordial anisotropies generated in the model of gauged hybrid inflation in which the complex waterfall field is charged under a $U(1)$ gauge field.
Primordial anisotropies are generated either actively during inflation or from inhomogeneities modulating the surface of end of inflation during waterfall transition. Using our consistent $\delta N$ mechanism in the previous chapter, we calculate the anisotropic power spectrum and bispectrum.
We show that the primordial anisotropies generated at the surface of end of inflation do not depend on the number of e-folds and therefore do not produce dangerously large anisotropies associated with the IR modes. Furthermore, one can find the parameter space that the anisotropies generated from the surface of end of inflation cancel the anisotropies generated during inflation, therefore relaxing the constrains on model parameters imposed from
IR anisotropies. We also show that the gauge field fluctuations induce a red-tilted power spectrum so the averaged power spectrum from the gauge field can change the total power spectrum from blue to red. Therefore, hybrid inflation, once gauged under a $U(1)$ field, can be consistent with the cosmological observations.
	
\end{quote}

\vspace{0.1cm}  \hrule
\vspace{0.5cm}

\section{Introduction}
Having presented the consistent $\delta N$ formalism for the anisotropic universe in the previous chapter, here we consider one application of this setup for a very specific inflationary model, i.e. Gauged Hybrid Inflation, in which the waterfall is a complex scalar field coupled to the $U(1)$ gauge field. The interesting property of this model is that primordial anisotropies can be generated both during inflation and also at the surface of end of inflation via waterfall phase transition. In addition, in this scenario, the surface of end of inflation is controlled by the waterfall field which is coupled to a $U(1)$ gauge field. \\
It was argued in \cite{Yokoyama:2008xw} that the inhomogeneities sourced by the gauge field fluctuations $\delta A_\mu$ at the surface of end of inflation can lead to statistical anisotropies. 
Here we ue our consistent $\delta N$ formalism, as was developed in the previous chapter, and revisit the generation of the anisotropies at the end of inflation. This also allows us to revisit the original idea of Yokoyama-Soda in \cite{Yokoyama:2008xw}. One novel feature of generating anisotropies at the surface of end of inflation is that the large unwanted anisotropy sourced by the cumulative IR contributions of super-horizon modes \cite{Bartolo:2012sd}, scaling like $N^2$ as a function of the total number of e-folds $N$, does not show up. \\
This chapter follows the material of \cite{Abolhasani:2013bpa}.
\section{Gauged Hybrid Inflation}
\label{hybrid}

In this section we present our setup of anisotropic inflation, the gauged hybrid inflation model.\\
As we already saw in chapter \ref{Chapter3}, the system contains the inflaton field $\phi$, the complex waterfall field $\psi$ and the gauge field $A_\mu$. As in standard hybrid inflation scenarios \cite{Linde:1993cn, Copeland:1994vg}, the waterfall field is responsible to terminate inflation abruptly. In this model, both inflaton and the gauge field are coupled to waterfall field so the surface of end of inflation is controlled both by the inflaton field and the gauge field. The gauge field fluctuations source anisotropy at the surface of end of inflation as advocated in  \cite{Yokoyama:2008xw}. Here we study anisotropies generated at the surface of end of inflation as well as anisotropies generated  actively during inflation  
using a consistent $\delta N$ formalism following the method employed in \cite{Abolhasani:2013zya}. \\
Since the background behavior of this model has been considered in previous chapters, we skip considering this in detail here and go directly to the final presented results in the previous chapters,
\ba
\label{cons-eq-7}
\dot A^2 f^2 e^{-2 \alpha} &=& I \, \epsilon V \, , \\
\label{klinGordon-7}
M_P^{-2}\frac{d \phi}{d \alpha} &\simeq& -\frac{V_\phi}{ V} + \frac{c-1}{c}\frac{V_\phi}{ V} \, .
\ea


\subsection{Solving for N}

Having specified the form of $f(\phi)$ we are ready to solve $N$ in terms of background fields dynamics. First, let us calculate $N$ as a function of the scalar field $\phi$. As described before, during the attractor phase the gauge field's back-reaction on inflaton dynamics is not negligible. The evolution of $\phi$ as a function of number of e-folds is given by
\ba
\label{phi-eq1}
\frac{d \phi}{d \alpha} &&= \left( I -1 \right) \frac{ M_{P}^{2}V_{, \phi}}{V}  \nonumber\\
&&= \left( I -1 \right) \frac{2 \phi }{p_c} \, .
\ea
Note that the effects of gauge field back-reaction on inflaton dynamics is captured by
the factor $I$ in Eq. (\ref{phi-eq1}). If we set $I=0$, then the gauge field has no effect on inflaton dynamics and one recovers the standard formula for $N(\phi)$ as in conventional hybrid inflation scenarios. Note that this is the critical limit in which the gauge field plays the role of an iso-curvature field during inflation. In other words, in the critical case we have turned on the conformal factor $f(\phi)$ such that  the gauge field fluctuations barely survive the background expansion and becomes exactly scale-invariant.

One can easily solve Eq. (\ref{phi-eq1}) to obtain
\ba
\label{N-phi-hybrid}
N(\phi) \simeq \frac{p_c}{2 \left(I- 1  \right) } \ln \left( \frac{\phi}{\phi_{f}} \right)  \, .
\ea
Note that we have replaced $\alpha = N$ and used the convention that $N$ is counted from the time of end of inflation $N_f=0$ such that  at the start of inflation where $\phi=\phi_i$, we have $N=N_i \simeq -60$. Eq. (\ref{N-phi-hybrid}) is one of our starting key
formula to find the final $\delta N$ expression.

Now we calculate $N$ as a function of the background gauge field dynamics. The key information is that during the attractor phase $R$ reaches a constant value as given by
Eq. (\ref{R-app-6}).  Plugging Eq. (\ref{R-app-6}) for definition of $R$ in Eq. (\ref{R-def-5}) and noting that
$f \simeq (a/a_f)^{-2} = e^{-2N}$ we obtain
\ba
\label{A-N}
A =  \sqrt{\frac{2R}{3}} M_P \left( e^{3 N} -1 \right) + A_{f} \, ,
\ea
in which $ A_f$ denotes the value of the gauge field at the end of inflation. From Eq. (\ref{A-N}) we see that the gauge field is completely  negligible during the most of the period of inflation  in which  $N<0$ and grows exponentially towards the end of inflation when it approaches  its final value $ A_f $.  Note that in the critical case in which $R \rightarrow 0$, we have $A \simeq A_f$ as expected for an iso-curvature field.

Alternatively, one can invert Eq. (\ref{A-N}) to find $N$ as a function of $A$
\ba
\label{N-A-hybrid}
N=  \frac{1}{3} \ln  \left( \sqrt{\frac{3}{2 R} } \left( \frac{ A -  A_{f} }{M_{P}} \right) + 1 \right) \, .
\ea
This is our second key formula to express the final formula of $\delta N$ in terms of the background fields perturbations.

Our last task in finding $\delta N$ is to take into account the contributions from the surface of the end of inflation  in $\delta N$. Here we follow closely the method developed in multi-brid inflation \cite{Sasaki:2008uc, Naruko:2008sq}. We parameterize the surface of end of inflation in Eq. (\ref{transition4}) via \cite {Emami:2011yi}
\ba
\label{gamma}
\phi_{f} = \phi_c \cos \gamma \quad , \quad
A_{f} =  \frac{g\,  \phi_c}{\mathbf{e}} \sin \gamma \, .
\ea
We note that the angle $\gamma$ should be treated as an independent variables so when we calculate
$\delta N$, we have to take into account the perturbations in $\gamma(\phi, A)$ which represents the contribution of the surface of the end of inflation in $\delta N$.

\subsection{ Calculating $\delta N$}
\label{deltaN }

Our goal is to calculate $\delta N$ up to second order in  $\delta \phi$ and $\delta A$ perturbations. We relegate the details of the analysis to
Appendix \ref{deltaN-app}. However, before using $\delta N$ formalism in this setup, one has to verify the validity of the gradient expansion and the separate Universe approach as studied in \cite{ Lyth:2004gb,  Naruko:2012fe, Naruko:2012um, Sugiyama:2012tj}. We demonstrate the validity of the separate Universe approach for our setup to second order in perturbation theory  in Appendix \ref{separate-universe}. This will allow us to use $\delta N$ formalism to
calculate the power spectrum and the bispectrum.

Calculating $\delta N( \phi, A)$ up to second order we have
\begin{align}
\label{delta N total}
\delta N &= N_{{\phi}} \delta {\phi} + N_{\dot{A}} \delta \dot{A} + N_{{A}} \delta {A}+ \frac{N_{{\phi} {\phi}}}{2} {\phi}^2  +  \frac{N_{\dot{A} \dot{A}}}{2} \delta \dot{A}^2 + \frac{N_{{A}{A}}}{2} \delta {A}^2\nonumber\\
&~~~+ N_{{\phi} \dot{A}} \delta {\phi} \delta \dot{A_x} +  N_{{\phi} {A}} \delta {\phi} \delta {A_x} + N_{{A} \dot{A}} \delta \dot{A} \delta {A} \, ,
\end{align}
in which to leading order in $R \ll 1$, for linear terms we have
\begin{align}
\label{N-phi}
N_{{\phi}} & =   \frac{p_{c}}{2(I-1)} \frac{1}{{\phi}}   + \frac{\e M_{P}p_{c}}{ 6 \phi_c \, g \, (I-1)}\frac{\tan{\gamma}}{\cos{\gamma}} \frac{f_{\phi}}{f} \sqrt{ 6 R} \\
\label{N-dotA}
N_{\dot {A}} & = \left(-\frac{2NI}{(I-1)} +  \frac{ \e \, p_{c}}{6 g (I-1)}\frac{\tan{\gamma}}{\cos{\gamma}} \frac{M_{P}}{\phi_{c}} \sqrt{ 6 R} \right) \left( \frac{1}{\dot {A}}\right) \\
\label{N-A}
N_{{A}} & = \frac{\e \, p_{c}}{2 g \phi_c (I-1)}\frac{\tan{\gamma}}{\cos{\gamma}} \, .
\end{align}
We note that in above expressions, the first term in $N_\phi$ represents the usual contribution in curvature perturbations from the inflaton field except with the small correction from the gauge field back-reaction given by the factor $I$ in the denominator.  The second term in
$N_\phi$, proportional to $\e$, represents the contribution in isotropic power spectrum from the inhomogeneities generated at the surface of end of inflation. On the other hand, the first term in $N_{\dot A}$, containing $ N I$, represents the anisotropies generated actively during inflation from $\delta \dot A_i$ fluctuations. This is the crucial contribution first derived in the analysis of  \cite{Abolhasani:2013zya} in the $\delta N$ formalism. Furthermore, the second term in $N_{\dot A}$ containing $\e$, represents the anisotropies generated purely at the surface of end of inflation again from the $\delta \dot A_i$ fluctuations. This is obtained  for the first time in this work in the context of $\delta N$ formalism. As we shall see, this is the  term which  captures the anisotropies generated from the surface of end of inflation as envisaged by Yokoyama and Soda in \cite{Yokoyama:2008xw}.  Finally, the term $N_A$ represents the anisotropies generated purely from the surface of end of inflation from  $\delta A_i$ fluctuations. This is the term which was included in the analysis of \cite{Emami:2011yi}. As we shall see below, the contribution from $N_A$ term is exponentially suppressed compared to the contribution from $N_\phi$ and $N_{\dot A}$, consistent with the  analysis of \cite{Emami:2011yi}. However, in the analysis of \cite{Emami:2011yi} the crucial contribution of $N_{\dot A}$, Eq. (\ref{N-dotA}),  was not included which has led to the conclusion different than the result obtained in \cite{Yokoyama:2008xw}.

Correspondingly, for the non-linear terms we get
\ba
N_{{\phi} {\phi}} & =&  - \frac{p_{c}}{2(I-1)} \frac{1}{{\phi}^2}  - \frac{2NI}{(I-1)} \left( \left(\frac{f_{\phi}}{f}\right)^2+ \frac{f_{,\phi \phi}}{f}\right) + \frac{\e^2 M_{P}^2 p_{c}}{6 g^2 \phi_c^2 (I-1)}\frac{(\sin^2{\gamma}+1)}{\cos^4{\gamma}} \left(\frac{f_{\phi}}{f} \right)^2 I \epsilon \nonumber \\
&&~~  -\frac{2 \e Ip_{c}}{3 g (I-1)^2}\frac{\tan{\gamma}}{\cos^2{\gamma}} \frac{ M_{P}f_{\phi}^2}{\phi_c f^2} \sqrt{ 6 R}+
\frac{\e p_{c}}{2g (I-1)}\frac{\tan{\gamma}}{\cos{\gamma}} \frac{ M_{P}f_{\phi \phi}}{f \phi_c} \sqrt{\frac{ 2R }{3}} \\
N_{{A} {A}} & =&  \frac{\e^2 p_{c}}{2 g^2 \phi_c^2(I-1)}\frac{(\sin^2{\gamma}+1)}{\cos^4{\gamma}} \\
N_{\dot {A}\dot {A}} & = &\left(\frac{-2NI}{(I-1)} + \frac{ \e^2 \, R\,  p_{c}}{3 g^2 (I-1)}\frac{(\sin^2{\gamma}+1)}{\cos^4{\gamma}} \frac{M_{P}^2}{\phi_{c}^2} -\frac{2 \e I p_{c}}{3g (I-1)^2}\frac{\tan{\gamma}}{\cos{\gamma}} \frac{M_{P}}{\phi_{c}}
\sqrt{6 R} \right) \frac{1}{\dot {A}^2}
\ea
\ba
N_{{A} {\phi}} & = & \frac{\e^2 p_{c}}{6 g^2 (I-1)}\frac{(\sin^2{\gamma}+1)}{\cos^4{\gamma}}\frac{M_{P}f_{\phi}}{\phi_c f} \sqrt{ 6 R} - \frac{ \e I p_{c}}{g (I-1)^2}\frac{\tan{\gamma}}{\cos{\gamma}} \frac{f_{\phi}}{\phi_c f} \\
N_{{A} \dot{A}} & =& \left(\frac{\e^2 p_{c}}{6 g^2 (I-1)}\frac{(\sin^2{\gamma}+1)}{\cos^4{\gamma}}\frac{M_{P}}{\phi_{c}} \sqrt{6 R} - \frac{ \e I p_{c}}{g (I-1)^2}\frac{\tan{\gamma}}{\cos{\gamma}}\right)  \left( \frac{1}{\dot {A}}\right) \\
N_{\dot{A} {\phi}} & =& - \frac{I p_{c}}{(I-1)^2} \frac{1}{{\phi}}  - \frac{4NI}{(I-1)} \frac{f_{\phi}}{\phi_c f} + \frac{\e^2 M_{P}^2 p_{c}}{6 g^2 (I-1)\phi_{c}^2}\frac{(\sin^2{\gamma}+1)}{\cos^4{\gamma}} \frac{f_{\phi}}{f} I \epsilon \nonumber \\
\label{N-dotA-phi}
& &~~  + \frac{ \e p_{c}}{6 g (I-1)}\frac{\tan{\gamma}}{\cos{\gamma}} \frac{ M_{P}f_{\phi}}{\phi_c f} \sqrt{6 R} \, ( 1- 4 I)  \, .
\ea

For the future reference, we note that if we set $I=0$ in the above expressions, we obtain
the results in the critical limit $p=p_c$ in which the gauge field does not back-react on the inflaton dynamics and its contributions into anisotropies are generated purely from the surface of end of inflation controlled by the gauge coupling $\e$.

\subsection{Seed Quantum Fluctuations}

Having calculated $\delta N$ to second order in fields perturbations we are ready to calculate the power spectrum and the bispectrum. Before doing that, we need some information about the statistical properties of $\delta \phi$ and $\delta A$ fluctuations necessary for
$\delta N$ analysis. As usual, we take $\delta \phi$ to be a near scale-invariant perturbations with amplitude $H/2\pi$ at the time of horizon crossing such that
\ba
\label{dis phi}
\left<\delta  \phi_{\bfk} \delta \phi_{\bfk'}\right> \equiv (2\pi)^{3} P_{\delta  \phi}(k)~\delta^3(\bfk+\bfk') ~~,~~ {\cal P}_{\delta  \phi}\equiv \frac{k^{3}}{2 \pi^{2}}P_{\delta  \phi}(k) = \left( \frac{H_*}{2 \pi} \right)^2 \, .
\ea
Note that  this is calculated at the time when the mode of interest leaves the horizon, denoted by  $*$,  when $k=a_* H_*$.

To find the power spectrum of gauge field fluctuations  we solve the mode function for $\delta A_i$ with the initial Bunch-Davies vacuum when the mode of interest is deep inside the horizon.   The canonically normalized gauge field quantum fluctuations are decomposed via
\cite{Bartolo:2012sd}
\ba
\label{A-V-eq}
\delta A_i =   \sum_{\lambda = \pm} \int \frac{d^3k}{(2\pi)^{3/2}}e^{i \bfk.\mathbf{x}}\vec{\epsilon}_{\lambda}(\bfk) \frac{\widehat{V}_i}{f}  \, ,
\ea
where
\ba
\widehat{V} = a_{\lambda}(\bfk)V_{\lambda}(k) + a_{\lambda}^{\dagger}(-\bfk)V_{\lambda}^{*}(k) \, .
\ea
As usual  $a_{\lambda}(\bfk)$ and $a_{\lambda}^{\dagger}(\bfk)$
represent the annihilation and the creation operators. Furthermore,  $\epsilon_\lambda$ represents the circular polarization for $\lambda =\pm$,
satisfying  $\vec k \,  . \,   \vec \epsilon_\pm(\bfk) =0 \, , \,  \bfk \, \times \,  \vec \epsilon_\pm(\bfk) =
\mp i k \, \vec \epsilon_\pm(\bfk)  \, , \vec \epsilon_\lambda (-\bfk) = \vec \epsilon_\lambda\, (\bfk)^*$, normalized such that
$\vec \epsilon_\lambda(\bfk)  \, . \,  \vec \epsilon_{\lambda'}(\bfk)^*= \delta_{\lambda \lambda'}$ and
\ba
\sum_{\lambda}\epsilon_{\lambda,i}({\bfk})\epsilon^{*}_{\lambda,j}({\bfk}) = \delta_{ij} - \hat{k}_{i}\hat{k}_{j} \, .
\ea
As we have seen in the previous chapter, the mode function for the gauge field would be, (\ref{mode-deltaA-6})
\begin{align}
\label{canonical gauge field}
V_{\lambda}(k) \simeq \frac{1+ i k \eta}{\sqrt{2}k^{3/2}\eta} e^{-ik\eta} \, .
\end{align}
As a result, on super-horizon scales we obtain \cite{ Abolhasani:2013zya}
\ba
\label{mode-deltaA}
\frac{\delta \vec{\dot{A}}}{\dot{A}} = \frac{1}{M_P} \sqrt{\frac{3}{2 R}}
\frac{H}{ \sqrt{2 k^3}}
\sum_{\lambda}\vec{\epsilon}_{\lambda}     \quad \quad (k < a H)
\ea
Eq. (\ref{mode-deltaA}) shows that  $ \delta \dot A/\dot A $ is scale-invariant on super-horizon scales which is crucial for our analysis below. This is a consequence of turning on the
conformal factor $f(\phi)$ with the specific form given in Eq. (\ref{f-scale5}).

One crucial point to mention is that although $V_\lambda(k)$ has the profile of a massless scalar field fluctuations as given in Eq. (\ref{canonical gauge field}), but the profile of gauge field
$\delta A$ is related to $V_\lambda(k)$ by additional factor of $f(\phi)^{-1}$ as shown in
Eq. (\ref{A-V-eq}). As a result, the power spectrum of $\delta A$, $\calP_{\delta A}$, is hugely suppressed compared to $\calP_{\delta \phi}$. More specifically
\ba
\label{P-A-less}
\calP_{\delta A_*} = \frac{1}{f(\phi_*)^2} \calP_{\delta \phi} \simeq e^{4 N_*}  \calP_{\delta \phi_*} \, .
\ea
Noting that $N_* \simeq -60$, we conclude that $\calP_{\delta A_*} \ll \calP_{\delta \phi}$.
This was the key argument in  \cite{Emami:2011yi} as why the  method employed in \cite{Yokoyama:2008xw} does not work. As we discussed in previous sub-section, in the paragraph after Eq. (\ref{N-A}),  in order to reproduce the results of \cite{Yokoyama:2008xw} correctly, we have to take into account
the contribution of $\delta \dot A/\dot A$, represented by $N_{\dot A}$ in Eq. (\ref{N-dotA}),
which is scale-invariant as given in Eq. (\ref{mode-deltaA}) following the method developed  in \cite{ Abolhasani:2013zya}. This was the crucial missing piece in the analysis of \cite{Emami:2011yi}.


\section{Power Spectrum}
\label{power-spec}

Having calculated $\delta N$ to second order in Eq. (\ref{delta N total}) we can calculate the power spectrum and the bispectrum using the standard formula
\ba
\calR(\mathbf{x},t) = \delta N( \phi, A) \, .
\ea
The power spectrum analysis are presented here while the bispectrum analysis are given in next Section.

The quantities $N_\phi, N_A$ and $N_{\dot A}$ are given in Eqs. (\ref{N-A}), (\ref{N-dotA})
and (\ref{N-A}). As a result
\ba
\calP_\calR = N_\phi^2 \calP_{\delta \phi} + \dot A^2 N_{\dot A}^2 \calP_{\delta \dot A/\dot A}
+ N_{A}^2 \calP_{\delta A}
\ea
As argued in Eq. (\ref{P-A-less}), one can safely neglect the contribution of $\calP_{\delta A}$ in the above formula since it is exponentially suppressed compared to $\calP_{\delta \phi}$.
This is reminiscent of the argument used in \cite{Emami:2011yi}. However, as will show, the source of anisotropy is actually from $\calP_{\delta \dot A/\dot A}$. Neglecting the contribution of $\calP_{\delta A}$ we obtain
\ba
\calP_\calR \simeq  \calP_{\calR }^{(0)}  + \dot A^2 N_{\dot A}^2 \calP_{\delta \dot A/\dot A} \, ,
\ea
in which the isotropic power spectrum is calculated to be
\ba
\label{power}
\calP_{\calR }^{(0)} = N_{\phi_*}^2 \calP_{\delta \phi_*} = \left( \frac{p_c H_*}{4 \pi \phi_*}
\right)^2 \left[ 1+ \frac{\e p_c M_P}{g \phi_c} \frac{\tan \gamma}{\cos \gamma} \sqrt{\frac{2 R}{3}} \right]^2   \, ,
\ea
in which an asterisk represents the time of horizon crossings, $k= a_* H_*$, and
the relation $f_\phi/f \simeq p_c/\phi$ has been used. Furthermore, it is assumed that
$I \ll 1 $ which follows from the observational constraint as we shall see below,
see also  \cite{ Watanabe:2010fh, Abolhasani:2013zya}.

Now we calculate the anisotropy generated during and at the surface of end of inflation coming from $\calP_{\delta \dot A/\dot A}$. First, from Eq. (\ref{mode-deltaA}),  we note that
\ba
\calP_{\delta \dot A/\dot A} = \frac{k^3}{2 \pi^2 }
\left\langle  \frac{\delta \dot A_x(k_1)}{A_x}
\frac{\delta \dot A_x(k_2)}{A_x} \right\rangle = \frac{3 H_*^2 \sin^2 \theta }{8 \pi^2 M_P^2 R} \, ,
\ea
in which $\cos \theta = \hat {\bf n}. \hat{\bf k}$ represents the angle between the preferred direction (in our notation the $x$-direction) and the wave-number.  Using the expression for
$N_{\dot A}$ given in Eq. (\ref{N-dotA}), the anisotropy  in power spectrum becomes
\ba
\Delta \calP_\calR \equiv \dot A^2 N_{\dot A}^2 \calP_{\delta \dot A/\dot A}  =
\left(-2NI  +  \frac{ \e \, p_{c}}{6 g }\frac{\tan{\gamma}}{\cos{\gamma}} \frac{M_{P}}{\phi_{c}} \sqrt{ 6 R} \right)^2  \frac{3 H_*^2 \sin^2 \theta }{8 \pi^2 M_P^2 R} \, .
\ea
On the other hand, using the relation $R= I \epsilon/2 $ and
\ba
\label{epsilon-eq}
\epsilon \simeq \frac{2 \phi_*^2}{p_c^2 M_P^2}
\ea
and assuming $\phi_* \simeq \phi_f$ which is a good approximation in hybrid inflation,
the fractional change in power spectrum due to anisotropies from the gauge fields is
\ba
\frac{\Delta \calP_\calR}{\calP_{\calR }^{(0)}} =  \left( \frac{\sqrt{24 I N^2} - \frac{\e}{g} \tan \gamma}{1+ \frac{\e p_c M_P}{g \phi_c} \frac{\tan \gamma}{\cos \gamma} \sqrt{\frac{2}{3} R}} \right)^2  \sin^2 \theta \, .
\ea

Now, in the limit in which $R= I \epsilon/2 \ll1 $ (assuming the pre-factor in denominator above does not diverge for typical value of $\e$ and $\gamma$) we obtain
\ba
\label{g-eq}
g_* \simeq  -\left( \sqrt{24 I N^2} - \frac{\e}{g} \tan \gamma \right)^2 \, .
\ea
This is one of the main result of this paper.

Eq. (\ref{g-eq}) is very interesting. The first term in the bracket captures the effects of anisotropy generated actively during inflation \cite{ Watanabe:2010fh, Bartolo:2012sd,
	Abolhasani:2013zya, Emami:2013bk}.
As is well-known the anisotropy generated during inflation is given by
\ba
\label{g-IR}
g_*|_{IR} = -24 I N^2
\ea
in which the subscript IR indicates the accumulative contributions of the IR modes which leave the horizon at  $N$ e-folds towards the end of inflation. As mentioned in \cite{Bartolo:2012sd} the IR contribution can become
too large so in order not to produce too much anisotropy from IR accumulations one requires
that $N$ is not too large. With $N=60$  to solve the flatness and the horizon problem, and $| g_* | \lesssim 10^{-2}$ in order to satisfy the observational bounds, one requires $I \lesssim 10^{-7}$.

The second term in Eq. (\ref{g-eq}) represents the anisotropy generated purely from the surface of end of inflation. This is the contribution advocated originally in
\cite{Yokoyama:2008xw}. Denoting this contribution by $g_*|_{\e}$ we have
\ba
\label{ge}
g_*|_{\e} = - \frac{\e^2}{g^2} \tan^2 \gamma \, .
\ea
Note that this term exists even we set $I=0$. This is the critical limit in which the gauge field becomes an iso-curvature field which does not contribute into power spectrum
during inflation but can generate additional anisotropic curvature perturbations from the
inhomogeneities generated at the surface of end of inflation a la Lyth \cite{Lyth:2005qk}.
We can also compare our expression for $g_*|_{\e}$ with the value for $g_*$ obtained in
\cite{Yokoyama:2008xw}, see also \cite{ Emami:2011yi} and \cite{Lyth:2012vn}.
Interestingly, we find that in the physical limit in which $| g_*| \ll 1$, our value for $g_*|_{\e}$ agrees with $g_*$ obtained in  \cite{Yokoyama:2008xw}.
This may seem somewhat surprising, since in the analysis of \cite{Yokoyama:2008xw} the exponential evolution of the gauge field during inflation and the the effects of $N_{\dot A}$ in $\delta N$ expansion is not taken into account. However, the analysis of \cite{Yokoyama:2008xw} relies crucially on the idea of \cite{Lyth:2012vn} in which the inhomogeneities at the surface of end of inflation are generated from a scale-invariant perturbation, which in our
case it is $\delta \dot A/\dot A$. The fact that the fluctuations $\delta \dot A/\dot A$ are massless and scale-invariant guarantees that the method proposed in \cite{Lyth:2005qk} still applies in the analysis of \cite{Yokoyama:2008xw}. This is the reason why the results obtained in \cite{Yokoyama:2008xw} are correct as a matter of principle independent of
their somewhat ad-hoc $\delta N$ analysis.

One curious conclusion from  Eq. (\ref{g-eq}) is that  one can choose the parameter space such that  the anisotropies generated actively during inflation are cancelled from the anisotropies generated from the surface of end of inflation, yielding $g_*=0$. For this to happen one requires
\ba
\label{tuning}
\tan \gamma = \frac{g}{\e} \sqrt{24 I N^2} \, .
\ea
For a given value of number of e-foldings $N$, this determines how one should approach the surface of end of inflation in order to obtain $g_* =0$. Of course, in this limit, one should take into account the sub-leading terms containing factors of $I$ in $1-I $ expressions and so on
in $\delta N$ expression in previous sub-section. Nonetheless, it may be possible to achieve $g_*=0$ to higher order in $I$ too.

One interesting implication of  Eq. (\ref{g-eq}) is that one does not need to impose the strong fine-tuning $I \lesssim 10^{-7}$ anymore. The smallness of $g_*$ is now controlled by the near cancellation of the two terms in  Eq. (\ref{g-eq}) instead of requiring the first term, containing $\sqrt {I N^2}$, itself to be small.

Now let us look at the spectral tilt, $n_s$.  Similar to \cite{Ohashi:2013qba} we have
\ba
\label{ns}
n_s -1= \frac{d \ln \calP_\calR}{d \ln k} = ( n_s^{(0)} -1  ) + \frac{\cos^2 \theta }{1+ g_* \cos^2 \theta } \frac{d g_*}{d \ln k} \, ,
\ea
in which $n_s^{(0)}$ represents the spectral index obtained from the isotropic power spectrum. In conventional models of hybrid inflation in which the waterfall is a real field and is not gauged under the $U(1)$ field $n_s^{(0)}>1$ so the power spectrum is blue-tilted. As a result, the simple models of hybrid inflation are ruled out from the PLANCK data which requires $n_s <1$. On the other hand,  with $g_*$ given in Eq. (\ref{g-eq}), the change in spectral index induced by anisotropies, $\Delta n_s$, in the limit $| g_*| \ll 1$ is obtained to be
\ba
\label{delta-ns}
\Delta n_s \simeq - \sqrt{-96 I g_*}  \, \langle \cos^2 \theta \rangle =- \frac{ \sqrt{-96 I g_*}}{3} \, ,
\ea
in which we have taken the average value $\langle \cos^2 \theta \rangle = 1/3$.

Interestingly, anisotropies, once averaged over the whole sky, generate a red-tilted power spectrum. This may help to generate a red-tilted power spectrum for hybrid inflation.  As an example, suppose we take $g_* \sim -10^{-2}$ and $I \sim 10^{-2}$, then we obtain $\Delta n_s \sim - 0.03$
which can bring the value of $n_s$ in models of  gauged hybrid inflation within the desired range of PLANCK data. Having this said, this value of $I$ may seem too large compared to results obtained in \cite{ Watanabe:2010fh, Bartolo:2012sd,
	Abolhasani:2013zya, Emami:2013bk}. However, in those works anisotropies are generated during inflation so $g_*$ is given in Eq. (\ref{g-IR}), yielding $I \sim 10^{-7}$. However, in our model, as we mentioned before, anisotropies generated at the surface of end of inflation can nearly cancel the anisotropies generated during inflation so one can saturate the observational bound on $g_*$ without requiring very small value of $I$. Therefore, taking
a value of $I$ much larger than the value required in \cite{ Watanabe:2010fh, Bartolo:2012sd,  Abolhasani:2013zya, Emami:2013bk} is easy in our model.

Before ending this section we comment that we have neglected the contributions of the waterfall quantum fluctuations and the longitudinal excitation in curvature perturbations. The reason is that the waterfall is very massive during inflation and its power spectrum is highly blue-tilted. As studied in great details in  \cite{Abolhasani:2010kr, Abolhasani:2011yp,      Fonseca:2010nk, Gong:2010zf, Lyth:2010ch, Lyth:2012yp} the contributions of the waterfall on large scale (i.e. CMB scale) curvature perturbations are exponentially suppressed. Similarly, the longitudinal mode is very massive and its excitations are highly suppressed compared to the transverse modes. As demonstrated in \cite{Emami:2013bk} in a similar model,
the contributions of the longitudinal excitation in curvature perturbations are exponentially suppressed compared to the transverse mode.

\section{Bispectrum}
\label{bi-spec}

In this Section we calculate the Bispectrum of our model. For similar analysis in anisotropic inflation background of \cite{Watanabe:2009ct} see \cite{Bartolo:2012sd, Shiraishi:2013vja}.
As demonstrated in \cite{Abolhasani:2013zya}
the main contribution in bispectrum comes from the $N_{\dot A \dot A}$ terms. The contributions from terms containing $N_{\phi \phi}$ are slow-roll suppressed as demonstrated by Maldacena \cite{Maldacena:2002vr}. Furthermore, similar to power spectrum case, terms containing $N_{AA}, N_{A \dot A}$  and $N_{A \phi}$ are suppressed as argued in Eq. (\ref{P-A-less}). In addition, one can check that the contribution from the term
$N_{\dot A \phi}$ is suppressed by a factor of $1/N$  compared to the contribution from
the  $N_{\dot A \dot A}$ term.  As a result, in our analysis below we consider the contributions from $N_{\dot A \dot A}$ term.

Calculating the three point function, and taking $I \ll 1$,   we obtain
\ba
\label{leading three0}
\left \langle \calR (\bfk_{1}) \calR (\bfk_{2}) \calR (\bfk_{3})   \right \rangle  &\simeq&
N_{, \dot{A}\dot{A}}( k_{1}) N_{, \dot{A}}( k_{2}) N_{, \dot{A}} ( k_{3}) \int \frac{d^3\bfp}{2(2\pi)^3} \left \langle \delta \dot{A}_{x} (\bfk_{1})  \delta \dot{A}_{x} (\bfk_{2})
\delta \dot{A}_{i} (\bfp) \delta \dot{A}_{i} (\bfk_{3} - \bfp)  \right \rangle \nonumber\\
 && + 2 \mathrm{perm.}=   \left(N_{k_3} I - \frac{\e^2 \,  R \, p_{c}}{6 g^2}\frac{(\sin^2{\gamma}+1)}{\cos^4{\gamma}} \frac{M_{P}^2}{\phi_{c}^2} - \frac{ \e\,  I p_{c}}{3 g }\frac{\tan{\gamma}}{\cos{\gamma}} \frac{M_{P}}{\phi_{c}}\sqrt{3I \epsilon} \right) \nonumber \\
&& \times \left(2N_{k_1} I - \frac{\e \, p_{c}}{6 g }\frac{\tan{\gamma}}{\cos{\gamma}} \frac{M_{P}}{\phi_{c}} \sqrt{3I \epsilon} \right)\Bigg{(}2N_{k_2} I - \frac{\e\,  p_{c}}{6 g }  \frac{\tan{\gamma}}{\cos{\gamma}} \frac{M_{P}}{\phi_{c}} \sqrt{3I \epsilon} \Bigg{)} \nonumber\\
&&\times \int \frac{d^3\bfp}{(2\pi)^3}  \left\langle \frac{\delta \dot{A}_{x} (\bfk_{1})}{\dot{A}}  \frac{\delta \dot{A}_{x} (\bfk_{2})}{\dot{A}}\frac{\delta \dot{A}_{i} (\bfp)}{\dot{A}}
\frac{\delta \dot{A}_{i} (\bfk_{3} -\bfp)}{\dot{A}} \right\rangle  + 2 \mathrm{perm.} \, ,
\ea
in which $N_{k_i}$ represents the time when the mode $\bfk_i$ leaves the horizon. Calculating the convolution integrals and defining the bispectrum $B_{\calR} (\bfk_1, \bfk_2, \bfk_3)$
via
\ba
\label{bi- def}
\langle \calR (\bfk_{1}) \calR (\bfk_{2}) \calR (\bfk_{3}) \rangle &\equiv& \left( 2 \pi \right)^3 \delta^3 \left( \bfk_{1} +  \bfk_{2} +  \bfk_{3}\right) B_{\calR}(\bfk_{1}, \bfk_{2}, \bfk_{3})
\ea
we get
\ba
\label{leading three}
B_{\calR}(\bfk_{1}, \bfk_{2}, \bfk_{3})   &\simeq& \frac{18 \epsilon^2}{R^2}
\left(2N_{k_1} I - \frac{\e \, p_{c}}{6 g }\frac{\tan{\gamma}}{\cos{\gamma}} \frac{M_{P}}{\phi_{c}} \sqrt{6 R} \right)\Bigg{(}2N_{k_2} I - \frac{\e\,  p_{c}}{6 g }  \frac{\tan{\gamma}}{\cos{\gamma}} \frac{M_{P}}{\phi_{c}} \sqrt{6 R} \Bigg{)}
\times  \nonumber \\
&& \left(N_{k_3} I - \frac{\e^2 \,  R \, p_{c}}{6 g^2}\frac{(\sin^2{\gamma}+1)}{\cos^4{\gamma}} \frac{M_{P}^2}{\phi_{c}^2} - \frac{ \e\,  I p_{c}}{3 g }\frac{\tan{\gamma}}{\cos{\gamma}} \frac{M_{P}}{\phi_{c}}\sqrt{6R} \right) \times \nonumber\\
&&\Big( C(\bfk_{1}, \bfk_{2})P_0(k_{1})P_0(k_{2}) + 2 \mathrm{perm.} \Big) \, , \nonumber\\
\ea
in which $P_0(k)= (2 \pi^2/k^3) \calP_{\calR}^{(0)}(k)$ and
the momentum shape function $C(\bfk_1, \bfk_2) $ is defined via  \cite{ Bartolo:2012sd,  Abolhasani:2013zya}
\ba
C(\bfk_{1}, \bfk_{2})\equiv\bigg{(}1 -   (\widehat \bfk_1.\widehat {\bf{n}} )^2  -   (\widehat \bfk_2.\widehat {\bf n})^2 +
(\widehat \bfk_1.\widehat {\bf n} ) \,  (\widehat \bfk_2.\widehat {\bf n}) \,  (\widehat \bfk_1.\widehat \bfk_2)  \bigg{)} \, .
\ea
As in the power spectrum case, the bispectrum is generated both during inflation and at the end of inflation. The bispectrum generated during inflation in Eq. (\ref{leading three}) are represented
by $N_{k_i} I $ factor while the bispectrum generated from inhomogeneous end of inflation effect are controlled by the gauge coupling $\e$.

Now let us look at the bispectrum generated purely during inflation by setting $\e=0$ in Eq. (\ref{leading three}). In this limit, we obtain
\ba
\label{B-active}
B_{\calR}(\bfk_{1}, \bfk_{2}, \bfk_{3}) |_{IR} = 288 I\, N_{k_1} N_{k_2} N_{k_3}
\Big( C(\bfk_{1}, \bfk_{2})P_0(k_{1})P_0(k_{2}) + 2 \mathrm{perm.} \Big),
(\e=0) \, .
\ea
This is in exact agreement with the result obtained in  \cite{Bartolo:2012sd, Abolhasani:2013zya}.

Now let us look at the critical case in which the gauge field plays the role of an iso-curvature field, $I=0$, and the bispectrum is generated purely at the surface of end of inflation.  In this limit, we get
\ba
\label{B-cric}
&& B_{\calR}(\bfk_{1}, \bfk_{2}, \bfk_{3}) |_{\e} =\nonumber\\
&& -\frac{2\e^4 }{ p_c g^4} \frac{\sin^2 \gamma (1+ \sin^2 \gamma)}{\cos^4 \gamma} \Big( C(\bfk_{1}, \bfk_{2})P_0(k_{1})P_0(k_{2}) + 2 \mathrm{perm.} \Big), (p=p_c) \, .
\ea
Comparing Eqs. (\ref{B-active}) and (\ref{B-cric}), we find that the bispectrum generated at
the surface of end of inflation always has an opposite sign compared to  the bispectrum generated during inflation.

It is instructive to look at the non-Gaussian parameter $f_{NL}$ defined in the squeezed limit $k_1 \ll k_2 \simeq k_3$ via
\ba
f_{NL} (\bfk_1, \bfk_2, \bfk_3) = \lim_{k_1 \rightarrow 0} \frac{5}{12}
\frac{B_\zeta(\bfk_1, \bfk_2, \bfk_3)}{P_\zeta(k_1) P_\zeta(k_2)} \, .
\ea
If we further assume $N_{k_1} \simeq N_{k_2} \simeq N_{k_3}$ and noting that
$R = I \epsilon/2$ and using Eq. (\ref{epsilon-eq}), from  Eq. (\ref{leading three}) we obtain
\ba
\label{fNL-g*}
f_{NL} = -\frac{5 \epsilon g_*}{R} \left[ N I - \frac{\e I}{g}\sqrt{\frac{2 I}{3}} \tan \gamma
- \frac{\e^2 R}{3  p_c \epsilon g^2} \frac{1+ \sin^2 \gamma}{\cos^2 \gamma }
\right] C(\bfk_1, \bfk_2) \, .
\ea
This is an interesting result. Specially, consider the situation in which Eq. (\ref{tuning}) is satisfied so $g_*=0$. Then Eq. (\ref{fNL-g*}) also  indicates that $f_{NL}=0$. In other words, the anisotropies generated actively during inflation are canceled by the anisotropies generated from the surface of end of inflation both at the level of power spectrum and bispectrum.

In the limit $\e=0$ one obtains the known result that
\ba
\label{fnl squeezed}
f_{NL} (\bfk_1, \bfk_2)|_{IR}&=& 240 I N(k_{1}) N(k_{2})^2 C(\bfk_{1}, \bfk_{2}) ,  \quad
( k_{1} \ll k_{2} \simeq k_{3} ), \quad (\e=0) \\
&\simeq& 10 N\,  |g_*| \, C(\bfk_1, \bfk_2)  \, . \nonumber
\ea
On the other hand, for the critical case with $I=0$ we obtain
\ba
\label{fNL-e0}
f_{NL}(\bfk_1, \bfk_2)|_\e &=&   -\frac{5\e^4}{3p_cg^4}  \frac{\sin^2 \gamma (1+ \sin^2 \gamma)}{\cos^4 \gamma}  C(\bfk_1, \bfk_2) \nonumber\\
&=&  -\frac{5 g_*^2}{3p_c} \frac{1+ \sin^2 \gamma}{\sin^2 \gamma }  C(\bfk_1, \bfk_2), 
\quad
( k_{1} \ll k_{2} \simeq k_{3} ), 
\quad (I=0)
\ea

To satisfy the observational bound from the Planck data, one should not produce too much non-Gaussianity. However, we note that the anisotropic non-Gaussianity given by the
shape function $C(\bfk_1, \bfk_2)$ is quite different than the known non-Gaussian shapes.
In the critical case in which $I=0$ and all anisotropies are generated at the surface of end of inflation, this is easy to satisfy. Indeed, Eq. (\ref{fNL-e0}) indicates that if $\gamma $ is not very close to zero, then $f_{NL}$ is quite negligible with small $g_*$. On the other hand, if  anisotropy is generated predominantly during inflation, then Eq. (\ref{fnl squeezed}) indicates that $f_{NL}$ is at the order of few which can be detectable.

Similar to the case of  power spectrum, Eqs. (\ref{B-cric}) and (\ref{fNL-e0}) have the interesting property that they doe not depend on the length of the duration of inflation. As a result, the IR  anisotropy problem associated with the model such as \cite{ Watanabe:2010fh, Bartolo:2012sd,  Abolhasani:2013zya, Emami:2013bk} in which
primordial anisotropies are generated during inflation does not show up  in models in which anisotropies are generated exclusively at the surface of end of inflation.

It is also constructive to compare our expression for $f_{NL}|_\e$ with $f_{NL}$ obtained
in \cite{Yokoyama:2008xw}. Taking the limit in which the direction-dependence in their formula collapses to our $C(\bfk_1, \bfk_2)$ their formula for $f_{NL}$ scales like
(in our notation)  $(\e^6/g^6) \tan^4 \gamma (1+ \tan^2 \gamma)$ which is different than our result obtained in Eq. (\ref{fNL-e0}).

\section{Summary}
\label{summary}

In this chapter, we have studied primordial quadrupole asymmetry in models of gauged hybrid inflation. As we mentioned there are two mechanisms to generate primordial anisotropies from gauge fields: either actively from IR contributions during inflation or from the inhomogeneities
generated at the surface of end of inflation via the waterfall mechanism. The anisotropies generated during inflation suffers from the IR problem in which $g_*$ grows as $N^2$ which
leads to too much anisotropies to be compatible with the observation. Similarly, in this mechanism, $f_{NL}$ scales like $N^3$ which is again problematic in the light of recent data
indicating no detectable non-Gaussianity. On the other hand, the mechanism of generating primordial anisotropy at the surface of end of inflation has the appealing feature that it does not suffer from the above mentioned IR problem. In this mechanism, the gauge field is an iso-curvature field which has negligible contribution in total energy density  so it does not affect the inflaton dynamics. However, it modulates the waterfall mechanism yielding to inhomogeneities at the surface of end of inflation. As a result, this generates primordial anisotropies purely at the end of inflation as pioneered by Yokoyama and Soda \cite{Yokoyama:2008xw}. \\
In this chapter we have employed our consistent $\delta N$ mechanism, presented in chapter \ref{Chapter5}, to calculate the curvature perturbations up to second order. We have clearly identified the contributions in anisotropic power spectrum from the above two mechanisms as given in Eq. (\ref{g-eq}). The  two limiting cases
are given by Eqs. (\ref{g-IR}) and (\ref{ge}). Similarly, the different contributions in bispectrum are identified as given by Eq. (\ref{leading three}) with the two limiting cases
given in Eqs. (\ref{fnl squeezed}) and (\ref{fNL-e0}).\\
The combined effects of generating anisotropies during and at the surface of end of inflation have some interesting observational implications. First, one can choose the parameter space such that these two source of anisotropies cancel each other's contributions so there is no net primordial anisotropies both at the level of power spectrum and bispectrum. Second, the anisotropic power spectrum are red-tilted. After averaging the anisotropic power spectrum over the sky, the total power spectrum can become red-tilted. This is an interesting observation which can save models of hybrid inflation which in the absence of gauge fields predict a blue-tilted  power spectrum in contrast with cosmological observations.


\chapter{The TT, TB, EB and BB  correlations in  anisotropic inflation } 

\label{Chapter8} \begin{flushleft}
	
\end{flushleft}

\lhead{Chapter 8. \emph{The TT, TB, EB and BB  correlations in  anisotropic inflation}} 

\vspace{0.5cm}
\hrule \vspace{0.3cm}

\begin{quote}
\textbf{Abstract:} In this chapter, we study the TT, TB, EB and BB correlations associated with the B-mode polarization of CMB map in models of charged anisotropic inflation, presented in \ref{Chapter3}. We calculate the statistical anisotropies generated in the power spectra of the curvature perturbation, the tensor perturbation and their cross-correlation. It is shown that the asymmetry in tensor power spectrum is a very sensitive probe of the gauge coupling. While the level of statistical  anisotropy in temperature power spectrum can be small and satisfy the observational bounds, the interactions from the gauge coupling can induce large directional dependence
in tensor modes. This will leave interesting anisotropic fingerprints in various correlations involving the B-mode polarization such as the TB cross-correlation which may be detected in upcoming Planck polarization data. In addition, the TT correlation receives an anisotropic contribution from the tensor sector which naturally decays after $l \gtrsim 100$. We expect that the mechanism of using tensor sector to induce  asymmetry at low $l$ to be generic which can also be applied to address other low $l$ CMB anomalies.
\end{quote}

\vspace{0.1cm}  \hrule
\vspace{0.5cm}

\section{Introduction}
Having presented the signatures of the primordial anisotropic universe in the temperature spectrum, in this chapter, we look for its unique fingerprint in the polarization map of the cosmic microwave background. \\ 
There has been some experiments looking for the imprints of the Gravitational waves in the CMB. There was a report from the BICEP2 collaboration, \cite{Ade:2014xna}, for a strong detection of the Inflationary Gravitational Waves, IGWs. However, a recent joint analysis of the BICEP2-Keck-Planck collaboration, \cite{Ade:2015tva}, demonstrated that at least half of the signal is due to the polarization emission from the galactic dust and the residual power is also consistent with zero IGW signal. There are also other types of the CMB experiments looking for the GW, e.g. 100 square degrees of SPTPOL, \cite{Keisler:2015jua}. However, so far there has not been enough sensitivity to detect the IGW from these kind of experiments. \\
In addition, the forthcoming generation of the tensor experiments such as SPTPol \cite{Austermann12}, ACTPol \cite{Niemack10}, PolarBear \cite{PolarBear} and CLASS \cite{Eimer12}, can probe some detailed properties of the tensor mode. \\
For example, when the statistical features of the primordial perturbations are not isotropic, a large number of new observables arises, including the correlation functions with different multipole moment $l$, and TB and EB cross-correlations, which have been forbidden by the isotropic statistics of the primordial fluctuations \cite{Watanabe:2009ct, Chen:2013eaa, Chen:2014vja}.\\
Motivated by the future data, in this chapter, we study inflationary dynamics producing statistical
anisotropies in details. We investigate the anisotropic inflation scenario with a charged scalar inflaton coupled to the gauge field, presented in \ref{Chapter3}. We show that the gauge coupling induces large statistical anisotropies in tensor perturbations as compared to model of anisotropic inflation with no charge coupling \cite{Ohashi:2013qba}. We show that while statistical anisotropies in temperature power spectrum can be small as required by the Planck data \cite{Kim:2013gka}, the tensor mode can develop significant statistical anisotropies. Therefore it is important for the forthcoming experiments to look for the statistical anisotropies in the B-mode polarization even though the statistical anisotropies in temperature map is well-constrained. For this purpose, in this chapter, we calculate the primordial correlations of the curvature perturbation $\zeta$, and the two tensor modes $h_+$ and $h_\times$. The CMB temperature and polarization correlations TT, TE, EE, TB, EB and BB are then calculated from these primordial perturbations, for the same and different multipole moments. \\
A novel result in our analysis is that the anisotropies in the tensor sector also contribute to the TT correlation on the CMB. However, the transfer function from primordial tensor to CMB temperature decays towards large $l$. Thus the TT anisotropy coming from the primordial tensor has a decaying amplitude and is highly suppressed after $l \gtrsim 100$. As a result, we naturally obtain anisotropies at low multipoles of TT without modifying the high multipoles. This scale-dependent anisotropies will have a better fit to the CMB anomalies. \\
This chapter is based on \cite{Chen:2014eua}.
\section{Anisotropic Inflation}
\label{model}
As we already discussed, the goal of this chapter is to study the fingerprint of the primordial anisotropies in the gravitational waves. \\
The background behavior of this model has been already discussed in the previous chapters, So we skip presenting the model and go directly into the perturbation level of this model. 

\subsection{Perturbations}

Here we present cosmological perturbations in anisotropic inflation. 
The general form of the metric and matter perturbations has been studied in chapter \ref{Chapter4} where we verified that the leading contributions in the final action come from the matter sector. 
To simplify the situation further, we go to flat gauge where the curvature perturbations is given by the
inflaton perturbations $\zeta = - \frac{H}{\dot \phi} \delta \phi$. As a result, the metric has no scalar perturbations and we are left with the simple form of metric perturbations
\ba
\label{dynamical metric}
ds^2 = a(\eta)^2 \left(-d\eta^2 + \left[\delta_{ij}+ h_{ij} \right]dx^i dx^j \right ) \, .
\ea
Note that since we work in small anisotropy limit, we can set $a=b$ to leading order in Eq. (\ref{metric4}).
Here the perturbations $h_{ij} $ represents the tensor modes subject to the transverse and traceless conditions  $\partial_i h_{ij} = 0$ and $ h_{ii}=0$ where the repeated indices are summed over.  We denote the two independent polarizations of the metric by $h_\times$ and  $h_+$.\\
As for the perturbations in gauge field  sector  there is one non-dynamical degree of freedom,  $\delta A_{0}$, which must be integrated out from the action. However, similar to the case of   non-dynamical degrees of freedom from the metric perturbations, it turns out that the new terms from integrating out $\delta A_{0}$ are also sub-leading. As a result, the leading interaction terms in the total Lagrangian come from  the dynamical degrees of freedom $\delta A_i$.

In order to simplify the analysis further, we can use the remaining  two-dimensional rotational symmetry on the $y-z$ plane to set $k_{z} \equiv 0$ so $\overrightarrow{k} = (k_x , k_{y} , 0) =
k \, (\cos \theta, \sin \theta, 0 )$ in Fourier space. In addition, since the gauge field  has three polarizations in the unitary gauge,
two transverse and one longitudinal polarizations, we  can choose the following ansatz for the gauge field perturbations, i.e. $\delta A_i$,
\ba
\delta A_\mu^{(S)} = (\delta A_0, \delta A_1, \partial_y M, 0) \quad \quad , \quad \quad
\delta A_\mu^{(V)} =(0, 0, 0, D) \, .
\ea
where we have defined, $\delta A_1, \delta A_2$ and $\delta A_3$ to be $\delta A_x, \partial_y M$ and $D$ respectively. Here $A_\mu^{(V)}$ refers to one transverse polarizations in the vector sector  while $A_\mu^{(S)}$
represents the two polarizations in the scalar sector.
Furthermore, we can decompose the two polarizations in $A_\mu^{(S)}$ into  one transverse and one longitudinal polarizations
as follows \cite{Emami:2013bk}
\begin{align}
\label{D12}
D_{1}&\equiv \delta A_{1} -ik \cos{\theta}M \\
D_{2}&\equiv \cos{\theta} \delta A_{1} + ik \sin^2{\theta}M \, .
\end{align}
In this decomposition $D_{1}$ represents  the transverse polarization while $D_{2}$ refers to the longitudinal polarization of the gauge field.  However, as it has been demonstrated in \cite{Emami:2013bk}, the interactions containing the longitudinal mode are  exponentially suppressed during inflation and can be neglected from the analysis. Physically, this is understandable since the interactions containing the
longitudinal mode $D_2$ originate from the ``Higgs mechanism'' via the interaction $\e^2 A_\mu A^\mu \phi^2$ which are exponentially suppressed during much of the period of inflation as discussed before.

We can quantize the curvature perturbation and the gauge field perturbations as usual.
For the curvature perturbation, note that we work in the flat gauge so
\ba
\label{zeta-phi}
\zeta = - \frac{H}{\dot \phi} \delta \phi  = \frac{\delta \phi}{M_P \sqrt{2 \epsilon_H}}.
\ea
Expanding the quantum operator $\widehat \zeta$ in terms of the annihilation and the creation operator
$a(\bfk)$ and $a^{\dagger}(\bfk)$ we have
\ba
\widehat \zeta (\bfx , \eta) =  \int \frac{d^3k}{(2\pi)^{3/2}} e^{i \mathbf{k}.\mathbf{x}}
\widehat\zeta(\mathbf{k},\eta) \quad , \quad
\widehat\zeta(\mathbf{k},\eta) = \zeta(k, \eta) a(\bfk) +   \zeta^*(k, \eta)  a^{\dagger}(- \bfk)
\ea
where the creation and the annihilation operators satisfy the usual commutation relation
$ [ a(\bfk),    a^\dagger  (\bfk')] = \delta^{(3)} (\bfk-\bfk' )$.

The wave function  of the curvature perturbation has the standard form of  the excitations of
a  massless scalar field on a dS background
\ba
\label{zeta-wave}
\zeta_k(\eta) = \frac{i H \eta }{M_{P}\sqrt{2\epsilon_{H}k} }  \left( 1 - \frac{i}{k\eta} \right) e^{-ik\eta} \, .
\ea
The power spectrum of the curvature perturbations is given by
\ba
\langle   \widehat{\zeta}(\mathbf{k_1})  \widehat{\zeta} (\mathbf{k_2}) \rangle
= (2 \pi)^3 \delta^{(3)} (\bfk_1 + \bfk_2) P_\zeta (k_1) \quad , \quad
\calP_\zeta \equiv   \frac{ k_1^3  }{ 2 \pi^2}  P_\zeta(k_1)
\ea
In particular, the power spectrum for the free isotropic theory is
\ba
\label{calP0}
\calP_\zeta^{(0)} = \frac{H^2}{8 \pi^2 \epsilon_H M_P^2}\, .
\ea

Similarly, the quantum excitations of the gauge field perturbations $\widehat D_{1\bf k}(\eta)$ and $\widehat D_\bfk(\eta)$ can be expanded in terms of their annihilation and creation operators with the wave functions
\ba
\label{D-wave}
\sin{\theta} D_{1 k}(\eta)  = D_k(\eta) =
\frac{i}{f \sqrt{2k}} \left( 1 - \frac{i}{k\eta} \right) e^{-ik\eta} \, .
\ea

Now we present our decomposition of the tensor perturbations  $h_{ij}$ into $h_\times$ and $h_+$ polarizations following the method of  \cite{Ohashi:2013qba}.  Decomposing
$h_{ij}$ into $e_{ij}^{(s)}(\bfk)$ in Fourier space, the traceless  and transverse conditions, $h_{ii} = h_{ij,j} =0$,
yields
\ba
e_{i i}^{(s)}(\bfk) = 0 \quad , \quad  k_j e_{i j}^{(s)}(\bfk) = 0  \, ,
\ea
with $s= \times, +$ representing the two polarizations. In addition, we choose the following normalization
\ba
e^{(s)}_{ij}(\mathbf{k}) e^{*(s')}_{ij}(\mathbf{k}) = \delta_{ss'} \, ,
\ea
where $*$ represents the complex-conjugation. Note that we also have $e^{(s)}_{ij}(\mathbf{k}) = e^{*(s)}_{ij}(\mathbf{-k})$.

The  quantum operators $\widehat{h}_{ij}(\mathbf{k},\eta) $ in Fourier space
are represented in terms  of the annihilation and creation operators by
\ba
\label{tensor}
\widehat{h}_{ij}(\mathbf{k},\eta) = \sum _{s=+,\times}  \widehat{h}_{s}(\mathbf{k},\eta)
e_{ij}^{(s)}(\bfk)
\quad , \quad \widehat{h}_{s}(\mathbf{k},\eta)=
h_{s}(k, \eta)a_{s}(\mathbf{k})+ h^{*}_{s}(k, \eta)a^{\dag}_{s}(-\mathbf{k}) \, ,
\ea
with  the commutation relations  $ [ a_{s}(\bfk),    a_{s}^\dagger  (\bfk')] = \delta_{s s'} \delta^{(3)} (\bfk-\bfk' )$.

As we mentioned before, we chose the convention that
$\mathbf{k} = k(\cos{\theta}, \sin{\theta}, 0)$. With this choice, the polarizations
$e^{+}_{ij}(\mathbf{k})$ and $e^{\times}_{ij}(\mathbf{k})$ become
\begin{align}
\label{polarization2}
e^{+}_{ij}(\mathbf{k}) =\frac{1}{\sqrt{2}} \left( \begin{array}{ccc}
\sin^2{\theta} & -\sin{\theta}\cos{\theta} & 0 \\
-\sin{\theta}\cos{\theta} & \cos^2{\theta} & 0 \\
0 & 0 & -1 \\
\end{array} \right) ~,~
e^{\times}_{ij}(\mathbf{k}) = \frac{i}{\sqrt{2}}\left( \begin{array}{ccc}
0 & 0 & -\sin{\theta} \\
0 & 0 & \cos{\theta} \\
-\sin{\theta} & \cos{\theta} & 0 \\
\end{array} \right).
\end{align}
Using Eq. (\ref{tensor}) and Eq. (\ref{polarization2}), we find the following expression for the Fourier mode of the tensor field
\begin{align}
\label{polarization3}
\widehat{h}_{ij}(\mathbf{k}) =\frac{1}{\sqrt{2}} \left( \begin{array}{ccc}
\widehat{h}_{+}\sin^2{\theta} & -\widehat{h}_{+}\sin{\theta}\cos{\theta} & -i\widehat{h}_{\times}\sin{\theta} \\
-\widehat{h}_{+}\sin{\theta}\cos{\theta} & \widehat{h}_{+}\cos^2{\theta} & i\widehat{h}_{\times}\cos{\theta} \\
-i\widehat{h}_{\times}\sin{\theta} &  i\widehat{h}_{\times}\cos{\theta}& -\widehat{h}_{+} \\
\end{array} \right) \, .
\end{align}
We will use this expression later on when  calculating the cross-correlation between the tensor mode and the curvature as well as the gauge field.

The profile of the tensor excitations has the standard form
\ba
\label{hs-wave}
{h}_{s}(k,\eta) = \frac{2 i H\eta }{M_{P}\sqrt{2 k}}\left(1-\frac{i}{ k\eta} \right)e^{-ik\eta} ~~~,~~~ (s= +, \times) \, .
\ea
The power spectrum of the tensor perturbations is given by
\ba
\langle   \widehat{h}_{ij}(\mathbf{k_1})  \widehat{h}_{ij}(\mathbf{k_2}) \rangle
= (2 \pi)^3 \delta^{(3)} (\bfk_1 + \bfk_2) P_h(k_1) \quad , \quad
\calP_h \equiv   \frac{ k_1^3  }{ 2 \pi^2}  P_h(k_1)
\ea
In the absence of anisotropy the power spectrum has the standard form
\ba
\label{calPh0}
\calP_h^{(0)} = \frac{2 H^2}{\pi^2 M_P^2}  = 16 \epsilon_H  \calP_\zeta^{(0)} \, .
\ea
Therefore, defining the tensor-to scalar ratio $r \equiv \calP_h/\calP_\zeta  $ we have $r = 16 \epsilon_H$
for the isotropic theory.


\subsection{The Interaction Lagrangian}
\label{interactions}

Having presented the perturbations in some details, here we separate the Lagrangian into the free field part and interaction part. Here and below we call the latter the interaction Lagrangian.
The starting Lagrangian from the action (\ref{action4}) is
\ba
\label{interaction Lagrangian1}
L = -\frac{a^4}{4}f(\phi)^2 F_{\mu \nu} F^{\mu \nu} - \frac{a^4}{2}  \e^2 \phi^2 A_{\mu} A^{\mu} \, .
\ea
Expanding the above action around the background values,  neglecting the contributions of the non-dynamical field $\delta A_0$ which are sub-leading as discussed before, and using the relation
$\left(\frac{\partial f^2}{\partial \phi}\right)\delta \phi = 4 f^2 \zeta$,
the interaction Lagrangians in the Fourier space is calculated as
\ba
\label{Lzetah}
L_{\zeta h_+} &=&
- \frac{3\sqrt{2}}{2} I \epsilon_{H} M_{P}^2\sin^2{\theta}a^2\left(-\eta\right)^{-2}\left( \zeta^{*} {h}_{+} + c.c.  \right)  + \frac{\e^2 \sqrt{2}}{6}  I \epsilon_{H}M_{P}^4 \sin^2{\theta} \times \nonumber\\
&& \left(\frac{a ^4}{f^2}\right) \left( \zeta^{*} {h}_{+} + c.c. \right)  \\
\label{LzetaA}
L_{\zeta D_1} &=& -2 M_{P}\sqrt{3I \epsilon_{H}}\sin^2{\theta} \left(\frac{a f}{\eta}\right) \left( \zeta ^{*} D'_{1} + c.c.\right) - 2 \e^2 M_{P}^3 \sqrt{\frac{I \epsilon_{H}}{3}}\sin^2{\theta} \times \nonumber\\&& \left(\frac{a^3}{f}\right) \left( \zeta ^{*} D_{1} + c.c.\right) \\
\label{Lh+D1}
L_{h_+  D_1} &=& \frac{M_{P}}{2} \sqrt{\frac{3I\epsilon_{H}}{2}}\sin^2{\theta}\left( \frac{fa}{\eta}\right) \left( D^{'*}_{1} {h}_{+} +  c.c.\right) + \sqrt{\frac{I}{6\epsilon_{H}}} \e^2 M_{P}^3 \sin^2{\theta}\times \nonumber\\&&\left( \frac{a^3}{f}\right)\left( D^{*}_{1} {h}_{+} + c.c.\right) \\
\label{LhxzD}
L_{h_\times  D} &=&
\frac{M_{P}}{2} \sqrt{\frac{3I \epsilon_{H}}{2}}\sin{\theta} \left(\frac{fa}{\eta}\right)\left( i D^{'} {h}^{*}_{\times} + c.c.\right) + \sqrt{\frac{I}{6\epsilon_{H}}}\e^2 M_{P}^3 \times \nonumber\\&&\sin{\theta} \left(\frac{a^3}{f}\right)\left( i D{h}^{*}_{\times} + c.c. \right)
\ea
where c.c stands for complex conjugation.

The above interaction Lagrangians are needed in order to calculate the anisotropy corrections in
$\langle \zeta \zeta \rangle,  \langle h_s  h_{s'} \rangle$ and the cross-correlations $\langle \zeta h_s \rangle$.  Note that in the free (isotropic) theory with $I=\e =0$ there is no anisotropy corrections in power spectra and $\langle \zeta h_s \rangle$ as expected.

\section{Anisotropic Correlations  }
\label{correlations}

Having calculated the interaction Lagrangians as given in Eqs. (\ref{Lzetah})-(\ref{LhxzD})
now we are ready to calculate the anisotropic correlation functions by using the in-in formalism.
For this purpose, we need to obtain the interaction Hamiltonian from the interaction Lagrangian.
One should notice that $H_{int} = -L_{int}$ is not necessary true with kinetically coupled interactions.
So it is worth to check it in this model before proceeding with the in-in calculation of the correlation functions. We have calculated it in the Appendix \ref{int-hamilton}. It turns out that the above formula is true for the whole of the interactions except $H_{\zeta h_{+}}$. So in the following we use $H_{int} = -L_{int}$ everywhere except that in $H_{\zeta h_{+}}$, special care is taken of.

We are interested in anisotropic contributions in $\langle \zeta_\bfk \zeta_\bfk^* \rangle  $, $\langle h_{s \, \bfk} h_{s' \, \bfk}^* \rangle  $  and $ \langle \zeta_{ \bfk} h_{s' \, \bfk}^* \rangle$. We calculate each term in turn.
Note that the wave function of the free theory for $\zeta_k, D_k, D_{1 k}$ and $h_{s k}$ are given in Eqs. (\ref{zeta-wave}),  (\ref{D-wave}) and (\ref{hs-wave}).

\subsection{Anisotropies in curvature power spectrum}
\label{zeta-power}
Here we calculate the anisotropic contributions in curvature perturbation power spectrum $\langle \zeta_\bfk \zeta_\bfk^* \rangle $, taking into account the tensor contribution.
We denote the change in curvature perturbation power spectrum from the anisotropic sources by $\delta \langle \zeta_\bfk \zeta_\bfk^* \rangle  $.  This analysis has been performed in chapter \ref{Chapter4} in detail and for the comparison purposes, we just review them here. \\
Using the in-in formalism, the leading order corrections in curvature perturbation power spectrum are given by
\ba
\label{delta-P-zeta}
\delta \langle \zeta_\bfk \zeta_\bfk^* \rangle  = - \int_{\eta_{0}}^{\eta_{e}} d\eta_{1} \int_{\eta_{0}}^{\eta_{1}}d\eta_{2} \left \langle
\bigg{[} L_{I}(\eta_{2}) , \bigg{[} L_{I}(\eta_{1}) ,
\zeta_\bfk (\eta_e) \zeta_\bfk^*(\eta_e)  \bigg{]}\bigg{]} \right \rangle  \, ,
\ea
where $L_I$ represents the interaction Lagrangian.  The lower limit of the integral should be set
$k \eta_0 \rightarrow -\infty$  corresponding to initial  modes being deep inside the horizon. However, as studied in \cite{Bartolo:2012sd, Emami:2013bk}, the interactions responsible for anisotropies operate on super-horizon scales so to a good approximation one can safely take $k \eta_0 =-1$ corresponding to the time when the mode leaves the horizon.
The upper limit of the above integral as usual corresponds to $k \eta_e \simeq 0$.

The interaction Lagrangians relevant to
$\delta \langle \zeta_\bfk \zeta_\bfk^* \rangle$ are $L_{\zeta h_+}$ and  $L_{\zeta D_1}$  given
in Eqs. (\ref{Lzetah}) and (\ref{LzetaA}).  A look at these two equations show that $L_{\zeta h_+}$
is suppressed compared to $L_{\zeta D_1}$ by the factor $\sqrt{I \eH} \ll 1$. Therefore, the leading order anisotropic corrections in  curvature perturbation power spectrum comes from  $L_{\zeta D_1}$.
In addition, $L_{\zeta D_1}$ has two independent terms denoted by $L^{(1)}_{\zeta D_1}$ and
$L^{(2)}_{\zeta D_1}$:
\ba
L^{(1)}_{\zeta D_1} &\equiv&
-2 M_{P}\sqrt{3I \epsilon_{H}}\sin^2{\theta} \left(\frac{a f}{\eta}\right) \left( \zeta ^{*} D'_{1} + c.c.\right) \\
L^{(2)}_{\zeta D_1} &\equiv&
- 2 \e^2 M_{P}^3 \sqrt{\frac{I \epsilon_{H}}{3}}\sin^2{\theta} \left(\frac{a^3}{f}\right) \left( \zeta ^{*} D_{1} + c.c.\right)
\ea
Depending on whether one chooses either $L^{(1)}_{\zeta D_1}$ or $L^{(2)}_{\zeta D_1}$ in place of  $L_I(\eta_1)$ and $L_I(\eta_2)$ in the   integral in Eq. (\ref{delta-P-zeta}), there are four possible contributions in  $\delta \langle \zeta_\bfk \zeta_\bfk^* \rangle$ denoted by $\delta \langle \zeta_\bfk \zeta_\bfk^* \rangle_{ij}$ where $i, j=1, 2$ with the  assumption  that
$L_I(\eta_1) = L^{(i)}_{\zeta D_1}$  and  $L_I(\eta_2) = L^{(j)}_{\zeta D_1}$.
For example,    $\delta \langle \zeta_\bfk \zeta_\bfk^* \rangle_{12}$ means
$L_I(\eta_1) = L^{(1)}_{\zeta D_1}$  and  $L_I(\eta_2) = L^{(2)}_{\zeta D_1}$. With this identification we have
\ba
\label{correction zeta}
\delta  \bigg{\langle} \zeta_{\bfk}(\eta_{e}) \zeta_{\bfk}^*(\eta_{e})\bigg{\rangle} &=&
\delta \bigg{\langle} \zeta_{\bfk}(\eta_{e}) \zeta_{\bfk}^*(\eta_{e})\bigg{\rangle} _{11}
+\delta \bigg{\langle} \zeta_{\bfk}(\eta_{e}) \zeta_{\bfk}^*(\eta_{e})\bigg{\rangle} _{12}
\nonumber\\
&+& \delta \bigg{\langle} \zeta_{\bfk}(\eta_{e}) \zeta_{\bfk}^*(\eta_{e})\bigg{\rangle} _{21} + \delta \bigg{\langle} \zeta_{\bfk}(\eta_{e}) \zeta_{\bfk}^*(\eta_{e})\bigg{\rangle} _{22}
\ea
The details of the in-in analysis are presented in Appendix \ref{in-in}. As a sample analysis, here we present the integral form of
$\delta  \bigg{\langle} \zeta_{\mathbf{k}_{1}}(\eta_{e}) \zeta_{\mathbf{k}_{1}}(\eta_{e})^*\bigg{\rangle}_{11} $
which is  (here and hence after, the momentum conservation $\delta$-function is omitted to save writing)
\ba
\delta  \bigg{\langle} \zeta_{\mathbf{k}_{1}}(\eta_{e}) \zeta_{\mathbf{k}_{1}}^*(\eta_{e})\bigg{\rangle}_{11}  =&&  384 I \eH M_P^2  \sin^4{\theta} \
\int_{\eta_{0}}^{\eta_{e}} d\eta_{1} \left(\frac{a f}{\eta} \right)_{\eta_1}
\im \left[ \zeta_k (\eta_{1})\zeta_k^{*}(\eta_{e})\right] \times \nonumber\\
&& \int_{\eta_{0}}^{\eta_{1}}d\eta_{2}
\left(\frac{a f}{\eta} \right)_{\eta_2}
\im \left[ \zeta_k(\eta_{2})\zeta_k^{*}(\eta_{e})  D_{1k}^{' *} (\eta_{1}) D_{1k}^{'}(\eta_{2})
\right] \, .
\ea
Expanding the integrand for small $k \eta$ arguments and assuming $k \eta_0 = -1$ and
$k \eta_e =0$ as explained above, the above integral yields
\ba
\delta  \bigg{\langle} \zeta_{\mathbf{k}_{1}}(\eta_{e}) \zeta_{\mathbf{k}_{1}}(\eta_{e})\bigg{\rangle}_{11}&=& \frac{6 I N^2}{k_{1}^3 \epsilon_{H}} \left(\frac{H}{M_{P}} \right)^2 \sin^2{\theta}
\ea
in which $N= - \ln (-k \eta_e)$ represents the number of e-folds when the mode
$k$ has left the horizon. Taking $k$ to be the CMB scales we need $N \sim 60$ in order to solve
the flatness and the horizon problem.

Performing the same procedure for other integrals, we obtain  \cite{Emami:2013bk}
\ba
\delta  \bigg{\langle} \zeta_{\mathbf{k}}(\eta_{e}) \zeta_{\mathbf{k}}^*(\eta_{e})\bigg{\rangle}_{12}&=& -\frac{31 }{490}  \frac{\e^2 I}{k^3 {\epsilon_{H}}} \sin^2{\theta} \\
\delta  \bigg{\langle} \zeta_{\mathbf{k}}(\eta_{e}) \zeta_{\mathbf{k}}^*(\eta_{e})\bigg{\rangle}_{21}&=& -\frac{ I \e^2 N}{7 k^3 \epsilon_H}   \sin^2{\theta} \\
\delta  \bigg{\langle} \zeta_{\mathbf{k}}(\eta_{e}) \zeta_{\mathbf{k}}^*(\eta_{e})\bigg{\rangle}_{22}&=& \frac{9}{2156}  \frac{\e^4 I}{k_{1}^3}\left(\frac{M_{P}}{m} \right)^2 \sin^2{\theta}
\ea
So combining these four contributions, and assuming $N \gg 1$,  we have
\ba
\label{correction zeta final}
\delta  \bigg{\langle} \zeta_{\mathbf{k}}(\eta_{e}) \zeta_{\mathbf{k}}^*(\eta_{e})\bigg{\rangle}
&\simeq&
\left( \frac{6 I N^2}{ \epsilon_{H}} \frac{H^2}{M_{P}^2}
-\frac{ I \e^2 N}{7  \epsilon_H}
+\frac{9 \, \e^4 I}{2156} \frac{M_{P}^2}{m^2}  \right) \left(\frac{\sin^2{\theta}}{k^3}\right)\\
&=&   \frac{6 I N^2}{ \epsilon_{H}} \frac{H^2}{M_{P}^2}    \frac{\sin^2{\theta}}{k^3} F (\beta)
\ea
where we have defined
\ba
\label{beta-def}
\beta \equiv \frac{\e^2 }{42 N}  \left(\frac{M_P}{H} \right)^2  \quad , \quad
F(\beta) \equiv 1- \beta + \frac{9}{22} \beta^2 \, .
\ea
We have also written $m$ in terms of $H$ and $\epsilon_H$ by using $m^2 = \left(3 \epsilon_H  H^2 \right)$.\\
Note that $\beta$ is a measure of the gauge field coupling $\e^2$. In particular, in the model of \cite{Watanabe:2010fh, Bartolo:2012sd, Ohashi:2013qba} with $\e=0$ we have  $F(\beta)=1$.
With $M_P/H \sim 10^{5}$, and with $\e \gtrsim 10^{-4}$ we obtain $\beta \gtrsim 1$. For larger value of $\e$ we see that  $F(\beta)$ grows  like $\beta^2$.

The anisotropic power spectrum $\delta \calP_\zeta= \frac{k^3}{2 \pi^2} \delta {\langle} \zeta_{\mathbf{k}}(\eta_{e}) \zeta_{\mathbf{k}}^*(\eta_{e}) {\rangle}$ therefore is
\ba
\label{delta-calP-zeta}
\delta \calP_\zeta= \frac{3 I N^2 H^2 }{\eH \pi^2 M_P^2} F(\beta) \sin^2 \theta \, .
\ea
Correspondingly, the total anisotropic power spectrum  $\calP_\zeta$ is
\ba
\calP_\zeta =  \calP_\zeta^{(0)} \left[ 1+ 24 I N^2  F(\beta)  \sin^2 \theta
\right] \, ,
\ea
where $\calP_\zeta^{(0)}$ represents the isotropic power spectrum for the free theory. Note that in the limit  when $\e=\beta=0$ so $F(\beta)=1$, our result for  $\delta \calP_\zeta$ agrees with the result
in  \cite{Watanabe:2010fh, Bartolo:2012sd, Ohashi:2013qba}.

Now defining the anisotropy estimator $g_*$ via
\ba
\label{g*}
\calP_\zeta =  \calP_\zeta^{(0)} \left[ 1+  g_* \left( \widehat\bfk \cdot \widehat\bfp \right)^2 \right]
\ea
where $\bfp$ is the preferred anisotropic direction in the sky (the $x$-direction in our example), we
obtain
\ba
\label{g*2}
g_* = -24 I N^2 F(\beta) \,
\ea
It is important to note that the form of $g_*$ we have defined here is with respect to the primordial curvature power spectrum. In the TT and other correlation functions, the anisotropy not only comes from the $g_*$ here, but also comes from an ``effective $g_*$'' contribution from the tensor sector. Such an ``effective $g_*$'' has a scale dependence on the TT and other correlations because the tensor mode is decaying after it returns to the horizon. We will return to this issue later.
Also we see that the gauge coupling $\e$ appears  in $g_*$ via the parameter $\beta$.
Taking $| g_*| \lesssim 10^{-2}$ from the Planck data  constraint  \cite{Kim:2013gka} we require
$I F(\beta) \lesssim 10^{-6}$.

\begin{figure}[!t]
	\centering
	\includegraphics[width=0.6\textwidth]{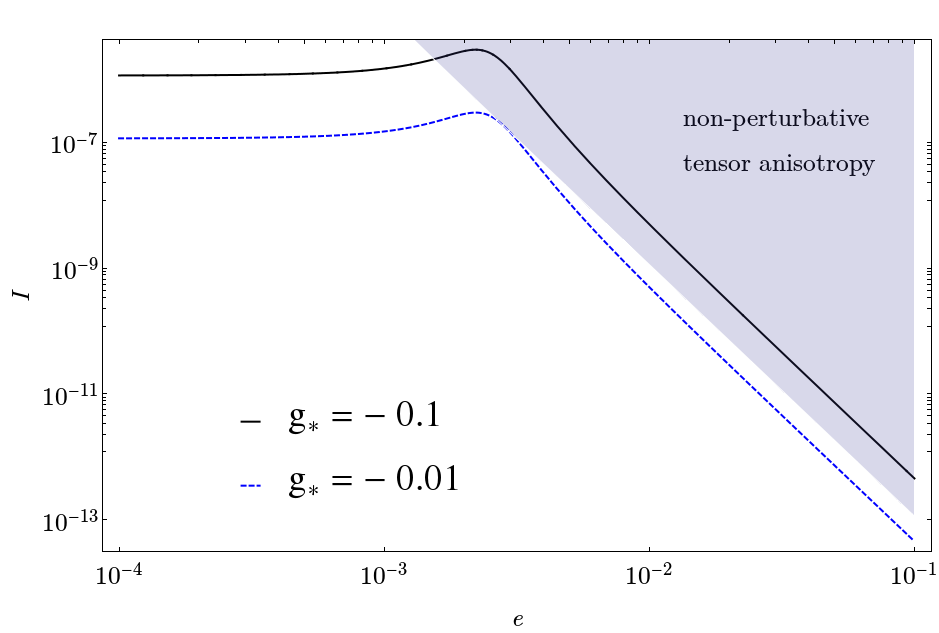}
	\caption{The  allowed range of $I$ and $\e$ for different values of $g_*$.  The shaded region
		is beyond our current scope because a large value of $\e$ induces too much anisotropy in tensor power spectrum
		which undermines our perturbative approach, i.e. $\delta \calP_h $ in Eq. (\ref{delta-Ph})
		becomes comparable to the   isotropic power spectrum $\calP_h^{(0)}$. In addition, note that we require $\e < 0.1$ in order to get large enough number of e-folds. As a result the range  $\e < 0.1$ is consistent both at the background and perturbation levels.}
	\label{Ie-plot}
\end{figure}

One may ask what the theoretical limits on the value of $\e$ or the parameter $\beta$ are. First, we have to make sure that we get enough number of e-folds of inflation at the background level. As we mentioned before, the number of e-folds depends logarithmically on $\e$
so as studied in \cite{Emami:2010rm} one can take say $\e <0. 1$ to get  a long enough
period of inflation. In addition, our assumption in parametrizing the anisotropy was that
the anisotropic  power spectrum  is smaller than the isotropic power spectrum, i.e.
$|g_*| < 1$ so our perturbative approach using the leading order in-in formalism is valid.
Therefore, demanding $|g_*| < 1$ we need  $I F(\beta) < 10^{-4}$. We have presented the contour plot of the allowed range of $I$ and $\e$ in Fig. \ref{Ie-plot}. As can be seen, we need $\e \lesssim 10^{-3}$ in order not to produce too much anisotropy in tensor perturbations (to be discussed in next subsection). Therefore, with $\e \lesssim 10^{-3}$, and $M_P/H \sim 10^5$ we have $\beta \lesssim 3$.

Checking the behavior of the function $F(\beta)$ for the approximate allowed range
$\beta \lesssim 3$ indicates that  $g_*$ has  a weak dependence on $\e$.
One can check that  for $0\le  \beta \le 3$, $F(\beta)$
takes the value in the range $0.4 \lesssim    F(\beta) \lesssim 1.7 $.  As we shall see this conclusion plays important roles for the predictions of our model for various cross-correlations.  While the anisotropy in curvature perturbation is under control for the above range  of $\beta$, the tensor perturbations become highly anisotropic when $\beta \gtrsim 1$.

Here we pause to mention one conceptual problem associated with anisotropic inflation. As seen from
Eq. (\ref{g*2}) the amplitude of anisotropy in scalar power spectrum scales like $N^2$. If inflation is prolonged in the past, this yields a large value of $g_*$ and the system becomes highly anisotropic. Therefore, our treatment of taking the anisotropies as small corrections to the isotropic
FRW background will be invalid. As pointed out in  \cite{Bartolo:2012sd} this corresponds to IR gauge field fluctuations which have left the horizon in the past inflationary history and contributed to the classical background trajectory. Therefore, a prolonged period of inflation will bring more contributions from these anisotropic IR modes which can destroy the near isotropy of the background.
In order to prevent this to happen, we demand that the total period of anisotropic inflation is under control, say less than few hundreds of e-folds.

\subsection{Correction to the tensor power spectrum}

Now we calculate the anisotropy in tensor power spectra
$\langle {h}_{\bfk \times } {h}_{\bfk \times }^*\rangle$ and $\langle {h}_{\bfk + } {h}_{\bfk + }^*\rangle$.

Let us  start with  $\langle {h}_{\times } {h}_{\times }^*\rangle$.  The relevant  interaction Lagrangian is
$L_{D {h}_{\times } } =  L_{D {h}_{\times } }^{(1)} +  L_{D {h}_{\times } }^{(2)} $ where
$ L_{D {h}_{\times } }^{(1)}$ and  $ L_{D {h}_{\times } }^{(2)}$ respectively are the first and the second terms in Eq. (\ref{LhxzD}).   Following the same convention as  in our analysis for
anisotropies in curvature perturbation power spectrum we have
\ba
\delta \bigg{\langle} {h}_{\times } {h}_{\times }^*\bigg{\rangle} &=& - \int_{\eta_{0}}^{\eta_{e}} d\eta_{1}\int_{\eta_{0}}^{\eta_{1}} d\eta_{2}\bigg{[}L_{D {h}_{\times }  } ,\bigg{[}L_{D {h}_{\times}}, {h}_{\times\mathbf{k}} {h}_{\times\mathbf{k}}\bigg{]}\bigg{]} \nonumber\\
&=& \delta \bigg{\langle} \widehat{h}_{\times }\widehat{h}_{\times }\bigg{\rangle}_{11} +  \delta \bigg{\langle} \widehat{h}_{\times }\widehat{h}_{\times }\bigg{\rangle}_{12} + \delta \bigg{\langle} \widehat{h}_{\times }\widehat{h}_{\times }\bigg{\rangle}_{21} + \delta \bigg{\langle} \widehat{h}_{\times }\widehat{h}_{\times }\bigg{\rangle}_{22}
\ea
The details of the in-in analysis are presented in Appendix \ref{in-in}. The results for each contribution are
\ba
\label{tensor times 1}
\delta \bigg{\langle} {h}_{\times } {h}_{\times }^*\bigg{\rangle}_{11}
&=& \left(\frac{12}{k^3}\right)\left(\frac{H}{M_{P}}\right)^2 I \epsilon_{H} N^2\sin^2{\theta}
\\
\label{tensor times 2}
\delta \bigg{\langle} {h}_{\times}  {h}_{\times}^*\bigg{\rangle}_{12}&=&  -\left(\frac{4}{7 k^3}\right)N I\e^2\sin^2{\theta}\\
\label{tensor times 3}
\delta \bigg{\langle} {h}_{\times}  {h}_{\times}^*\bigg{\rangle}_{21}&=&  -\left(\frac{62}{245 k^3}\right) I \e^2 \sin^2{\theta}
\\
\label{tensor times 4}
\delta \bigg{\langle} {h}_{\times}  {h}_{\times}^*\bigg{\rangle}_{22}
&=& \left(\frac{6 }{539 k^3 }\right) \left(\frac{M_{P}}{H}\right)^2 \left(\frac{I \e^4}{\epsilon_{H}}\right) \sin^2{\theta}
\ea

Summing up those four terms we get
\ba
\label{correction hh times final}
\delta \bigg{\langle} \widehat{h}_{\times\mathbf{k}_{1}}  \widehat{h}_{\times\mathbf{k}_{2}}\bigg{\rangle}
&\simeq& \left( 12I \epsilon_{H}N^2 \frac{H^2}{M_{P}^2} -\frac{4}{7}N I\e^2 + \frac{6 I \e^4}{539 \eH} \frac{M_{P}^2 }{H^2 } \right) \frac{\sin^2{\theta}}{k^3} \nonumber\\
&=&  12I \epsilon_{H}N^2 \frac{H^2}{M_{P}^2} F(\hat \beta)  \frac{\sin^2{\theta}}{k^3},
\ea
where the function $F(x)$ is defined in Eq. (\ref{beta-def}) and $\hat \beta$ is related to
$\beta$ via
\ba
\hat \beta \equiv \frac{2 \beta}{\epsilon_H} \, .
\ea
Note the crucial point that $\hat \beta$ is enhanced compared to $\beta$ by the factor $1/\epsilon_H$.
As we have discussed before, from the observational and the theoretical constraints on $\beta$ we have
$\beta \lesssim 1$. Now, from the above relation between $\hat \beta$ and $\beta$, we see that
$\hat \beta$ can be as large as $100$ for $\epsilon_H \sim 0.01$. As we shall see shortly, the anisotropy in tensor power spectra becomes very strong for large $\e$ so there will be  upper bound on $\e$ and
$\hat \beta$.

Now we calculate   $\langle {h}_{+ }{h}_{+}^*\rangle$. In this case the relevant interaction Lagrangians are  $L_{D_{1} {h}_{+ } }$ and $ L_{\zeta {h}_{+ }  }$ so we have
\ba
\label{tensor plus}
\delta \bigg{\langle} {h}_{+} {h}_{+}\bigg{\rangle} &=& - \int_{\eta_{0}}^{\eta_{e}} d\eta_{1}\int_{\eta_{0}}^{\eta_{1}} d\eta_{2}\bigg{[}L_{D_{1} {h}_{+ }  } ,\bigg{[}L_{D_{1} {h}_{+}}, {h}_{+} {h}_{+}\bigg{]}\bigg{]} - \nonumber\\
&&\int_{\eta_{0}}^{\eta_{e}} d\eta_{1}\int_{\eta_{0}}^{\eta_{1}} d\eta_{2}\bigg{[}L_{\zeta {h}_{+ }  } ,\bigg{[}L_{\zeta {h}_{+}}, {h}_{+} {h}_{+}\bigg{]}\bigg{]}
\ea
Comparing $L_{D_{1} {h}_{+ } }$ and $ L_{\zeta {h}_{+ }  }$ we see that $ L_{\zeta {h}_{+ }  }$ is suppressed compared to $L_{D_{1} {h}_{+ } }$ by a factor $\sqrt{I} \ll 1$
so to leading order in anisotropy we can neglect the contribution from  $ L_{\zeta {h}_{+ }  }$ in
$\langle {h}_{+ }{h}_{+}^*\rangle$. Now the analysis is exactly the same as what we performed
in  $\langle {h}_{\times }{h}_{\times}^*\rangle$ and therefore
\ba
\label{tensor plus final}
\delta \bigg{\langle} {h}_{+} {h}_{+}\bigg{\rangle} =
\delta \bigg{\langle} {h}_{\times} {h}_{\times}\bigg{\rangle} \, .
\ea

To summarize,  the anisotropy in total tensor power spectrum is
\ba
\label{delta-Ph}
\delta \calP_h &=& 2 \left( \frac{k^3}{2 \pi^2}   \right)
\delta \bigg{\langle} {h}_{\times} {h}_{\times}^*\bigg{\rangle} \nonumber\\
&=&   24 I \epsilon_{H}N^2 \frac{H^2}{M_{P}^2} F(\hat \beta)  \sin^2{\theta} ~,
\ea
so the total tensor power spectrum is
\ba
\calP_h = \calP_h^{(0)} \left( 1 +  6   I \epsilon_{H}N^2 F(\hat \beta) \sin^2 \theta
\right)
\ea
This is an interesting formula indicating that the effects of the gauge coupling is very strong
in tensor power spectrum anisotropy. This is because $\hat \beta = 2 \beta/\epsilon_H$ so with
$\beta \sim 1$ we gain  $\hat \beta \sim 100$  and therefore
$\delta \calP_h /\calP_h^{(0)} \simeq 24 I N^2 (\beta^2/\epsilon_H) = |g_*| \beta^2/\epsilon_H$.
With $| g_*| \sim O(\epsilon_H)$, which is consistent with the observational constraints and with
$\beta \sim 1$, one easily gets to the regime in which  $\delta \calP_h /\calP_h^{(0)} \simeq1 $.
Note that the epsilon enhancement for the charged interaction is a very special feature of this model, from the specific form of the potential. Explicitly, since our interaction includes $\e^2 \phi^2 A^2$, the charged contribution in $\langle \zeta \zeta\rangle$ comes from $(\e^2 \phi A \delta\phi \delta A )^2$, while the charged contribution in $\langle hh\rangle$ comes from $(\e^2 \phi^2 A h_{ij} \delta A)^2$. So we see that the ratio between these two effects are controlled by $\phi^2$.
In the chaotic inflation, $\phi^2$ is proportional to $1/\epsilon_H$. \footnote{On the other hand, for the symmetry breaking potential, the ratio would be $\epsilon_H$, so no hope to see any enhancement for this case.}
This signals the strong dependence of the tensor anisotropies to the gauge coupling. Of course, we can not trust our analysis when we approach the limit $\delta \calP_h /\calP_h^{(0)} \simeq1 $. This is because we have followed a perturbative approach and only kept the leading interaction terms in
our in-in analysis. Our situation  is in contrast to models of anisotropic inflation with a real inflaton field, as studied in  \cite{Ohashi:2013qba}, where $\e= \beta=0$,  so $F(\beta) =1$ and
$\delta \calP_h /\calP_h^{(0)} = - g_* \eH /4$ which is highly suppressed.

Demanding that $\delta \calP_h < \calP_h^{(0)}$ so our theoretical analysis is under perturbative control, we obtain the following upper bound
on the parameter $\beta$
\ba
\label{beta-cons}
\beta \lesssim  \sqrt{\frac{\eH}{ |g_* | }} \, .
\ea
With $\epsilon_H \sim 10^{-2}$  and $| g_*| \lesssim 10^{-2}$ we conclude that $\beta \lesssim 1$
in order for the anisotropic contribution in tensor power spectrum to be under control, corresponding to  $\e \lesssim 10^{-3}$.  The contour plot of  $I $ versus $\e$ is shown in Fig. \ref{Ie-plot}. The strong constraints on the allowed range of $\e$ comes from the tensor power spectrum.

\subsection{Cross-Correlation between $\zeta$ and $h_{ij}$}

Here we calculate the cross-correlation $\zeta$ and $h_{ij}$. We should calculate the following terms,
\ba
\bigg{\langle} \zeta_{\mathbf{k}_{1}}(\eta_{e}) {h}_{+\mathbf{k}_{2}}(\eta_{e}) \bigg{\rangle} &=& i \int_{\eta_{0}}^{\eta_{e}} d\eta_{1} \bigg{\langle}   \bigg{[} H_{\zeta {h}_{+}} , \zeta_{\mathbf{k}_{1}}{h}_{+\mathbf{k}_{2}}\bigg{]}   \bigg{\rangle}
- \int_{\eta_{0}}^{\eta_{e}} d\eta_{1}\int_{\eta_{0}}^{\eta_{1}} d\eta_{2}
\bigg{\langle} \bigg{[}L_{\zeta D_{1}} ,\bigg{[}L_{D_{1} {h}_{+}}, \nonumber\\&& \zeta_{\mathbf{k}_{1}}{h}_{+\mathbf{k}_{2}}\bigg{]}\bigg{]}  \bigg{\rangle}  -\int_{\eta_{0}}^{\eta_{e}} d\eta_{1}\int_{\eta_{0}}^{\eta_{1}} d\eta_{2}
\bigg{\langle} \bigg{[} L_{D_{1} {h}_{+}},\bigg{[} L_{\zeta D_{1}}, \zeta_{\mathbf{k}_{1}}{h}_{+\mathbf{k}_{2}}\bigg{]}\bigg{]}  \bigg{\rangle}
\nonumber\\
&\equiv& \bigg{\langle} \zeta_{\mathbf{k}_{1}}(\eta_{e}) {h}_{+\mathbf{k}_{2}}(\eta_{e})\bigg{\rangle}_{1} + \bigg{\langle} \zeta_{\mathbf{k}_{1}}(\eta_{e}) {h}_{+\mathbf{k}_{2}}(\eta_{e})\bigg{\rangle}_{2} + \bigg{\langle} \zeta_{\mathbf{k}_{1}}(\eta_{e}) {h}_{+\mathbf{k}_{2}}(\eta_{e})\bigg{\rangle}_{3} \nonumber\\
\ea
where the indices 1, 2 and 3 indicate the above three integrals respectively. The nested integrals 2 and 3 each have four different contributions as in previous analysis so in total we have nine contributions in the above cross correlation.  The details of the analysis are given in Appendix \ref{in-in} and here we present the final result:
\ba
\label{final cross correlation}
\bigg{\langle} \zeta_{\mathbf{k}_{1}}(\eta_{e}) {h}_{+\mathbf{k}_{2}}(\eta_{e})\bigg{\rangle} &\simeq&
I \left(- 6\sqrt{2} \frac{ N^2 H^2 }{M_{P}^2}  + \frac{  \sqrt{2} \e^2 N}{7 \epsilon_{H} }  - \frac{3\sqrt{2} \e^4 }{1078 \epsilon_{H} } \frac{M_{P}^2}{H^2}  \right) \frac{\sin^2{\theta}}{k^3}
\ea
Eq. (\ref{final cross correlation}) is the final result for the cross-correlation.

Finally one can easily check that
\ba
\bigg{\langle} \zeta_{\mathbf{k}_{1}}(\eta_{e}) {h}_{\times\mathbf{k}_{2}}(\eta_{e})\bigg{\rangle}=0 \, .
\ea
This is because at the second order level $\zeta $ does not see ${h}_{\times}$.

The power spectrum of $\langle \zeta h \rangle$ cross-correlation is therefore
\ba
\label{calP-zh}
\calP_{\zeta h} &=& \frac{k^3}{2 \pi^2 } \bigg{\langle} \zeta_{\mathbf{k}}(\eta_{e}) {h}_{+\mathbf{k}^*}(\eta_{e})\bigg{\rangle}  \nonumber\\
&=& -24 \sqrt{2}  I N^2 \eH \calP_\zeta^{(0)} G(\beta) \sin^2 \theta
\ea
where the function $G(\beta)$ is defined via
\ba
\label{G-beta}
G(\beta) \equiv 1- \frac{\beta }{\epsilon_H} + \frac{9}{11} \frac{\beta^2 }{\epsilon_H} \, .
\ea
For typical value of $\epsilon_H \ll 1$, the function  $G(\beta)$ has two positive roots $\beta_1$ and $\beta_2$ where  $\beta_1  \ll 1$ and $\beta _2 \gtrsim  1$. Thus the function $G(\beta)$ is negative in the range $\beta_1 < \beta < \beta_2$ while it is positive beyond this region. Therefore, with appropriate
choice of $\e$ or $\beta$ the cross-correlation $\calP_{\zeta h} $ can have both signs, i.e. $\zeta$ and $h$
can be either correlated or anti-correlated.

Alternatively, one can also write $\calP_{\zeta h} $ in terms of $g_*$ as (note that $g_* <0$)
\ba
\calP_{\zeta h} =  \sqrt{2} g_*  \eH \calP_\zeta^{(0)} \frac{G(\beta)}{F(\beta)} \, .
\ea
Note that in the limit where $\e= \beta=0$ so $F(\beta) = G(\beta) =1$, the above expression coincides with the result obtained in  \cite{Ohashi:2013qba}. With $\beta \sim1$ the ratio $\calP_{\zeta h}/\calP_{\zeta}^{(0)} $ in our model is about one or two orders of magnitude bigger than the result in
\cite{Ohashi:2013qba} in which $\beta =0$.

Before closing this Section and presenting our numerical results for various correlations, let us summarize the main results of our model.
The anisotropy in curvature perturbation power spectrum is given by Eq. (\ref{delta-calP-zeta})
with $g_*$ given in Eq. (\ref{g*}). As discussed below  Eq. (\ref{g*}), $\delta \calP_\zeta$ and $g_*$ depend weakly on $\beta$ so we do not get strong constraints on the value of $\e$ from the constraints on curvature perturbations anisotropies. On the other hand, the anisotropic tensor power spectrum is given in Eq. (\ref{delta-Ph}). The crucial point is that $\delta \calP_h$ scales
with $F (\hat \beta)$ where $\hat \beta = 2 \beta/\epsilon_H$. With $\beta \sim 1$  we get
$\hat \beta \sim 100$ which yields  an enhancement $\sim 10^4$ from the function  $F (\hat \beta)$.
This is a novel effect indicating that while the scalar perturbations are well-constrained to be statistically symmetric, the tensor perturbations show strong directional dependence. This is the motivation for careful scrutiny of B-mode polarizations for the TB, EB and BB correlations in the upcoming Planck polarization maps.
Finally the cross-correlation of scalar-tensor,  $\calP_{\zeta h}$,  is given in Eq. (\ref{calP-zh}). The situation here is a hybrid of the above two limits of $\delta \calP_h$ and $\delta \calP_\zeta$.
For large enough value of $\e$, i.e. with $  \beta \sim 1$, we get an enhancement of order 10-100 from
the function $G(\beta)$.



\section{From the primordial fluctuations to the CMB}
\label{numerical}
In this section, we shall relate the calculation of the above sections to CMB anisotropies, using the same method as in \cite{Watanabe:2010bu}. We use the spin weighted spherical harmonics in our analysis, \cite{Kamionkowski:1996zd, Hu:1997hp}.

To calculate the CMB anisotropies, one has to project the three-momentum onto two dimensional spherical harmonics. The $\int d^3k$ integral breaks up into two pieces with the presence of statistical anisotropy -- the radius part and the angular part. The correlators of CMB observables takes the form
\begin{align}\label{eq:aXaX}
\langle a^{X_1}_{l_1, m_1} a^{X_2}_{l_2, m_2}\rangle = 4 \pi
\int \frac{dk}{k} \Delta_{l_1}^{i_1 X_1}(k) \Delta_{l_2}^{i_2 X_2}(k)
\int  d\Omega ~  [{}_{i_1}Y^*_{l_1m_1}(\theta, \phi)] [{}_{i_2}Y_{l_2m_2}(\theta, \phi)] P^{i_1, i_2}(k, \theta, \phi)~,
\end{align}
where $X^i$ takes value (T, E, B), which are the temperature anisotropy, the E-mode and B-mode respectively. Here although the dimensionless power spectrum in principle depends on $k$, we approximate it as scale invariant. The effects on the plots are tinny\footnote{On the other hand, the scale dependence is needed once data analysis is to be performed for this model and the modification is straightforward if the scale dependence can be written in a factorizable form $P^{i_1, i_2}(k, \theta, \phi) = \sum_n f_n(k)g_n(\theta,\phi)$.}. Also, in our case there is no $\phi$-dependence for the power spectrum and the $\theta$-dependence takes the shape $P^{i_1, i_2} = P^{i_1, i_2}(\sin^2 \theta)$. In the absence of $\phi$-dependence, the momentum along the $\phi$-rotation is conserved and thus $\langle a^{X_1}_{l_1, m_1} a^{X_2}_{l_1, m_1}\rangle$ is non-vanishing only when $m_1 = m_2$.

However, the rotational symmetry on the $\theta$ direction is broken. As a result, the correlation $\langle a^{X_1}_{l_1, m_1} a^{X_2}_{l_1, m_1}\rangle$ is not restricted to $l_1=l_2$. Instead, other than those diagonal correlations, we also have $l_1=l_2\pm1$ for TB and EB, and $l_1=l_2\pm 2$ for TT, TE, EE and BB respectively.

In \eqref{eq:aXaX}, the $i_1$ and $i_2$ are indices indicating the spin of the component. And the ${}_{i}Y^*_{lm}(\theta, \phi)$ is the spin-$i$-weighted spherical harmonics. On the $P^{i_1, i_2}(\sin^2 \theta)$ side, the correlation functions on the spin bases takes the form
\begin{align}
P^{0,0} =& P^{\zeta\zeta}~, \quad P^{0,\pm2} = P^{\pm2,0} = \frac{1}{\sqrt2} P^{0,+}~, \quad
P^{\pm2, \pm2} = \frac{1}{2} \left( P^{++} + P^{\times\times}\right)~, \nonumber\\
 P^{\pm2, \mp2} =& \frac{1}{2} \left( P^{++} - P^{\times\times}\right)~.
\end{align}
Note that  $P^{\zeta\zeta} \equiv P_{\zeta}$, $P^{0,+} \equiv P_{\zeta h_+}$, $P^{++}\equiv P_{h_+}$ and $P^{\times\times} \equiv P_{h_\times}$ components are given by  equations \eqref{correction zeta final}, \eqref{final cross correlation}, \eqref{correction hh times final} and \eqref{tensor plus final} respectively.

Among those angular integrals $\int  d\Omega ~  [{}_{i_1}Y^*_{l_1m_1}(\theta, \phi)] [{}_{i_2}Y^*_{l_2m_2}(\theta, \phi)] P^{i_1, i_2}(\sin^2\theta)$, we would like to emphasize a particularly interesting integral:
\begin{align}
\int  d\Omega ~  [{}_{0}Y^*_{lm}(\theta, \phi)] [{}_{2}Y_{lm}(\theta, \phi)] \sin^2\theta ~.
\end{align}
Note that this term picks up cross correlation of $\langle\zeta h_+\rangle$, and maps it onto TT or other CMB anisotropies. \\
The transfer function part encodes very complicated late time physics. Fortunately this part does not have angular dependence thus one can use the standard Boltzmann code for the calculation. Here ``the Cosmic Linear Anisotropy Solving System'' (CLASS) \cite{Blas:2011rf} is used for solving those transfer functions, where the cosmological parameters are chosen to be the same as those used by the BICEP2 group.

Now for the purpose of studying statistical anisotropies, we are not to sum over $m_1$ and $m_2$ (unlike the case above) because the summation would average away some signals of the anisotropy. Rather, we leave $m_1=m_2=m$ free and look into $m=0$ and $m=l$ cases respectively for illustration.

Three sets of parameters are examined numerically:
\begin{itemize}
	\item Case I: Real inflaton field with no electric gauge coupling:  In this class of models we plot $I=10^{-7}$ and $\e=0$. This case is the same as previous studies on the chargeless scalar field \cite{Watanabe:2010bu}. On the plots, Case I is shown in blue color.
	\item Case II: Balanced: In this class of models we plot $I=10^{-7}$ and $\e=10^{-3}$. Such choice of parameters does not significantly modify $g_*$ of the scalar sector. However, the gravitational sector is largely modified because the two point correlation functions with tensors are more sensitive to the charge of the complex inflaton, as we have seen in the previous sections. On the plots, Case II is shown in black color.
	\item Case III: Charge coupling dominated: In this class of models we plot $I=10^{-11}$ and $\e=0.025$. In this case  $g_*$, which is a measure of the scalar power spectrum anisotropy,
	is considerably smaller than Cases I and II since $g_*$
	is mostly sensitive to $I$ than $\e$. However, the charge contribution is large in the tensor sector and is marginally under control to calculate the statistical anisotropy perturbatively (where the anisotropic term is about 34\% of the isotropic term). On the plots, Case III is shown in red color.
\end{itemize}

By choosing the above parameters, we have taken into consideration that the anisotropies in the temperature correlations cannot be large, making use of Planck data in range $2 \leq l \leq 2000$ \cite{Kim:2013gka}. On the other hand, if scale dependence of the anisotropy is allowed, the constraint may not be so strong. In that case, we would have larger effects from anisotropies and our plots are shifted upwards.

The figures are plotted in logarithm scales because sometimes different cases differ by order-of-magnitude. However, we note that in various cases the correlations could go negative. To represent as detailed information as possible, we shall use solid lines to denote the logarithm of a quantity, where the quantity is positive (for example, $\log C_{l, l+2}^{TT}$) and dashed lines for multiplying $-1$ before taking logarithm (for example, $\log (- C_{l, l+2}^{TT})$), where the quantity is negative.

\subsection{The TT and BB modes}

\begin{figure}[!h]
	\centering
	\includegraphics[width=0.45\textwidth]{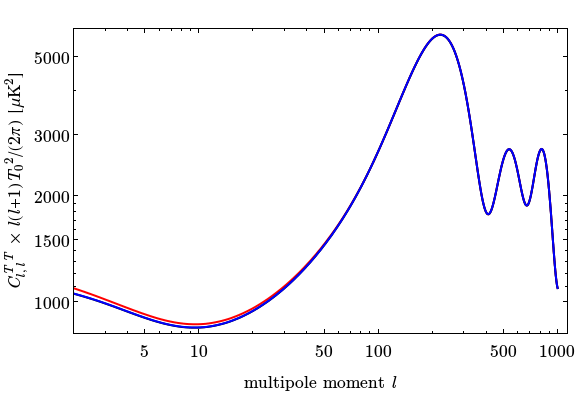}
	\hspace{0.05\textwidth}
	\includegraphics[width=0.45\textwidth]{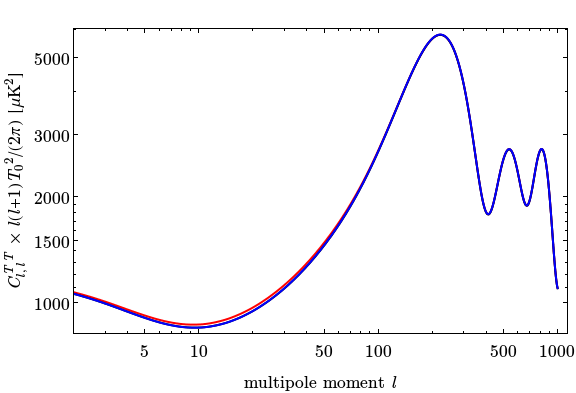}
	\caption{\label{fig:TT0} The TT correlation at $l_2=l_1$. Here and hence after, the blue curve denotes Case I, with $I=10^{-7}$ and $\e=0$; the black curve denotes Case II, with $I=10^{-7}$ and $\e=10^{-3}$. The red curve denotes Case III, with $I=10^{-11}$ and $\e=0.025$. The left panel is for $m=0$ and the right panel is for $m=l_1$. It is important to note that the anisotropic contribution in Case III is considerably larger than the anisotropic contribution in case I and II at low $l$. But the contribution decays at high $l$. Here the black and blue curves are not visibly distinguishable from each other, which also coincides with the isotropic power spectrum, because of the tight constraint on the $g_*$ of the scalar sector.}
\end{figure}

\begin{figure}[!h]
	\centering
	\includegraphics[width=0.45\textwidth]{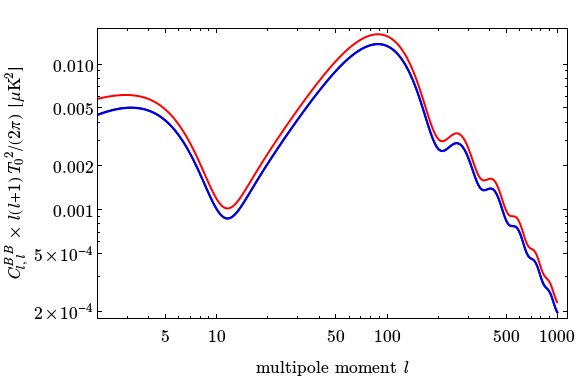}
	\hspace{0.05\textwidth}
	\includegraphics[width=0.45\textwidth]{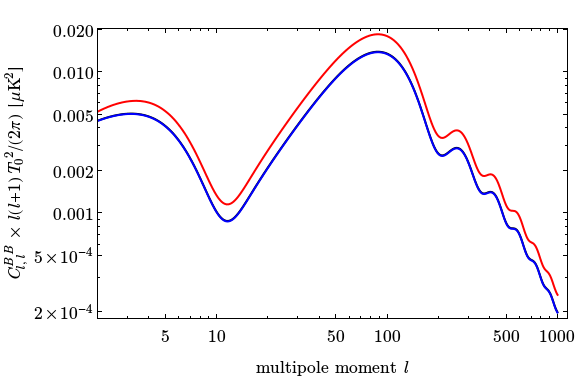}
	\caption{\label{fig:BB0} The $m=0$ (left) and $m=l_1$ (right) plots for BB correlation with $l_2=l_1$. }
\end{figure}

\begin{figure}[!h]
	\centering
	\includegraphics[width=0.45\textwidth]{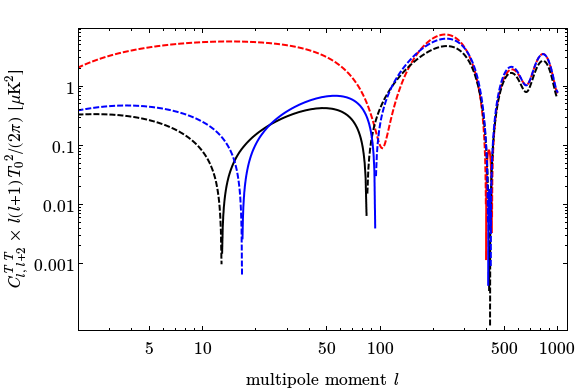}
	\hspace{0.05\textwidth}
	\includegraphics[width=0.45\textwidth]{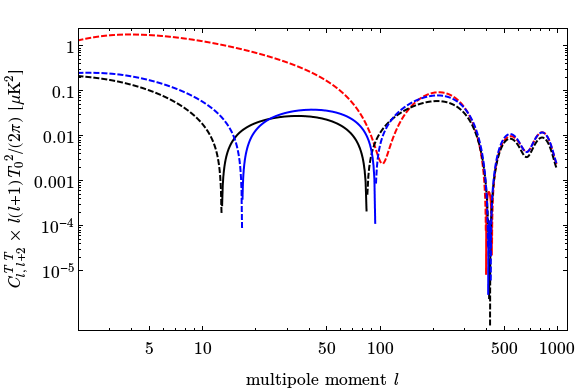}
	\caption{\label{fig:TT2} The $m=0$ (left) and $m=l_1$ (right) plots for TT correlation with $l_2=l_1+2$. Here and hence after, the dashed lines denote the plotted quantity (here $C_{l, l+2}^{TT}$) is negative along this line segment, and thus we plot $-C_{l, l+2}^{TT}$ on the logarithm scales.}
\end{figure}

\begin{figure}[!h]
	\centering
	\includegraphics[width=0.45\textwidth]{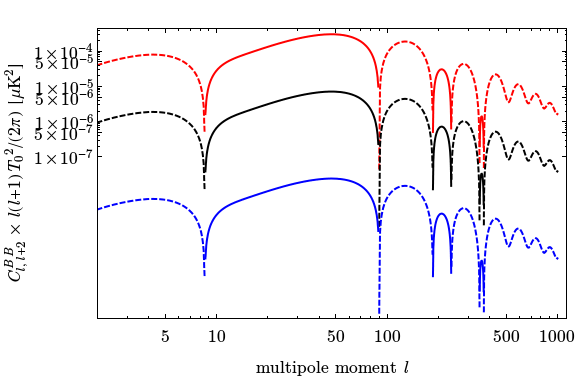}
	\hspace{0.05\textwidth}
	\includegraphics[width=0.45\textwidth]{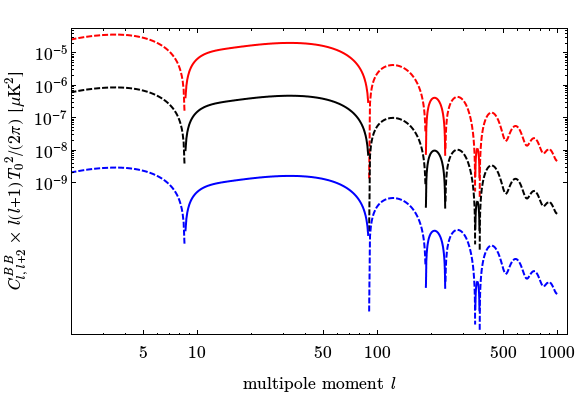}
	\caption{\label{fig:BB2} The $m=0$ (left) and $m=l_1$ (right) plots for BB correlation with $l_2=l_1+2$.}
\end{figure}

With the above choice of parameters and conventions, here we plot the TT and BB power spectra in Figs.~\ref{fig:TT0}, \ref{fig:BB0}, \ref{fig:TT2} and \ref{fig:BB2}. In Figs.~\ref{fig:TT0} and \ref{fig:BB0} the $l_1=l_2$ part of the correlation function is plotted and in Figs.~\ref{fig:TT2} and \ref{fig:BB2} with $l_2 = l_1 +2$. The left and right panels corresponds to $m=0$ and $m=l_1$ respectively. From Figs.~\ref{fig:TT0}, \ref{fig:BB0}, one observes that the anisotropic modification to the standard TT and BB power spectra are slight but still visible.

\subsection{TB and EB correlations}

\begin{figure}[!h]
	\centering
	\includegraphics[width=0.45\textwidth]{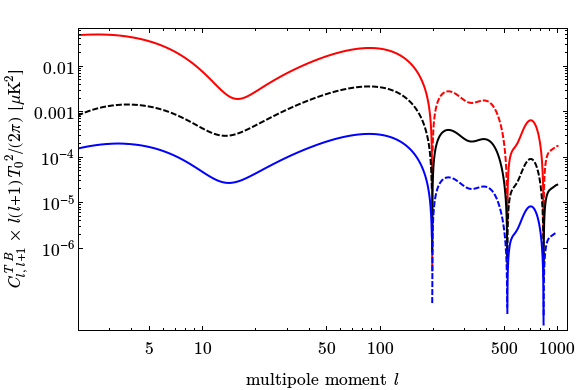}
	\hspace{0.05\textwidth}
	\includegraphics[width=0.45\textwidth]{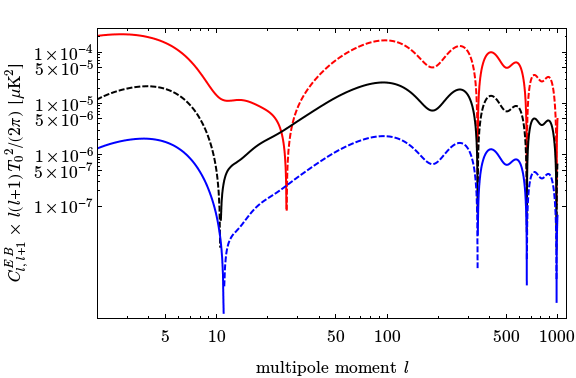}
	\caption{\label{fig:TBEB1} The $m=l_1$ plots for TB (left) and EB (right) correlation with $l_2=l_1+1$.}
\end{figure}

In the anisotropic case, the TB and EB correlations are opened up, with $l_2 = l_1 \pm 1$. In Fig.~\ref{fig:TBEB1}, those cross correlations are plotted.

\subsection{TE and EE correlations}

\begin{figure}[!h]
	\centering
	\includegraphics[width=0.45\textwidth]{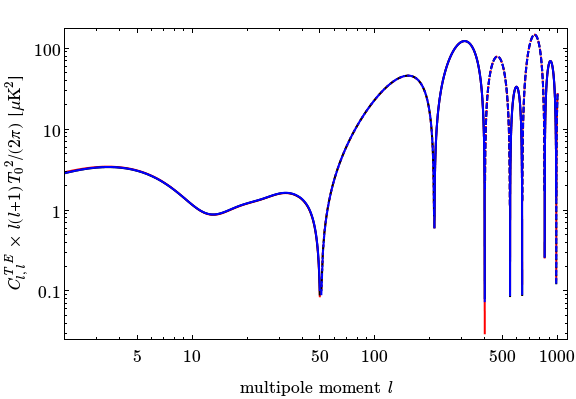}
	\hspace{0.05\textwidth}
	\includegraphics[width=0.45\textwidth]{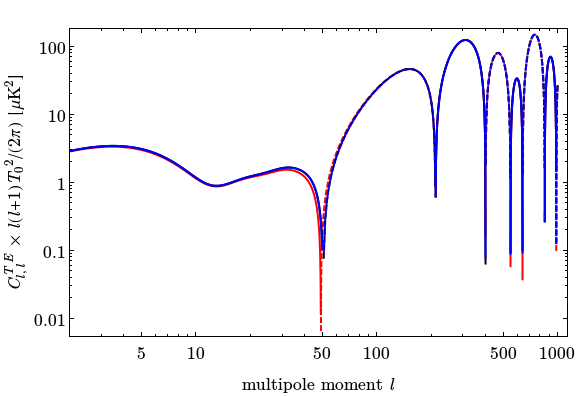}
	\caption{\label{fig:TE0} The $m=0$ (left) and $m=l_1$ (right) plots for TE correlation with $l_2=l_1$.}
\end{figure}

\begin{figure}[!h]
	\centering
	\includegraphics[width=0.45\textwidth]{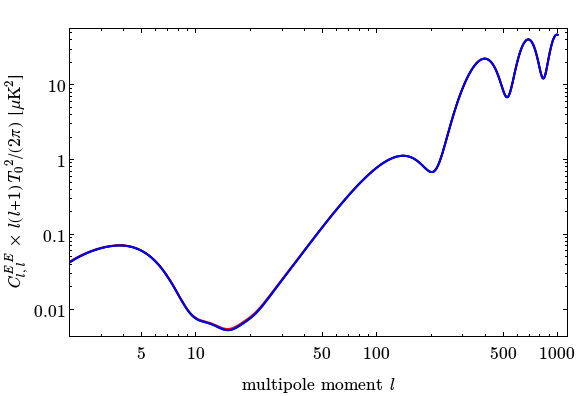}
	\hspace{0.05\textwidth}
	\includegraphics[width=0.45\textwidth]{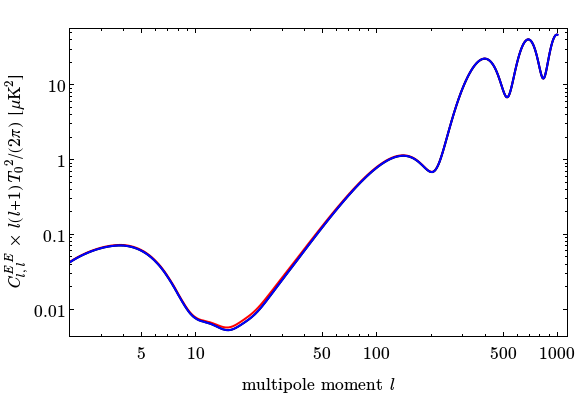}
	\caption{\label{fig:EE0} The $m=0$ (left) and $m=l_1$ (right) plots for EE correlation with $l_2=l_1$.}
\end{figure}

\begin{figure}[!h]
	\centering
	\includegraphics[width=0.45\textwidth]{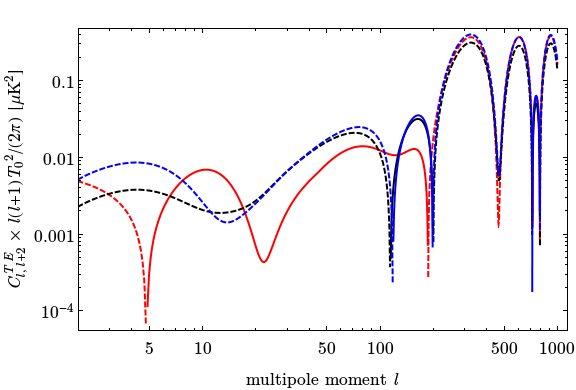}
	\hspace{0.05\textwidth}
	\includegraphics[width=0.45\textwidth]{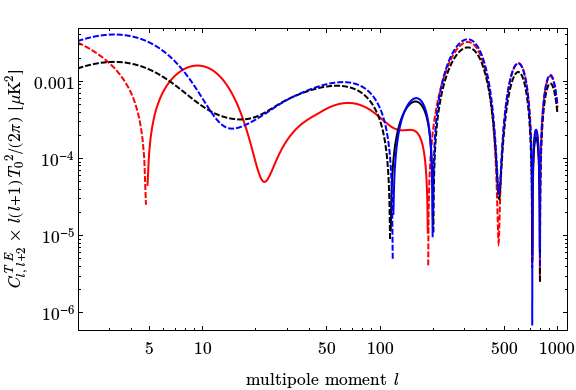}
	\caption{\label{fig:TE2} The $m=0$ (left) and $m=l_1$ (right) plots for TE correlation with $l_2=l_1+2$.}
\end{figure}

\begin{figure}[!h]
	\centering
	\includegraphics[width=0.45\textwidth]{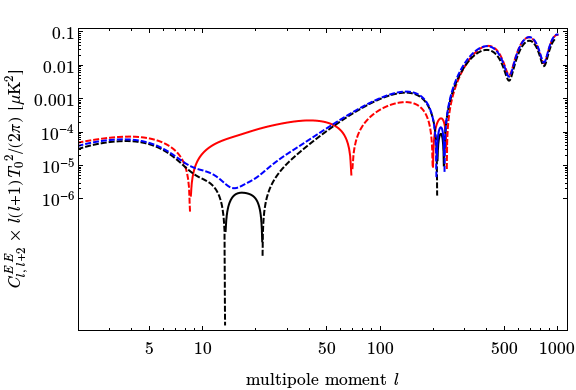}
	\hspace{0.05\textwidth}
	\includegraphics[width=0.45\textwidth]{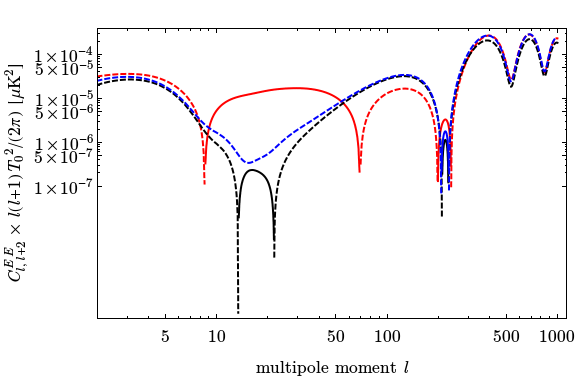}
	\caption{\label{fig:EE2} The $m=0$ (left) and $m=l_1$ (right) plots for EE correlation with $l_2=l_1+2$.}
\end{figure}

In Figs. \ref{fig:TE0}, \ref{fig:EE0}, \ref{fig:TE2} and \ref{fig:EE2}, the TE and EE correlations for $l_2=l_1$ and $l_2=l_1+2$ are plotted respectively.

\section{Summary}

In this chapter, we have studied  statistical anisotropies in model of anisotropic inflation from a charged inflaton field. More specifically, we work in large field inflation model with the chaotic potential  $V(\varphi) = \frac{1}{2}m^2 |\varphi|^2$ with the conformal coupling $f $ given
in Eq. (\ref{f-scale5}).  It is worth to emphasize that our results, especially the enhancement of the tensor mode, is because we work in large field model. We have calculated the anisotropies generated in $\langle \zeta \zeta \rangle $ and $\langle h_{s}  h_{s} \rangle$ with the tensor polarizations $s=+, \times$. In addition, we also calculated the cross-correlation $\langle \zeta  h_{s} \rangle$ which is sourced in the  anisotropic inflationary background. Our general conclusion is that
the anisotropy in curvature perturbation power spectrum $\delta \calP_\zeta$ can be small so one can easily satisfy the observational bound  \cite{Kim:2013gka} $|g_*| \lesssim 10^{-2} $.  However, the effects of gauge coupling $\e$ appears strongly in tensor perturbations. In particular, we have shown that while $g_*$ is small enough to be within the observational bound, $\delta \calP_h$ induced from anisotropies  can be comparable to the background tensor power spectrum $\calP_h$. \\
The induced anisotropy of TT spectrum from the primordial tensor fluctuations can obtain an ``effective $g_*$''  as large as a few percents at low $l$. This effective $g_*$ contribution is decaying towards high $l$, characterized by the tensor to temperature transfer function. It remains interesting to explore CMB anisotropies with this new profile of TT anisotropy.\\
We would also like to mention that the mechanism we have used to generate small $l$ CMB anomalies from the tensor sector is quite general. It may also be applied to other kinds of CMB anomalies such as non-Gaussianities, hemispherical asymmetry, etc. \\

Based on the correlations $\langle \zeta \zeta \rangle $, $\langle h_{s}  h_{s} \rangle$
and $\langle \zeta h_s \rangle $ obtained in this model, we have calculated the observational signatures on the CMB temperature and polarization maps. Especially, we have shown that when the gauge interaction dominates over the interaction between the inflaton and the gauge field, the tensor sector can be more anisotropic than the scalar sector, resulting in enhanced correlations involving B-modes. For example, in Fig~\ref{fig:TBEB1} one can observe that the TB correlation in the model with the
charged coupling can be order-of-magnitude larger compared to model with no electric charge coupling,  which is also considerably larger than the size of the BB correlation near the reionization bump at $l\lesssim 10$.


\chapter{CMB statistical anisotropies of classical and quantum origins } 

\label{Chapter9} 

\lhead{Chapter 9. \emph{CMB statistical anisotropies of classical and quantum origins }} 

\vspace{0.5cm}
\hrule \vspace{0.3cm}

\begin{quote}
\textbf{Abstract:} In this chapter, we examine the impact of different anisotropic relics on inflation, in particular the predictions on the density perturbations. These relics can be the source of the large scale anomalies in the cosmic microwave background. There are two different types of background relics, one from the matter sector and the other purely from the metric. Although the angular-dependence of the statistical anisotropy in both cases are degenerate, the scale-dependence are observationally distinctive. In addition, we demonstrate that non-Bunch-Davies vacuum states can extend the statistical anisotropy to much shorter scales, and leave a scale-dependence that is insensitive to the different backgrounds but sensitive to the initial quantum state.
\end{quote}

\vspace{0.1cm}  \hrule
\vspace{0.5cm}
\section{Introduction}
In the last chapters, we mostly focus on the primordial anisotropies which arise from the vector realization of the anisotropic inflation where the scale dependence had a very specific form, logarithmic behavior, and the initial quantum state had been assumed to be in Bunch Davis state. Here in order to complete the previous study, we consider the case where both of the above conditions are relaxed. The importance of this case study is that it can be thought as a way to classify different relic models and their predictions. 
These predictions include the scale dependence and angular dependence of the anomalies, together with other possible predictions on such as spatial curvature and non-Gaussianities. When comparing with data, these predictions provide theoretical templates which may provide a unified explanation for several anomalies. Some related new anomalies
may be predicted and verified, substantially increasing the statistical significance.
In addition, systematic studies of different anomalies in model-building can tell us not only why they are present, but also which fundamental physics we are able to probe.

With these motivations in mind, we note that there are two classes of models with initial anisotropics relics. The source of the initial anisotropy can either be matter fields, or solely from the gravitational sector\footnote{There is also a large class of models where the anisotropy has an active source. For example, inflation supported by an attractor vector field \cite{Watanabe:2009ct, Emami:2010rm}, see also  \cite{Chen:2014eua, Ohashi:2013qba, Thorsrud:2013kya} and the references therein, or bifurcation of inflationary trajectory \cite{Li:2009sp, Afshordi:2010wn, Wang:2013zz}. Alternatively, the anisotropy may not be efficiently diluted when the inflationary dynamics is modified \cite{Endlich:2012pz, Bartolo:2013msa, Akhshik:2014zz}. To distinguish, we do not call them the {\em relics} models.}.
The main goal of this chapter is to compare the predictions of these two classes of models.
For the first type of models, an example of relic vector field has been studied analytically and numerically in Ref.~\cite{Chen:2013eaa}. This model gives a specific prediction on the form and scale-dependence of the statistical anisotropy of the CMB. The dependence of the prediction on the initial quantum fluctuation state is also studied. For the second type, a Bianchi-type inflationary background model has been studied in \cite{Gumrukcuoglu:2007bx}. The density perturbations in this study was done only numerically. To properly compare them with the first type of models and to make the prediction more relevant to the data analyses, we use the same perturbative method as in \cite{Chen:2013eaa,Chen:2013tna} to solve these models analytically. We examine the angular and scale-dependence of the statistical anisotropy in these two types of models. In addition, we study the effects of the initial quantum state on these predictions following \cite{Chen:2013eaa,Chen:2013tna}, and emphasize how the resulting distinctive scale-dependence can be used as a probe of non-Bunch-Davies (non-BD) vacuum. 

This chapter follows the material of \cite{Chen:2014vja}

\section{Background evolution}
\label{model}

We start with the minimal model of inflation based on a scalar field minimally coupled to gravity
\ba
\label{action} S= \int
d^4 x  \sqrt{-g} \left [ \frac{M_P^2}{2} R - \frac{1}{2} \partial_\mu \phi
\,  \partial^\mu \phi - V(\phi) \right] \, ,
\ea
in which $M_P$ is the reduced Planck mass.

Before inflation reaches its attractor isotropic FRW phase, the expansion rates along different spatial directions may be different. The difference can be modeled by the type I Bianchi Universe, with the metric
\ba
\label{bian0}
ds^2 = - dt^2 + a^2 d x^2 + b^2(d y^2 +d z^2) \, .
\ea
Note that to simplify the analysis, we have assumed that there is a remnant two-dimensional symmetry in
$y-z$ plane. Later on we consider the most general case in which all three directions are anisotropic.

Considering the following ansatz for the scale factors $a$ and $b$, $a \equiv e^{\alpha(t)}$ and $b \equiv e^{\alpha(t)+3\sigma(t)}$, the metric (\ref{bian0}) becomes
\ba
\label{bian01}
ds^2 &=& - dt^2 + e^{2\alpha(t)}\left(d x^2
+e^{6\sigma(t)}(d y^2 +d z^2) \right) \, .
\ea
With the metric in this form, the background field equations are
\ba
\label{back-rho-eq}
\ddot\phi+3\left(\dot \alpha + 2 \dot \sigma \right)\dot \phi+ V_\phi &=&0  \\
\label{Ein1-eq}
3 M_P^2 \left(\dot \alpha^2+ 4\dot \alpha \dot \sigma + 3\dot \sigma^2 \right) &=& \frac{1}{2}\dot
\phi^2+V(\phi) \\
\label{Ein2-eq}
M_P^2 \left( \ddot \alpha + 3\dot \alpha \left( \dot \alpha + 2 \dot \sigma \right) \right) &=& V(\phi) \\
\label{anisotropy-eq}
\ddot \sigma +3\dot \sigma \left( \dot \alpha + 2 \dot \sigma \right)&=& 0\, ,
\ea
in which a dot indicates derivative with respect to $t$.

One can integrate the above equations and to leading order in slow-roll expansion obtain (the details can be found in appendix \ref{slow-roll} )
\ba
\label{app.a}
a & \simeq & H_{0}^{-1} \left( -\eta \right)^{-1} \\
\label{app.b}
b & \simeq & H_{0}^{-1} \left( -\eta \right)^{-1}\left( 1 + \left(\frac{\dot{\sigma_{0}}}{H_{0}}\right) \left( \mathcal{H}_{0} \eta \right)^3 \right) ~,
\ea
in which the subscript $0$ represents the values of the corresponding quantities at the start of inflation
$\eta =\eta_0$, $H= \dot \alpha$ is the leading order Hubble expansion rate and $\mathcal{H}  \equiv a H$.

\section{Perturbations}

Here we study perturbations in this model. The perturbation in this model is solved numerically in \cite{Gumrukcuoglu:2007bx}. However, in order to compare the results with a different model presented in \cite{Chen:2013tna, Chen:2013eaa}, here we solve the model analytically as in \cite{Chen:2013tna, Chen:2013eaa}. In principle one should take into account the perturbations in both of the matter and metric sectors. This can be achieved by integrating out the non-dynamical degrees of freedom. We leave the details of this analysis to appendix \ref{metric pert}. However, due to the slow-roll approximation, it turns out that the additional terms from integrating out the metric degrees of freedom are sub-leading compared to the typical terms coming from the matter sector and in order to read off the leading corrections we can neglect them all together \cite{Emami:2013bk}.

Neglecting the metric perturbations,  the second order action for $\delta \phi$ is then well approximated by (see Appendix \ref{quad-action} for the total form of the quadratic action)
\ba
L_{\phi \phi} \simeq \frac{b^2}{2}\mid \delta \phi_{k} '\mid^2
- \left[ \frac{b^2}{2} k_x^2 + \frac{a^2}{2} (k_y^2 + k_z^2) \right]
\mid \delta \phi_{k}\mid^2
- \frac{a^2 b^2}{2} V_{,\phi\phi}\mid \delta \phi_{k}\mid^2 ~,
\ea
where the last term is also slow-roll suppressed and can be neglected.
Throughout the paper we use the
prime to indicate the derivative with respect to the conformal time defined in terms of the scale factor $a(t)$, $d \eta = dt/a(t)$.
The equation of motion for  $\delta \phi$ in Fourier space is
\ba
\label{KG1}
\delta \phi_{k}'' + 2 \frac{b'}{b}\delta \phi_{k}' + \left( k_{x}^2 + \frac{a^2}{b^2}\left(k_{y}^2 + k_{z}^2 \right)\right)\delta \phi_{k} =0 \, .
\ea
We can expand $\delta \phi $ in terms of the usual creation and annihilation operators as
\ba
\label{a-adag}
\delta \phi = \int \frac{d^3k}{(2\pi)^3}\bigg{[}u_{k}a_{\kk} + u_{k}^{*}a_{-\kk}^{\dag}\bigg{]}e^{i\mathbf{k.x} } \equiv  \int \frac{d^3k}{(2\pi)^3} \delta \phi_{\kk} e^{i \kk \cdot {\bf x} } \, .
\ea
By using Eqs. (\ref{app.a}-\ref{app.b}) and the above expansion, the perturbed scalar field equation (\ref{KG1}) is written as
\ba
\label{mode function}
u_{k}'' -\frac{2}{\eta}\left(1 - 3 \left(\frac{\dot{\sigma_{0}}}{H_{0}}\right)\left(\mathcal{H}_{0}\eta\right)^3  \right)u_{k}' + \left( k_{x}^2 + \left( 1 - 2 \left(\frac{\dot{\sigma_{0}}}{H_{0}}\right) \left(\mathcal{H}_{0}\eta\right)^3 \right)\left(k_{y}^2 + k_{z}^2 \right)\right)u_{k} =0 ~.
\ea

In this paper we are interested in small anisotropies so we can solve the above equation perturbatively. Since the effect of Bianchi anisotropy has been parameterized by $\frac{\sigma_{0}'}{\mathcal{H}_{0}}\left(\mathcal{H}_{0}\eta\right)^3$, we would expect that all modes, either near the horizon or well inside the horizon, are affected by the anisotropy of this order. However, as shown in \cite{Chen:2013tna}, in order to see this explicitly
a proper change of variables in (\ref{mode function}) is necessary. In the following, first we solve equation \eqref{mode function} using the original  variable $u_k$. As we will see, the expansion breaks down for modes deep inside the horizon. We improve our expansion scheme by changing to a new variable and present an expansion which is suitable for both near horizon and UV modes.

\subsection{Near Horizon expansion}
Now we would like to solve the equation of motion for perturbations. Following \cite{Chen:2013tna, Chen:2013eaa}, we can expand $u_{k}$ as
\ba
\label{uk}
u_{k} = \mathcal{C}_{+}u_{k(0)} + u_{k(1)} ~,
\ea
in which the zeroth order isotropic wave function is given by
\ba
u_{k(0)} = \frac{H_{0}}{\sqrt{2k^3}}\left( 1+ ik\eta \right)e^{-ik\eta} ~.
\ea
One can interpret $\mathcal{C}_{+}$ as the correction in wave function normalization and $u_{k(1)}$ as the corrections in the profile of wave function in the presence of anisotropy.

The next order $u_{k(1)}$ can be solved perturbatively from the following equation,
\ba
u_{k(1)}'' -\frac{2}{\eta}u_{k(1)}' + k^2 u_{k(1)} = - \frac{6}{\eta}\left(\frac{\dot{\sigma_{0}}}{H_{0}}\right)
\left(\mathcal{H}_{0}\eta\right)^3u_{k(0)}'+ 2\left(\frac{\dot{\sigma_{0}}}{H_{0}}\right) \left(\mathcal{H}_{0}\eta\right)^3 \left(k_{y}^2 + k_{z}^2 \right)u_{k(0)} ~.
\ea
Using the ansatz
\ba
u_{k(1)} = \frac{\dot{\sigma_{0}}}{\sqrt{2k^3}}\mathcal{H}_{0}^3 \sum_{n=3}^{5} \alpha_{n}\eta^{n} e^{-ik\eta}
\ea
we get
\ba
\alpha_{3} &=& - \frac{1}{4k^2}\left( 4k_{x}^2 + k_{y}^2 + k_{z}^2 \right) \\
\alpha_{4} &=& - \frac{i}{4k}\left( 4k_{x}^2 + k_{y}^2 + k_{z}^2 \right) \\
\alpha_{5} &=& - \frac{1}{4}\left( k_{y}^2 + k_{z}^2 \right)
\label{alpha5value}
\ea
We see that for $k\eta > \left(\frac{\dot{\sigma_{0}}}{H_{0}}\right)^{-1}\left( \mathcal{H}_{0} \eta\right)^{-3}$ the above expansion breaks down, as we discussed.
We will come back to this point soon.

We determine $\mathcal{C}_{+}$ by using the following normalization condition
\ba
[\delta \phi_{\mathbf{q}}, \delta \pi_{\mathbf{p}}] = i (2\pi)^3 \delta^3(\mathbf{p}+ \mathbf{q}) \, ,
\ea
where $\delta \pi_{\mathbf{p}}$ is the momentum conjugate associate with
$\delta \phi_{\mathbf{p}}$,  $\delta \pi_{\mathbf{p}} = b^2 \delta \phi'_{\mathbf{p}}$. The above condition leads to the following equation
\ba
\frac{1}{H_{0}^2\eta^2}\left( 1 + 2 \left(\frac{\dot{\sigma_{0}}}{H_{0}}\right)\left( \mathcal{H}_{0} \eta\right)^3 \right) \left( u_{q} u_{q}^{'*} - u_{q}^{*} u_{q}^{'} \right) = i \, .
\ea
Since $\eta \to 0 $, we just keep the leading constant term. It turns out that only $\alpha_{3}$ plays role while the other higher terms are exponentially suppressed.
We get
\ba
\label{c plus0}
|\mathcal{C}_{+}|^2 =  1 + \frac{3\mathcal{H}_{0}^2}{4k^3}\sigma_{0}' \left(1 + 3 \cos^2{\Theta} \right)  \, ,
\ea
in which the amplitude of momentum $k$ and the angle $\Theta $ are defined as
\ba
k^2 \equiv k_x^2 + k_y^2 + k_z^2 ~, ~~~
\cos\Theta \equiv k_x/k ~.
\label{k_definition}
\ea

\subsection{UV safe expansion}

One might have some doubts in the above expansion scheme because
it breaks down for short wavelength modes $k\eta > \left(\frac{\dot{\sigma_{0}}}{H_{0}}\right)^{-1}\left( \mathcal{H}_{0} \eta\right)^{-3}$ due to the last term (\ref{alpha5value}).
Physically we do not expect this to happen. This problem is especially important if we would like to study the effect of anisotropic relics on the short wavelength modes. So to demonstrate explicitly the validity of our method, we have to elaborate the expansion scheme. It turns out that this can be fixed by properly choosing the variable used in the perturbative method. The expansion will be perturbative for all modes if we choose to perturbatively expand the exponent in the variable $u_k$ \cite{Chen:2013tna}.
Defining
\ba
\label{psi u}
\psi_{\mathbf{k}}(\eta)\equiv \log{\left(u_{\mathbf{k}}(\eta)\right)} \, ,
\ea
we expand $\psi_{\mathbf{k}}(\eta)$ in orders of $\sigma_{0}'$
\ba
\psi_{\mathbf{k}}(\eta) = \psi_{\mathbf{k}(0)}(\eta)+ \psi_{\mathbf{k}(1)}(\eta)+ ... ~.
\ea
One can then solve this perturbatively (see Appendix \ref{psi solution}) and get
\begin{align}
\label{psi zeroth}
\psi_{\mathbf{k}(0)}(\eta) &= \log{\left(u_{\mathbf{k}(0)}(\eta)\right)}
\\
\psi_{\mathbf{k}(1)}(\eta) &= \frac{3i}{8k^3} \mathcal{H}_{0}^2\sigma_{0}' \left(1 + 3 \cos^2{\Theta}\right) +\frac{ \mathcal{H}_{0}^2\sigma_{0}' }{1+ik\eta}\sum_{n=3}^{5} \alpha_{n}\eta^{n}\nonumber\\
&\equiv \frac{\mathcal{H}_{0}^2 \sigma_{0}'}{1+ik\eta}\sum_{n=0}^{5} \beta_{n}\eta^{n} \, ,
\end{align}
where $\beta_{n}$ are given by
\ba
\beta_{0} &=& \frac{3i}{8k^3}\left(1 + 3 \cos^2{\Theta}\right) \\
\beta_{1} &=& -\frac{3}{8k^2}\left(1 + 3 \cos^2{\Theta}\right) \\
\beta_{2} &=& 0 \\
\beta_{m} &=& \alpha_{m} ~~,~~ (m=3,4,5) \, .
\ea
For UV modes, $\psi_{k(0)} \sim -ik\eta$ and
$\psi_{k(1)} \sim (\sigma'_0 \mathcal{H}_0^2 \eta^3) k\eta$.
So the anisotropic corrections remain small for all modes.

At late time, the conserved curvature perturbation approaches to the attractor single field expression, $\zeta \approx - H_{0} \left(\frac{\delta \phi}{\dot \phi_{0}}\right)$, in the gauge used here. Therefore we can use this time-delay formula to compute the power spectrum by evaluating the variables at their attractor values. The statistical anisotropy in the finite result shows up through the coefficient ${\cal C}_+$ we just computed. We thus have
\ba
\label{power-anis}
\left( \frac{k^3}{2 \pi^2} \right)  \big \langle \zeta^2   \big \rangle = P_{\zeta0}\left( 1 + \frac{3\mathcal{H}_{0}^2}{4k^3}\sigma_{0}' \left(1 + 3 \cos^2{\Theta} \right) \right) \, ,
\ea
where the isotropic power spectrum is defined via  $P_{\zeta0} \equiv \frac{H_{0}^4}{\left(2\pi\dot{\phi_{0}} \right)^2}$.

Now we can compare this result with that in the model of relic vector field \cite{Chen:2013tna,Chen:2013eaa}. In both models, the anisotropy is axial symmetric, so as expected they have the same angular-dependence. But due to the different sources, the anisotropies in these two types of models have different scale-dependence. In the relic vector case the anisotropy decays towards smaller scales as $\sim 1/k^4$. However, here in the Bianchi type cases in which  anisotropy is generated from anisotropic scale factors, it decays as $\sim 1/k^3$. These two different behavior are related to the different decay speeds of the background relics in the models.


\section{An non-BD example: Gaussian state}

For inflation with minimal number of e-folds, the initial state of quantum fluctuations also do not have to be in their attractor vacuum states. It is therefore a sensible question to consider the effects of the non-BD states, and see how the initial quantum states of the universe leave their imprints in the statistical anisotropy of the CMB \cite{Chen:2013tna, Emami:2014tpa}. Conversely any distinctive predictions can then be used as a probe of the initial quantum state of the Universe.
In the following, we use a specific example for the non-BD vacuum, namely the Gaussian state [\cite{Polarski:1995jg},\cite{Chen:2013tna}]\footnote{There are other proposals and methods to model \cite{Barrow:1997sy, Barrow:1998ih} and probe the initial non-BD states \cite{Brandenberger:2000wr,Easther:2001fi,Chen:2006nt,Holman:2007na,Meerburg:2009ys,Chen:2009bc,Agullo:2010ws,Ganc:2011dy,Chialva:2011hc,Berezhiani:2014kga}.}. As we will see, there are two types of scale-dependence. One has an oscillatory behavior while the other is not oscillatory.
To start, let us write down the quadratic Hamiltonian for the quantum fluctuations of the inflaton field, $\delta \phi$, in a canonical form
\ba
v_\mathbf{k} &=& b \delta \phi_\mathbf{k} \\
\pi_\mathbf{k} &=& v_\mathbf{k}' - \frac{b'}{b}v_\mathbf{k}
\ea
The Hamiltonian is
\ba
H_2 &=& \int \frac{d^3 \mathbf{k}}{(2\pi)^3} \frac{1}{2} \bigg{[} \left(\pi_\mathbf{k} \pi^{*}_\mathbf{k}\right) + \frac{\left(k_y^2 + k_z^2 + k_x^2 \left( 1 + \sigma_{0}' \mathcal{H}_{0}^2 \eta^3\right)^2\right)}{\left( 1 + \sigma_{0}' \mathcal{H}_{0}^2 \eta^3\right)^2} \left(v_\mathbf{k} v^{*}_\mathbf{k} \right)+ \frac{\left(-1 + 2\sigma_{0}'\mathcal{H}_{0}^2\eta^3\right)}{\eta \left( 1 + \sigma_{0}' \mathcal{H}_{0}^2 \eta^3 \right)} \nonumber\\
&&\times \left(\pi_\mathbf{k} v^{*}_\mathbf{k} + \pi^{*}_\mathbf{k} v_\mathbf{k}  \right) \bigg{]} \nonumber\\
\ea
Using the Schrodinger picture to quantize the fields as
\ba
\label{wavefunction}
v_\mathbf{k} &=& f_\mathbf{k}(\eta) a_\mathbf{k}(\eta_0) + f^{*}_\mathbf{k}(\eta) a^{\dag}_{-\mathbf{k}}(\eta_0), \nonumber\\
\pi_\mathbf{k} &=& -i \left(g_\mathbf{k}(\eta) a_\mathbf{k}(\eta_0) - g^{*}_\mathbf{k}(\eta) a^{\dag}_{-\mathbf{k}}(\eta_0)\right),
\ea
where the creation and annihilation operators satisfy the usual commutation relations and
\ba
f_\mathbf{k}(\eta) = u_{0\mathbf{k}}b(\eta)\left(C_{+0}e^{\psi_{1\mathbf{k}}} + C_{-0}e^{\psi^{*}_{1\mathbf{k}}} \right)
\ea
is proportional to mode function with $C_{+0}$ and $C_{-0}$ being initial  constants
while
\ba
g_{\mathbf{k}}(\eta) = i \left( f'_{\mathbf{k}} - \mathcal{H}_{b}f_{\mathbf{k}} \right) \, .
\ea
Now we can define the Gaussian state at $\eta_0$ as,
\ba
a_{\mathbf{k}}(\eta_0) \mid 0, \eta_0 \rangle =0 ~.
\ea
Through this condition, the initial quantum state acquires an anisotropic component due to the anisotropic background.
By using Eq. (\ref{wavefunction}), we get
\ba
\label{condition1}
(g_{\mathbf{k}} - i \widehat{k} v_{\mathbf{k}})\mid_{\eta_{0}} =0 \, ,
\ea
where we have defined,
\ba
\label{hatk}
\widehat{k}^2 \equiv \frac{\left(k_y^2 + k_z^2 + k_x^2 \left( 1 + \sigma_{0}' \mathcal{H}_{0}^2 \eta^3\right)^2\right)}{\left( 1 + \sigma_{0}' \mathcal{H}_{0}^2 \eta^3\right)^2}
\ea
On the other hand, we may also use the normalization condition for $C_{+0}$ and $C_{-0}$ as
\ba
\label{condition2}
\mid C_{+0}\mid ^2 - \mid C_{-0}\mid ^2 =1.
\ea
Now by using Eqs. (\ref{condition1}) and (\ref{condition2}), the power spectrum is proportional to
\ba
\mid C_{+0} + C_{-0} \mid ^2 e^{2\psi_{\mathbf{k}(1)}\mid_{\eta\rightarrow 0}} = 1 + \frac{1}{2k^2\eta_0^2} + \sigma_{0}' \mathcal{H}_{0}^2 \left( - \frac{\eta_0}{2k^2} - \frac{5\eta_0}{2k^2}\cos^2{\Theta} \right) + {\rm oscillation ~~ terms},
\label{non-BD_results}
\ea
where we have the following expression for the oscillation terms
\ba
{\rm oscillation~~ terms} &=& \left[ -\frac{1}{2k^2\eta_0^2} + \sigma_{0}' \mathcal{H}_{0}^2 \left( \frac{3}{4k^4\eta_0} + \frac{\eta_0}{2k^2} + \frac{\eta_{0}^3}{2} + \cos^2{\Theta}\left( \frac{9}{4k^4 \eta_{0}} + \frac{5\eta_{0}}{2k^2}-\frac{\eta_{0}^3}{2}\right) \right) \right]  \nonumber\\
&&\times \cos{2k\eta_{0}} 
+\left[ - \frac{1}{k\eta_0} + \sigma_{0}' \mathcal{H}_{0}^2 \left( \frac{-3 + 2k^4\eta_{0}^4}{8k^5 \eta_{0}^2} + \cos^2{\Theta} \left( \frac{ -9 + 22 k^4 \eta_{0}^4}{ 8 k^5 \eta_{0}^2} \right)
\right) \right]\nonumber\\
&&\times \sin{2k\eta_{0}} 
\ea
We are mostly interested in the non-oscillatory anisotropic terms because such terms are sensitive probes of the initial quantum states \cite{Chen:2013tna, Chen:2013eaa}.\footnote{If $\eta_0$ is at the beginning of inflation, the frequency of the oscillatory components is high and approaches the ultimate resolution of CMB.} Interestingly, while in the BD cases, the scale-dependence of the statistical anisotropies are different for the relic vector field model  ($\sim 1/k^4$) \cite{Chen:2013tna, Chen:2013eaa} and the Bianchi model ($\sim 1/k^3$) as we obtained in previous Section, the effect of the non-BD Gaussian state on both models are the same. Such a state generically extends the anisotropy to much smaller scales and the scale-dependence
for both cases are  $\sim 1/k^2$. This can also be readily understood. In the BD case, the background evolution plays the dominant role in the final results. However, in the non-BD case the initial quantum states play more important roles enhancing the anisotropy of the shorter wave-length modes.
We have given just one example of non-BD state here. It is plausible that the enhancement caused by other non-BD states can have different scale-dependence.


\section{Generalization: full anisotropy in all 3 spatial directions}
\label{total Anisotrop}

In the previous sections we have reduced the three-dimensional spatial translational symmetry to the
two-dimensional translational symmetry. In the rest of the paper, we generalize these results to the maximally anisotropic case in which all three scale factors are different.
We expect the scale-dependence of the anisotropic power spectrum
to be the same as we studied above. However, we expect the angular-dependence to be different.

\subsection{Background }

In this case the background metric is given by
\ba
\label{biantotal}
ds^2 = - dt^2 + a^2 d x^2 + b^2d y^2 + c^2d z^2 \, .
\ea
Considering the following anasatz for the scale factors $a$, $b$ and $c$,
\ba
\label{abc}
a(t)&=& e^{\alpha(t)} \\
b(t)&=& e^{\alpha(t)+ 3\sigma(t)}\\
c(t)&=& e^{\alpha(t)+ 3\delta(t)} \, ,
\ea
the background equations of motion is
\ba
\label{back-rho-eq-to}
\ddot\phi+3\left(\dot \alpha + \dot \sigma + \dot \delta \right)\dot \phi+ V_\phi &=&0  \\
\label{Ein1-eq-to}
3 M_P^2 \left(\dot \alpha^2+ 2\dot \alpha \left(\dot \sigma + \dot \delta \right) + 3\dot \sigma \dot \delta\right) &=& \frac{1}{2}\dot
\phi^2+V(\phi) \\
\label{Ein2-eq-to}
M_P^2 \left( \ddot \alpha + 3\dot \alpha \left( \dot \alpha +  \dot \sigma + \dot \delta \right) \right) &=& V(\phi) \\
\label{anisotropy-eqy-to}
\ddot \sigma +3\dot \sigma \left(\dot \alpha +  \dot \sigma + \dot \delta  \right)&=& 0\, ,
\\
\label{anisotropy-eqz-to}
\ddot \delta +3\dot \delta \left(\dot \alpha +  \dot \sigma + \dot \delta  \right)&=& 0\, .
\ea
\begin{figure}[t]
	\includegraphics[width=0.6\textwidth]{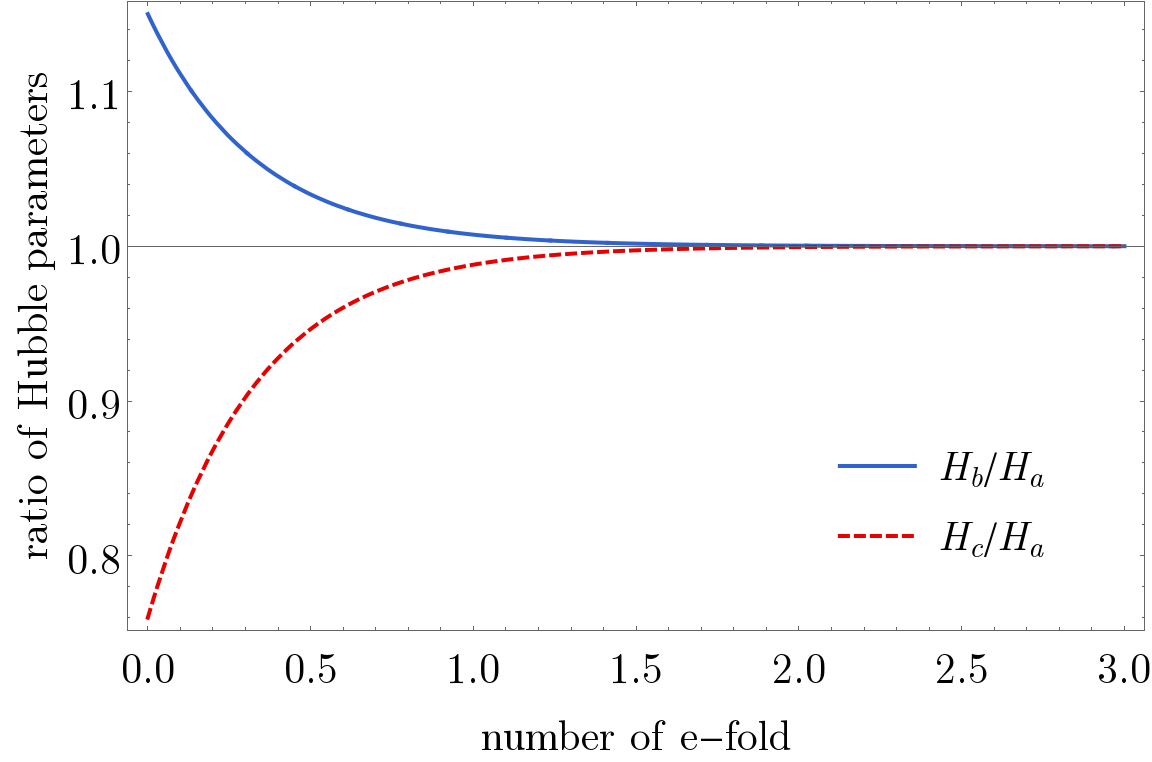}
	\caption{Here we plot the evolution of $H_{b}$ and $H_{c}$. As we expect, the attractor solution is FRW as the system approaches to it very rapidly. The parameters are chosen such that at an initial time $t_0$, $\dot \sigma_{0} = 0.05 \dot \alpha_{0}$ and $\dot \delta_{0} = -0.08 \dot \alpha_{0}$.}
	\vspace{0.5cm}
	\label{Fattractor}
\end{figure}
Although the above equations seem to be complicated, they can be simplified by using the slow-roll approximation. The situation is similar to the previous case where both of $\dot \sigma$ and $\dot \delta$ decay like $a^{-3}$ and our system approaches to its attractor   FRW phase. We present the attractor solutions in Fig.~\ref{Fattractor}.

As in our previous case, we can integrate the above equations and find the following approximate solutions
\ba
\label{app.ato}
a & \simeq & H_{0}^{-1} \left( -\eta \right)^{-1 } \\
\label{app.bto}
b & \simeq & H_{0}^{-1} \left( -\eta \right)^{-1}\left( 1 + \left(\frac{\dot{\sigma_{0}}}{H_{0}}\right) \left( \mathcal{H}_{0} \eta \right)^3 \right) \\
\label{app.cto}
c & \simeq & H_{0}^{-1} \left( -\eta \right)^{-1}\left( 1 + \left(\frac{\dot{\delta_{0}}}{H_{0}}\right) \left( \mathcal{H}_{0} \eta \right)^3 \right) \, .
\ea


\subsection{Perturbations of the fully anisotropic background}

Now we consider the perturbations of our full anisotropic background. As we have justified before, we can safely neglect the metric perturbations and only consider the inflaton fluctuations. Then the second order action is
\ba
L_{\phi \phi} = \frac{bc}{2}\mid \delta \phi_{k} '\mid^2  - \left(\frac{bc}{2}k_{x}^2 + \frac{a^2c}{2b}k_{y}^2 +\frac{a^2b}{2c}k_{z}^2 \right)\mid \delta \phi_{k}\mid^2 ~.
\ea
Again we have neglected the terms that are slow-roll suppressed. Now the equation of motion for $\delta \phi$ is
\ba
\delta \phi_{k}'' + \left(\frac{b'}{b} + \frac{c'}{c}\right)\delta \phi_{k}' + \left( k_{x}^2 + \frac{a^2}{b^2}k_{y}^2 + \frac{a^2}{c^2}k_{z}^2\right)\delta \phi_{k} =0 \, .
\ea
Expanding $\delta \phi $ in terms of usual creation and annihilation operators as in Eq. (\ref{a-adag}),  the perturbed scalar field equation becomes
\ba
\label{mode functionto}
u_{k}''  &-&\frac{1}{\eta}\left[ 2 - 3 \left(\frac{\dot{\sigma_{0}}}{H_{0}}\right)\left(\mathcal{H}_{0}\eta\right)^3  - 3 \left(\frac{\dot{\delta_{0}}}{H_{0}}\right)\left(\mathcal{H}_{0}\eta\right)^3 \right]u_{k}' \nonumber\\
&&+ \bigg{[}k_{x}^2 + \left( 1 - 2 \left(\frac{\dot{\sigma_{0}}}{H_{0}}\right) \left(\mathcal{H}_{0}\eta\right)^3 \right)k_{y}^2
+ \left( 1 - 2 \left(\frac{\dot{\delta_{0}}}{H_{0}}\right) \left(\mathcal{H}_{0}\eta\right)^3 \right)k_{z}^2 \bigg{]}u_{k} =0\, .
\ea
Parallel to what we did for the axial-symmetric case,
first we solve the above equation for the ``near to horizon modes" and then we improve our expansion  by changing the variable. Subsequently, we  present an expansion which is suitable for both of near horizon and UV modes.

\subsection{Generalized near horizon expansion}

Following our previous procedure, we can expand $u_{k}$ as in Eq.~(\ref{uk}). Then the goal is finding $C_{+}$ and $u_{k(1)}$.
Let us start with the differential equation of motion for $u_{k(1)}$,
\ba
u_{k(1)}'' -\frac{2}{\eta}u_{k(1)}' + k^2 u_{k(1)} = - \frac{3}{\eta}\left(\frac{\dot{\sigma_{0}}+\dot{\delta_{0}}}{H_{0}}\right)
\left(\mathcal{H}_{0}\eta\right)^3u_{k(0)}'+ 2\left(\frac{\dot{\sigma_{0}}}{H_{0}}k_{y}^2 + \frac{\dot{\delta_{0}}}{H_{0}}k_{z}^2 \right) \left(\mathcal{H}_{0}\eta\right)^3 u_{k(0)} ~,
\ea
from which we get
\ba
u_{k(1)} = \frac{\mathcal{H}_{0}^3}{\sqrt{2k^3}}\sum_{n=3}^{5} \Omega_{n}\eta^{n}
e^{-ik\eta} \, ,
\ea
where
\ba
\Omega_{3} &=& - \frac{1}{4k^2}\left( 2k^2 \left(\dot{\sigma_{0}}+ \dot{\delta_{0}} \right)-3\left(\dot{\sigma_{0}} k_{y}^2 + \dot{\delta_{0}}k_{z}^2 \right) \right)\\
\Omega_{4} &=& - \frac{i}{4k}\left( 2k^2 \left(\dot{\sigma_{0}}+ \dot{\delta_{0}} \right)-3\left(\dot{\sigma_{0}} k_{y}^2 + \dot{\delta_{0}}k_{z}^2 \right) \right) \\
\Omega_{5} &=& - \frac{1}{4}\left(\dot{\sigma_{0}} k_{y}^2 + \dot{\delta_{0}}k_{z}^2 \right)  \, .
\ea
As it has been discussed before, we determine $\mathcal{C}_{+}$ by using the normalization condition which leads to
\ba
\label{c plus}
|\mathcal{C}_{+}|^2 =  1 + \frac{3\mathcal{H}_{0}^2}{4k^3}\bigg{[} 2 \left(\sigma'_{0}+ \delta'_{0}\right)-3 \sin^2{\Theta} \left(\sigma'_{0} \cos^2{\Phi} + \delta'_{0} \sin^2{\Phi}\right)\bigg{]}
\ea
where we have chosen the wave number $\kk$ as
\ba
\kk = (k\cos{\Theta}\, , k \sin{\Theta} \cos{\Phi}\, , k \sin{\Theta} \sin{\Phi} ) \, .
\ea
Note that as before $ \Theta $ is the angle of $\hat \kk$ with respect to the $x$ axis while
$\Phi$ is the azimuthal angle of $\hat \kk$  in $y-z$ plane.

\subsection{Generalized UV safe expansion}

As in the previous cases, we expect that all modes, including the near-horizon and UV modes, receive the same order of anistropic corrections. So we improve the expansion scheme by changing the variable as Eq. (\ref{psi u}).
Expanding $\psi_{\mathbf{k}}(\eta)$ in orders of $\dot{\sigma}_{0}$ and  $\dot{\delta}_{0}$ leads us to the following expression for $\psi_{\mathbf{k}(1)}(\eta)$
\ba
\psi_{\mathbf{k}(1)}(\eta) &=& \frac{3i\mathcal{H}_{0}^2}{8k^3}\bigg{[} 2 \left(\sigma'_{0}+ \delta'_{0}\right)-3 \sin^2{\Theta}  \left(\sigma'_{0}\cos^2{\Phi} + \delta'_{0}\sin^2{\Phi}\right)\bigg{]}
+\frac{\mathcal{H}_{0}^2 a_{0}}{1+ik\eta}\sum_{n=3}^{5} \Omega_{n}\eta^{n}\nonumber\\
&=& \frac{\mathcal{H}_{0}^2}{1+ik\eta}\sum_{n=0}^{5} \Xi_{n}\eta^{n} \, ,
\ea
where $\Xi_{n}$ are given by
\ba
\Xi_{0} &=& \frac{3i}{8k^3}\bigg{[} 2 \left(\sigma'_{0}+ \delta'_{0}\right)-3 \sin^2{\Theta} \left(\sigma'_{0} \cos^2{\Phi} + \delta'_{0} \sin^2{\Phi}\right)\bigg{]} ~,
\\
\Xi_{1} &=& -\frac{3}{8k^2}\bigg{[} 2 \left(\sigma'_{0}+ \delta'_{0}\right)-3 \sin^2{\Theta} \left(\sigma'_{0} \cos^2{\Phi} + \delta'_{0} \sin^2{\Phi}\right)\bigg{]} ~,
\\
\Xi_{2} &=& 0 ~,
\\
\Xi_{m} &=& a_{0}\Omega_{m} ~~,~~ (m=3,4,5) ~.
\ea

Using the same time-delay formula, $\zeta \approx - H_{0} \left(\frac{\delta \phi}{\dot \phi_{0}}\right)$, and the above formulas for $\delta \phi$, we can calculate the power spectrum of curvature perturbation as,
\ba\label{generalgstar}
\left( \frac{k^3}{2 \pi^2} \right)\langle \zeta^2  \rangle = P_{\zeta0}\left( 1 + \frac{3\mathcal{H}_{0}^2}{4k^3}\bigg{[} 2 \left(\sigma'_{0}+ \delta'_{0}\right)-3 \sin^2{\Theta} \left(\sigma'_{0}  \cos^2{\Phi} + \delta'_{0} \sin^2{\Phi}\right)\bigg{]}\right) \, .
\ea
Eq. (\ref{generalgstar}) is  the main result of this section and shows the non-trivial shape of anisotropic power spectrum as a function of the angles $\Theta$ and $\Phi$. In the limit where
$\sigma'_{0} = \delta'_{0}$ and $\Phi=0$, the above result coincides with the result in Eq. (\ref{power-anis}) as expected.

\begin{figure}[t]
	\includegraphics[width=\textwidth]{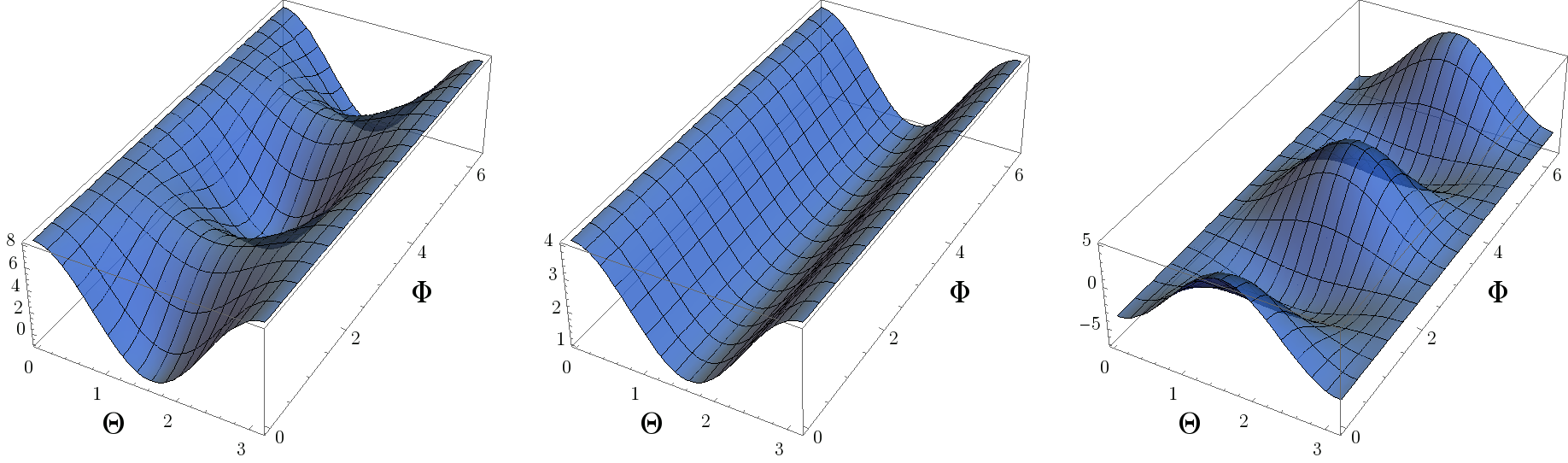}
	\caption{Here we plot $\bigg{[} 2 \left(\sigma'_{0}+ \delta'_{0}\right)-3\sin^2{\Theta} \left(\sigma'_{0} \cos^2{\Phi} + \delta'_{0} \sin^2{\Phi}\right)\bigg{]}$ for different choices of $\sigma'_{0}$ and $\delta'_{0}$. The left, middle and right panels correspond to $\sigma'_{0} = 3 \delta'_{0} $, $\sigma'_{0} = \delta'_{0} $ and $\sigma'_{0} = -3 \delta'_{0} $ respectively.}
	\vspace{0.5cm}
	\label{2d angular}
\end{figure}

Since the above formula is somewhat complicated, in Fig.~\ref{2d angular} we draw  a few diagrams to illustrate this two-dimensional angular patterns. We perform this for different choices of $\sigma'_{0}$ and $\delta'_{0}$.

\section{Statistical anisotropies on the CMB}

In this section, we expand the anisotropies derived from previous sections in terms of the spherical harmonics basis. The correlation functions of the expansion coefficients are the observables on the CMB. The anisotropic corrections in Eq. (\ref{generalgstar}) are
\begin{align} \label{eq:dp}
\Delta P_\zeta & = \frac{3\mathcal{H}_0^2}{4k^3} \left[
2\left(\sigma_0'+\delta_0'\right) - 3  \sin^2{\Theta} \left( \sigma_0' \cos^2\Phi + \delta_0' \sin^2\Phi \right)
\right]
\nonumber\\ &
= \frac{3\mathcal{H}_0^2}{4k^3} \left[
\frac{1}{2} (1+3\cos^2\Theta) \left(\delta_0'+\sigma_0'\right)  - \frac{3}{2} (1-\cos^2\Theta)(1-2\cos^2\Phi) \left(\delta_0'-\sigma_0'\right)
\right]~.
\end{align}
In terms of the $g_{L, M}$ parameters \cite{Pullen:2007tu} defined via
\ba
\Delta P_\zeta = \sum_{L, M} g_{L, M} Y_{LM} (\Theta, \Phi)
\ea
the  anisotropic correction can be written as
\begin{align}
\Delta P_\zeta = P_{\zeta_0} \left[ g_{0,0}Y_{0,0} + g_{2,0}Y_{2,0} + g_{2,2}Y_{2,2} + g_{2,-2}Y_{2,-2}\right],
\end{align}
where
\begin{align}
g_{0,0} = \frac{3\mathcal{H}_0^2}{4k^3}2\sqrt\pi \left( \delta_0'+ \sigma_0' \right)~, \quad
g_{2,0} = \frac{3\mathcal{H}_0^2}{4k^3} \frac{2\sqrt\pi}{\sqrt 5} \left( \delta_0'+ \sigma_0' \right)~, \quad
g_{2,2} = g_{2,-2} =  \frac{3\mathcal{H}_0^2}{4k^3} \sqrt{\frac{6\pi}{5}} \left( \delta_0'- \sigma_0' \right)~.
\end{align}
In terms of $a_{lm}$, the anisotropy can be expressed as
\begin{align} \label{eq:alm}
a_{lm} = 4\pi (-i)^l \int \frac{d^3 k}{(2\pi)^3} ~ g_l (k) \zeta_\mathbf{k} Y^*_{lm}(\hat k)~, \qquad
C_l = \frac{1}{2l+1} \sum_m \langle a_{lm}a^*_{lm}\rangle  \, ,
\end{align}
where $g_l(k)$ is the radiation transfer function.

Inserting the anisotropic corrections from \eqref{eq:dp},  for the anisotropic corrections  in $C_\ell$ we obtain
\begin{align}
\Delta C_l =&
3\pi \mathcal{H}_0^2 P_{\zeta_0} \left[ \int dk ~ \frac{g_l^2(k)}{k^{4}}  \right]
\Big\{   \frac{1}{2l+1}  \sum_m \int d\Omega_k ~ Y_{lm}(\mathbf{k}) Y^*_{lm}(\mathbf{k})
\nonumber\\ &
\times \left[
\frac{1}{2} (1+3\cos^2\Theta) \left(\delta_0'+\sigma_0'\right)  - \frac{3}{2} (1-\cos^2\Theta)(1-2\cos^2\Phi) \left(\delta_0'-\sigma_0'\right)
\right] \Big\} \, .
\end{align}
Note that the following integrals are $l$-independent:
\begin{align} \label{eq:aniso-integ}
& \frac{1}{2l+1}  \sum_m \int d\Omega_k ~ Y_{lm}(\mathbf{k}) Y^*_{lm}(\mathbf{k}) \cos^2\Theta = \frac{1}{3} ~,
\nonumber\\ &
\frac{1}{2l+1}  \sum_m \int d\Omega_k ~ Y_{lm}(\mathbf{k}) Y^*_{lm}(\mathbf{k}) \cos^2\Phi = \frac{1}{2} ~,
\nonumber\\ &
\frac{1}{2l+1}  \sum_m \int d\Omega_k ~ Y_{lm}(\mathbf{k}) Y^*_{lm}(\mathbf{k}) \cos^2\Theta\cos^2\Phi = \frac{1}{6} ~.
\end{align}
The detail of the above calculation can be found in Appendix \ref{sec:summ-rules-spher}. As a result, the correction $\Delta C_l$ does not obtain additional $l$-dependence other than from the radiation transfer function:
\begin{align}
\Delta C_l =
3\pi \mathcal{H}_0^2 P_{\zeta_0} \left(\delta_0'+\sigma_0'\right) \int dk ~ \frac{g_l^2(k)}{k^{4}} ~.
\end{align}

In  the presence of statistical anisotropies, non-diagonal couplings of $\langle a_{l_1m_1}a_{l_2m_2}\rangle$ with $l_1 \neq l_2$ are turned on \cite{Ma:2011ii, Kim:2013gka, Chen:2013eaa, Watanabe:2010bu, Book:2011na}. Making use of the Gaunt's formula
\begin{align}
& \int_0^\pi  d\Theta \int_0^{2\pi} d\Phi  ~ Y^*_{l_1m_1}Y_{l_2m_2} Y_{l_3m_3}
\nonumber\\ =& ~
(-1)^{m_1} \sqrt{\frac{(2l_1+1)(2l_2+1)(2l_3+1)}{4\pi}}
\left( \begin{array}{ccc}
l_1 & l_2 & l_3\\
0 & 0 & 0
\end{array} \right)
\left( \begin{array}{ccc}
l_1 & l_2 & l_3\\
-m_1 & m_2 & m_3
\end{array} \right) ~,
\end{align}
the $Y_{2,m}$ anisotropy introduces
\begin{align} \label{eq:lmlmc}
\langle a_{l_1m_1}a^*_{l_2m_2}\rangle_{2,m}
=&
g_{2,m}P_{\zeta_0}\frac{8\pi}{3} i^{l_2-l_1}(-1)^{m_1} \sqrt{(2l_1+1)(2l_2+1)}
\left( \begin{array}{ccc}
l_1 & l_2 & 2\\
0 & 0 & 0
\end{array} \right)
\left( \begin{array}{ccc}
l_1 & l_2 & 2\\
-m_1 & m_2 & m
\end{array} \right)  \nonumber\\
& \times \int \frac{dk}{k^{4}}~ g_{l_1}(k)g_{l_2}(k)~,
\end{align}
where $m$ takes values in  $\{0, 2, -2\}$. The non-zero elements of $\langle a_{l_1m_1}a^*_{l_2m_2}\rangle_{2,m}$ are those with $l_1=l_2$ or $l_1 = l_2 \pm 2$ which also have comparable amplitudes.


\section{Summary}

In this chapter, we have studied statistical anisotropies in a model of inflation with a relic background anisotropy of the Bianchi I type. We have compared the predictions in the density perturbations in this model with another type of relic anisotropy model where the source is a vector field in the matter sector. We also considered the effect of a non-BD Gaussian state in such a model, as an illustrating example of how the non-BD states in minimal inflation models can extend the effects of the anisotropy to shorter scales in density perturbations.

\begin{figure}[htbp]
	\centering
	\includegraphics[width=1.0\textwidth]{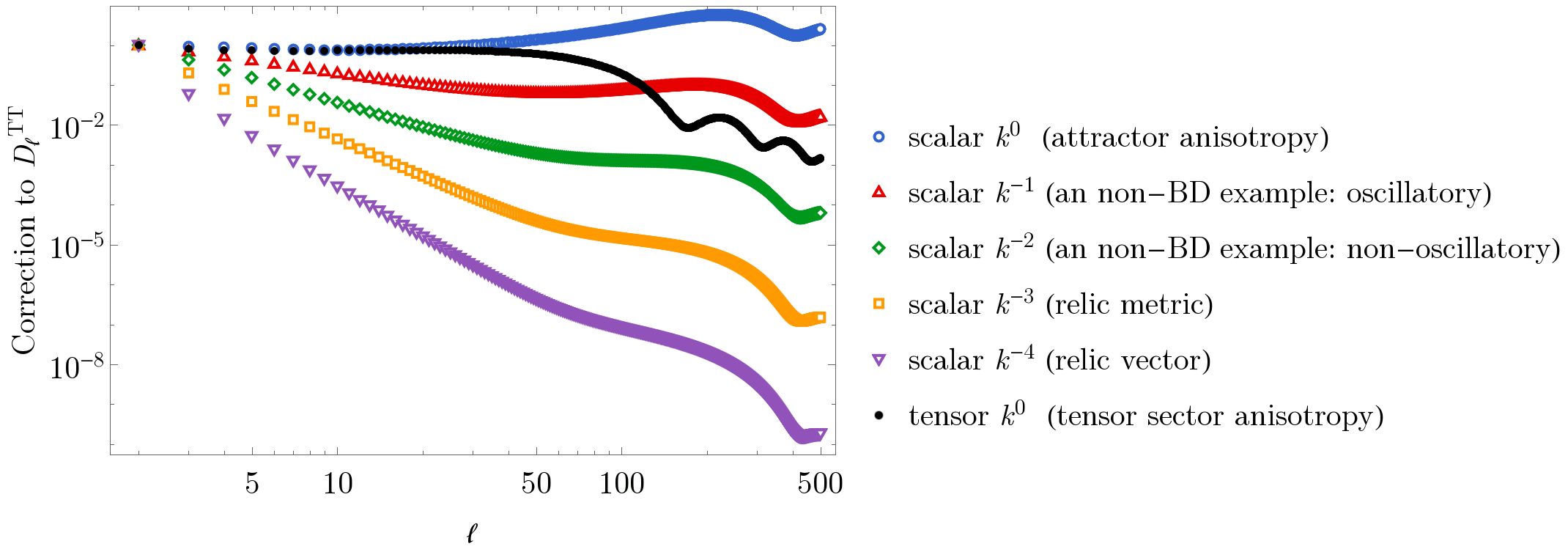}
	\caption{\label{fig:plotPower} Summary of scale-dependence of the anisotropic component in power spectrum for various models. The $k$-dependence of the primordial power spectrum, and the examples are listed in the plot legend. The correction to the CMB temperature anisotropy is plotted in the figure. For the oscillatory result in the non-BD example, only the envelop is plotted.}
\end{figure}

In Fig.~\ref{fig:plotPower} we have summarized the scale-dependence of the anisotropic power spectrum of these relic scenarios. As expected, the angular dependence of the statistical anisotropy in density perturbations are the same for all three models, because we have set up the same axial symmetric initial condition. Nonetheless, interestingly, the differences in the underlying physics of the models still lead to distinctive observable differences. In the BD vacuum case, the statistical anisotropy in the vector relic field model decays as $\sim 1/k^4$ while in the Bianchi model it decays as $\sim 1/k^3$. These behaviors are determined by the different background sources of the anisotropy. In the example of non-BD Gaussian state, the scale-dependence in both models become the same, $\sim 1/k^2$ for the non-oscillatory part, dominated by the similar initial quantum states in both cases. Such a quantum state enhances the anisotropy in much shorter scales, and becomes an interesting probe of the initial quantum state of the universe.

For comparison, in Fig.~\ref{fig:plotPower} we also listed the predictions from the models of  anisotropic inflation
based on attractor gauge field dynamics such as in \cite{Watanabe:2009ct, Emami:2010rm} in which
anisotropies are generated actively during entire period of inflation. To leading order,
the anisotropic power spectrum in these models are given by $\delta P_\zeta \propto P_{\zeta 0} N(k)^2$ in which $N(k)$ represents the number of e-folds when the mode of interest $k$ leaves the horizon. To leading order (neglecting the logarithmic scale-dependence of $N(k)$ to $k$) the attractor anisotropic models predict nearly scale-invariant anisotropic power spectrum, as we discussed in previous chapters. As another class of scale dependence, when the anisotropies are originated from the tensor sector, the scale dependence is characterized by  the CMB transfer function from the primordial tensor mode into temperature \cite{Chen:2014eua}.

Finally we also generalized the anisotropy from the axial symmetry to arbitrary angular dependence and calculated the corresponding anisotropic power spectrum.


\chapter{Conclusion} 

\label{Conclusion} 

\lhead{Conclusion} 

In modern cosmology, inflationary paradigm is regarded as a successful scenario in explaining some of the big-bang shortcomings without extreme fine tuning. This theory is very well consistent with the recent
observations from Planck satellite. From the theoretical point of view, we can introduce any fields with arbitrary spins during the early universe. However, until very recently most of the attentions was on the models with pure scalar fields. The reason was two folds. On the one hand it is much easier to work with the scalar fields where we have the rotational symmetry in the system and on the other hand, due to the lack of the resolution, there was not any motivations to take this difficulties into account. However, this picture has been changed by the WMAP, where this satellite announced a possible deviation from the statistical isotropy in the CMB data. Right after this, there were considerable attentions on primordial anisotropies and people started building inflationary models that contain not only the scalar field but also include higher spin fields like vectors, spinors etc. Moreover since some of the usual formalisms have been made for the statistical isotropic case, it is also required to revisit them and consider their extensions when it is necessary. Among these new opportunities, the vector fields seems a minimalistic extensions on the top of the well known scalar fields models. \\
The main focus of this thesis was on considering the vector realization of the anisotropic inflation. First of all, we tried to build up a consistent inflationary model for taking into account the primordial anisotropies. We then tried to extend some of the very well-known formalisms in our case and finally we tried to go ahead and consider the unique signatures of anisotropic inflation in the cosmic microwave background radiation. \\
More precisely, we considered the following items during this thesis, \\  
During the first few chapters, We reviewed the standard model of cosmology as well as its shortcoming in explaining the current universe. \\
In chapter 3, we reviewed the cosmological perturbation theory. We then introduced some physical observables during this phase and used the current observations to figure them out. \\
In chapter 4, we studied anisotropic inflation in models with charged scalar field. We have shown that the system reaches an attractor solution sometimes during inflation where the anisotropies produced during this phase becomes comparable to the slow-roll parameters. \\
In chapter 5, we studied the curvature perturbations in our charged scalar field inflationary model. We verified that the main contribution to anisotropies does come from the matter sector and one can neglect the metric degrees of freedom at the leading order. In addition, we proved that the gauge field's longitudinal mode is exponentially suppressed as compared to the transverse mode. We obtained an upper bound on the gauge coupling in order for the model to be consistent with the observations. \\
In chapter 6, we presented a consistent $\delta N$ formalism for calculating the curvature perturbations in anisotropic cosmological backgrounds. We then used our formalism to calculate the power spectrum, bispectrum and the trispectrum in our anisotropic setup. \\
In chapter 7, we studied anisotropies generated in the model of gauged hybrid inflation in which the complex waterfall field is charged under a $U(1)$ gauge field. We showed that the primordial anisotropies are generated either actively during inflation or from inhomogeneities modulating the surface of end of inflation. We showed that the gauge field fluctuations induce a red-tilted power spectrum so the averaged power spectrum from the gauge field can change the total spectral tilt from blue to red. \\
In chapter 8, we studied the TT, TB, EB and BB correlations associated with the B-mode polarization of the CMB map in models of charged anisotropic inflation. We showed that the asymmetry in the tensor power spectrum is a very sensitive probe of the gauge coupling.  While the level of statistical  anisotropy in temperature power spectrum can be small and satisfy the observational bounds, the interactions from the gauge coupling can induce large directional dependence
in tensor modes. Since the tensor transfer function would decay after $l = 100$, we can explain the low $l$ anomalies in the CMB data. \\
In chapter 9, we examined the impact of different anisotropic relics during inflation. These relics can be the source of the large scale anomalies in cosmic microwave background. In addition, we demonstrate that non-Bunch-Davies vacuum states can extend the statistical anisotropy to much shorter scales, and leave a scale-dependence that is insensitive to the different backgrounds but sensitive to the initial quantum state.\\
At the end, it is important to mention that while the CMB searches are near to their completion, it is important to use the future large scale surveys to probe the statistical anisotropies, \cite{Emami:2015uva}.   
\appendix

\chapter{}
\section{Metric Perturbations}
\label{appendix1}

Here we study the metric perturbations in Bianchi I background and their transformation properties  under a general coordinate transformation. Consider the general coordinate transformation
\ba
\label{xi}
x^\mu \rightarrow x^{\mu} + \xi^{\mu} \quad \quad ,  \quad \quad
\xi^\mu = \left(  \xi^0 \, ,\,  \partial_i \lambda  \, ,\, \partial_i \Lambda + \xi_\perp^i
\right)
\ea
in which $\xi^0, \lambda$ and $\Lambda$ are scalars and $\xi_\perp^i$ is vector subject to
$\partial_i {\xi_\perp^i}$=0.  For the future reference note that by appropriate choice of $\xi^0, \lambda$ and $\Lambda$   one can remove three scalar degrees of metric perturbations in Eq. (\ref{deltag}) while the freedom from $\xi_\perp^i$ can remove only one  vector degree of freedom.

Under the coordinate transformation Eq. (\ref{xi}) we have
\ba
\delta g_{\mu \nu} \rightarrow \delta g_{\mu \nu} -{^{(0)}} g_{ \mu \nu, \kappa}\, \xi^{\kappa} -
{^{(0)}}g_{\alpha \nu}\,  \partial_\mu \xi^\alpha -  {^{(0)}}g_{\alpha \mu}\,  \partial_\nu \xi^\alpha
\ea
in which ${^{(0)}}g_{\alpha \mu}$ is the background Bianchi metric given in Eq. (\ref{Bianchi-metric}).

More explicitly, one can check that
\ba
A && \rightarrow A - \frac{1}{a} \left( a \xi^0 \right)'   \\
\beta&& \rightarrow \beta + \xi^0 - \lambda'\\
B &&\rightarrow  B+ \frac{a}{b} \xi^0 - \frac{b}{a} \Lambda'\\
\bar \psi && \rightarrow \bar \psi +  \frac{a'}{a} \xi^0 + \partial_x^2 \lambda\\
\gamma && \rightarrow \gamma - \frac{b}{a}  \Lambda - \frac{a}{b}   \lambda\\
\psi  && \rightarrow \psi  + \frac{b'}{b} \xi^0 \\
E && \rightarrow E - \Lambda
\ea
and
\ba
B_i && \rightarrow B_i - \frac{b}{a}\,  {\xi_\perp^{i'}} \\
\Gamma_i && \rightarrow  \Gamma_i  - \frac{b}{a}\,  {\xi_\perp^i} \\
E_i && \rightarrow E_i - \xi_\perp^i
\ea
Using the above transformation properties one can check that the following two scalar variables are gauge invariant
\ba
\label{delta-phi-psi}
\delta \phi_\psi &&\equiv \delta \phi + \frac{\dot \phi}{H_b} \psi  \\
\hat \gamma &&\equiv \gamma - \frac{b}{a} E + \frac{a}{b} \partial^{-2}_{1} \left( \bar \psi - \frac{H_a}{H_b} \psi
\right)
\ea
In this view $\delta \phi_\psi$ represents the inflaton perturbations on  $\psi=0$ surface which
reduces to  inflaton perturbations on flat slice in FRW background while $\hat \gamma$ is identically zero in FRW background.

In our analysis we adopt the following gauge
\ba
\label{gauge}
\psi= \bar \psi = E = E_i =0 \, ,
\ea
which one can check is a consistent gauge.  Note that the three scalar conditions $\psi= \bar \psi = E =0$ fixes three scalar freedoms $\xi^0, \lambda $ and $\Lambda$ while the vector condition $E_i=0$ fixes the remaining one degree of freedom $\xi_\perp$.
The advantage in choosing the gauge in Eq. (\ref{gauge}) is that it reduces to the flat gauge in
the isotropic limit where $\psi =\bar \psi$.

We also note that after inflation ends and the universe becomes isotropic the scalar perturbation $\gamma$ and the vector perturbation $\Gamma_i$ combine to furnish two polarizations of the tensor perturbations. Note that in anisotropic background, the condition
$\partial_i \Gamma_i=0$ leaves only one degree of freedom. As a result $\Gamma_i$ can count only for one tensor polarization and the remaining polarization is taken care of by
$\gamma$ as mentioned.

\chapter{}
\section{Integrating out non-Dynamical Fields}
\label{reduced}

In this appendix we present the detail analysis of integrating out the non-dynamical fields $\delta A_0, \beta, A$ and $B$ in terms of the dynamical fields $\delta \rho, \gamma, \delta A_1$ and $M$. The second order action is given in Eq. (\ref{deltag5}).  Correspondingly, the second order action for the scalar perturbations in Fourier space is

\ba
\label{S2-scalar-k}
&&S_2 = \int d \eta d^3 k \left[ b b' k_x^2 (A^* \beta + A \beta^*) + \frac{a b}{2} (\frac{a'}{a} + \frac{b'}{b}) k_y^2 (A^* B + A B^*) +  \frac{a b}{2} k_x^2 k_y^2 (\gamma^* A + \gamma A^*) 
\right. \nonumber \\ &&\left.
- a^2 b^2 V(\rho_0) |A|^2
- \frac{\e^2}{2} b^2 \rho_0^2 A_x^2 |A|^2 - \frac{a b}{4} k_x^2 k_y^2 (\beta^* B + \beta B^*) + \frac{a' b}{2} k_x^2 k_y^2 (\gamma^* \beta + \gamma \beta^*) + \frac{a^2}{4} k_x^2 k_y^2 |\beta|^2
\right. \nonumber \\ &&\left.
+ \frac{a b}{4} k_x^2 k_y^2 (\gamma^* \beta' + \gamma \beta'^*)
+ \frac{\e^2}{2} b^2 \rho_0^2 A_x^2 k_x^2 |\beta|^2
- \frac{b^2}{4} k_x^2 k_y^2 (B^* \gamma' + B \gamma'^*)+ \frac{b^2}{4} (\frac{b'}{b} - \frac{a'}{a}) k_x^2 k_y^2 (\gamma^* B + \gamma B^*)
\right. \nonumber \\ &&\left.
  + \frac{b^2}{4} k_x^2 k_y^2 |B|^2 + \frac{b^2}{4} k_x^2 k_y^2 | \gamma'|^2 - \frac{\e^2}{2}b^2 A_x^2 \rho_0^2 k_x^2 k_y^2 |\gamma|^2  + \frac{f^2 b^2}{2 a^2} A_x'^2 k_x^2 k_y^2 |\gamma|^2 - \frac{b^2}{4} (\frac{b''}{b} - \frac{a''}{a}) k_x^2 k_y^2 |\gamma|^2
\right. \nonumber \\ &&\left.
+  \frac{b^2}{2} |\delta \rho'|^2 -  \frac{b^2}{2} \rho_0' (A^* \delta \rho' + A \delta \rho'^*) - \frac{b^2}{2} \rho_0' k_x^2 (\beta^* \delta \rho + \beta \delta \rho^*) - \frac{a b}{2} \rho_0' k_y^2 (B^* \delta \rho + B \delta \rho^*)- \frac{b^2}{2} k_x^2 | \delta \rho|^2
\right. \nonumber \\ &&\left.
  - \frac{a^2}{2} k_y^2 |\delta \rho|^2 + \frac{\e^2 b^2}{2}  \rho_0^2 |\delta A_0|^2 + i k_x \frac{\e^2 b^2}{2} \rho_0^2 A_x  (\beta^* \delta A_0 - \beta \delta A_0^*)
  -\frac{\e^2 b^2}{2} \rho_0^2 |\delta A_1|^2 - \frac{\e^2 b^2}{2}A_x^2 |\delta \rho|^2
\right. \nonumber \\ &&\left.
- \e^2 b^2 \rho_0 A_x (\delta \rho^* \delta A_1 + \delta \rho \delta A_1^*) + i k_x k_y^2 \frac{\e^2 a b}{2} \rho_0^2 A_x (\gamma M^* - \gamma^* M) - \frac{\e^2 a^2}{2} \rho_0^2 k_y^2 |M|^2 + \frac{b^2}{2 a^2} f^2  |\delta A_1'|^2
\right. \nonumber \\ &&\left.
 - \frac{\e^2 b^2}{2} \rho_0^2 A_x (A^* \delta A_1 + A \delta A_1^*)
- \frac{\e^2  b^2}{2} \rho_0 A_x^2 (A^* \delta \rho + A \delta \rho^*) - i k_x \frac{b^2 f^2}{2 a^2}  (\delta A_1'^* \delta A_0 - \delta A_1' \delta A_0^*)
\right. \nonumber \\ &&\left.
+  \frac{b^2}{2 a^2} f^2  k_x^2 |\delta A_0|^2
-  \frac{b^2 f^2}{2 a^2} A_x' (A^* \delta A_1' + A \delta A_1'^*) +i k_x  \frac{b^2 f^2}{2 a^2} A_x' (A^* \delta A_0 - A \delta A_0^*) + \frac{f^2}{2} k_y^2 |M'|^2
\right. \nonumber \\ &&\left.
 - i k_x k_y^2 \frac{b f^2 A_x'}{2a}  (\gamma M'^* - \gamma^* M')   + i k_x k_y^2 \frac{b f^2}{2 a} A_x'   (\gamma \delta A_0^* - \gamma^* \delta A_0) -\frac{b f^2}{2 a} A_x' k_y^2 (B \delta A_1^* + B^* \delta A_1)
\right. \nonumber \\ &&\left.
  +  i k_x\frac{b f^2}{2 a} A_x' k_y^2 (B^* M - B M^*)  + \frac{f^2}{2} k_y^2| \delta A_0 |^2 - \frac{f^2}{2} k_y^2 (M'^* \delta A_0 + M' \delta A_0^*) - \frac{f^2}{2} k_y^2 |\delta A_1 |^2
\right. \nonumber \\ &&\left.
- \frac{f^2}{2} k_x^2 k_y^2 |M |^2 + i k_x \frac{f^2}{2} k_y^2 (\delta A_1^* M - \delta A_1 M^*) + \frac{b^2 f f_{, \rho}}{a^2}  A_x' (\delta A_1'^* \delta \rho + \delta A_1' \delta \rho^*) + \frac{b^2  f_{, \rho}^2}{2 a^2} A_x'^2 |\delta \rho |^2
\right. \nonumber \\ &&\left.
+ i k_x \frac{b^2 f f_{, \rho}}{a^2}  A_x' (\delta A_0^* \delta \rho - \delta A_0 \delta \rho^*) - \frac{b^2 f f_{, \rho}}{2 a^2} A_x'^2 ( A^* \delta \rho +  A \delta \rho^*)  +  \frac{b^2 f f_{, \rho \rho}}{2 a^2} A_x'^2 |\delta \rho |^2 - \frac{a^2 b^2}{2} V_{, \rho \rho} |\delta \rho |^2 
\right. \nonumber \\ &&\left.
 -  \frac{a^2 b^2}{2} V_{, \rho } (\delta \rho A^* + \delta \rho^* A)
\right] \, .
\ea

We have to integrate out the non-dynamical variables $\{\delta A_0, \beta, A, B\}$ from the action Eq. (\ref{S2-scalar-k}). The analysis are simple but tedious. To outline the analysis, here we demonstrate how to integrate out $\beta$. The action expanded in powers of $\beta$ is
\ba
\label{L-beta}
{\cal L } = c_1 \beta \beta^* + c_2 \beta^* + c_2^* \beta  +... \, ,
\ea
in which the dots indicates the rest of the action containing the dynamical fields $\{\delta \rho, \delta A_1, M, \gamma \}$ and $\{ \delta A_0, A, B\}$ and $c_{1, 2}$ are functions which can be read off from the action Eq. (\ref{S2-scalar-k})
\ba
c_1 = \frac{a^2}{4} k_x^2 k_y^2 + \frac{\e^2}{2} b^2 \rho^2 k_x^2 A_x^2
\ea
and
\ba
c_2 = b \, b' k_x^2 A - \frac{ab}{4} k_x^2 k_y^2 \, B + \frac{a' b}{2} k_x^2 k_y^2 \, \gamma
- k_x^2 k_y^2 \left( \frac{ab}{4} \right)' - \frac{b^2}{2} \rho' k_x^2 \delta \rho
+ i k_x \frac{\e^2 b^2}{2} \rho^2 A_x \delta A_0 \, .
\ea
Varying the action with respect to $\beta^*$ yields $\beta = - c_2/c_1$. Plugging this into the action yields
\ba
{\cal L} = -\frac{|c_2|^2}{c_1} + ... .
\ea
Following the same steps to integrate out $\delta A_0, A$ and $B$ we can write the dynamical action as
\ba
L_{(2)} =  L_{\rho \rho} + L_{\gamma \gamma} + L_{MM}  + L_{A_1 A_1} + L_{\rho \gamma} + L_{\rho M} + L_{\rho A_1} + L_{\gamma M} + L_{\gamma A_1}
+ L_{M A_1}  + ... \, ,
\ea
in which the dots indicate the rest of the action coming from the dynamical fields $\{\delta \rho, \delta A_1, M, \gamma \}$. Here we have defined

\ba
\label{L_{rho-rho}}
L_{\rho \rho}  &&= \left(\frac{b^2}{2} - \frac{1}{\lambda_8}\frac{b^4}{4}\rho_{0}^{'2}- \frac{\lambda_6^2}{\lambda_{13} \lambda_8^2}\frac{b^4}{4}\rho_{0}^{'2}  \right) \Big| \delta \rho ^{'}\Big|^2+ \bigg{(}-\frac{b^2}{2}k^2 - \frac{b^2}{2}\e^2 A_{x}^2+ \frac{b^2}{2a^2}A_{x}^{'2}f_{,\rho}^2
+ \frac{b^2}{2a^2}A_{x}^{'2}f f_{,\rho\rho} \nonumber \\
&&
- \frac{a^2 b^2}{2}V_{, \rho \rho}
-\frac{1}{\lambda_1}\frac{b^4}{4}\rho_{0}^{'2}k_{x}^4 - \frac{|\lambda_5|^2}{\lambda_2}- \frac{|\lambda_9|^2}{\lambda_8}- \frac{|\lambda_{14}|^2}{\lambda_{13}}\bigg{)}\Big| \delta \rho \Big|^2 + 
\bigg{(}  \frac{\lambda_9}{\lambda_8}\frac{b^2}{2}\rho_{0}^{'} - \frac{\lambda_6 \lambda_{14}}{\lambda_8 \lambda_{13}}\frac{b^2}{2}\rho_{0}^{'}\bigg{)}\Big( \delta \rho \delta \rho^{*} \Big)^{'} \nonumber\\
\label{L_{gamma-gamma}}
\ea
\ba
L_{\gamma \gamma} & =& \left(\frac{b^2}{4}k_{x}^2 k_{y}^2 -\frac{a^2 b^2}{16}\frac{1}{\bar \lambda_2}k_{x}^4 k_{y}^4- \frac{a^2 b^2}{16}\frac{\bar \lambda_3^2}{\lambda_{8}\bar \lambda_2^2 }k_{x}^4 k_{y}^4 - \frac{|\lambda_{11}|^2}{\lambda_{13}} \right) \Big| \gamma ^{'}\Big|^2+ \bigg{(}-\frac{b^2}{2}\e^2 A_{x}^2 \rho_{0}^2 k_{x}^2 k_{y}^2 + \frac{b^2}{2a^2}A_{x}^{'2}f^2 k_{x}^2 k_{y}^2 \nonumber \\
& - &\frac{b^2}{4}(\frac{b^{''}}{b}-\frac{a^{''}}{a})k_{x}^2 k_{y}^2 - \frac{b^2}{4a^2}\frac{1}{\bar \lambda_1}f^4 A_{x}^{'2} k_{x}^2 k_{y}^4 - \frac{(\bar \lambda_4)^2}{\bar \lambda_2} - \frac{\lambda_7^2}{\lambda_8} -\frac{\lambda_{12}^2}{\lambda_{13}} \bigg{)}\Big| \gamma \Big|^2 +
\bigg{(} \frac{a b}{4}\frac{\bar \lambda_4}{\bar \lambda_2}k_{x}^2 k_{y}^2 - \frac{a b}{4}\frac{\lambda_7 \bar \lambda_{3}}{\lambda_8 \bar \lambda_{2}}k_{x}^2 k_{y}^2 \nonumber \\
&- &\frac{\lambda_{11} \lambda_{12}}{\lambda_{13} }\bigg{)}\Big( \gamma \gamma^{*} \Big)^{'}
\ea
\ba
\label{L_{M-M}}
L_{MM} & =& \left(\frac{f^2}{2}k_{y}^2 -\frac{f^4}{4}\frac{1}{\lambda_2}k_{y}^4- \frac{f^4}{4}\frac{ |\lambda_3|^2}{\lambda_{8}\lambda_2^2 }k_{y}^4 - \frac{|\lambda_{17}|^2}{\lambda_{13}} \right) \Big| M ^{'}\Big|^2+ \bigg{(}-\frac{a^2}{2}\e^2\rho^2 k_{y}^2 - \frac{f^2}{2}k_{x}^2 k_{y}^2
- \frac{b^2}{4a^2\lambda_{13}} 
\nonumber \\
&&
\times f^4 A_{x}^{'2} k_{x}^2 k_{y}^4 \bigg{)}\Big| M \Big|^2  + \bigg{(} \frac{b}{2a\lambda_{13}}f^2 A_{x}^{'}k_{x} k_{y}^2\bigg{)}\Big(i\lambda_{17} M^{*} M^{'}- i\lambda_{17}^{*} M^{'*} M \Big)
\ea
\ba
\label{L_{A1-A1}}
L_{A_{1}A_{1}}& = &\left(\frac{b^2}{2a^2}f^2-\frac{b^4}{4a^4\lambda_2}f^4k_{x}^2- \frac{ (\lambda_{10})^2}{\lambda_{8}} - \frac{(\lambda_{16})^2}{\lambda_{13}} \right) \Big| \delta A_{1} ^{'}\Big|^2+ \bigg{(}-\frac{b^2}{2}\e^2\rho^2 - \frac{f^2}{2}k_{y}^2 - \frac{b^4}{4\lambda_{8}}e^4 \rho^4 A_{x}^{2} \nonumber \\
&&
-\frac{(\lambda_{15})^2}{\lambda_{13}} \bigg{)}\Big| \delta A_{1} \Big|^2 +\bigg{(} \frac{b^2\lambda_{10}}{2\lambda_{8}}\e^2 \rho^{2} A_{x} - \frac{\lambda_{15}\lambda_{16}} {\lambda_{13}}\bigg{)}\Big( \delta A_{1} \delta A_{1}^{*} \Big)^{'}
\ea
\ba
\label{L_{rho-gamma}}
L_{\rho \gamma }& =& \Big{(}\frac{b^3}{2a^3\bar \lambda_{1}}f^3 f_{,\rho}A_{x}^{'2}k_{x}^2 k_{y}^2 - \frac{\bar \lambda_4 \bar \lambda_5}{\bar \lambda_2}- \frac{\lambda_7 \lambda_9}{\lambda_8}-\frac{\lambda_{12} \lambda_{14}}{\lambda_{13}}\Big{)} \Big{(}\delta \rho \gamma^{*} + c.c. \Big{)}+ \Big{(} \frac{ab}{4}\frac{\bar \lambda_{5}}{\bar \lambda_{2}}k_{x}^2 k_{y}^2 - \frac{ab}{4}\frac{\bar \lambda_{3}\lambda_{9}}{\bar \lambda_{2}\lambda_{8}}k_{x}^2 k_{y}^2 \nonumber \\
-&& \frac{\lambda_{11}\lambda_{14}}{\lambda_{13}} \Big{)}\Big{(} \delta \rho \gamma^{'*}+c.c. \Big{)}
+ \Big{(} \frac{b^2}{2}\frac{\lambda_{7}}{\lambda_{8}}\rho^{'} - \frac{b^2}{2}\frac{\lambda_{6}\lambda_{12}}{\lambda_{8}\lambda_{13}} \rho^{'}\Big{)}\Big{(}\delta \rho^{'} \gamma^{*} + c.c. \Big{)}+ \Big{(} \frac{ab^3}{8}\frac{\bar \lambda_3}{\bar \lambda_2 \lambda_8} \rho^{'} k_{x}^2 k_{y}^2\nonumber\\
&-& \frac{b^2}{2}\frac{\lambda_6 \lambda_{11}}{\lambda_8 \lambda_{13}} \rho^{'}\Big{)}
\Big{(}\delta \rho^{'} \gamma^{'*} + c.c. \Big{)}
\ea
\ba
\label{L_{rho-M}}
L_{\rho M }& =& \Big{(}\frac{b}{2a}\frac{\lambda_{14}}{\lambda_{13}}f^2 A_{x}^{'}k_{x}k_{y}^2\Big{)} \Big{(}i\delta \rho M^{*} + c.c. \Big{)}+ \Big{(} \frac{f^2}{2\lambda_{2}}k_{y}^2\Big{)}\Big{(}\lambda_{5}\delta \rho M^{'*} + c.c. \Big{)}-\Big{(}\frac{\lambda_{9}}{2\lambda_{8}\lambda_{2}} f^2 k_{y}^2\Big{)}\Big{(}\lambda_{3}\delta \rho M^{'*}\nonumber\\
&+& c.c. \Big{)}-\Big{(}\frac{\lambda_{14}}{\lambda_{13}}\Big{)}\Big{(}\lambda_{17}\delta \rho^{*} M^{'}+ c.c. \Big{)}+ \Big{(}\frac{b^3}{4a}\frac{\lambda_{6}}{\lambda_{13}\lambda_{8}}f^2 A_{x}^{'}\rho^{'}k_{x}k_{y}^2\Big{)} \Big{(}i\delta \rho^{'} M^{*} + c.c. \Big{)}+\Big{(}\frac{b^2}{4\lambda_{8}\lambda_{2}} f^2 k_{y}^2\rho^{'} \Big{)}\nonumber\\
&&\Big{(}\lambda_{3}\delta \rho^{'} M^{'*}+ c.c.\Big{)}- \Big{(}\frac{b^2\lambda_{6}}{2\lambda_{8}\lambda_{13}}\rho^{'} \Big{)}\Big{(}\lambda_{17}\delta \rho^{'*} M^{'}+ c.c.\Big{)}
\ea
\ba
\label{L_{rho-A1}}
L_{\rho A_{1} }& = &\Big{(}-\e^2b^2 \rho A_{x} + \frac{b^2\lambda_{9}}{2\lambda_{8}}\e^2 \rho^2 A_{x} - \frac{\lambda_{14}\lambda_{15}}{\lambda_{13}}\Big{)} \Big{(}\delta \rho \delta A_{1}^{*} + c.c. \Big{)}+ \Big{(} \frac{b^2}{a^2}A_{x}^{'}f f_{,\rho}- \frac{b^2}{2a^2}\frac{i\lambda_{5}^{*}}{\lambda_{2}}k_{x}f^2\nonumber \\
&-& \frac{\lambda_{9}\lambda_{10}}{\lambda_{8}}-\frac{\lambda_{14}\lambda_{16}}{\lambda_{13}} \Big{)}\Big{(} \delta \rho^{*} \delta A_{1}^{'}+c.c. \Big{)}+ \Big{(} -\frac{b^4}{4\lambda_{8}}\e^2\rho^{'}\rho^{2}A_{x} -\frac{b^2}{2}\frac{\lambda_{6}\lambda_{15}}{\lambda_{8}\lambda_{13}}  \rho^{'}\Big{)}\Big{(}\delta \rho^{'} \delta A_{1}^{*} + c.c. \Big{)}\nonumber \\
&+& \Big{(} \frac{b^2}{2}\frac{\lambda_{10}}{\lambda_8} \rho^{'}- \frac{b^2}{2}\frac{\lambda_6 \lambda_{16}}{\lambda_8 \lambda_{13}} \rho^{'}\Big{)}
\Big{(}\delta \rho^{'*} \delta A_{1}^{ '} + c.c. \Big{)}
\ea
\ba
\label{L_{gamma-M}}
L_{\gamma M }& =& \Big{(}\frac{ab}{2}\e^2\rho^{2} A_{x}k_{x}k_{y}^2 + \frac{b\lambda_{12}}{2a\lambda_{13}}f^2 A_{x}^{'}k_{x}k_{y}^2 \Big{)} \Big{(}i\gamma  M^{*} + c.c. \Big{)} - \Big{(} \frac{b}{2a}A_{x}^{'}f^{2}k_{x}k_{y}^2\Big{)}\Big{(}i\gamma  M^{'*} + c.c. \Big{)}\nonumber\\
&+&\frac{1}{2\lambda_{2}}f^2k_{y}^2\Big{(}\lambda_{4}\gamma  M^{'*} + c.c. \Big{)}-\frac{\lambda_{7}}{2\lambda_{2}\lambda_{8}}f^2k_{y}^2\Big{(}\lambda_{3}\gamma  M^{'*} + c.c. \Big{)}-\frac{\lambda_{12}}{\lambda_{13}}\Big{(}\lambda_{17}^{*} \gamma  M^{'*} + c.c. \Big{)}\nonumber\\
&+&\Big{(}\frac{b\lambda_{11}}{2a\lambda_{13}}f^2 A_{x}^{'}k_{x}k_{y}^2\Big{)}\Big{(}i\gamma^{'}  M^{*} + c.c. \Big{)}-\Big{(}\frac{ab^3}{16\lambda_{1}\lambda_{2}}\e^2\rho^{2}f^2 A_{x}k_{x}^{3}k_{y}^4 \Big{)}\Big{(}i\gamma^{'}  M^{'*} + c.c. \Big{)} \nonumber\\ &-&\Big{(}\frac{ab\bar\lambda_{3}}{8\lambda_{2}\bar\lambda_{2}\lambda_{8}}f^2 k_{x}^{2}k_{y}^4 \Big{)}\Big{(}\lambda_{3}\gamma^{'}  M^{'*} + c.c. \Big{)}
-\Big{(}\frac{\lambda_{11}}{\lambda_{13}} \Big{)}\Big{(}\lambda_{17}^{*}\gamma^{'}  M^{'*} + c.c. \Big{)}
\ea
\ba
\label{L_{gamma-A1}}
L_{\gamma A_{1} }& =& \Big{(}\frac{b^2\lambda_{7}}{2\lambda_{8}}\e^2\rho^{2} A_{x}-\frac{\lambda_{12}\lambda_{15}}{\lambda_{13}}\Big{)} \Big{(}\gamma  \delta A_{1}^{*} + c.c. \Big{)} +\Big{(}\frac{b^2}{2a^2}\frac{i\lambda_{4}}{\lambda_{2}}f^{2}k_{x}-\frac{\lambda_{7}\lambda_{10}}{\lambda_{8}} -\frac{\lambda_{12}\lambda_{16}} {\lambda_{13}}\Big{)}\Big{(}\gamma \delta A_{1}^{'*} + c.c. \Big{)}\nonumber\\
&+&\Big{(}\frac{ab^3\bar\lambda_{3}}{8\bar\lambda_{2}\lambda_{8}}\e^2\rho^{2} A_{x}k_{x}^2k_{y}^2-\frac{\lambda_{11}\lambda_{15}}{\lambda_{13}}\Big{)}
\Big{(}\gamma^{'} \delta A_{1}^{*} + c.c. \Big{)}+\Big{(}\frac{b^5}{16a\lambda_{1}\lambda_{2}}\e^2\rho^{2} A_{x}f^2k_{x}^4k_{y}^2-\frac{ab\bar\lambda_{3}\lambda_{10}}{4\bar\lambda_{2}\lambda_{8}}k_{x}^2k_{y}^2\nonumber\\
&-&\frac{\lambda_{11}\lambda_{16}}{\lambda_{13}} \Big{)}\Big{(}\gamma^{'} \delta A_{1}^{'*} + c.c. \Big{)}
\ea
and
\ba
\label{L_{M-A1}}
L_{M A1}& =& \Big{(}-\frac{1}{2}f^2k_{x}k_{y}^2 + \frac{b\lambda_{15}}{2a\lambda_{13}}f^2 A_{x}^{'}k_{x}k_{y}^2 \Big{)} \Big{(}i\delta A_{1} M^{*} + c.c. \Big{)} - \Big{(} \frac{b^2}{4a^2\lambda_{2}}f^{4}k_{x}k_{y}^2\Big{)}\Big{(}i\delta A_{1}^{'*} M^{'} + c.c. \Big{)}\nonumber\\
&-&\frac{\lambda_{10}}{2\lambda_{2}\lambda_{8}}f^2k_{y}^2\Big{(}\lambda_{3}\delta A_{1}^{'} M^{'*} + c.c. \Big{)}-\Big{(}\frac{\lambda_{16}}{\lambda_{13}} \Big{)}\Big{(}\lambda_{17}^{*}\delta A_{1}^{'}  M^{'*} + c.c. \Big{)}
+\Big{(}\frac{b^2}{4\lambda_{2}\lambda_{8}}\e^2\rho^{2}A_{x}f^2k_{y}^2\Big{)}\nonumber\\
&&\Big{(}\lambda_{3}\delta A_{1} M^{'*} + c.c. \Big{)}-\frac{\lambda_{15}}{\lambda_{13}}\Big{(}\lambda_{17}^{*}\delta A_{1} M^{'*} + c.c. \Big{)}
+\frac{b\lambda_{16}}{2a\lambda_{13}}f^2A_{x}^{'}k_{x}k_{y}^2\Big{(}i\delta A_{1}^{'} M^{*} + c.c. \Big{)} \nonumber\\
\ea
The parameters $\lambda_i$ and $\bar \lambda_i$ which are introduced after integrating out the non-dynamical fields  are defined via

\ba
\label{lambdak1}
\bar \lambda_1 &&= \frac{b^2}{2 a^2} k^2 f^2 + \frac{\e^2 }{2} b^2 \rho^2 \\
\bar \lambda_2 &&= \frac{a^2}{4} k_x^2 k_y^2 + \frac{\e^2 b^4 k_x^2 k^2}{4 a^2 \bar \lambda_1} f^2  \rho^2 A_x^2  \\
\bar \lambda_3 &&=  b b' k_x^2 - \frac{\e^2 b^4  k_x^2}{4 a^2 \bar \lambda_1}f^2 \rho^2 A_x A_x' \\
\bar \lambda_4 &&=  \frac{a b}{4} \left(\frac{a'}{a}-\frac{b'}{b} \right) k_x^2 k_y^2
+ \frac{\e^2 b^3 k_x^2 k_y^2}{4 a \bar \lambda_1} f^2 \rho^2 A_x A_x' \\
\bar \lambda_5 &&=- \frac{b^2}{2} \rho' k_x^2 + \frac{\e^2 b^4 k_x^2 }{2 \bar \lambda_1 a^2}
f f_{,\rho} \rho^2 A_x A_x'
\\
\lambda_1 &&=   \frac{a^2}{4} k_x^2 k_y^2 + \frac{\e^2 b^2 }{2}k_x^2 \rho^2 A_x^2 \\
\lambda_2 &&=   \frac{b^2 f^2}{2 a^2} k^2 + \frac{\e^2 b^2 a^2}{8 \lambda_1} \rho^2 k_x^2 k_y^2\\
\lambda_3 &&= - \frac{i k_x b^2}{2 a^2} f^2 A_x' + \frac{i \, \e^2  b^3 b'}{2\lambda_1}k_x^3 \rho^2 A_x \\
\lambda_4 &&= -\frac{a \lambda_3}{b} k_y^2 + \frac{i\, \e^2 a b^3}{8 \lambda_1}
\left(\frac{a'}{a}+3\frac{b'}{b} \right)  k_x^3 k_y^2 \rho^2 A_x \\
\lambda_5 &&=  \frac{i b^2}{a^2} k_x f f_{,\rho} A_x' - \frac{i\, \e^2 b^4}{4 \lambda_1}k_x^3\rho^2 \rho' A_x \\
\lambda_6 && = \frac{a b}{2}  \left(\frac{a'}{a}+\frac{b'}{b} \right) k_y^2 + \frac{ a b\bar \lambda_3}{4 \bar \lambda_2} k_x^2 k_y^2
\\
\lambda_7 && =  \frac{a b}{2} k_x^2 k_y^2 - \frac{a b^2 b'}{4 \lambda_1} \left(\frac{a'}{a}-\frac{b'}{b} \right) k_x^4 k_y^2 - \frac{\lambda_3^* \lambda_4}{\lambda_2}
\ea
\ba
\lambda_8 && = - a^2 b^2 V - \frac{b^2 b'^2}{\lambda_1} k_x^4 - \frac{|\lambda_3|^2}{\lambda_2} - \frac{\e^2 b^2}{2} \rho^2 A_x^2
\\
\lambda_9 && =  -\frac{a^2 b^2}{2} V_{,\rho} - \frac{b^2}{2 a^2} f f_{,\rho} A_x'^2 +
\frac{b^3 b'}{2 \lambda_1}\rho' k_x^4 - \frac{\lambda_3^* \lambda_5}{\lambda_2}
- \frac{\e^2 b^2}{2} \rho A_x^2
\\
\lambda_{10} && = - \frac{b^2}{2 a^2} f^2 A_x' - \frac{i  \, b^2 \lambda_3^* }{2 a^2 \lambda_2} f^2 k_x \\
\lambda_{11} && = -\frac{b^2}{4} k_x^2 k_y^2 - \frac{a^2 b^2}{16 \bar \lambda_2} k_x^4 k_y^4 - \frac{a b \bar \lambda_3 \lambda_6^*}{ 4 \bar \lambda_2\lambda_8} k_x^2 k_y^2
\\
\lambda_{12} && =  -\frac{b^2}{4} \left(\frac{a'}{a}-\frac{b'}{b} \right) k_x^2 k_y^2
+ \frac{a b \bar \lambda_4}{4 \bar \lambda_2} k_x^2 k_y^2 - \frac{\lambda_7 \lambda_6^*}{\lambda_8}
\\
\lambda_{13} && = \frac{b^2}{4} k_x^2 k_y^2 - \frac{a^2 b^2}{16 \bar \lambda_2} k_x^4 k_y^4 - \frac{| \lambda_6  |^2}{\lambda_8} \\
\lambda_{14} && = -\frac{a b}{2} \rho' k_y^2 + \frac{a b \bar \lambda_5}{4 \bar \lambda_2}
k_x^2 k_y^2 - \frac{\lambda_9 \lambda_6^*}{\lambda_8} \\
\lambda_{15} && =  - \frac{b}{2 a} f^2 A_x' k_y^2 + \frac{\e^2 b^2 \lambda_6^*}{2 \lambda_8} \rho^2 A_x \\
\lambda_{16} && =  -\frac{\lambda_6^* \lambda_{10}}{\lambda_8}
+ \frac{\e^2  b^5 }{16 a \bar \lambda_1 \bar \lambda_2 }f^2  \rho^2 A_x  k_x^4 k_y^2 \\
\label{lambda17}
\lambda_{17} && =  -\frac{\lambda_3^* \lambda_6^*}{2\lambda_2 \lambda_8} f^2 k_y^2
+\frac{i\, \e^2 a b^3}{16 \bar \lambda_1 \bar \lambda_2} f^2 A_x \rho^2 k_x^3 k_y^4
\ea
One can see that $\bar \lambda_i$ are determined by $\bar \lambda_1$ while $\lambda_i$
depends on both $\bar \lambda_1$ and $\lambda_1$. One can check from the detail processes of integrating out the non-dynamical fields that $\bar \lambda_1$ is obtained from integrating out $\delta A_0$. On the other hand, as we have seen 
in Section \ref{matter-leading}, the leading interactions originate from integrating out the matter sector. As a result, it is expected that $\bar \lambda_1$ plays the dominant role in determining the earliest time in which the interaction $\e^2 \rho^2 A_\mu A^\mu$ becomes comparable to $f^2 F^2$ interaction.  

Here we justify this conclusion specifically. To see this, let us look at $\lambda_1$. Dividing the second term in $\lambda_1$ to the first term in $\lambda_1$ yields 
\ba
\label{lambda1-ratio}
\frac{\e^2 \rho^2}{M_P^2 k^2} A_x^2  \, .
\ea
On the other hand, during most of period of inflation $\partial_t (\dot A f^2 e^{\alpha})=0 $
so $\dot A_x \sim e^{-\alpha } f^{-2} \sim e^{3\alpha}$. As a result $A_x \sim \dot A_x/3 H$.
Using the relation $\dot A_x^2 \sim (I \epsilon_H) e^{2 \alpha } f^{-2}$ from the attractor solution  we obtain  $A_x^2 \sim (I \epsilon_H) b^2 M_P^2/f^2$. Plugging this value of $A_x^2$ in the ratio Eq. (\ref{lambda1-ratio}) above yields 
\ba
(I \epsilon_H) \frac{\e^2 b^2 \rho^2}{k^2 f^2} \, .
\ea
Up to the pre-factor $I \epsilon_H \ll 1$ this ratio is the same as the ratio one obtains in comparing the second term in $\bar \lambda_1$ to the first term in $\bar \lambda_1$.
Now if we define ${\eta_c'}$ as the time when the second term in $\lambda_1$ becomes comparable  to the first term in $\lambda_1$, then $\eta_c \simeq (I \epsilon_H)^{-1/6}  \eta_c'$. Noting that $\eta<0$, we conclude that  $\eta_c \ll \eta_c' $. As a result the interaction $\e^2 \rho^2 A_\mu A^\mu$ becomes comparable to $f^2 F^2$ sooner in $\bar \lambda_1$ than in $\lambda_1$. Now, since the rest of $\lambda_i$ and $\bar \lambda_i$ are controlled by  either $\bar \lambda_1$ or $\lambda_1$, then we conclude that the earliest time when $\e^2 \rho^2 A_\mu A^\mu$  becomes comparable to $f^2 F^2$ is determined by $\bar \lambda_1$ as we used to fix $\eta_c$. \\ 
\chapter{}
\section{Second Order Slow-roll Action}
\label{second slow}

In this appendix we calculate the second order action for the canonical fields in the slow roll approximation. Following the discussions  in Section \ref{sec-ac} we divide the dynamical action into two different regions depending on whether the charge $\e$ is important or not.
In the first phase, $\eta < \eta_c$, the charge effect is sub-dominant while in the
second phase, $\eta > \eta_c$, its effect is dominant and the longitudinal mode, as we will introduce it in Eq. (\ref{D12}), has the same contribution as the transverse mode. In the following, first we write the action in the first phase and then we go to the second phase.

\subsubsection{First Phase}

Our goal is to write down the action in terms of the free Lagrangians plus the interaction terms.
Using the slow-roll approximation given in Eq. (\ref{slow roll5}),  and taking $I \ll 1$ as mentioned in the main text, to leading orders in slow parameters and $I$ we have
\begin{align}
\label{Total scalar action1}
S_{2}^{(1)} = \int d\eta d^3k \bigg{(} L_{\rho \rho} + L_{\gamma\gamma} + L_{MM} + L_{A_{1}A_{1}} + L_{\rho\gamma} +   L_{\rho M} +  L_{\rho A_{1}} +  L_{\gamma A_{1}} +  L_{\gamma M} + L_{ A_{1} M} \bigg{)}~,
\end{align}
Where
\begin{align}
\label{Lrhorho}
L_{\rho \rho} & = \left(\frac{b^2}{2} \right) \Big| \delta \rho ^{'}\Big|^2 + \left(\frac{b^2}{2} \right)\bigg{(}-k^2+ (-\eta)^{-2}\bigg{(}6\epsilon_{H}-6\frac{\eta_{H}}{1-I}-12\frac{I}{1-I}(1-2\sin^2{\theta})\nonumber\\
&~~~+3\frac{\e^2 \cM^4}{k^2\lambda^2}\frac{a^2}{f^{2}}I\epsilon_{H}\cos^2{\theta}\bigg{)}\bigg{)}\Big| \delta \rho \Big|^2~~,\\
\label{Lgammagamma}
L_{\gamma \gamma} & = \left(\frac{b^2}{2a}k^2 \sin^2{\theta}\cos^2{\theta}\right)^2\bigg{(} 1 + 6\epsilon_{H}\bigg{)} \Big| \gamma ^{'}\Big|^2 -\left(\frac{b^2}{2a}k^3 \sin^2{\theta}\cos^2{\theta}\right)^2 \Big| \gamma \Big|^2~~,\\
\label{Lrhogamma}
L_{\rho \gamma} & = \left( \frac{\e^2}{8}\frac{b^3 \eta_{e}^4}{\eta^5}\frac{\cM^2}{\lambda^2 M_{P}}k^2 \cos^2{\theta} \sin^4{\theta} \sqrt{3\lambda} I \epsilon_{H}^{3/2} \right) \left( \delta \rho^{*} \gamma + c.c. \right) - \frac{3\sqrt{2}}{2} \frac{b^3}{a\eta}k^2 \cos^2{\theta} \sin^4{\theta} I\sqrt{\epsilon_{H}}\left( \delta \rho^{*} \gamma' + c.c. \right) \nonumber\\
&~~~-\frac{\sqrt{2}}{4} \frac{b^3}{a} k^2 \cos^2{\theta} \sin^2{\theta} \sqrt{\epsilon_{H}} \left( \delta \rho^{'*} \gamma' + c.c. \right) 
-\left( \frac{\e^2}{8}\frac{b^3 \eta_{e}^4}{\eta^6}\frac{\cM^2}{M_{P}\lambda^2} \cos^2{\theta} \sin^4{\theta} \sqrt{3\lambda}\left( 1 + 3\sqrt{2} \right) I \epsilon_{H}^{3/2} \right) \nonumber\\
&~~~\left( \delta \rho^{*} \gamma' + c.c. \right)
\end{align}
\begin{align}
\label{LMM}
L_{MM} & = \left(\frac{b^2}{2a^2}k^2f^2 \sin^2{\theta} \cos^2{\theta} + \frac{\e^2 \cM^4}{16\lambda^2}b^2 \sin^4{\theta}\frac{\epsilon_{H}}{1-I}\right)\Big| \cM^{'}\Big|^2 +\bigg{(}-\frac{b^2}{2a^2}k^4f^2 \sin^2{\theta} \cos^2{\theta}\nonumber\\
&~~~-\frac{\e^2 \cM^4}{16\lambda^2}b^2 k^2 \sin^2{\theta}\frac{\epsilon_{H}}{1-I}\bigg{)}\bigg{)}\Big| M \Big|^2~~,
\\
\label{LA1A1}
L_{A1 A1} & = \left(\frac{b^2}{2a^2}f^2 \sin^2{\theta} + \frac{\e^2 \cM^4}{16k^2\lambda^2}b^2 \cos^2{\theta}\frac{\epsilon_{H}}{1-I}\right)\Big| \delta A_{1}^{'}\Big|^2 +\bigg{(}-\frac{b^2}{2a^2}k^2f^2 \sin^2{\theta}\nonumber\\
&~~~-\frac{\e^2 \cM^4}{16\lambda^2}b^2\frac{\epsilon_{H}}{1-I}\bigg{)}\bigg{)}\Big|\delta A_{1} \Big|^2~~,
\\
\label{LrhoA1}
L_{\rho A1} & = \left(\frac{1}{\eta}\right) \left(\frac{fb^2}{a}\sqrt{6I} \sin^2{\theta} + \frac{\e^2 \cM^4}{8k^2\lambda^2}\frac {ab^2}{f} \epsilon_{H}\cos^2{\theta}\right)\Big{(} \delta \rho^{*}\delta A_{1}^{'} + c.c.\Big{)}-\e^2b^2 \rho A_{x}\Big{(}\delta \rho \delta A_{1}^{*} + c.c. \Big{)}~~,\\
\label{L_{rho-M}}
L_{\rho M} & = \left(\frac{1}{\eta}\right) \left(-\frac{fb^2}{a}\sqrt{6I} \sin^2{\theta}(k\cos{\theta}) + \frac{\e^2 \cM^4}{8k\lambda^2}\frac {ab^2}{f} \epsilon_{H} \sin^2{\theta}\cos{\theta}\right)\Big{(} \delta \rho^{*} M^{'} + c.c.\Big{)}~~\\
\label{LMgamma}
L_{\gamma M} & = -\left(\frac{fb^3}{4a^2}k^5\sqrt{3I\epsilon_{H}}\sin^4{\theta}\cos^3{\theta}\right)\Big{(}i\gamma^{*}M + c.c.)\Big{)}-\left(\frac{b^3}{2a^2}\frac{f}{\eta}k^3\sqrt{3I\epsilon_{H}}\sin^2{\theta}\cos^3{\theta}\right)\nonumber\\
&\Big{(}i\gamma^{*}M^{'} + c.c.)\Big{)}
-\left(\frac{b^3}{4a^2}\frac{f}{\eta}k^3\sqrt{3I\epsilon_{H}}\sin^2{\theta}\cos^3{\theta}(1+\cos^2{\theta})\right)\Big{(}i\gamma^{'*}M + c.c.)\Big{)}\nonumber\\
&+\left(\frac{fb^3}{4a^2}k^3\sqrt{3I\epsilon_{H}}\sin^4{\theta}\cos^3{\theta}\right)\Big{(}i\gamma^{'*}M^{'} + c.c.)\Big{)}~~,\\
\label{LA1gamma}
L_{\gamma A_{1}} & =  \left(\frac{fb^3}{4a^2}k^4\sqrt{3I\epsilon_{H}}\sin^4{\theta}\cos^2{\theta}\right)\Big{(}\gamma^{*}\delta A_{1} + c.c.)\Big{)}+\left(\frac{b^3}{2a^2}\frac{f}{\eta}k^2\sqrt{3I\epsilon_{H}}\sin^2{\theta}\cos^2{\theta}\right)\nonumber\\
&\Big{(}\gamma^{*}\delta A_{1}^{'} + c.c.)\Big{)}
+\left(\frac{b^3}{4a^2}\frac{f}{\eta}k^2\sqrt{3I\epsilon_{H}}\sin^2{\theta}\cos^2{\theta}(1+\cos^2{\theta})\right)\Big{(}\gamma^{'*}\delta A_{1} + c.c.)\Big{)}\nonumber\\
&-\left(\frac{fb^3}{4a^2}k^2\sqrt{3I\epsilon_{H}}\sin^4{\theta}\cos^2{\theta}\right)\Big{(}\gamma^{'*}\delta A_{1}^{'} + c.c.)\Big{)}~~,\\
\label{LA1M}
L_{A_{1}M} & = \left(\frac{f^2b^2}{2a^2}k^3\sin^2{\theta}\cos{\theta}\right)\Big{(}i\delta A_{1}^{*}M + c.c.)\Big{)}+
\Big{(}-\frac{f^2b^2}{2a^2}k\sin^2{\theta}\cos{\theta}\nonumber\\
&+b^2\frac{\e^2 \cM^4}{16k\lambda^2}\frac{\epsilon_{H}}{1-I}\sin^2{\theta}\cos{\theta}\Big{)}\Big{(}iM^{'}\delta A_{1}^{'*} + c.c.)\Big{)}~~,
\end{align}
Now looking at  Eq. (\ref{LA1M}) we can easily see that this term is not as small as the other interaction terms. Actually this term seems to be of the same order as our free field action. This means that  $M $ and $\delta A_{1}$ are not the physical fields. One should consider a rotation in $ \{M, \delta A_{1} \}$ space such that all of the interaction terms become small compared to the free field action.
One can easily check that the following two new fields $D_1$ and $D_2$ work for us in the sense that they do not mix with each other and all of the interaction terms would be small:
\ba
\label{D12}
D_{1}&\equiv \delta A_{1} -ik \cos{\theta}M \\
D_{2}&\equiv \cos{\theta} \delta A_{1} + ik \sin^2{\theta}M \, .
\ea

One can check that  $D_{1}$ represents  the transverse polarization while $D_{2}$ is for the longitudinal polarization of the gauge field perturbations $\delta A_\mu$. Now the different parts of the action can be rewritten in terms of these two new fields  as
\begin{align}
L_{MM} +L_{\delta A_{1}\delta A_{1}} +L_{M\delta A_{1}} &= \frac{b^2 f^2}{2 a^2}\sin^2{\theta}\left( |D_{1}'|^2 -\left( k^2 + \frac{\e^2 \cM^4}{8 \lambda^2 } \frac{a^2}{f^2}\frac{\epsilon_{H}}{1-I} \right)|D_{1}|^2 \right)\nonumber\\
&+ \frac{\e^2 \cM^4}{16 \lambda^2 k^2}\frac{\epsilon_{H}b^2}{1-I}\left( |D_{2}'|^2 - k^2 |D_{2}|^2 \right)~~,\\
\label{intD1-rho}
L_{\delta \rho M}+L_{\delta \rho \delta A_{1}} &= \left(\frac{1}{\eta}\right)\frac{b^2}{a} \sqrt{6I} \sin^2{\theta} f \Big{(} \delta \rho^{*} D_{1}^{'} + c.c.\Big{)} - \left( \frac{a^2}{f\eta} \right) \e^2 \sqrt{\frac{I \epsilon_{H}^2}{\lambda}}M_{P} \sin^2{\theta}\Big{(} \delta \rho^{*} D_{1} + c.c.\Big{)}\nonumber\\
& - \left( \frac{a^2}{f\eta} \right) \e^2 \sqrt{\frac{I \epsilon_{H}^2}{\lambda}}M_{P} \cos{\theta}\Big{(} \delta \rho^{*} D_{2} + c.c.\Big{)}~~,
\end{align}
\begin{align}
\label{intD1-gamma}
L_{\gamma M}+L_{\gamma \delta A_{1}} &=
\left(\frac{b^3}{4a^2}k^4 f\sqrt{3I\epsilon_{H}}\sin^4{\theta}\cos^2{\theta}\right)\Big{(}\gamma^{*} D_{1} + c.c.)\Big{)}+\left(\frac{b^3}{2a^2}\frac{1}{\eta}k^2f\sqrt{3I\epsilon_{H}}\sin^2{\theta}\cos^2{\theta}\right)\nonumber\\
&\Big{(}\gamma^{*} D_{1}^{'} + c.c.)\Big{)}
+\left(\frac{b^3}{4a^2}\frac{1}{\eta}k^2f\sqrt{3I\epsilon_{H}}\sin^2{\theta}\cos^2{\theta}(1+\cos^2{\theta})\right)\Big{(}\gamma^{'*}D_{1} + c.c.)\Big{)}\nonumber\\
&-\left(\frac{b^3}{4a^2}k^2f\sqrt{3I\epsilon_{H}}\sin^4{\theta}\cos^2{\theta}\right)\Big{(}\gamma^{'*}\delta A_{1}^{'} + c.c.)\Big{)}~~ \, ,
\end{align}
where in Eq. (\ref{intD1-rho}) and Eq. (\ref{intD1-gamma}) only the leading terms have been written. It can easily be seen that since the charge effect is not important in this phase, the longitudinal mode is sub-leading.
Now we can define the canonical variable as,
\begin{align}
\label{canonical variables}
\overline{\delta \rho} &\equiv b \delta \rho \\
\overline{\gamma}&\equiv \frac{a}{\sqrt{2}}\frac{k_{x}^2 k_{y}^2}{k^2} \gamma \\
\overline{D_{1}}&\equiv \frac{b}{a}f \sin{\theta} D_{1} \\
\overline{D_{2}}&\equiv \frac{\e \cM^2}{2\sqrt{2}\lambda M_{P} k}\sqrt{\frac{\epsilon_{H}}{1-I}} bD_{2}
\end{align}
One can write the action in terms of the canonical variables. Due to technical reasons we only express the free Lagrangians in terms of the canonical variables while the
interaction terms may be expressed in terms of the old variables
\begin{align}
\label{rho rho action}
L_{\rho \rho} &= \frac{1}{2}|\overline{\delta \rho}'|^2 + \frac{1}{2}\bigg{[} -k^2 + (-\eta)^{-2}\bigg{(}2+9\epsilon_{H}-6\frac{\eta_{H}}{1-I}
-12\frac{I}{1-I}(1-2\sin^2{\theta})\bigg{)}\bigg{]}|\overline{\delta \rho}|^2\\
\label{gamma gamma action}
L_{\gamma \gamma} &= \frac{1}{2}|\overline{\gamma}^{'}|^2 + \frac{1}{2}\bigg{[} -k^2 + (-\eta)^{-2}\bigg{(}2 + 15\epsilon_{H} +I\epsilon_{H}(-6+11\sin^2{\theta} + \cot^2{\theta})\bigg{)}\bigg{]}|\overline{\gamma}|^2 \\
\label{D1 D1  action}
L_{D_{1} D_{1}} &= \frac{1}{2}|\overline{D_{1}}^{'}|^2 + \frac{1}{2}\bigg{[} -k^2 + (-\eta)^{-2}\bigg{(}2+9\epsilon_{H}-3\frac{\eta_{H}}{1-I}\bigg{)}\bigg{]}|\overline{D_{1}}|^2\\
\label{D2 D2  action}
L_{D_{2} D_{2}} &= \frac{1}{2}|\overline{D_{2}}^{'}|^2 + \frac{1}{2}\bigg{[} -k^2 + (-\eta)^{-2}\bigg{(}2+3\epsilon_{H} + I\epsilon_{H}\bigg{)}\bigg{]}|\overline{D_{2}}|^2\\
\end{align}
These are the final results for the action in the first phase used in Eqs. (\ref{rhorho action1})-
(\ref{D2D2 action1}).

\subsubsection{Second Phase}
In this part we  write the leading order action in the second phase from which we can read off the canonical fields. Considering the dominant effects of $\e$ in $\bar \lambda_1$ as mentioned in Section {\ref{sec-ac}} yields
\begin{align}
\label{rho rho action2}
L_{\rho \rho} &= \frac{b^2}{2}|\delta \rho'|^2 -\left(\frac{\e^2 I \epsilon_{H} \lambda }{\cM^4}\right) \left(\frac{b^2}{f^{2}\eta^{2}}\right)|\delta \rho|^2\\
\label{D1 D1 action2}
L_{D_{1} D_{1}} &= \frac{b^2}{2a^2}f^2\sin^2{\theta}|D_{1}'|^2 -\left(\frac{ a^2 \e^2 \epsilon_{H} \cM^4}{16\lambda^2 f^2}\right)\left(\frac{b^2}{a^2}f^2\sin^2{\theta}\right)|D_{1}|^2\\
\label{D2 D2 action2}
L_{D_{2} D_{2}} &= \frac{b^2}{2a^2}f^2|D_{2}'|^2 -\left(\frac{ a^2 \e^2 \epsilon_{H}\cM^4}{16\lambda^2 f^2}\right)\left(\frac{b^2}{a^2}f^2\right)|D_{2}|^2\\
\label{rhoD1 action2}
L_{\rho D_{1}} &= \left(\frac{1}{\eta}\right)\frac{b^2}{a} \sqrt{6I} \sin^2{\theta} f \Big{(} \delta \rho^{*} D_{1}^{'} + c.c.\Big{)} - \left( \frac{a^2}{f\eta} \right) \e^2 \sqrt{\frac{I \epsilon_{H}^2}{\lambda}}M_{P} \sin^2{\theta}\Big{(} \delta \rho^{*} D_{1} + c.c.\Big{)}\\
\label{rhoD2 action2}
L_{\rho D_{2}} &= \left(\frac{1}{\eta}\right)\frac{b^2}{a} \sqrt{6I} \cos{\theta} f \Big{(} \delta \rho^{*} D_{2}^{'} + c.c.\Big{)}- \left( \frac{a^2}{f\eta} \right) \e^2 \sqrt{\frac{I \epsilon_{H}^2}{\lambda}}M_{P} \sin^2{\theta}\Big{(} \delta \rho^{*} D_{2} + c.c.\Big{)}\\
L_{\rho \gamma} & = \left( \frac{\e^2}{8}\frac{b^3 \eta_{e}^4}{\eta^5}\frac{\cM^2}{\lambda^2 M_{P}}k^2 \cos^2{\theta} \sin^4{\theta} \sqrt{3\lambda} I \epsilon_{H}^{3/2} \right) \left( \delta \rho^{*} \gamma + c.c. \right) - \frac{3\sqrt{2}}{2} \frac{b^3}{a\eta}k^2 \cos^2{\theta} \sin^4{\theta} I\sqrt{\epsilon_{H}}\times  \nonumber\\
&~~~ \times \left( \delta \rho^{*} \gamma' + c.c. \right)-\frac{\sqrt{2}}{4} \frac{b^3}{a} k^2 \cos^2{\theta} \sin^2{\theta} \sqrt{\epsilon_{H}} \left( \delta \rho^{'*} \gamma' + c.c. \right) 
\end{align}
Now from the above equations, we can find the canonical variables in the second phase as,
\begin{align}
\label{canonical variables2 appen}
\overline{\delta \rho}_{k} &\equiv b \delta \rho_{k} \\
\overline{D_{1k}}&\equiv \frac{b}{a} \sin{\theta} f D_{1k} \\
\overline{D_{2k}}&\equiv \frac{b}{a} f D_{2k}
\end{align}
as used in Eq. (\ref{canonical variables2}).

\chapter{}
\section{Outgoing Mode Functions}
\label{outgoing modes}

In this appendix we calculate the mode function during the second phase, $\eta< \eta_c < \eta_e$.

The canonical inflaton field mode function, $\overline{\delta \rho}_{\textbf{k}}$, associated
with the free inflaton Lagrangian given in Eq. (\ref{rhorho action5})  is
\ba
\label{mode function-u}
\overline{\delta \rho}_{\textbf{k}} &=& u^{(m)}_{\textbf{k}}a_{\rho\textbf{k}} + u^{(m)*}_{(-\textbf{k})}a^{\dag}_{\rho(-\textbf{k})} \nonumber \\
u^{m}_{\textbf{k}} &=&\sqrt{-\eta}\left[ c_{1k}H_{3/4}^{(1)}\left( \frac{\sqrt{\Omega}}{2\eta^2 }\right) + c_{2k}H_{3/4}^{(2)}\left( \frac{\sqrt{\Omega}}{2\eta^2 }\right)\right]~~~,~~~ \Omega \equiv \frac{2\e^2I\epsilon_{H}\lambda M_P^4}{\cM^4}\eta_{e}^4.
\ea
Here $H_{3/4}^{(1, 2)} ( x)$ are the Hankel functions with index $\frac{3}{4}$ and $c_{i \,  k}$ are constant of integrations to be found by matching conditions at $\eta =\eta_c$.

Similarly the canonical transverse mode function,  $\overline{D_{1\textbf{k}}}$, associated with the Lagrangian $L_{D_1 D_1}$ in Eq. (\ref{D1D1 action5}) is
\ba
\label{mode function-v}
\overline{D_{1\textbf{k}}} &=& \left(\frac{b}{a} \sin{\theta} f\right) \left(v^{(m)}_{\textbf{k}}b_{\rho\textbf{k}} + v^{(m)*}_{(-\textbf{k})}b^{\dag}_{\rho(-\textbf{k})} \right) \nonumber \\
v^{m}_{\textbf{k}} &=&\left( \frac{a \sqrt{-\eta}}{b\sin{\theta} f}\right)\left[ d_{1k}H_{3/4}^{(1)}\left( \frac{\sqrt{\Delta}}{2\eta^2 }\right) + d_{2k}H_{3/4}^{(2)}\left(\frac{\sqrt{\Delta}}{2\eta^2 }\right)\right]~~~,~~~ \Delta \equiv \frac{3\e^2\epsilon_{H}}{2\lambda}\eta_{e}^4.
\ea
Finally the $\overline{D_{2\textbf{k}}}$ mode function is
\ba
\label{mode function-w}
\overline{D_{2\textbf{k}}} &=&\left(\frac{b}{a}f\right) \left( w^{(m)}_{\textbf{k}}c_{\rho\textbf{k}} + w^{(m)*}_{(-\textbf{k})}c^{\dag}_{\rho(-\textbf{k})} \right)\nonumber \\
w^{(m)}_{\textbf{k}} &=&\left(\frac{a \sqrt{-\eta}}{bf}\right) \left[ e_{1k}H_{3/4}^{(1)}\left( \frac{\sqrt{\Delta}}{2\eta^2 }\right) + e_{2k}H_{3/4}^{(2)}\left( \frac{\sqrt{\Delta}}{2\eta^2 }\right)\right].
\ea
Our goal is to find the constants of integration $c_{ik},d_{ik},e_{ik}~ ,~ i=1,2$ with the appropriate matching conditions at $\eta = \eta_c$.
To impose the matching conditions we require that the original fields $\delta \rho, M, A_1$ and their derivatives  to be continuous at $\eta= \eta_c$. With the incoming mode functions given by Eq. (\ref{mode function0}) and after  imposing the matching conditions one obtains
\begin{align}
\label{c12}
c_{1,2} &= \frac{\pi}{8i}\frac{e^{-ik\eta_{c}}}{\sqrt{-2k\eta_{c}}}\left(\pm H^{(2,1)}_{3/4}\left(\frac{\sqrt{\Omega}}{2\eta_{c}^2 }\right)\left(3-\frac{3i}{k\eta_{c}}+ ik\eta_{c}\right)\mp H^{(2,1)}_{-1/4}\left(\frac{\sqrt{\Omega}}{2\eta_{c}^2 }\right)\frac{\sqrt{\Omega}}{\eta_{c}^2 }\left(1-\frac{i}{k\eta_{c}}\right)\right),\\
\label{d12}
d_{1,2}& = \frac{\pi}{8i}\frac{e^{-ik\eta_{c}}}{\sqrt{-2k\eta_{c}}}\left(\pm H^{(2,1)}_{3/4}\left(\frac{\sqrt{\Delta}}{2\eta_{c}^2 }\right)\left(3-\frac{3i}{k\eta_{c}}+ ik\eta_{c}\right)\mp H^{(2,1)}_{-1/4}\left(\frac{\sqrt{\Delta}}{2\eta_{c}^2 }\right)\frac{\sqrt{\Delta}}{\eta_{c}^2 }\left(1-\frac{i}{k\eta_{c}}\right)\right),\\
\label{e12}
e_{1,2} &= \frac{\pi}{8i}\frac{e^{-ik\eta_{c}}}{\sqrt{-2k\eta_{c}}}\left(\pm H^{(2,1)}_{3/4}\left(\frac{\sqrt{\Delta}}{2\eta_{c}^2 }\right)\left( ik\eta_{c}\right)\mp H^{(2,1)}_{-1/4}\left(\frac{\sqrt{\Delta}}{2\eta_{c}^2 }\right)\frac{\sqrt{\Delta}}{\eta_{c}^2 }\left(1-\frac{i}{k\eta_{c}}\right)\right).
\end{align}
This fixes the form of the outgoing mode functions. However, as discussed at the end of Section \ref{sec-ac}, during most of the period of the second phase the arguments of the Hankel functions above are smaller than unity so the mode functions to a good approximation follow the profile of a massless scalar field. As a result, in our In-In integrals we can use the mode functions as given in Eq. (\ref{mode function0}).

\chapter{}
\section{Gradient Expansion Ordering of Perturbations}
\label{App-B}

In this section we estimate the ordering of $\beta^i$ and  $\delta q^\mu$ and calculate the contributions of
$\delta q^\mu$ and $\delta \pi^{\mu}_{\nu}$ in the energy conservation equation,
Eq. (\ref{cont}).

First of all let us check the transverse conditions on the heat flow and anisotropic pressure. By definition one has
\ba
\label{trans-cond}
u^{\mu} q_{\mu} =0 \qquad  , \quad  u^{\mu} \pi^{\nu}_{\, \, \mu} =0.
\ea
The fluid's 4-velocity  can be read as 
\ba
\label{u-up-dn}
u^{\mu}=\left[\dfrac{1}{\cal N}, \vec{0} \right]+{\cal O}(\epsilon^2)  \quad ,\qquad u_{\mu}=\left[-{\cal N}, \dfrac{\beta_i}{\cal N}\right]+{\cal O}(\epsilon^2).
\ea
From the background equations we conclude  that $\bar q^\mu$ and $\bar \pi^{0}_{\mu}$ are zero. Now using the transverse condition \eqref{trans-cond} one concludes that to all order
\ba
\delta q_0 =0 \quad , \quad \delta \pi^{\nu}{}_{0} =0
\ea
This also yields  $\delta q^{0} = {\cal O} (\delta^2)$.
For the ordering of $\delta \pi^{0}{}_{i}$ one has
\ba
\label{pi-0-i}
\delta \pi^{0}{}_{i}= a_i^2 \delta \pi^i{}_{0} + \beta_i \pi^{i}{}_{i}  = \beta_i \pi^{i}{}_{i}  \, .
\ea
We will use this equation later in order to find the ordering of the gradient expansion of perturbations.
 
Let us now look at the gradient expansion ordering of $\beta^{i}$. 
For this we look at $\delta G^{0}{}_{i}$ and $\delta G^i{}_0$ components of Einstein equations. With some efforts one can show that 
\ba
\delta G^{0}{}_{i} &=& \epsilon {\cal O} (\delta) + \beta {\cal O}(\delta) 
\\
\nonumber
\delta G^i{}_0 &=& \beta^i \left(3 \bar H_i \bar H- \sum_i \bar H_i^2 - 3\dot{\bar H}+ \dot{\bar H}_i  \right) + \epsilon {\cal O} (\delta) + \beta {\cal O}(\delta) 
\\
&=&\beta^i \left(\bar{R}^i{}_{i} -\bar{R}^0{}_{0} \right) + \epsilon {\cal O} (\delta) + \beta {\cal O}(\delta)  .
\ea
The easiest way to see this is to adopt the local inertial frame in which $\Gamma^\alpha_{\beta \gamma} =0$. 
Therefore, the corresponding Einstein equations yield
\ba
\label{G-0-i}
  \delta T^{0}{}_{i}&=& \epsilon {\cal O} (\delta) + \beta {\cal O}(\delta) 
\\
\label{G-i-0}
 \delta T ^i{}_{0}&=& \beta^i \left(\bar{R}^i{}_{i} -\bar{R}^0{}_{0} \right) + \epsilon {\cal O} (\delta) + \beta {\cal O}(\delta)  .
\ea
Similarly, using the spatial components of energy momentum conservation one can put limits on $\delta T^{0}{}_i$. The continuity equation $\nabla_{\mu} T^{\mu}_i$ to leading order yields
\ba
\label{T-i-0-con}
\nonumber
(\partial_0 +3H-H_i) \delta T^{0}{}_i -a_i^2 H_i \delta T^{i}{}_0 &=& -H_i \beta_i \left(\bar{T}^i{}_i - \bar{T}^0{}_0  \right) +\epsilon {\cal O} (\delta) + \beta {\cal O}(\delta)
\\
&=&- H_i \beta_i \left(\bar{R}^i{}_i - \bar{R}^0{}_0  \right)+\epsilon {\cal O} (\delta) + \beta {\cal O}(\delta) \, .
\ea
One can show that the expression, $\bar{T}^i{}_i - \bar{T}^0{}_0  = \bar{R}^i{}_i - \bar{R}^0{}_0  $  is non-vanishing in general and is of the order of $\dot{H}$. 

Plugging Eqs. (\ref{G-0-i}) and (\ref{G-i-0}) into continuity equation Eq. (\ref{T-i-0-con}) yields 
 \ba
\label{T-0-i-con}
(\partial_0+3H-H_i) \delta T^{0}{}_{i} &=\epsilon {\cal O} (\delta) + \beta {\cal O}(\delta) .
\ea
This equation shows that  $\delta T^{0}{}_{i}$ has decaying  solutions approximately like $1/a^{2}$. So one can readily deduce that  $\delta T^{0}{}_{i}$ should be higher order in gradient expansion as
\ba
\label{T-0-i-order}
\delta T^{0}{}_{i} &=\epsilon^2 {\cal O} (\delta) + \epsilon \beta {\cal O}(\delta)  \, .
\ea
Before discussing about the consequences of the above equation it is more convenient to rephrase Eq. (\ref{G-i-0}) as follows
\ba
\label{G-i-0-2}
-\delta q_i &=& \beta_i \left(\bar{T}^i{}_{i} -\bar{T}^0{}_{0} \right)   + \epsilon {\cal O} (\delta) + \beta {\cal O}(\delta) ,
\ea
Furthermore  for   $\delta T^{0}{}_{i}$  we have
\ba
\label{G-0-i-2}
\delta T^{0}{}_{i}&=& (\bar{\rho}+\bar{p}) \beta_i -\delta q_i +\delta \pi^{0}{}_{i}+\epsilon {\cal O} (\delta) + \beta {\cal O}(\delta)  
\ea
By using Eq. \eqref{G-i-0-2} to eliminate $\delta q^i$ in favor of $\beta^i$ and Eq. \eqref{pi-0-i} to express $\delta \pi^{0}{}_{i}$ as a function of $\beta^i$  one can show that the leading order terms of $\beta^i$ cancel each other and one obtains
\ba
\label{G-0-i-3}
\delta T^{0}{}_{i}&=& \beta^i \, {\cal O}(\delta)+\epsilon \,{\cal O} (\delta) + \beta\, {\cal O}(\delta)  \, .  
\ea
On the other hand, comparing Eq. (\ref{T-0-i-order}) with Eq. (\ref{G-0-i-3}), one obtains the following result for the ordering of $\beta^i$
\ba
\beta^{i} &={\cal O} (\epsilon) ,
\ea
This also yields 
\ba
\label{ordering}
\delta q_i= {\cal O} (\epsilon) 
\\
\label{ordering2}
\delta \pi^{0}{}_{i}= {\cal O} (\epsilon) 
\ea
Now we investigate the contribution of heat flow in continuity equation Eq. \eqref{cont}. Using Eq. \eqref{ordering} and Eq. \eqref{ordering2} one finds that $\delta q^{\mu} \sim {\cal O} (\epsilon )$ and by noting that the background value of $q^{\mu}$ is also zero, we get
\be
\label{heat-con} 
-u_\mu u^\nu \nabla_\nu q^\mu + \nabla_\mu q^\mu = {\cal O} (\epsilon\delta,\epsilon^2 )
\ee
As a result one can deduce that heat conduction can be ignored in the continuity equation at the first order of perturbations and gradient expansion.
\\
Now it is time to calculate the contribution of anisotropic pressure on the continuity equation
\ba
u^{\mu} \nabla_{\nu} \pi^{\nu}{}_{\mu} &=& \bar{u}^{\mu} \bar{\nabla}_{\nu} \bar{\pi}^{\nu}{}_{\mu} +\delta (u^{\mu} \partial_{\nu} \pi^{\nu}{}_{\mu}+u^{\mu} \Gamma^{\nu}_{\nu \rho} \pi^{\rho}{}_{\mu} - u^{\mu} \Gamma^{\rho}_{\mu \nu } \pi^{\nu}{}_{\rho}) 
\ea
Noting that $u^{\mu}=\left[1/{\cal N}, \vec{0} \right]+{\cal O}(\epsilon^2)$ and  $\delta \pi^{\nu}{}_{0} =0$ to the all orders of perturbations $\delta$, one has
\ba
\label{aniso-pres-con}
u^{\mu} \nabla_{\nu} \pi^{\nu}{}_{\mu} = \dfrac{-1}{\cal N} H_i (\mathbf{x},t)  \pi^i_i (\mathbf{x},t).
\ea

\chapter{}
\section{$(i\neq j)$ Components of Einstein equation}
\label{off-Ein}

In this Appendix we look into off-diagonal components of spatial Einstein equations
$M_P^2 \delta G^i_j = \delta T^i_J$ for $i \neq j$. These equations are trivial at the background level. The leading order perturbation equations lead
\ba
\label{gamma-1}
{}^{(1)}\ddot{\gamma}_{ij} + 3H {}^{(1)}\dot{\gamma}_{ij}+ \left[3 H_i H_j-3({\cal H}^2+\dot{H}) -2\sum_{i} H_i^2 - |\epsilon^{ijk}|\dot{H}_k  \right]{}^{(1)} \gamma_{ij} &=& \dfrac{2}{a_i a_j M_P^2} {}^{(1)}\delta \pi_{ij},
\qquad (i\neq j) \nonumber\\
\ea
in which $\epsilon^{ijk}$ is the Levi-Civita symbol. As one can see the above equation has decaying solutions. This is due to the fact that the background metric does not admit off-diagonal spatial components.  Weinberg has argued that the anisotropic stress  for a wide class of theories to be some linear combinations of $\delta u$, $\delta p$  and $\delta \rho$ \cite{Weinberg:2003sw}. We partially extend this assumption and  assume that the anisotropic stress can also obtain contributions from gauge fields, $A_{\mu}$. So anisotropic stress tensor $\pi_{ij}$ for $i \neq j$ can be some linear combination of $\partial_i \partial_j p$, $\partial_i \partial_j \rho$, $\partial_i  u_j$, $\partial_i  A_j$, $\partial_i q_j$, $u_i u_j$, $u_i A_j$, $u_i q_j$ and finally $A_i q_j$ . As $\dot{A}_i \dot{A}_j $ for $i \neq j$ is forbidden by the background equations, this term does not contribute to the off-diagonal part of $\pi_{ij}$. These contributions are at least at the first order of gradient expansion $\epsilon$. So one readily deduces that
\ba
\pi_{ij} ={\cal O} (\epsilon).
\ea
Now,  Eq. \eqref{gamma-1} can be rephrased as \ba
\label{gamma-1b}
{}^{(1)}\ddot{\gamma}_{ij} + 3H {}^{(1)}\dot{\gamma}_{ij}+ m^2_{ij} \gamma_{ij} = {\cal O}(\epsilon) \, ,
\ea
 with $m^2_{ij} \sim H^2$ so $\gamma_{ij}$  has decaying solutions scaling  approximately as $a^{3/2}$. This is a consequence of the fact  that the background equations do not admit $\gamma_{ij} \neq 0$ for $i \neq j$. At the second order in perturbation variables $\delta$, the homogeneous equation has the same form, but it can be verified that all  possible source terms  are at least at the first order in gradient expansion $\epsilon$. This argument can be repeated for all orders of perturbations. This argument leads to the conclusion that in the $n$-th order
of perturbation theory, the off diagonal spatial part of metric,  after the decaying solutions become negligible, are of the first order of gradient expansion. As a result, one deduces
\ba
\gamma_{ij} = \cal {O} (\epsilon)
\ea
This equation is important for gradient expansion of Einstein equations.  

\chapter{}
\section{The $\delta N$ analysis}
\label{deltaN-app}

In this Appendix we outline the details of the $\delta N$ analysis leading to Eqs.
(\ref{N-A})-(\ref{N-dotA-phi}). The starting equations are
\ba
\label{N-phi-Ab}
N &&= \frac{p_c}{2 \left( I -1 \right) } \ln \left( \frac{\phi}{\phi_{f}} \right)  \\
\label{N-phi-A2b}
&& = \frac{1}{3} \ln \left(\sqrt{\frac{3}{ 2 R} }  \left( \frac{ A -  A_{f} }{M_{P}} \right) + 1 \right) \, .
\ea
In addition during inflation, from the attractor condition we have
\ba
\label{R-defb}
R \equiv \frac{\dot A_x^2 f(\phi)^2 e^{-2 \alpha}}{2 V}  \simeq \frac{I \epsilon}{2}
\ea
in which the last approximate equality is for the general case when $p>p_c$ and $I \neq 0$.
In this view $R$ is treated as a free parameter which should be varied when calculating $\delta N$. This is because in general $N$ is defined in the phase space as a function of $(\phi, \dot \phi)$ and $(A_x, \dot A_x)$. Because of the  slow-roll conditions we can solve for $\dot \phi$ in terms of $\phi$ so we do not need to vary $\delta \dot \phi$ as an independent variable. However, for the gauge field, because of the  gauge invariance,
it is $\dot A_x$ (i.e. $F_{0 x}$) which is physical and not $A_x$ itself. The contribution of
gauge field $A_x$ (and not its derivative) only appears at the surface of end of inflation in which the gauge symmetry is spontaneously broken due to Higgs mechanism and the gauge field  becomes massive.  Therefore, $\delta N$ has contributions both from $\delta \dot A_x$
and $\delta A_x$.

Finally, the surface of end of inflation is parameterized by the angle $\gamma$ via
\ba
\label{phif-b}
\phi_f = \phi_c \cos \gamma \quad , \quad
A_f = \frac{g \phi_c}{\e} \sin \gamma \, .
\ea
Expressing the surface of end of inflation in this way, the dependent variables $\phi_f$ and
$A_f$ are traded in terms of the independent variable $\gamma$.

Now our goal is to use the constraint Eqs. (\ref{R-defb}) and (\ref{phif-b}) to find $\delta N$ in terns of initial fluctuations $\delta \phi, \delta A$ and $\delta \dot A$. In the expressions below, we keep $N$ general, but it is understood that we evaluate the initial perturbations
at the time of horizon crossing corresponding to $N=-60$.

Varying Eq. (\ref{N-phi-Ab}) with respect to variables $\phi, \phi_f$ and $I$ (or $R$)  yields
\ba
\label{deltaN-phi-2}
\delta N = N_{, X_I} \delta {X_I} + \frac{1}{2} N_{, X_I X_J} \delta X_I \delta X_J
\ea
in which $X_I$ collectively  represents the variables $\{ \phi, \phi_f, I \}$. Note that  when $I \neq 0$ and $R \simeq I \epsilon/2$, we can use $\delta I$ and $\delta R$
interchangeably. Calculating the derivatives, we have
\ba
N_{,\phi} &=& \frac{p_c }{2(I-1) \phi}   \quad , \quad  N_{,\phi_f} = \frac{-p_c }{2(I-1) \phi_f}
\quad , \quad
N_{, I} = -\frac{p_c}{2 (I-1)^2} \ln \frac{\phi}{\phi_f} = -\frac{N}{I-1}  \nonumber\\
N_{,\phi \phi } &=& -\frac{p_c }{2(I-1) \phi^2}   \quad , \quad
N_{,\phi_f \phi_f } = \frac{p_c }{2(I-1) \phi_f^2}   \quad , \quad
N_{, I I } = \frac{2 N}{(I-1 )^2} \nonumber\\
N_{,I \phi } &=&   -\frac{p_c}{2 (I-1)^2 \phi}  \quad , \quad
N_{,I \phi_f } =   \frac{p_c}{2 (I-1)^2 \phi_f}  \quad , \quad
N_{,\phi \phi_f } =0 \, .
\ea
Varying the attractor condition, Eq. (\ref{R-defb}), yields
\begin{align}
\label{attractor-per}
\frac{\delta R}{R} = \frac{\delta I}{I}&=  \frac{2 f_{,\phi}}{f}\delta \phi +  \frac{2 \delta \dot{A_x}}{\dot{A}} + \left( \frac{f_{,\phi \phi}}{f} + \bigg{(}\frac{f_{,\phi}}{f}\bigg{)}^2    \right)\delta \phi^2
+ \bigg{(}\frac{\delta \dot{A}}{\dot{A}}\bigg{)}^2 +  \frac{4 f_{,\phi}}{f} \frac{\delta \dot{A_x}}{\dot{A}} \delta \phi  \nonumber\\
& -2 \delta N \bigg{(}1+  \frac{2 f_{,\phi}}{f}\delta \phi +  \frac{2 \delta \dot{A_x}}{\dot{A}}   \bigg{)} + 2 \delta N^2
\end{align}
On the other hand, varying Eq. (\ref{phif-b}) with respect to the independent variable $\gamma$ to second order yields
\ba
\label{phif-gamma}
\delta \phi_f = \phi_c \left[ - \sin \gamma \left( \delta_1 \gamma + \delta_2 \gamma \right)
-\frac{1}{2} \cos \gamma \left( \delta_1 \gamma \right)^2 \right] \, ,
\ea
in which $\delta_1\gamma$ and $\delta_2\gamma$ respectively show the first and second order perturbations in $\gamma$.

Plugging Eqs. (\ref{phif-gamma}) and (\ref{attractor-per}) in Eq. (\ref{deltaN-phi-2}), and
neglecting the sub-leading terms $\delta N^2$ and $N I^2$ for $I \ll 1$,  yields
\ba
\label{deltaN-phi-a}
\delta N &\simeq& -\left[  \frac{p_c }{2 \phi} + 2 N I \frac{f_{, \phi}}{f}  \right] \delta \phi
+ 2 N I \frac{\delta \dot A_x}{\dot A_x} -\frac{p_c}{2} \tan \gamma \,  \delta_1 \gamma
-\frac{p_c}{2} \tan \gamma \,  \delta_2 \gamma  \nonumber\\
&+& \left[ N I  \left( \frac{f_{,\phi \phi}}{f} + \left(\frac{f_{,\phi}}{f} \right)^2    \right) -  \frac{p_c I \, f_{, \phi}}{ \phi f} + \frac{p_c}{4 \phi^2} \right] \delta \phi^2
+ N I  \left( \frac{\delta \dot A}{\dot A} \right)^2
+\left[ 4 N I  \frac{f_{, \phi} }{f} - \frac{I p_c }{\phi} \right]  \frac{\delta \dot A_x}{\dot A} \delta \phi \nonumber\\
&-& \frac{p_c}{4} \frac{(\delta_1\gamma)^2}{\cos^2 \gamma } -I p_c \tan \gamma
 \left[ \frac{ f_{,\phi}}{f} \delta \phi + \frac{\delta \dot A_x}{\dot A_x}
\right] \delta_1 \gamma
\ea

Similarly, varying Eq. (\ref{N-phi-A2b}) to second order yields
\ba
\label{deltaN-phi-2b}
\delta N = N_{, Y_I} \delta {Y_I} + \frac{1}{2} N_{, Y_I Y_J} \delta Y_I \delta Y_J
\ea
in which $Y_I$ collectively  represents the variables $\{ \phi, A_f, R \}$. Note that in formula for $N$ given in Eq. (\ref{N-phi-A2b}) it is $R$ and not $I$which  appears. This is crucial for the critical case in which $I=0$ but $R$ is still a free parameter containing information for $\dot A_x$.   Calculating the derivatives, we have
\ba
N_{, A} &=& \frac{e^{-3 N}}{M_P \sqrt{6 R}} \quad , \quad
N_{, A_f} = - \frac{e^{-3 N}}{M_P \sqrt{6 R}} \quad , \quad
N_{, R} = -\frac{1}{6 R}  \left( 1- e^{-3 N}  \right) \nonumber\\
N_{, A A} &=&   N_{, A_f A_f}= -\frac{e^{-6 N}}{2 R M_P^2} \quad , \quad
N_{, R R} = \frac{1}{12 R^2} \left( 2- e^{-3 N} - e^{-6 N} \right) \nonumber\\
N_{, A A_f} &=& \frac{e^{-6 N}}{2 R M_P^2}  \quad , \quad
N_{, R A} = - N_{, R A_f}=  -\frac{e^{-6 N}}{2 M_P \sqrt{6 R^3}}
\ea
On the other hand
\ba
\label{Af-gamma}
\delta A_f = \frac{g \phi_c}{\e} \left[ \cos \gamma ( \delta_1 \gamma + \delta_2 \gamma)
- \frac{1}{2} \cos \gamma (\delta_1 \gamma)^2 \right]  \, .
\ea
Plugging Eqs. (\ref{Af-gamma}) and (\ref{attractor-per}) in Eq. (\ref{deltaN-phi-2b}) to leading order yields
\ba
\label{deltaN-A-a}
\delta N  &=&  -\frac{1}{3 }  \left( 1- e^{-3 N}  \right)  \left[ \frac{f_{, \phi}}{f} \delta \phi + \frac{\delta \dot A}{\dot A} \right] +
\frac{e^{-3 N}}{\sqrt{6 R}}\frac{\delta A}{M_P}
-\frac{e^{-3 N}}{ \sqrt{6 R}} \frac{g \phi_c}{\e M_P}  \cos \gamma  \,
( \delta_1 \gamma  + \delta_2 \gamma)  \nonumber\\
&-&\frac{e^{-6 N}}{4 R }\frac{\delta A^2}{M_P^2}
+ \left[  \frac{g \phi_c \sin \gamma}{2 \e M_P} \frac{e^{-3 N}}{\sqrt{6 R}}
- \frac{g^2 \phi_c^2 \cos^2 \gamma }{2 \e^2  M_P^2} \frac{e^{-6 N}}{2 R}
\right] (\delta_1 \gamma)^2 + \frac{g \phi_c \cos \gamma}{2 R \e M_P^2} e^{-6 N}
\delta_1 \gamma \delta A\nonumber\\
&-& \frac{1}{6 }  \left( 1- e^{-3 N}  \right) \left[   \left( \left(\frac{f_{,\phi}}{f}\right)^2 + \frac{f_{, \phi \phi}}{f} \right) \delta \phi^2 + \left( \frac{\delta \dot A}{\dot A} \right)^2 + 4 \frac{f_{, \phi}}{f} \frac{\delta \dot A}{\dot A} \delta \phi \right] + \frac{1}{6}  \left( 2- e^{-3 N} - e^{-6 N} \right) \nonumber\\
&\times&
\left( \frac{f_{, \phi}}{f} \delta \phi + \frac{\delta \dot A}{\dot A}   \right)^2 + \frac{e^{-6 N}}{3 M_P} \sqrt{\frac{3}{2 R}}  \left( \frac{f_{, \phi}}{f} \delta \phi + \frac{\delta \dot A}{\dot A}   \right) \left( - \delta A   + \frac{g \cos \gamma}{\e} \delta_1 \gamma
\right)
\ea

Now we have two formulas for $\delta N$, given by Eqs. (\ref{deltaN-A-a}) and  (\ref{deltaN-phi-a}). Similar to \cite{Sasaki:2008uc} we can use these two equations to eliminate
$\delta_1 \gamma$ and $\delta_2 \gamma$ in terms of initial perturbations $\delta \phi, \delta A$ and $\delta \dot A$.
We find
\ba
\label{gamma1}
\delta_1 \gamma &=& \frac{\e}{g \phi_c  \cos \gamma} \left( \delta A
+ \sqrt{\frac{2 R}{3}} M_P \left(  \frac{\delta \dot A_x}{\dot A} + \frac{f_{, \phi}}{f} \delta \phi \right)  \right) \nonumber\\
\delta_2 \gamma &=&  \frac{\e M_P \sqrt{6 R}}{g \cos \gamma \, \phi_c}
\left[ \frac{\e \sin \gamma }{2g \cos^2 \gamma \, \phi_c }  \left( \frac{\delta A^2}{M_P \sqrt{6 R}}  + \frac{M_P}{3} \sqrt{\frac{2R}{3}} \left( \frac{\delta \dot A}{\dot A} \right)^2
+ \frac{2  \delta A . \delta \dot A}{3 \dot A}
+ \frac{2 f_{, \phi}}{3 f} \delta \phi \delta A \right) \right.  \nonumber\\
&+&  \left. \left( \frac{f_{, \phi \phi}}{6 f}
   + \frac{\e M_P \sin \gamma }{18 g \cos^2 \gamma \, \phi_c } \sqrt{6 R}
\left( \frac{f_{, \phi}}{f} \right)^2  \right) \delta \phi^2
+ \frac{ f_{, \phi}}{3 f} \left(1+  \frac{\e \sin \gamma M_P }{3 g \cos^2 \gamma \, \phi_c } \sqrt{6 R} \right) \delta \phi \frac{\delta \dot A}{\dot A}
\right] \nonumber\\
\ea
Now we plug  back $\delta \gamma$ to either of Eqs. (\ref{deltaN-A-a}) or  (\ref{deltaN-phi-a}) yielding  our final formula for $\delta N$ given in Eqs.
(\ref{N-phi})-(\ref{N-dotA-phi}).

\chapter{}
\section{Investigating the Validity of Separate Universe Assumption}
\label{separate-universe}

Before using the $\delta N$ formalism, one has to verify the validity of the
gradient expansion and the separate universes approach in our model. For isotropic FRW background containing interacting scalar fields this was studied in \cite{ Lyth:2004gb,  Naruko:2012fe, Naruko:2012um, Sugiyama:2012tj}. In this picture, each local Hubble patch behaves as a background FRW universe with the effects of a very long mode to rescale the background quantities such as $a(t), H, \rho $ and $p$ appropriately.
For the model of anisotropic inflation this was first demonstrated in \cite{Abolhasani:2013zya}. In this Appendix, we prove  the validity of separate universe approach in our current model of gauged hybrid inflation. Because of  the complexity of waterfall dynamics, we demonstrate this up to second order in perturbation theory which enables us to calculate the  power spectrum and the bispectrum. In principle, one can prove the validity of  $\delta N$ to all orders in perturbation theory, but this is beyond the scope of this work.

Let us start with the general Bianchi I background with three different background scale factors
\ba
\label{Bianchi-metric1}
ds^2 = -dt^2 + a_1(t)^2 d x^2 +a_2(t)^2 d y^2 +a_3(t)^2 d z^2 \, .
\ea
Following the notations used in \cite{Miedema:1993} we define
\ba
H_i(t)= \dfrac{\dot{a_i}}{a_i} \qquad , \qquad H \equiv \dfrac{1}{3} \sum_{i=1}^3 H_i \, .
\ea
Here $H_i$ represents the Hubble expansion rate for the $i$-th  direction with  $i=1,2,3$ and a dot indicates the derivative with respect to cosmic $t$.

To solve the background fields  equations one has  to specify the form of  energy momentum tensor. The general  form of energy momentum tensor $T_{\mu \nu}$ for an imperfect fluid is given by  \cite{ellis98}
\ba
\label{eq:stress}
T_{\mu \nu} = (\rho + p)\,u_{\mu}\,u_{\nu}+ p\,g_{\mu \nu}+ q_{\mu}\,u_{\nu} + u_{\mu}\,q_{\nu} +  \pi_{\mu \nu}
\ea
with the  conditions
\ba
 q_{\mu}\,u^{\mu} = 0 \quad , \quad   \pi^{\mu}{}_{\mu} = 0  \quad , \quad  ~\pi_{\mu \nu} = \pi_{\nu \mu} \quad , \quad
~\pi_{\mu \nu}\,u^{\nu} = 0 \, . \nonumber
\ea
Here $u^\mu$ represents  the four-vector associated with the fluid,  $\rho$ is the  energy density, $p $ represents  the isotropic pressure, $\pi_{\mu \nu} $ stands for the trace-free  anisotropic pressure (stress) and $q^\mu$  is the  heat conduction.

The background Einstein equations are
\ba
\label{00-back}
3  {\cal H}^2&\equiv& \sum_{i > j} \bar H_j \bar H_{j} =  \frac{\bar \rho}{M_P^2}
\\
\bar T^0{}_i &=& \bar q_i =0
\\
\label{i=j-back}
M_P^2 \dot{\bar H}_i&=&-3  M_P^2 \bar H \bar H_i + \dfrac{1}{2} (\bar \rho-\bar p) + \bar \pi^{i}{}_{i} \, .
\ea
In this notation, $\bar H_i$ represents the background Hubble expansion rates while  $\bar \rho, \bar p$ and  so on are the background values of the fluid's physical parameters.  Also note that we have defined ${\cal H}$ as the effective Hubble expansion rate appearing in Friedmann equation, Eq. (\ref{00-back}), which should not be mistaken with the conformal Hubble expansion rate usually used in literature.

On the other hand,  the energy conservation equation $u_{\mu} \nabla_{\nu} T^{\mu \nu} = 0$ yields
\ba
\label{cont.-back}
-u_{\mu} \nabla_{\nu} T^{\mu \nu} = \dot{\bar \rho} + 3 H (\bar \rho + \bar p) + \bar H_j \bar \pi
^{i}{}_{j} \delta^{j}_i=0 .
\ea
where  $\bar H = \sum_{i} \bar H_i/3$.

Let us now look at the perturbations.  We follow the notation used in \cite{Sugiyama:2012tj}
in our $\delta N$ analysis.  We denote the order of spatial derivative or the gradient expansion  by $\epsilon=k/aH$ while the  perturbations are denoted by $\delta$. In general, one has to allow three different gradient expansion parameters $\epsilon_i$ for three different spatial directions $\epsilon_i = k/a_i H_i$. However, in order to simplify the analysis we assume $\epsilon_i \sim \epsilon$ but the extension  to the general case will be easy.

Using the standard ADM formalism the perturbations in metric are parameterized as
\ba
\label{ADM}
ds^2 = -{\cal N}^2 dt^2 + \gamma_{ij} \left( dx^i + \beta^i dt \right) \left( dx^j + \beta^j dt\right) \, ,
\ea
in which, as usual, ${\cal N}$ is the lapse function, $\beta_i$ represents the shift vectors and $\gamma_{ij}$ is the three-dimensional  spatial metric.  Furthermore,  it is instructive to decompose the spatial metric into
\ba
\label{gamma-ij}
\gamma_{ij} = a_i(t) a_j(t) e^{\psi_i (\mathbf{x},t)+\psi_j (\mathbf{x},t)} \tilde{\gamma}_{ij} \, ,
\ea
in which  $a_i(t)$ represents  the scale factor for the $i$-th spatial direction and $\psi_i(\mathbf{x},t)$ is similar  to curvature perturbation $\psi$ for the isotropic background.

As studied in \cite{Abolhasani:2013zya},  one important step in  the analysis of the gradient expansion for the  Einstein equations  is the ordering of the shift functions $\beta^i$. In \cite{Abolhasani:2013zya}, for the model of anisotropic inflation, it was shown that $\beta^{i} = {\cal O} (\epsilon)$ to all orders in perturbation theory.  Here we demonstrate that, although quantum back-reactions induce non-zero shift function  $\beta^i$, these corrections can still be neglected as the shift function is at the second order in perturbation theory $\beta^{i} = {\cal O} (\delta^2)$. In the following we demonstrate this point in details.

In order to  find an estimation for the shift function $\beta^{i}$, it is enough to use  $ (0,i)$ component of the  Einstein equation. Using the definition of energy-momentum tensor for an imperfect fluid, one  finds
\ba
\delta T^i_0  = -{\cal N} \delta q^i + \epsilon {\cal O} (\delta) + \beta {\cal O} (\delta) \, .
\ea
On the other hand, employing the $ (i,0)$ component of Einstein equation, the heat transfer can be related to the shift function as
\ba
\delta q^i = - (\bar{\rho}+\bar{p}+\bar{\pi}^i_i) \beta^i + \epsilon {\cal O} (\delta) + \beta {\cal O} (\delta).
\ea
 Let us simplify the above equation in order to estimate the  amplitude of $\beta^i$. In our specific model one can simply show that $\bar{\pi}^i_i \sim I H^2 \ll \epsilon H^2$. As a result the shift function $\beta^i$ can be estimated as
\ba
\label{beta-i}
\beta^i \simeq \dfrac{-\delta  q^i}{ \epsilon H^2 M^2_{P}},
\ea
in which $\epsilon$ denotes the slow-roll parameter and should not be mistaken with the gradient expansion parameter. Now using the action, Eq. (\ref{action3}), the heat transfer can be read as
\ba
\label{heat-transfer}
\delta q^i = - \dfrac{1}{\cal N} \left[f^2(\phi) T^i _m {}^{em} + i e \left(\delta \psi^{\ast} \partial_0 \delta \psi -\delta \psi \partial_0 \delta \psi ^{\ast}\right) A^i + e^2 \delta \psi^2 A_0 A^i  \right] \, .
\ea
It has been previously shown in \cite{Abolhasani:2013zya} that the first term in the big bracket above, denoting the heat transfer of electromagnetism, is ${\cal O} (\epsilon)$. Below we find the order of  the gradient expansion of the  other two terms in
Eq. (\ref{heat-transfer}).

In the main text we chose unitary gauge by setting the complex phase of the waterfall field to be zero. In the following, it is more helpful to choose the Coulumb-radiation gauge in which  $A_0 =\partial_iA_i=0$. As a result, the third term in Eq. (\ref{heat-transfer}) is zero and we are left only with the second term in Eq. (\ref{heat-transfer}).

Note that the waterfall is extremely heavy during inflation so the background value of the waterfall field is pinned to zero before the waterfall phase transition. In addition, the quantum fluctuations of the waterfall field after horizon crossing are continuously  damped  till  the  time of waterfall instability  after which the modes start growing exponentially as a result of the tachyonic instability \cite{Lyth:2010ch, Abolhasani:2010kr, Abolhasani:2010kn, Abolhasani:2011yp, Fonseca:2010nk,
Gong:2010zf, Lyth:2012yp}.

Before doing any explicit calculation it is worth discussing how the waterfall dynamics can contribute to the heat transfer. For a moment suppose that every  waterfall quantum fluctuations  in the vicinity of the transition point has the following solution
\ba
\delta \psi_k(n)  = \delta \psi_k(0) ~ e^{-i \omega(t)\, t + \Omega(t)\, t} \, ,
\ea
in which $n\equiv N-N_c$  and $N_c$ represents the time of the waterfall phase transition.  Here $\omega(t)$ and $\Omega(t)$ are two real functions quantifying the frequency of the oscillations and the growth rate of each  mode respectively.

Looking at  Eq. \eqref{heat-transfer}, we conclude  that only the  oscillatory phase $\omega(t)$ of the solution can contribute into the second term in Eq. \eqref{heat-transfer}.  Therefore, only  modes which become classical but still have oscillatory behavior will contribute to this term. This takes place near the waterfall transition point when the modes   become classical but still may have weak oscillatory behaviors.

Let us now examine the above intuition more carefully. The evolution equation for the quantum mode $v_k \equiv a \delta \psi_k$ can be read as \cite{Abolhasani:2010kr}
\ba
v''_k+ \left(k^2 -\dfrac{2+\epsilon_{\psi}^2 n}{\tau^2} \right)v_k=0 \, ,
\ea
in which the prime denotes the derivation respect to conformal time $\tau$ and $\epsilon_\psi$
is a large number measuring the tachyonic mass of the waterfall quantum fluctuations \cite{Abolhasani:2010kr}.  In order to estimate the heat transfer one can assume that every mode becomes classical when $\omega_k \tau \sim \lambda \sim 1$ in  which $\omega_k$ is the time-dependent frequency  in the above equation $\omega_k^2 \equiv k^2 -(2+\epsilon_{\psi}^2 n)/\tau^2$. It is  evident  that there is a narrow band of momenta which can contribute to the heat transfer
\ba
\label{band}
k^2_{min} = \dfrac{\lambda+2+\epsilon_{\psi}^2}{\tau^2}<k^2<k^2_{max} = \dfrac{2+\epsilon_{\psi}^2}{\tau^2}.
\ea
As a starting point let us estimate the background value of the heat transfer
\ba
\label{heat-transfer-back}
\delta \bar{q}^i = - \dfrac{1}{\cal N} ~ i \e Im\left[\delta \psi^{\ast} \partial_0 \delta \psi \right]_{0} \bar{A}^i  \, ,
\ea
in which the charge density of waterfall field can be estimated as
\ba
\label{charge-back}
i \e  \, Im\left[\delta \psi^{\ast} \partial_0 \delta \psi \right]_{0} \simeq  \dfrac{e}{a}  \int_{k_{min}}^{k_{max}} d^3k~ \omega_{k}(t) |\delta \psi_k|^2 \, .
\ea
For simplicity, in the vicinity of the  transition point,  for the narrow band of momenta given in Eq. (\ref{band}), the frequency of the modes can be estimated as $\omega_k(t) \sim 1/\tau$ so one has
\ba
\label{charge-back}
i \e \, Im\left[\delta \psi^{\ast} \partial_0 \delta \psi \right]_{0} \sim  \e H  \dfrac{H^2_0}{4 \pi^2}  \, ,
\ea
in which we have  used the fact that the short modes around the transition point have the following amplitude \cite{Gong:2010zf}
\ba
\delta \psi_S (n)= \dfrac{H_0}{\sqrt{2k}k_c}e^{-n} \, .
\ea
Adding up these results, from Eq. (\ref{beta-i}),  one  finds the following relation for the background value of shift function $\bar \beta^i$
\ba
\bar \beta^i \sim \dfrac{eHA^i}{4\pi^2 \epsilon M^2_{P}} = {\cal P_R} \dfrac{\e A^i}{H} \, .
\ea
As one can see, the shift function has a small value at the background level. Although its value can be ignored in the background equation,  but still it can give rise to complications in the perturbation equations. Therefore, it is helpful to estimate the amplitude of $\beta^i$ induced from the waterfall effects.  Using the relation $a^{-2}\dot{A} = 2R ~V$, with $R$ being the  ratio of the gauge field energy density to the total energy density, one  finds that
\ba
\dfrac{\e A_f}{H} \simeq 3R \ll1.
\ea
Moreover, one has ${\cal P_R} = {\cal O}(\delta^2)$. As a result one concludes  that
\ba
\label{beta-order}
\bar{\beta}^i \sim R~ \delta^2  \ll  \delta^2 \, .
\ea
This indicates that $\bar \beta^i$ is not larger than the second order in perturbations.  Let us now estimate the amplitude of perturbations in shift function,  $\beta^i_k$,  for each mode $k$. Following the same method as above one can simply find that
\ba
\label{beta-pert}
\beta_k \sim  R \left(\delta \psi^2 \right)_{k}
\ea

Now Let us look at the $\delta N$ prescription for this model with a background value of $\beta^i$. In an FRW universe  the  local Hubble expansion rate for each direction in the presence of perturbations  is defined as \cite{Abolhasani:2013zya}
 \ba
H_i(\mathbf{x},t) = \dfrac{\bar{H}_i + \dot{\psi} (\mathbf{x},t)}{{\cal N}}.
 \ea
With some analysis one can  can show that in the case of non-zero shift function $\beta^i$, the above prescription is modified to
\ba
H_i(\mathbf{x},t) = \dfrac{\bar{H}_i + \dot{\psi} (\mathbf{x},t)+\partial_i \beta^i + \sum_{i \neq j} \beta_j \partial^j \psi^i -\beta_i \partial^i \psi^i }{{\cal N}}.
\ea
with no sum on repeated  $i$ indices.

The  $\delta N$ formalism is at hand noting that from the  equations above one has the following formula for the number of e-fold expansion for each direction
\ba
\label{Ni-int}
N_i(\mathbf{x},t_1,t_2) \equiv \int_{t_1}^{t_2} H_i (\mathbf{x},t) {\cal N} dt = \int_{t_1}^{t_2} \bar H_i  dt + \int_{t_1}^{t_2} \dot{\psi}_i  dt + \int  \partial_i \beta^i dt + \int \left( \beta_j \partial^j \psi^i -\beta_i \partial^i \psi^i \right) dt. \nonumber\\
\ea
The first two terms above are the same as in \cite{Abolhasani:2013zya} while the remaining terms originate from the presence  of non-zero $\beta^i$. With a simple reasoning one can show that these additional terms do not play any role for the correlation functions which we are interested in. From Eq. \eqref{beta-order}, one concludes that the contribution of the last term in Eq. (\ref{Ni-int}) is at the third order of perturbation theory while to calculate the  power spectrum and the bispectrum we need  $\delta N$ formula up to second order in perturbations.

Moreover, It can be shown that the contribution of  $\partial_i \beta^i$ in Eq. (\ref{Ni-int}) is  not important. To see this note that  any Fourier mode of the square of the waterfall perturbations $\left(\delta \psi^2 \right)_k$ can have non-zero correlations only with itself. In other words, the square of waterfall fluctuations $\left(\delta \psi^2 \right)_k$ can be treated as  an individual fluctuation similar to fluctuation of other primary fields such as $\delta \phi$. In this view, the only connected diagram associated with the two-point correlation function which contains $\left(\delta \psi \right)_k$ is $\langle  \left(\delta \psi^2 \right)_k~ \left(\delta \psi^2 \right)_{k'} \rangle $. It is vivid that this contribution to the two point correlation function is at the fourth order of perturbation theory while the leading order terms from other fields are at the  second order. Similarly, the contribution of the waterfall field in the bispectrum can just emerge from the contractions of  the form $\langle  \left(\delta \psi^2 \right)_k~ \left(\delta \psi^2 \right)_{k'}~ \left(\delta \psi^2 \right)_{k''} \rangle $ which is at the sixth order in perturbation theory, negligible compared to leading terms  which are at the third order.  Therefore, one  concludes that despite the presence of a non-zero shift function in the vicinity of waterfall transition, any corrections due to heat transfer modification are too small to disrupt the  separate Universe assumption.

Finally, from Eq. (\ref{Ni-int}) one concludes
\ba
N_i(\mathbf{x},t_1,t_2) - \bar{N}_i(t) = \psi_i(t_2) -  \psi_i(t_1) + {\cal O} (\epsilon^2, \delta^3) \, ,
\ea
which, up to  ${\cal O} (\epsilon^2, \delta^3)$,  is the same  $\delta N$ formula as in \cite{Abolhasani:2013zya}.  As it is shown in Appendix  \ref{deltaN-app}, to calculate the power spectrum and the bispectrum we need
$\delta N$ formula only up to second order in perturbations. Therefore,  we conclude  that it is legitimate to apply $\delta N$ formalism in our analysis of power spectrum and bispectrum.

Finally we comment that the waterfall dynamics is not expected to induce observable curvature perturbations on large scales as studied in  e.g. \cite{Abolhasani:2010kr, Abolhasani:2011yp,      Fonseca:2010nk, Gong:2010zf, Lyth:2010ch, Lyth:2012yp}, for related works see also
\cite{Abolhasani:2010kn,    Levasseur:2010rk, Martin:2011ib, Clesse:2010iz,  Bugaev:2011qt, Kodama:2011vs, Mulryne:2011ni, Abolhasani:2012px,   Barnaby:2006km,   Barnaby:2006cq}. The waterfall dynamics affect only small scales, modes which leave the horizon around the time of waterfall.

\chapter{}
\section{ Interaction Lagrangians}
\label{int-lagrang}

In this Appendix we present the interaction Lagranians in Eqs. (\ref{Lzetah}) - (\ref{LhxzD}) in more details. Our starting point is the total Lagrangian
\ba
\label{interaction Lagrangian1}
L_{int} = -\frac{a^4}{4}f(\phi)^2 F_{\mu \nu} F^{\mu \nu} - \frac{a^4}{2} \e^2 \phi^2 A_{\mu} A^{\mu}
\ea
Expanding the above action around the background to second order in perturbations we get
\ba
\label{interaction Lagrangian2}
L_{int} &=& -2 f^2 A^{'2}_{x} \zeta h_{xx} + 4 f^2 A'_{x} \zeta \delta A'_{1} - 4f^2 A'_{x} \zeta \partial_{x}\delta
A_{0} - f^2 A'_{x} \delta A'_{1} h_{xx} + f^2 A'_{x} \partial_{x}\delta A_{0} h_{xx} + \frac{1}{2} f^2 \left( \partial_{x} \delta A_{0} \right)^2 \nonumber\\
&& -f^2 \delta A'_{1}\partial_{x}\delta A_{0} - f^2 A'_{x} \partial_{y} M' h_{xy} + f^2 A'_{x} h_{xy}\partial_{y} \delta A_{0} - f^2 A'_{x}h_{xz}D' + \frac{1}{2}f^2 \left( \partial_{y} \delta A_{0}\right)^2 -f^2\partial_{y}M' \partial_{y}\delta A_{0} \nonumber\\
&&+ \frac{1}{2}a^2 \e^2  \phi^2 \left(\delta A_{0} \right)^2 + a^2 \e^2 \phi^2 A_{x}\delta A_{1}h_{xx}
+ a^2 \e^2 \phi^2 A_{x} \partial_{y}M h_{xy} + a^2 \e^2 \phi^2 A_{x} D h_{xz} + a^2 \e^2 \phi A_{x}^2 \delta \phi h_{xx} \nonumber\\
&& - 2 a^2 \e^2 \phi A_{x} \delta \phi \delta A_{1}
\ea
Where we have used the following expression,
\ba
\left(\frac{\partial f^2}{\partial \phi}\right)\delta \phi = 4 f^2 \zeta
\ea
As we discussed in the main text,  we should integrate out the non-dynamical field $\delta A_{0}$. However, as we mentioned before, the resulting terms are sub-leading so we can safely neglect the contribution of $\delta A_{0}$ in Eq. (\ref{interaction Lagrangian2}).

Using the following useful formula
\ba
fA'_{x} &=& M_{P}\sqrt{3I\epsilon_{H}} (-\eta)^{-1} a \\
\e \phi A_x &=& M_{P}^2\e \sqrt{\frac{2I}{3}}\frac{a}{f} \\
\phi &=& M_{P}\sqrt{\frac{2}{\epsilon_{H}}} \\
f&=& \left(\frac{\eta^2}{\eta_{e}^2}\right)
\ea
In order to obtain the above equations, we have used the background attractor solution results which is given in Chapter \ref{Chapter4}.

We calculate $L_{\zeta h_+}, L_{\zeta D_1}, L_{D_1 h_+}$ and $L_{D h_\times}$ in turn.
For   $L_{\zeta h_+}$ we have
\ba
L_{\zeta h_+} &=& -f^2 A^{'2}_{x} \left( \zeta^{*} h_{xx} + \zeta h^{*}_{xx} \right) - \frac{1}{2} \e^2 a^2 A_{x}^2\phi  \frac{\dot \phi}{H}  \left( \zeta^{*} h_{xx} + \zeta h^{*}_{xx} \right)\nonumber\\
&=& - \frac{3\sqrt{2}}{2} I \epsilon_{H} M_{P}^2\sin^2{\theta}a^2\left(-\eta\right)^{-2}\left( \zeta^{*} {h}_{+} + \zeta {h}^{*}_{+} \right)  + \frac{\e^2 \sqrt{2}}{6}  I \epsilon_{H}M_{P}^4 \sin^2{\theta} \left(\frac{a ^4}{f^2}\right)\left( \zeta^{*} {h}_{+} + \zeta {h}^{*}_{+} \right) \nonumber\\
\ea

For $L_{\zeta D_1}$ we have
\ba
L_{\zeta D_1} &\simeq &  L_{\zeta \delta A_1}=
2f^2 A'_{x} \left( \zeta^{*} \delta A'_{1} + \zeta \delta A^{'*}_{1}\right) + \e^2 a^2 A_{x}\phi  \frac{\dot \phi}{H}  \left( \zeta^{*} \delta A_{1} + \zeta \delta A^{*}_{1} \right)  \nonumber\\
&=& -2 M_{P}\sqrt{3I \epsilon_{H}}\sin^2{\theta} \left(\frac{a f}{\eta}\right) \left( \zeta ^{*} D'_{1} + c.c.\right) - 2 \e^2 M_{P}^3 \sqrt{\frac{I \epsilon_{H}}{3}}\sin^2{\theta} \left(\frac{a^3}{f}\right) \left( \zeta ^{*} D_{1} + c.c.\right) \nonumber\\
\ea
where we have neglected the longitudinal mode $D_2$ so  $\delta A_1 \simeq  D_1 \sin \theta^2$.

Similarly, to calculate $L_{D_1 h_+}$ we have to calculate $L_{h_{xx} \delta A_{1}}$  and $L_{h_{xy} \delta A_{2}}$
which respectively are
\ba
L_{h_{xx} \delta A_{1}} &=& -\frac{1}{2} f^2 A'_{x} \left( \delta A^{'*}_{1} h_{xx} +  \delta A^{'}_{1} h^{*}_{xx}\right) + \frac{1}{2}\e^2 a^2 \phi^2 A_{x} \left( \delta A^{*}_{1} h_{xx} +  \delta A_{1} h^{*}_{xx}\right) \nonumber\\
&=& \frac{M_{P}}{2} \sqrt{\frac{3I\epsilon_{H}}{2}}\sin^4{\theta}\left( \frac{fa}{\eta}\right) \left( D^{'*}_{1} {h}_{+} +  c.c. \right) + \sqrt{\frac{I}{6\epsilon_{H}}} \e^2 M_{P}^3 \sin^4{\theta}\left( \frac{a^3}{f}\right)\left( D^{*}_{1} {h}_{+} +  c.c. \right) \nonumber\\
\ea
and
\ba
L_{h_{xy} \delta A_{2}} &=& -\frac{1}{2} f^2 A'_{x} k_{y} \left( i M^{'} h^{*}_{xy} -i   M^{'*} h_{xy}\right) + \frac{1}{2}\e^2 a^2 \phi^2 A_{x} k_{y}\left( i M h^{*}_{xy} -i   M^{*} h_{xy}\right) \nonumber\\
&=& \frac{M_{P}}{2} \sqrt{\frac{3I\epsilon_{H}}{2}} \sin^2{\theta}\cos^2{\theta}\left( \frac{fa}{\eta}\right)\left(  D^{'}_{1} {h}^{*}_{+} + c.c.\right) + \sqrt{\frac{I}{6\epsilon_{H}}} \e^2 M_{P}^3 \sin^2{\theta}\cos^2{\theta}
\nonumber\\
&&
\times \left( \frac{a^3}{f}\right)\left(  D_{1} {h}^{*}_{+} + c.c. \right) 
\ea
where the relation $M \simeq (i/k) \cos \theta D_1$ have been used in the limit where
we neglect the longitudinal mode.  Combining  $L_{h_{xx} \delta A_{1}}$  and $L_{h_{xy} \delta A_{2}}$ we obtain   $L_{D_1 h_+}$
as in Eq. (\ref{Lh+D1}).

Finally, to calculate $L_{D h_\times}$ we have to calculate $L_{h_{xz} \delta A_3}$ which is
\ba
L_{D h_\times} &=&
L_{h_{xz} \delta A_{3}}= -\frac{1}{2} f^2 A'_{x} \left( D^{'} h^{*}_{xz} +   D^{'*} h_{xz}\right) +
\frac{1}{2}\e^2 a^2 \phi^2 A_{x} \left(D h^{*}_{xz} +   D^{*} h_{xz}\right) \nonumber\\
&=& \frac{M_{P}}{2} \sqrt{\frac{3I \epsilon_{H}}{2}}\sin{\theta} \left(\frac{fa}{\eta}\right)\left( i D^{'} {h}^{*}_{\times} + c.c.\right) + \sqrt{\frac{I}{6\epsilon_{H}}}\e^2 M_{P}^3 \sin{\theta} \left(\frac{a^3}{f}\right)\left( i D {h}^{*}_{\times} + c.c. \right) \nonumber\\
\ea

\chapter{}

\section{ Interaction Hamiltonian}
\label{int-hamilton}
In this appendix, we are going to calculate the interaction Hamiltonian in our model which is required to proceed with in-in formalism. Since in this model, we do have kinetically coupled fields, one can be worried about the relation $H_{int} = - L_{int}$. So it is worth to calculate it by bruce force. We skip the details and only mention the final result for the interaction Hamiltonian.
\begin{align}
\label{int Hamil}
H_{int} &=  H_{\zeta h_+} + H_{\zeta D_1} + H_{D_{1} h_{+}} + H_{D h_\times}
\end{align}
Where we have {\color{black}{
\begin{align}
\label{Hzetah}
H_{\zeta h_+} & = - \frac{3\sqrt{2}}{2} I \epsilon_{H} M_{P}^2\sin^2{\theta}a^2\left(-\eta\right)^{-2}\left( \zeta^{*} {h}_{+} + \zeta {h}^{*}_{+} \right)  -  \frac{\e^2 \sqrt{2}}{6}  I \epsilon_{H}M_{P}^4 \sin^2{\theta} \left(\frac{a ^4}{f^2}\right)\left( \zeta^{*} {h}_{+} + \zeta {h}^{*}_{+} \right)
\\
\label{HzetaD1}
H_{\zeta D_1} & = +2 M_{P}\sqrt{3I \epsilon_{H}}\sin^2{\theta} \left(\frac{a f}{\eta}\right) \left( \zeta ^{*} D'_{1} + c.c.\right) + 2 \e^2 M_{P}^3 \sqrt{\frac{I \epsilon_{H}}{3}}\sin^2{\theta} \left(\frac{a^3}{f}\right) \left( \zeta ^{*} D_{1} + c.c.\right) \nonumber\\
& = - L_{\zeta D_1}  \\
\label{HD1h}
H_{h_+  D_1} &= -\frac{M_{P}}{2} \sqrt{\frac{3I\epsilon_{H}}{2}}\sin^2{\theta}\left( \frac{fa}{\eta}\right) \left( D^{'*}_{1} {h}_{+} +  c.c.\right) - \sqrt{\frac{I}{6\epsilon_{H}}} \e^2 M_{P}^3 \sin^2{\theta}\left( \frac{a^3}{f}\right)\left( D^{*}_{1} {h}_{+} + c.c.\right)\nonumber\\
& = - L_{h_+ D_1}  \\
\label{HDh}
H_{h_\times  D} &= -\frac{M_{P}}{2} \sqrt{\frac{3I \epsilon_{H}}{2}}\sin{\theta} \left(\frac{fa}{\eta}\right)\left( i D^{'} {h}^{*}_{\times} + c.c.\right) - \sqrt{\frac{I}{6\epsilon_{H}}}\e^2 M_{P}^3 \sin{\theta} \left(\frac{a^3}{f}\right)\left( i D{h}^{*}_{\times} + c.c. \right)\nonumber\\
& = - L_{h_\times D}
\end{align}}}
As a result, we see that $H_{int} = - L_{int}$ is not generally true for all kinetically coupled interactions. Especially for $H_{\zeta h_+}$, we can not use $H_{int} = - L_{int}$. We should notice that in the above analysis, we have neglected the mass terms. Because it is shown in \cite{Emami:2013bk} to a very good approximation, all of the fields are nearly massless in this model.
\chapter{}

\section{ The in-in analysis}

\label{in-in}

Here we present the integral form of the in-in integrals in more details.

\subsection{In-In integrals for anisotropic power spectrum}

For the anisotropy corrections in power spectrum we have
\ba
\label{delta-P-zeta-b}
\delta \langle \zeta_\bfk \zeta_\bfk^* \rangle  = - \int_{\eta_{0}}^{\eta_{e}} d\eta_{1} \int_{\eta_{0}}^{\eta_{1}}d\eta_{2} \left \langle
\bigg{[} L_{I}(\eta_{2}) , \bigg{[} L_{I}(\eta_{1}) ,
 \zeta_\bfk (\eta_e) \zeta_\bfk^*(\eta_e)  \bigg{]}\bigg{]} \right \rangle  \, .
\ea
where the leading interaction Lagrangian is   $L_{\zeta D_1} =   L^{(1)}_{\zeta D_1} +L^{(2)}_{\zeta D_1}$
with
\ba
L^{(1)}_{\zeta D_1} &\equiv&
 -2 M_{P}\sqrt{3I \epsilon_{H}}\sin^2{\theta} \left(\frac{a f}{\eta}\right) \left( \zeta ^{*} D'_{1} + c.c.\right)
 \\
L^{(2)}_{\zeta D_1} &\equiv&
- 2 \e^2 M_{P}^3 \sqrt{\frac{I \epsilon_{H}}{3}}\sin^2{\theta} \left(\frac{a^3}{f}\right) \left( \zeta ^{*} D_{1} + c.c.\right)
\ea
As explained in the main text, depending on whether one chooses either $L^{(1)}_{\zeta D_1}$ or $L^{(2)}_{\zeta D_1}$ in place of  $L_I(\eta_1)$ and $L_I(\eta_2)$ in the   integral Eq. (\ref{delta-P-zeta-b}), there are four different terms  in  $\delta \langle \zeta_\bfk \zeta_\bfk^* \rangle$ denoted by $\delta \langle \zeta_\bfk \zeta_\bfk^* \rangle_{ij}$ where $i=1, 2$ and with the  assumption  that
$L_I(\eta_1) = L^{(i)}_{\zeta D_1}$  and  $L_I(\eta_2) = L^{(j)}_{\zeta D_1}$.
For example,    $\delta \langle \zeta_\bfk \zeta_\bfk^* \rangle_{12}$ means
$L_I(\eta_1) = L^{(1)}_{\zeta D_1}$  and  $L_I(\eta_2) = L^{(2)}_{\zeta D_1}$. In total  we have
\ba
\label{correction zeta}
\delta  \bigg{\langle} \zeta_{\bfk}(\eta_{e}) \zeta_{\bfk}^*(\eta_{e})\bigg{\rangle} &=&
\delta \bigg{\langle} \zeta_{\bfk}(\eta_{e}) \zeta_{\bfk}^*(\eta_{e})\bigg{\rangle} _{11}
+\delta \bigg{\langle} \zeta_{\bfk}(\eta_{e}) \zeta_{\bfk}^*(\eta_{e})\bigg{\rangle} _{12}
+ \delta \bigg{\langle} \zeta_{\bfk}(\eta_{e}) \zeta_{\bfk}^*(\eta_{e})\bigg{\rangle} _{21} \nonumber\\
&+& \delta \bigg{\langle} \zeta_{\bfk}(\eta_{e}) \zeta_{\bfk}^*(\eta_{e})\bigg{\rangle} _{22} 
\ea
We calculate each of them in turn.
\ba
\delta  \bigg{\langle} \zeta_{\mathbf{k}}(\eta_{e}) \zeta_{\mathbf{k}}^*(\eta_{e})\bigg{\rangle}_{11}  &&=  384 I \eH M_P^2  \sin^4{\theta} \
\int_{\eta_{0}}^{\eta_{e}} d\eta_{1} \left(\frac{a f}{\eta} \right)_{\eta_1}
\im \left[ \zeta_k (\eta_{1})\zeta_k^{*}(\eta_{e})\right]  \\
&& \times \int_{\eta_{0}}^{\eta_{1}}d\eta_{2}
 \left(\frac{a f}{\eta} \right)_{\eta_2}
\im \left[ \zeta_k(\eta_{2})\zeta_k^{*}(\eta_{e})  D_{1k}^{' *} (\eta_{1}) D_{1k}^{'}(\eta_{2}) \right]
\ea
As discussed in the main text, expanding the integrand for small $k \eta$ arguments and assuming $k \eta_0 = -1$ and $k \eta_e =0$ the above integral can be calculated analytically and we get
\ba
\delta  \bigg{\langle} \zeta_{\mathbf{k}}(\eta_{e}) \zeta_{\mathbf{k}}(\eta_{e})\bigg{\rangle}_{11}&=& \frac{6 I N^2}{k^3 \epsilon_{H}} \left(\frac{H}{M_{P}} \right)^2 \sin^2{\theta}
\ea
Similarly
\ba
\delta  \bigg{\langle} \zeta_{\mathbf{k}}(\eta_{e}) \zeta_{\mathbf{k}}^*(\eta_{e})\bigg{\rangle}_{12}  &=&  128 I \eH M_P^2  \e^2  \sin^4{\theta} \
\int_{\eta_{0}}^{\eta_{e}} d\eta_{1} \left(\frac{a f}{\eta} \right)_{\eta_1}
\im \left[ \zeta_k (\eta_{1})\zeta_k^{*}(\eta_{e})\right] \nonumber\\
&& ~~~~~~~~~~~~~~~~~~~~~~~\times \int_{\eta_{0}}^{\eta_{1}}d\eta_{2}
 \left(\frac{a^3}{f} \right)_{\eta_2}
\im \left[ \zeta_k(\eta_{2})\zeta_k^{*}(\eta_{e})  D_{1k}^{ '*} (\eta_{1}) D_{1k}(\eta_{2}) \right] \nonumber\\
&=& -\frac{31 }{490}  \frac{\e^2 I}{k^3 {\epsilon_{H}}} \sin^2{\theta}  \, ,
\ea
\ba
\delta  \bigg{\langle} \zeta_{\mathbf{k}}(\eta_{e}) \zeta_{\mathbf{k}}^*(\eta_{e})\bigg{\rangle}_{21}  &=&  128 I \eH M_P^2  \e^2  \sin^4{\theta} \
\int_{\eta_{0}}^{\eta_{e}} d\eta_{1}     \left(\frac{a^3}{f} \right)_{\eta_1}  \im \left[ \zeta_k (\eta_{1})\zeta_k^{*}(\eta_{e})\right]  \nonumber\\
&& ~~~~~~~~~~~~~~~~~~~~~~~\times \int_{\eta_{0}}^{\eta_{1}}d\eta_{2}
   \left(\frac{a f}{\eta} \right)_{\eta_2}
\im \left[ \zeta_k(\eta_{2})\zeta_k^{*}(\eta_{e})  D_{1k}^{ *} (\eta_{1}) D_{1k}'(\eta_{2}) \right] \nonumber\\
&=& -\frac{ I \e^2 N}{7 k^3 \epsilon_H}   \sin^2{\theta}  \, .
\ea
Finally
\ba
\delta  \bigg{\langle} \zeta_{\mathbf{k}}(\eta_{e}) \zeta_{\mathbf{k}}^*(\eta_{e})\bigg{\rangle}_{22}  &=& \frac{128}{3} I \eH M_P^6  \e^4  \sin^4{\theta} \
\int_{\eta_{0}}^{\eta_{e}} d\eta_{1}     \left(\frac{a^3}{f} \right)_{\eta_1}  \im \left[ \zeta_k (\eta_{1})\zeta_k^{*}(\eta_{e})\right]  \nonumber\\
&& ~~~~~~~~~~~~~~~~~~~~~~~\times \int_{\eta_{0}}^{\eta_{1}}d\eta_{2}
  \left(\frac{a^3}{f} \right)_{\eta_2}
\im \left[ \zeta_k(\eta_{2})\zeta_k^{*}(\eta_{e})  D_{1k}^{ *} (\eta_{1}) D_{1k}(\eta_{2}) \right] \nonumber\\
 &=& \frac{9}{2156}  \frac{\e^4 I}{k^3}\left(\frac{M_{P}}{m} \right)^2 \sin^2{\theta} \, .
\ea

\subsection{In-in for tensor power spectra}

Now we calculate the anisotropy in tensor power spectra
 $\langle {h}_{\bfk \times } {h}_{\bfk \times }^*\rangle$ and $\langle {h}_{\bfk + } {h}_{\bfk + }^*\rangle$.

Let us  start with  $\langle {h}_{\times } {h}_{\times }^*\rangle$.  The interaction Lagrangian is
$L_{D {h}_{\times } } =  L_{D {h}_{\times } }^{(1)} +  L_{D {h}_{\times } }^{(2)} $ where
$ L_{D {h}_{\times } }^{(1)}$ and  $ L_{D {h}_{\times } }^{(2)}$ respectively are the first term and the second term in Eq. (\ref{LhxzD}).   Following the same convention as  in anisotropy analysis for curvature perturbation in power spectrum we have
\ba
\delta \bigg{\langle} {h}_{\times } {h}_{\times }^*\bigg{\rangle} &=& - \int_{\eta_{0}}^{\eta_{e}} d\eta_{1}\int_{\eta_{0}}^{\eta_{1}} d\eta_{2}\bigg{[}L_{D {h}_{\times }  } ,\bigg{[}L_{D {h}_{\times}}, {h}_{\times\mathbf{k}} {h}_{\times\mathbf{k}}\bigg{]}\bigg{]} \nonumber\\
&=& \delta \bigg{\langle} \widehat{h}_{\times }\widehat{h}_{\times }\bigg{\rangle}_{11} +  \delta \bigg{\langle} \widehat{h}_{\times }\widehat{h}_{\times }\bigg{\rangle}_{12} + \delta \bigg{\langle} \widehat{h}_{\times }\widehat{h}_{\times }\bigg{\rangle}_{21} + \delta \bigg{\langle} \widehat{h}_{\times }\widehat{h}_{\times }\bigg{\rangle}_{22}
\ea
The results for each contribution are
\ba
\label{hhtimes correction-b}
\label{tensor times 1}
\delta \bigg{\langle} {h}_{\times\mathbf{k}}  {h}_{\times\mathbf{k}}^*\bigg{\rangle}_{11}&=& -\left(12 I \epsilon_{H} M_{P}^2 \right) \sin^2{\theta} \int_{\eta_{0}}^{\eta_{e}} d\eta_{1}  \left(\frac{a f}{\eta}\right)_{\eta_{1}}\int_{\eta_{0}}^{\eta_{1}} d\eta_{2} \left(\frac{a f}{\eta}\right)_{\eta_{2}} {\rm Im}\left( {h}_{\times}(\eta_{1}){h}^{*}_{\times}(\eta_{e})\right) \times \nonumber\\
&& \times {\rm Im}\left({h}_{\times}(\eta_{2}) {h}_{\times}^{*}(\eta_{e}) D'(\eta_{2})D^{'*}(\eta_{1})\right) \nonumber\\
&=& \left(\frac{12}{k_1^3}\right)\left(\frac{H}{M_{P}}\right)^2 I \epsilon_{H} N^2\sin^2{\theta} \\
\label{tensor times 2}
\delta \bigg{\langle} {h}_{\times\mathbf{k}}  {h}_{\times\mathbf{k}}^*\bigg{\rangle}_{12}&=& -\left(8 I M_{P}^4 \right) \e^2 \sin^2{\theta} \int_{\eta_{0}}^{\eta_{e}} d\eta_{1}  \left(\frac{a^3 }{f}\right)_{\eta_{1}}\int_{\eta_{0}}^{\eta_{1}} d\eta_{2} \left(\frac{a f}{\eta}\right)_{\eta_{2}}{\rm Im}\left( {h}_{\times}(\eta_{1}){h}^{*}_{\times}(\eta_{e})\right) \times \nonumber\\
&& \times {\rm Im}\left( {h}_{\times}(\eta_{2}){h}_{\times}^{*}(\eta_{e})D'(\eta_{2})D^{*}(\eta_{1})\right) \nonumber\\
&=& -\left(\frac{4}{7 k^3}\right)N I\e^2\sin^2{\theta}\\
\label{tensor times 3}
\delta \bigg{\langle} {h}_{\times\mathbf{k}}  {h}_{\times\mathbf{k}}^*\bigg{\rangle}_{21}&=& -\left(8 I M_{P}^4 \right) \e^2 \sin^2{\theta} \int_{\eta_{0}}^{\eta_{e}} d\eta_{1}\left(\frac{a f}{\eta}\right)_{\eta_{1}}  \int_{\eta_{0}}^{\eta_{1}} d\eta_{2}\left(\frac{a^3 }{f}\right)_{\eta_{2}} {\rm Im}\left( {h}_{\times}(\eta_{1}){h}^{*}_{\times}(\eta_{e})\right) \times \nonumber\\
&& \times {\rm Im}\left( {h}_{\times}(\eta_{2}){h}_{\times}^{*}(\eta_{e})D(\eta_{2})D^{'*}(\eta_{1})\right) \nonumber\\
&=& -\left(\frac{62}{245 k^3}\right) I \e^2 \sin^2{\theta}
\ea
\ba
\label{tensor times 4}
\delta \bigg{\langle} {h}_{\times\mathbf{k}}  {h}_{\times\mathbf{k}}^*\bigg{\rangle}_{22}&=& -\left(\frac{16}{3} I M_{P}^6 \right)\e^4 \sin^2{\theta} \int_{\eta_{0}}^{\eta_{e}} d\eta_{1}\left(\frac{a^3 }{f}\right)_{\eta_{1}}\int_{\eta_{0}}^{\eta_{1}} d\eta_{2} \left(\frac{a^3}{f}\right)_{\eta_{2}}{\rm Im}\left( {h}_{\times}(\eta_{1}){h}^{*}_{\times}(\eta_{e})\right) \times \nonumber\\
&& \times {\rm Im}\left({h}_{\times}(\eta_{2}) {h}_{\times}^{*}(\eta_{e}) D(\eta_{2}) D^{*}(\eta_{1}) \right)\nonumber\\
&=& \left(\frac{6}{539 k^3 }\right) \left(\frac{M_{P}}{H}\right)^2 \left(\frac{I \e^4}{\epsilon_{H}}\right) \sin^2{\theta}
\ea
So finally we get,
\ba
\label{correction hh times final-b}
\delta \bigg{\langle} {h}_{\times\mathbf{k}_{1}}  {h}_{\times\mathbf{k}_{2}}\bigg{\rangle} \simeq \left( \left(12I \epsilon_{H}N^2\right)\left(\frac{H}{M_{P}}\right)^2 -\left(\frac{4}{7}N I\e^2\right) + \left(\frac{6}{539}\right) \left(\frac{M_{P}}{H}\right)^2 \left(\frac{I \e^4}{\epsilon_{H}}\right)\right)\left( \frac{\sin^2{\theta}}{k^3}\right) \nonumber\\
\ea

In addition, we have to calculate   $\langle {h}_{+ }{h}_{+}^*\rangle$. In this case the relevant interaction Lagrangians are  $L_{D_{1} {h}_{+ } }$ and $ L_{\zeta {h}_{+ }  }$ so we have
\ba
\label{tensor plus}
\delta \bigg{\langle} {h}_{+} {h}_{+}\bigg{\rangle} &=& - \int_{\eta_{0}}^{\eta_{e}} d\eta_{1}\int_{\eta_{0}}^{\eta_{1}} d\eta_{2}\bigg{[}L_{D_{1} {h}_{+ }  } ,\bigg{[}L_{D_{1} {h}_{+}}, {h}_{+} {h}_{+}\bigg{]}\bigg{]} - \int_{\eta_{0}}^{\eta_{e}} d\eta_{1}\int_{\eta_{0}}^{\eta_{1}} d\eta_{2}\bigg{[}L_{\zeta {h}_{+ }  } ,\bigg{[}L_{\zeta {h}_{+}}, {h}_{+} {h}_{+}\bigg{]}\bigg{]} \nonumber\\
\ea
As mentioned in the main text, comparing $L_{D_{1} {h}_{+ } }$ and $ L_{\zeta {h}_{+ }  }$ we see that $ L_{\zeta {h}_{+ }  }$ is suppressed compared to $L_{D_{1} {h}_{+ } }$ by a factor $\sqrt{I} \ll 1$
so to leading order in anisotropy we can neglect the contribution from  $ L_{\zeta {h}_{+ }  }$ in
$\langle {h}_{+ }{h}_{+}^*\rangle$. As a result the analysis is exactly the same as in  the case of  $\langle {h}_{\times }{h}_{\times}^*\rangle$ and therefore
\ba
\label{tensor plus final-b}
\delta \bigg{\langle} {h}_{+} {h}_{+}\bigg{\rangle} =
\delta \bigg{\langle} {h}_{\times} {h}_{\times}\bigg{\rangle} \, .
\ea
\subsection{In-in for  scalar-tensor cross-correlation}
Here we present the in-in analysis for the cross-correlation $\langle \zeta h_{s} \rangle $
for $s =\pm$.  The corresponding in-in integrals are
\ba
\bigg{\langle} \zeta_{\mathbf{k}_{1}}(\eta_{e}) {h}_{+\mathbf{k}_{2}}(\eta_{e})\bigg{\rangle} &=& i \int_{\eta_{0}}^{\eta_{e}} d\eta_{1}\bigg{[} H_{\zeta {h}_{+}} , \zeta_{\mathbf{k}_{1}}{h}_{+\mathbf{k}_{2}}\bigg{]} - \int_{\eta_{0}}^{\eta_{e}} d\eta_{1}\int_{\eta_{0}}^{\eta_{1}} d\eta_{2}\bigg{[}L_{\zeta D_{1}} ,\bigg{[}L_{D_{1} {h}_{+}}, \zeta_{\mathbf{k}_{1}}{h}_{+\mathbf{k}_{2}}\bigg{]}\bigg{]} \nonumber\\
&& -\int_{\eta_{0}}^{\eta_{e}} d\eta_{1}\int_{\eta_{0}}^{\eta_{1}} d\eta_{2}\bigg{[} L_{D_{1} \widehat{h}_{+}},\bigg{[} L_{\zeta D_{1}}, \zeta_{\mathbf{k}_{1}}{h}_{+\mathbf{k}_{2}}\bigg{]}\bigg{]} \nonumber\\
&=& \bigg{\langle} \zeta_{\mathbf{k}_{1}}(\eta_{e}) {h}_{+\mathbf{k}_{2}}(\eta_{e})\bigg{\rangle}_{1} + \bigg{\langle} \zeta_{\mathbf{k}_{1}}(\eta_{e}) {h}_{+\mathbf{k}_{2}}(\eta_{e})\bigg{\rangle}_{2} + \bigg{\langle} \zeta_{\mathbf{k}_{1}}(\eta_{e}) {h}_{+\mathbf{k}_{2}}(\eta_{e})\bigg{\rangle}_{3}
\ea
In the following, we calculate the above cross-correlation step by step,
\ba
\label{cross1}
\bigg{\langle} \zeta_{\mathbf{k}}(\eta_{e}) {h}_{+\mathbf{k}}(\eta_{e})^*\bigg{\rangle}_{1} &=&2i M_{P}^2 \left(I \epsilon_{H} \sin^2{\theta} \right)\int_{\eta_{0}}^{\eta_{e}} d\eta_{1} \bigg{(}-3\sqrt{2}\left(\frac{a^2}{\eta^2} \right) -  \frac{\e^2M_{P}^2 }{3} \left(\frac{a^4}{f^2} \right) \bigg{)}_{\eta_{1}} \nonumber\\
&& \times {\rm Im}\left( \zeta(\eta_{1}) \zeta^{*}(\eta_{e}){h}_{+}(\eta_{1}){h}^{*}_{+}(\eta_{e})\right) \nonumber\\
&=& \frac{2I}{3k_1^3} \left( \frac{H}{M_{P}} \right)^2  \left( 3\sqrt{2} N + \frac{\e^2 M_{P}^2}{28 H^2} \right)\sin^2{\theta} \\
&&\nonumber\\
\label{cross2}
\bigg{\langle} \zeta_{\mathbf{k}}(\eta_{e}) {h}_{+\mathbf{k}}(\eta_{e})^*\bigg{\rangle}_{2} &=&
- \int_{\eta_{0}}^{\eta_{e}} d\eta_{1}\int_{\eta_{0}}^{\eta_{1}} d\eta_{2}\bigg{[} L_{\zeta D_{1}}
,\bigg{[}L_{D_{1} {h}_{+}}, \zeta_{\mathbf{k}_{1}}\widehat{h}_{+\mathbf{k}_{2}}\bigg{]}\bigg{]}\nonumber\\
&\equiv& \bigg{\langle} \zeta_{\mathbf{k}_{1}} {h}_{+\mathbf{k}_{2}}\bigg{\rangle}_{21} + \bigg{\langle} \zeta_{\mathbf{k}_{1}} {h}_{+\mathbf{k}_{2}}\bigg{\rangle}_{22} + \bigg{\langle} \zeta_{\mathbf{k}_{1}} {h}_{+\mathbf{k}_{2}}\bigg{\rangle}_{23} + \bigg{\langle} \zeta_{\mathbf{k}_{1}}
\widehat{h}_{+\mathbf{k}_{2}}\bigg{\rangle}_{24}\nonumber\\
&&\nonumber\\
\label{cross3}
\bigg{\langle} \zeta_{\mathbf{k}}(\eta_{e}) {h}_{+\mathbf{k}}(\eta_{e})^*\bigg{\rangle}_{3} &=&
- \int_{\eta_{0}}^{\eta_{e}} d\eta_{1}\int_{\eta_{0}}^{\eta_{1}} d\eta_{2}\bigg{[} L_{D_{1} \widehat{h}_{+}}
,\bigg{[}L_{\zeta D_{1}}, \zeta_{\mathbf{k}_{1}}{h}_{+\mathbf{k}_{2}}\bigg{]}\bigg{]}\nonumber\\
&\equiv& \bigg{\langle} \zeta_{\mathbf{k}_{1}} {h}_{+\mathbf{k}_{2}}\bigg{\rangle}_{31} + \bigg{\langle} \zeta_{\mathbf{k}_{1}} {h}_{+\mathbf{k}_{2}}\bigg{\rangle}_{32} + \bigg{\langle} \zeta_{\mathbf{k}_{1}} {h}_{+\mathbf{k}_{2}}\bigg{\rangle}_{33} + \bigg{\langle} \zeta_{\mathbf{k}_{1}}
{h}_{+\mathbf{k}_{2}}\bigg{\rangle}_{34} \nonumber\\
\ea
where we have defined $\bigg{\langle} \zeta_{\mathbf{k}_{1}} {h}_{+\mathbf{k}_{2}}\bigg{\rangle}_{2i}$ as,
\ba
\label{cross21}
\bigg{\langle} \zeta_{\mathbf{k}}(\eta_{e}) {h}_{+\mathbf{k}}(\eta_{e})^*\bigg{\rangle}_{21}&=& \left(24 I \epsilon_{H} M_{P}^2 \sqrt{2} \right) \sin^4{\theta} \int_{\eta_{0}}^{\eta_{e}} d\eta_{1}  \left(\frac{a f}{\eta}\right)_{\eta_{1}}\int_{\eta_{0}}^{\eta_{1}} d\eta_{2} \left(\frac{a f}{\eta}\right)_{\eta_{2}} {\rm Im}\left( \widehat{h}_{+}(\eta_{1})\widehat{h}^{*}_{+}(\eta_{e})\right)  \nonumber\\
&& \times {\rm Im}\left( \zeta(\eta_{2})\zeta^{*}(\eta_{e})D'_{1}(\eta_{2})D^{'*}_{1}(\eta_{1})\right) \nonumber\\
&=& - \left(\frac{3\sqrt{2}}{k_1^3}\right)\left(\frac{H}{M_{P}}\right)^2 I N^2\sin^2{\theta} \\
\label{cross22}
\bigg{\langle} \zeta_{\mathbf{k}}(\eta_{e}) {h}_{+\mathbf{k}}(\eta_{e})^*\bigg{\rangle}_{22}&=& \left(16 I \sqrt{2}M_{P}^4 \right) \e^2 \sin^4{\theta} \int_{\eta_{0}}^{\eta_{e}} d\eta_{1}  \left(\frac{a^3 }{f}\right)_{\eta_{1}}\int_{\eta_{0}}^{\eta_{1}} d\eta_{2} \left(\frac{a f}{\eta}\right)_{\eta_{2}}{\rm Im}\left( \widehat{h}_{+}(\eta_{1})\widehat{h}^{*}_{+}(\eta_{e})\right) \nonumber\\
&& \times {\rm Im}\left( \zeta(\eta_{2})\zeta^{*}(\eta_{e})D'_{1}(\eta_{2})D^{*}_{1}(\eta_{1})\right) \nonumber\\
&=& \left(\frac{4\sqrt{2}}{3 \epsilon_{H}k_1^3}\right) \left( \frac{3}{28}N - \frac{361}{3920} \right)I\e^2\sin^2{\theta} \\
\label{cross23}
\bigg{\langle} \zeta_{\mathbf{k}}(\eta_{e}) {h}_{+\mathbf{k}}(\eta_{e})^*\bigg{\rangle}_{23}&=& \left(8 I \epsilon_{H} \sqrt{2}M_{P}^4 \right)\e^2 \sin^4{\theta} \int_{\eta_{0}}^{\eta_{e}} d\eta_{1}\left(\frac{a f }{\eta}\right)_{\eta_{1}}\int_{\eta_{0}}^{\eta_{1}} d\eta_{2} \left(\frac{a^3}{f}\right)_{\eta_{2}}{\rm Im}\left( \widehat{h}_{+}(\eta_{1})\widehat{h}^{*}_{+}(\eta_{e})\right) \nonumber\\
&& \times {\rm Im}\left( \zeta(\eta_{2})\zeta^{*}(\eta_{e})D_{1}(\eta_{2})D^{'*}_{1}(\eta_{1})\right) \nonumber\\
&=& \left(\frac{31 \sqrt{2}}{980 k_1^3}\right) I \e^2 \sin^2{\theta} \\
\label{cross24}
\bigg{\langle} \zeta_{\mathbf{k}}(\eta_{e}) {h}_{+\mathbf{k}}(\eta_{e})^*\bigg{\rangle}_{24}&=& \left(\frac{16}{3} I \sqrt{2} M_{P}^6 \right)\e^4 \sin^4{\theta} \int_{\eta_{0}}^{\eta_{e}} d\eta_{1}\left(\frac{a^3 }{f}\right)_{\eta_{1}}\int_{\eta_{0}}^{\eta_{1}} d\eta_{2} \left(\frac{a^3}{f}\right)_{\eta_{2}}{\rm Im}\left( \widehat{h}_{+}(\eta_{1})\widehat{h}^{*}_{+}(\eta_{e})\right) \nonumber\\
&& \times {\rm Im}\left( \zeta(\eta_{2})\zeta^{*}(\eta_{e})D_{1}(\eta_{2})D^{*}_{1}(\eta_{1})\right)\nonumber\\
&=& -\left(\frac{3\sqrt{2}}{2156 k_1^3 \epsilon_{H}}\right) \left(\frac{M_{P}}{H}\right)^2 I \e^4 \sin^2{\theta}
\ea
Where we have defined $\ln{\left(-k_1\eta_{e}\right)} = -N$.\\

In addition, we have defined $\bigg{\langle} \zeta_{\mathbf{k}_{1}} {h}_{+\mathbf{k}_{2}}\bigg{\rangle}_{3i}$ as,

\ba
\label{cross31}
\bigg{\langle} \zeta_{\mathbf{k}}(\eta_{e}) {h}_{+\mathbf{k}}(\eta_{e})^*\bigg{\rangle}_{31}&=& \left(24 I \epsilon_{H} \sqrt{2} M_{P}^2\right) \sin^4{\theta} \int_{\eta_{0}}^{\eta_{e}} d\eta_{1}  \left(\frac{a f}{\eta}\right)_{\eta_{1}}\int_{\eta_{0}}^{\eta_{1}} d\eta_{2} \left(\frac{a f}{\eta}\right)_{\eta_{2}} {\rm Im}\left( \zeta(\eta_{1})\zeta^{*}(\eta_{e})\right) \nonumber\\
&&  {\rm Im}\left({h}_{+}(\eta_{2}){h}^{*}_{+}(\eta_{e})D'_{1}(\eta_{2})D^{'*}_{1}(\eta_{1})\right) \nonumber\\
&=& - \left(\frac{3\sqrt{2}}{k^3}\right)\left(\frac{H}{M_{P}}\right)^2 I N^2\sin^2{\theta} \\
\label{cross32}
\bigg{\langle} \zeta_{\mathbf{k}}(\eta_{e}) {h}_{+\mathbf{k}}(\eta_{e})^*\bigg{\rangle}_{32}&=& \left(16 I \sqrt{2}M_{P}^4 \right) \e^2 \sin^4{\theta} \int_{\eta_{0}}^{\eta_{e}} d\eta_{1}  \left(\frac{a f}{\eta}\right)_{\eta_{1}}\int_{\eta_{0}}^{\eta_{1}} d\eta_{2} \left(\frac{a^3 }{f}\right)_{\eta_{2}}{\rm Im}\left( \zeta(\eta_{1})\zeta^{*}(\eta_{e})\right)
 \nonumber\\
&& {\rm Im}\left( {h}_{+}(\eta_{2})\widehat{h}^{*}_{+}(\eta_{e}) D_{1}(\eta_{2})D^{'*}_{1}(\eta_{1})\right) \nonumber\\
&=& \left(\frac{31 \sqrt{2}}{490 k^3}\right) I \e^2 \sin^2{\theta} \\
\label{cross33}
\bigg{\langle} \zeta_{\mathbf{k}}(\eta_{e}) {h}_{+\mathbf{k}}(\eta_{e})^*\bigg{\rangle}_{33}&=& \left(8 I \epsilon_{H} \sqrt{2}M_{P}^4 \right)\e^2 \sin^4{\theta} \int_{\eta_{0}}^{\eta_{e}} d\eta_{1}\left(\frac{a^3 }{f}\right)_{\eta_{1}}\int_{\eta_{0}}^{\eta_{1}} d\eta_{2} \left(\frac{a f}{\eta}\right)_{\eta_{2}} {\rm Im}\left( \zeta(\eta_{1})\zeta^{*}(\eta_{e})\right) \nonumber\\
&&  {\rm Im}\left( {h}_{+}(\eta_{2}){h}^{*}_{+}(\eta_{e}) D_{1}^{'}(\eta_{2})D^{*}_{1}(\eta_{1})\right) \nonumber\\
&=& \left(\frac{2\sqrt{2}}{3 \epsilon_{H}k^3}\right) \left( \frac{3}{28}N - \frac{361}{3920} \right)I\e^2\sin^2{\theta}
\ea
\ba
\label{cross34}
\bigg{\langle} \zeta_{\mathbf{k}}(\eta_{e}) {h}_{+\mathbf{k}}(\eta_{e})^*\bigg{\rangle}_{34}&=& \left(\frac{16}{3} I \sqrt{2} M_{P}^6 \right)\e^2 \sin^4{\theta} \int_{\eta_{0}}^{\eta_{e}} d\eta_{1}\left(\frac{a^3 }{f}\right)_{\eta_{1}}\int_{\eta_{0}}^{\eta_{1}} d\eta_{2} \left(\frac{a^3}{f}\right)_{\eta_{2}}
{\rm Im}\left( \zeta(\eta_{1})\zeta^{*}(\eta_{e})\right) \nonumber\\
&& {\rm Im}\left( {h}_{+}(\eta_{2}){h}^{*}_{+}(\eta_{e}) D_{1}(\eta_{2})D^{*}_{1}(\eta_{1})\right) \nonumber\\
&=& -\left(\frac{3\sqrt{2}}{2156 k^3 \epsilon_{H}}\right) \left(\frac{M_{P}}{H}\right)^2 I \e^4 \sin^2{\theta}
\ea

So adding these nine terms and assuming $N \gg 1$ we obtain
\ba
\label{final cross correlation-b}
\bigg{\langle} \zeta_{\mathbf{k}_{1}}(\eta_{e}) {h}_{+\mathbf{k}_{2}}(\eta_{e})\bigg{\rangle} &\simeq&
I \left(- 6\sqrt{2} \frac{ N^2 H^2 }{M_{P}^2}  + \frac{  \sqrt{2} \e^2 N}{7 \epsilon_{H} }  - \frac{3\sqrt{2} \e^4 }{1078 \epsilon_{H} } \frac{M_{P}^2}{H^2}  \right) \frac{\sin^2{\theta}}{k^3}
\ea

As for the other cross-correlation we have
\ba
\bigg{\langle} \zeta_{\mathbf{k}_{1}}(\eta_{e}) \widehat{h}_{\times\mathbf{k}_{2}}(\eta_{e})\bigg{\rangle}=0 \, .
\ea
This is because at the second order level $\zeta $ does not see $\widehat{h}_{\times}$.
\chapter{}
\section{Details of slow roll approximations}
\label{slow-roll}

In this appendix we  calculate the approximate solution of Eqs. (\ref{back-rho-eq}-\ref{anisotropy-eq}).

We start with Eq.(\ref{anisotropy-eq}). Since there is not any source of the anisotropy from the matter sector the anisotropic expansion rate decays exponentially and the non-linear term anisotropic terms in Einstein equations are  not important. Using the conformal time, $d\eta \equiv \frac{dt}{a(t)}$, the solution of Eq. (\ref{anisotropy-eq}) is
\ba
\label{sigma evo}
\sigma' = \sigma'_{0} \left(\frac{a_{0}}{a}\right)^2 \, ,
\ea
in which prime refers to the derivative respect to the conformal time and the subscript $0$ means the initial values of the corresponding quantities. Since $\sigma'$ decays very rapidly, it will not change the evolution of $\alpha$. Now by using the definition of the slow-roll parameter,
\ba
\epsilon_{H} \equiv -\frac{\ddot{\alpha}}{\dot{\alpha}^2} \, ,
\ea
we have
\ba
\left(1-\epsilon_{H} \right) = \frac{\alpha''}{\alpha'^2} \Longrightarrow \alpha' = \frac{\mathcal{H}_{0}}{1- \left(1-\epsilon_{H}\right)\mathcal{H}_{0}\left(\eta - \eta_{0} \right)} \, ,
\ea
in which we have $\mathcal{H} = a H$. Since $\mathcal{H}_{0} \eta_{0} \simeq -1$, we have
\ba
\label{scalefactor evo}
\mathcal{H} &=& \frac{\mathcal{H}_{0}}{\epsilon_{H} + \left(\epsilon_{H}-1\right)\mathcal{H}_{0}\eta} \nonumber\\
& \simeq& \left(1+\epsilon_{H}\right)\left(-\eta\right)^{-1} \, .
\ea
Integrating the above equation, we can calculate $a(\eta)$ as,
\ba
\label{a appro}
a \simeq H_{0}^{-1}\left(-\eta\right)^{-(1+ \epsilon_{H})} \, .
\ea
Now by using the above equations we can also find the evolution of $\frac{b'}{b}$. From Eq. (\ref{bian01}), written in terms of $\alpha$ and $\sigma$, we have
\ba
\frac{a'}{a} - \frac{b'}{b} = -3 \sigma' \, .
\ea
Integrating the above equation, we obtain
\ba
\label{b appro}
b &\sim & H_{0}^{-1}\left( -\eta \right)^{-(1+ \epsilon_{H})}\left( 1 + \left(\frac{\dot{\sigma_{0}}}{H_{0}}\right)\left(\mathcal{H}_{0}\eta\right)^3\right) \, .
\ea
Dropping the slow-roll parameter $\epsilon_H$ we recover Eqs. (\ref{app.a}) and (\ref{app.b}).
\chapter{}
\section{Details of metric perturbations}
\label{metric pert}

In this appendix, we look at the perturbations of the action both from the metric and matter sectors. First  we consider the metric perturbations. Then we proceed by considering the matter sector and finally we show that, due to the hierarchy between the terms from the matter sector and the metric back-reactions, we can neglect metric perturbations and only consider the matter effects \cite{Emami:2013bk}.

\subsection{The metric perturbations}

Now we look at the perturbations of the background metric (\ref{bian0}). Since the metric components in the $x$-direction are different from the $y$ and $z$ directions, the three-dimensional rotation invariance is broken into a subset of two-dimensional rotation invariance in $y-z$ plane. Therefore, in order to classify our perturbations, one can look at the transformation properties of the physical fields under the rotation in $y-z$ plane. Therefore, we decompose all of the metric and matter perturbations into scalar and vector components with respect to the 2D rotation in the $y-z$ plane. We also note that there are no tensor perturbations in two dimensions. In order to simplify the analysis and by employing the remnant symmetry in $y-z$ plane, we put $k_{z} =0$.

With these discussions  the most general form of metric perturbations is
\ba
\label{deltag}
\delta g_{\alpha \beta} =   \left(
\begin{array}{c}
- 2 a^2 A~~~~~~~~~~~~  a^2 \partial_x \beta~~~~~~~~~~~~~~~~~~~~ a\, b \left( \partial_i B + B_i \right)
\\\\
   ~~~~~~~~~~~~~~~~~~~~~~~~      - 2 a^2 \bar \psi   ~~~~~~~~~~~~~~~~~~       a b\,  \partial_x \left( \partial_i \gamma + \Gamma_i \right)
\\\\
 ~~~~~~~~~~~~~~~~~~~~~~~~~~~~~~~~~~~~~~~~~~~~~~~~~~~~~~~~~~     b^2 \left( - 2 \psi \delta_{ij} + 2 E_{, ij} + E_{i,j} +E_{j,i} \right)
\end{array}
\right )  \, .
\ea
In this decomposition $A, \beta, B, \bar \psi, \gamma, \psi $ and $E$ are scalar perturbations while $B_i, \Gamma_i$ and $E_i$ are vector perturbations subject to transverse conditions
\ba
\label{transverse}
\partial_i E_i = \partial_ i B_i = \partial_i \Gamma_i =0 \, .
\ea
One can  choose the following gauge for the metric perturbations \cite{Emami:2013bk}:
\ba
\label{gauge0}
\psi= \bar \psi = E = E_i =0 \, .
\ea
The gauge in Eq. (\ref{gauge0}) is similar to the flat gauge in standard FRW background.

\subsection{The quadratic action }
\label{quad-action}

Here we present the quadratic action for the inflaton field and metric degrees of freedom.
Following the approach of \cite{Emami:2013bk},  the second order action for the scalar degrees of freedom  in Fourier space is
\ba
\label{S2-scalar-k}
&&S_2 = \int d \eta d^3 k \left[ b b' k_x^2 (A^* \beta + A \beta^*) + \frac{a b}{2} (\frac{a'}{a} + \frac{b'}{b}) k_y^2 (A^* B + A B^*) +  \frac{a b}{2} k_x^2 k_y^2 (\gamma^* A + \gamma A^*) 
 \right. \nonumber \\ &&\left.
 - a^2 b^2 V(\phi) |A|^2
 - \frac{a b}{4} k_x^2 k_y^2 (\beta^* B + \beta B^*) + \frac{a' b}{2} k_x^2 k_y^2 (\gamma^* \beta + \gamma \beta^*) + \frac{a b}{4} k_x^2 k_y^2 (\gamma^* \beta' + \gamma \beta'^*)
\right. \nonumber \\ &&\left.
+ \frac{a^2}{4} k_x^2 k_y^2 |\beta|^2 - \frac{b^2}{4} k_x^2 k_y^2 (B^* \gamma' + B \gamma'^*)
+ \frac{b^2}{4} (\frac{b'}{b} - \frac{a'}{a}) k_x^2 k_y^2 (\gamma^* B + \gamma B^*) + \frac{b^2}{4} k_x^2 k_y^2 |B|^2 
\right. \nonumber \\ &&\left.
+ \frac{b^2}{4} k_x^2 k_y^2 | \gamma'|^2 - \frac{b^2}{4} (\frac{b''}{b} - \frac{a''}{a}) k_x^2 k_y^2 |\gamma|^2 -  \frac{b^2}{2} \phi' (A^* \delta \phi' + A \delta \phi'^*)
+  \frac{b^2}{2} |\delta \phi'|^2 - \frac{b^2}{2} \phi' k_x^2 (\beta^* \delta \phi + \beta \delta \phi^*)
\right. \nonumber \\ &&\left.
 - \frac{a b}{2} \phi' k_y^2 (B^* \delta \phi + B \delta \phi^*) - \frac{b^2}{2} k_x^2 | \delta \phi|^2 - \frac{a^2}{2} k_y^2 |\delta \phi|^2
- \frac{a^2 b^2}{2} V_{, \phi \phi} |\delta \phi|^2  -  \frac{a^2 b^2}{2} V_{,\phi} (\delta \phi A^* + \delta \phi^* A)
\right] \, . \nonumber\\
\ea
We have to integrate out the non-dynamical variables $\{\beta, B, A\}$ from the action Eq. (\ref{S2-scalar-k}). The analysis is simple but tedious. It turns out that it would be much easier to first integrate out $\beta$, then $B$ and finally $A$. Performing the details of integrating out analysis, the final action for the remaining dynamical field is $L= L_{\phi\phi} + L_{\gamma-\gamma} + L_{\phi \, \gamma}$ in which
\ba
\label{L_{phi-phi}}
L_{\phi \phi} &=& \frac{b^2}{2} \Big| \delta \phi ^{'}\Big|^2 - \bigg{(}\frac{b^2}{2}k^2 + \frac{a^2 b^2}{2}V_{, \phi \phi}
+\frac{b^4 k_{x}^2}{a^2 k_{y}^2} \phi^{'2} + \frac{b^6 k^4}{a^6 k_{y}^6} \frac{\phi^{'2}}{\lambda^2}\left( a^4 k_{y}^2 V(\phi) + 4 b'^2 k_{x}^2\right) + \frac{b^4 k^2}{2 a^4 k_{y}^4} \frac{\phi'}{\lambda} \times \nonumber\\
& \times & \left( 2 a^4 k_{y}^2 V_{,\phi} - 8 b b' k_{x}^2 \phi' \right)\bigg{)}\Big| \delta \phi \Big|^2
+ \bigg{(}\frac{b^4 k^2}{2 a^2 k_{y}^2} \frac{\phi'^{2}}{\lambda} \bigg{)}^{'}\Big( \delta \phi \Big)^{2} \\
\label{L_{gamma-gamma}}
L_{\gamma \gamma} & =& \frac{b^4 k_{x}^4}{a^2\lambda^2}\left( \frac{b'^2}{b^2}+ \frac{\phi'^{2}}{2}\right) \Big| \gamma ^{'}\Big|^2 - \bigg{(} \frac{b^2}{2} \frac{k_{x}^4}{\lambda}\left(k_{y}^2 - 2\frac{a'b b'}{a^3}+ 2\frac{b^{'2}}{a^2}\right)\bigg{)}^{'}\Big| \gamma \Big|^2
\ea
\ba
\label{L_{phi-gamma}}
L_{\phi \gamma} & =& \left(-\frac{b^3}{2a}k_{x}^2 \phi'-\frac{b^5 k^2 k_{x}^2}{a^5 k_{y}^4} \frac{\phi'}{\lambda^2}\left( a^4 k_{y}^2 V(\phi) + 4 b'^2 k_{x}^2\right) - \frac{b^3 k_{x}^2}{2 a^4 k_{y}^2}\frac{1}{\lambda} \left( -2k^2 \phi' a b b' + V_{,\phi}k_{y}^2 a^5 - 4 a b b' k_{x}^2\phi'\right)\right) \nonumber\\
&& \times\left(\delta \phi^{*}\gamma ^{'} + c.c.\right)+
\bigg{(}\frac{b^3 k_{x}^2}{2a}\phi'\left(\frac{a^{'}}{a}-\frac{b^{'}}{b}\right)
-\frac{b^3 k_{x}^2}{2 a^4 k_{y}^2}\frac{1}{\lambda}\left( -a^3 k^2 k_{y}^2 \phi' + 2 a' b b'k^2 \phi'- 2 a b'^2 k^2 \phi'\right)\bigg{)} \nonumber\\
&&\times \bigg{(}\delta \phi^{*}\gamma + c.c.\bigg{)}
- \frac{b^3 k_{x}^2}{2 a}\frac{\phi'}{\lambda}\bigg{(}\delta \phi^{'*}\gamma^{'} + c.c.\bigg{)}
\ea
Note that  we have defined $k$ as, $k^2 \equiv k_{x}^2 + \frac{a^2}{b^2} k_{y}^2$. (Note that there is a clash of notation here, this definition of $k$ is different from those in the main text defined as Eq.~(\ref{k_definition}).)
In addition, $\lambda$ has been defined as
\ba
\lambda \equiv \frac{a'}{a} + \frac{b'}{b} + 2 \frac{b^2}{a^2}\frac{k_{x}^2}{k_{y}^2}\frac{b'}{b} \, .
\ea

\subsection{Leading Correction}
To see the leading corrections in the action let us take a look at Eq. (\ref{L_{phi-gamma}}). As we can see all of the terms are proportional to $\phi'$ which means that they are all slow-roll suppressed. These terms are due to the metric perturbations since in the original action Eq.~(\ref{S2-scalar-k}) there is not any mixing between $\phi$ and $\gamma$.  The situation for Eq.~(\ref{L_{phi-phi}}) is the same, terms that are not directly from the matter sector are proportional to $\phi'$ or $V_{,\phi\phi}$ and so are slow-roll suppressed.  Therefore, we conclude that the metric perturbations in quadratic action are sub-leading compared to the contributions from the matter sector
fluctuations.

\chapter{}
\section{Detail analysis of $\psi$}
\label{psi solution}

Here we write down the equation of motion for $\psi$, which is defined by Eq. (\ref{psi u}), and try to solve it perturbatively.
\ba
\label{psi eq}
\psi''_{k} + \psi'^2_{k} - \frac{2}{\eta}\left( 1 - 3\left(\frac{\dot{\sigma_{0}}}{H_{0}}\right) (\mathcal{H}_{0}\eta)^3 \right)\psi'_{k} + \left[ k^2 - 2\left(\frac{\dot{\sigma_{0}}}{H_{0}}\right)(\mathcal{H}_{0}\eta)^3 \left(k_{y}^2 + k_{z}^2\right) \right] \psi_k =0 \, .
\ea
Expanding $\psi$ as
\ba
\psi_{\mathbf{k}}(\eta) = \psi_{\mathbf{k}(0)}(\eta)+ \psi_{\mathbf{k}(1)}(\eta)+ ...
\ea
the first order equation of motion for $\psi$ is
\ba
\psi''_{k(1)} + 2 \psi'_{k(0)}\psi'_{k(1)} - \frac{2}{\eta}\psi'_{k(1)} + 2 \left(\frac{\dot{\sigma_{0}}}{H_{0}}\right)(\mathcal{H}_{0}\eta)^3 \left( \frac{3}{\eta}\psi'_{k(0)} - \left(k_{y}^2 + k_{z}^2\right) \right) = 0
\ea
We can solve this equation and use the normalization condition to fix the constant of integration. The final result is
\ba
\psi_{\mathbf{k}(1)}(\eta) &=& \frac{\mathcal{H}_{0}^2 \sigma_{0}'}{1+ik\eta}\sum_{n=0}^{5} {\beta_{n}}\eta^{n}
\ea
Where ${\beta_{+n}}$ are given by
\ba
{\beta_{0}} &=& \frac{3i}{8k^3}\left(1 + 3 \cos^2{\Theta}\right) \\
{\beta_{1}} &=& -\frac{3}{8k^2}\left(1 + 3 \cos^2{\Theta}\right) \\
{\beta_{2}} &=& 0 \\
{\beta_{m}} &=& \alpha_{m} ~~,~~ (m=3,4,5) \, .
\ea


\section{Summation rules of spherical harmonics}
\label{sec:summ-rules-spher}

In the following, we present  a general expression for the $l$-dependence of the diagonal part of $ C_{l}$ due to a general anisotropic model,
\begin{align} \label{summation}
\frac{1}{2l+1}  \sum_m \int d\Omega_k ~ Y_{lm}(\mathbf{k}) Y^*_{lm}(\mathbf{k})Y_{LM}(\mathbf{k}) = \sqrt{\frac{2L+1}{4\pi}}\left( \begin{array}{ccc}
      l & l & L\\
      0 & 0 & 0
    \end{array} \right)\sum_m (-1)^{m}
  \left( \begin{array}{ccc}
      l & l & L\\
      -m & m & M
    \end{array} \right) ~.
\end{align}
It is worth to simplify Eq. (\ref{summation}). We first note that, due to the conservation of angular momentum, we have $M=0$. In addition, we can use the following identity,
\begin{align}
\sum_m (-1)^{m}
  \left( \begin{array}{ccc}
      l & l & L\\
      -m & m & 0
    \end{array} \right) = (-1)^{l}\sqrt{2l+1}\delta_{L0} ~.
\end{align}
So
\begin{align}
\frac{1}{2l+1}  \sum_m \int d\Omega_k ~ Y_{lm}(\mathbf{k}) Y^*_{lm}(\mathbf{k})Y_{LM}(\mathbf{k}) = & \sqrt{\frac{2L+1}{4\pi}} \delta_{L0}(-1)^{l}\sqrt{2l+1}\left( \begin{array}{ccc}
      l & l & L\\
      0 & 0 & 0
    \end{array} \right)\nonumber\\
    =& \sqrt{\frac{1}{4\pi}}(-1)^{l}\sqrt{2l+1}\left( \begin{array}{ccc}
      l & l & 0\\
      0 & 0 & 0
    \end{array} \right) ~.
\end{align}
Finally by using the following identity,
\begin{align}
\left( \begin{array}{ccc}
      l & l & 0\\
      0 & 0 & 0
    \end{array} \right) = (-1)^{l}\frac{1}{\sqrt{2l+1}} ~,
\end{align}
we get
\begin{align} \label{summation1}
\frac{1}{2l+1}  \sum_m \int d\Omega_k ~ Y_{lm}(\mathbf{k}) Y^*_{lm}(\mathbf{k})Y_{LM}(\mathbf{k}) = \sqrt{\frac{1}{4\pi}} \delta_{L0}\delta_{M0} ~.
\end{align}

\end{document}